\documentclass[12pt]{article}
\usepackage{mathtext}
\usepackage{graphicx}
\usepackage[T2A]{fontenc}
\usepackage[utf8]{inputenc}
\usepackage[rightcaption]{sidecap}
\usepackage{amsfonts}
\usepackage{amssymb}
\usepackage{amsmath}
\usepackage{appendix}
\usepackage{braket}
\renewcommand{\a}{\alpha}
\usepackage{epstopdf}
\usepackage{float}
\usepackage{braket}
\usepackage{indentfirst}
\usepackage{array}
\newcommand{\nnn}{\bigskip}
\newcommand{\nn}{\medskip}
\newcommand{\n}{\smallskip}

\renewcommand{\d}{\delta}
\renewcommand{\b}{\beta}
\newcommand{\g}{\gamma}

\newcommand{\QFT}{{\rm QFT}}
\newcommand{\D}{\Delta}

\newcommand{\e}{\epsilon}

\newcommand{\ar}{\longrightarrow}
\newcommand{\w}{\omega}
\newcommand{\s}{\sigma}
\newcommand{\la}{\lambda}
\renewcommand{\a}{\alpha}
\usepackage[a4paper, margin=2.5cm]{geometry}
\usepackage{authblk}
\begin{document}
\title{Quantum computations (course of lectures)}

\author{Yuri I. Ozhigov}
\affil{\it Moscow State University of M.V.Lomonosov, Faculty of Computational Mathematics and Cebernetics, Moscow center of fundamental and applied mathematics, Institute of physics and technology of K.A.Valiev (RAS), e-mail: ozhigov@cs.msu.ru
}
\date{}
\maketitle




\newpage

\section*{Annotation}

This course of lectures has been taught for several years at the Lomonosov Moscow State University; its modified version in 2021 is read in the Zhejiang University (Hangzhou), in the framework of summer school on quantum computing. The course is devoted to a new type of computations based on quantum mechanics. Quantum computations are fundamentally different from classical ones in that they occur in the space of so-called quantum states, and not in ordinary binary strings. The physical implementation of quantum computing - a device called a quantum computer has already been partially created, and its technology continues to develop intensively. Quantum computing is a real process in which the mathematical description is inextricably linked with quantum physics. In particular, the quantum mechanics of complex systems, the model of which is a quantum computer, is currently only being created, so quantum computing is a fundamental direction to a greater extent than an applied one. Therefore, in the course - in general, mathematical, much attention is paid to the physical implementation of this new type of computations. Various forms of quantum computing are considered: the Feynman gate model, fermionic and adiabatic computations. A class of problems is described in which quantum computing is not only more efficient than classical ones, but also cannot be replaced by them. These are the most important tasks of describing complex processes at the predictive level. In particular, using the phenomenon of quantum nonlocality, discovered at the end of the 20th century. Estimates of the ultimate possibility of quantum computing are also given - lower estimates of quantum complexity. The course is designed for students of physics and mathematics and natural science specialties as well as all those interested in this subject. It requires familiarity with the basics of linear algebra and mathematical analysis in the first two courses of the university.

\tableofcontents

\newpage
\thispagestyle{plain} 

\ \ \ \ \ 

\newpage

\addcontentsline{toc}{section}{Introduction}
\section*{Introduction}

Natural science studies the patterns that real-world scenarios obey. Where there are no laws, chaos reigns. Classical mechanics was born out of the chaos of the medieval description of the world, but it came across the chaos of the microcosm at the beginning of the 20th century, from which quantum mechanics emerged, which gave us microelectronics and IT technologies. Further progress is associated with complex scenarios, primarily biological ones. Their description is currently dominated by the ideas of classical physics, and chaos also reigns for the time being. The geometric growth of biological databases such as Protein Data Bank and the demand for increasingly powerful computers for only external, shallow processing of accumulated information has not been accompanied by visible progress either in medicine or in fundamental biology, since the discovery of the DNA double helix in 1953. Restoring order in the chaotic field that modern biology represents requires the introduction of an accurate theory of the microcosm, that is, quantum methods. A quantum computer serves this purpose.

\begin{figure}
\caption{Quantum mechanics and complex processes}
\includegraphics[width=1.0\textwidth]{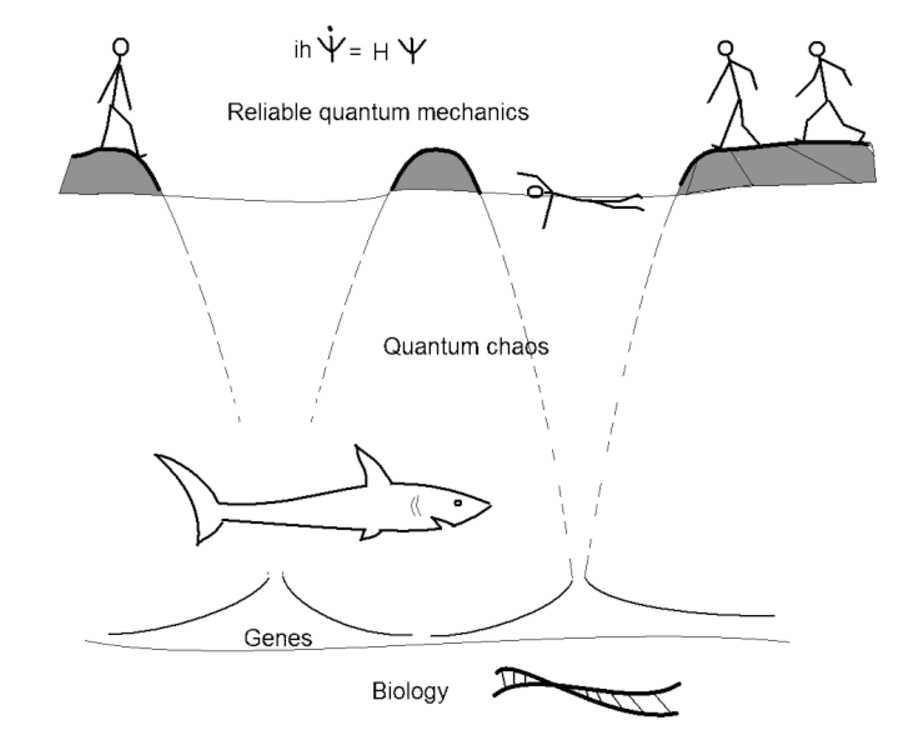} 
\label{fig:gener}
\end{figure}

The path to building a complete theory of the microcosm, including complex systems and processes, is very difficult and we are still, in fact, at its beginning (look at the Figure \ref{fig:gener}).

However, all the steps taken since 1982, when the idea of a quantum computer was first made public, were steps in the right direction. In this course, we will follow these steps from a mathematical point of view. 

\begin{figure}[h]
\centering
\caption{Andrei Markov - junior. Founder of constructive mathematics}
\includegraphics[width=0.3\textwidth]{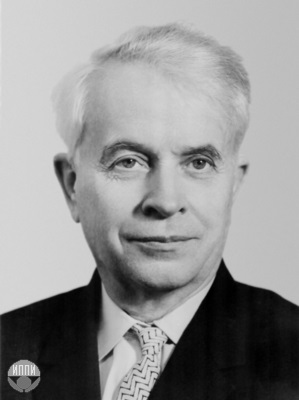} 
\label{fig:markov}
\end{figure}

Quantum computing is the mathematical support of a quantum computer. To understand the place of this subject, we must give an idea of the quantum computer itself and the need to create it.

\begin{figure}[h]
\caption{Lov Grover, author of quantum search}
\centering
\includegraphics[width=0.3\textwidth]{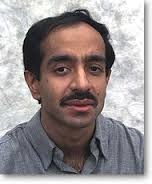} 
\label{fig:grover}
\end{figure}

The physics of a quantum computer is essentially the quantum physics of complex systems, which differs from the quantum theory of simple systems, the so-called Copenhagen theory. The difference is in the new, more stringent restrictions on the mathematical apparatus of infinitesimals, and also in the fact that for complex systems computations play a very special role, which was not in the old, Copenhagen theory. The name of this project originated in the distant times of the late 20th century, when we imagined its final product as a kind of computing machine standing on a desktop, or in a laboratory, and capable of calculating some things faster than classical supercomputers (Grover \ref{fig:grover} fast quantum search).

The last 30 years have shown the naivety of such an idea. In reality, we do not want to calculate something abstract, we want to manage complex natural processes that are necessary for our survival. And this control should be carried out at the quantum level, because the course of a complex process is determined in the microcosm, where the laws of quantum physics prevail.

The computations themselves are needed only to control the life. To do this, you need to know well what this or that move in control will lead to, that is, you need to be able to predict the behavior of a controlled system (a nanodevice, a living cell or an organism), and do it in real time, that is, faster than the controlled system itself responds to our control. This is the role that a quantum computer should play, no matter how it looks.

The first step towards the creation of this device has already been taken by our predecessors - the founders of quantum mechanics. This grandiose step in the knowledge of the world led to the appearance of computers as such. Of course, it is possible to count the history of computations from mechanical arithmometers of the time of Boole; historically, this has grounds, but still a computer is a microelectronic device - a set of microcircuits based on silicon-germanium heterostructures. And the principle of operation of such devices is based on the quantum representation of the state of electrons in a solid, that is, on quantum mechanics.

\thispagestyle{plain} 

All modern microelectronics is an achievement of quantum theory. And here it is used to control huge ensembles of identical particles - bosons, like photons, or fermions, like electrons. The methods of mathematical analysis work very well for such ensembles, which is the reason for the success at this stage, which covers almost the entire 20th century. We are able to manage such microelectronic systems well.

\thispagestyle{plain}

Today, we need to learn how to manage more complex systems of biological nature. Here we are talking about many atoms that have an individuality. Individual DNA links can no longer be combined into ensembles of identical particles, like helium-4 atoms in a liquid state, or like electrons in a semi-conducting layer of a heterostructure. This stage is designated by the term "quantum computer", and we have yet to pass it. We are talking here about the management of living things, which radically distinguishes the tasks of the present time from past epochs.
The role of analytical methods of the past today passes to computers, and the ideology of Computer Science becomes the main one in the physics of complex processes. Consequently, quantum computing becomes a mathematical form of controlling these processes.

A quantum computer is a method of penetrating into the depths of the microcosm, into the area where quantum theory itself must be transformed and adapted to the enormous complexity of living matter. The project of its creation is extensive and diverse, it is impossible to cover it in one lecture course.The role of analytical methods of the past is now being transferred to computers, and the ideology of Computer Science is becoming the main one in the physics of complex processes. Consequently, quantum computing becomes a mathematical form of controlling these processes.

\thispagestyle{plain} 

This lecture course presents only one side of it - mathematical, and from the biased point of view of the author, who himself deals with this topic. The listener can find a description of other aspects of this project in the constantly growing literature and by referring to the archive http:arxiv.org, the quant-ph section. The mathematical side of a quantum computer is very important, since here the analytical apparatus familiar to physicists of the 20th century is not quite adequate to reality. This was clearly demonstrated in the 90s by the example of the so-called fast quantum algorithms, the main examples of which we will consider in detail.

The development of the quantum computer project can have a serious impact on the development of natural science in the coming decades, so many works on it are not published, and their results are immediately used in the field of information technology. This applies to quantum cryptography-the part of the project that operates with one or two qubits or with precision quantum devices; both of these areas are widely used in practice. Here we will touch on these areas only very briefly. Our subject will be quantum computing and its connection with the most important general scientific tasks.

\thispagestyle{plain} 

There are excellent physical monographs on this subject, starting with the canonical book by Lev Landau and Yevgeny Lifshitz \cite{LLi} and many other equally excellent books. However, quantum computing dictates a slightly different, more formal and concise style of presentation based on linear algebra. This approach allows you to quickly master the formal language in which quantum computing processes will be described. This will allow us to formulate an abstract Feynman model of a quantum computer-with a user interface in the form of quantum gates (see \cite{Fe2}; the first detailed definition can be found in \cite{Deu}), and move on to quantum algorithms.

We will describe in more of less details three such algorithms that solve mathematical problems: the Grover algorithm, the fast quantum Fourier transform and - briefly - the associated Shor integer factorization algorithm, as well as the Zalka - Wiesner algorithm. The first two illustrate the property of fast concentration of the amplitude on the target state, which allows us to achieve the so-called quantum speedup of classical computations. Here we will also discuss computations with an external object - an oracle, and its quantum form. Attention will also be paid to the lower bounds of quantum complexity - the limiting capabilities of quantum computing. The third algorithm is designed for predictive modeling of the real evolution of a complex system at the quantum level. The choice of these algorithms is dictated by the fact that they fully reveal the essence of quantum computing and their possible applications in the field of mathematical problems.

\thispagestyle{plain} 

In the first two lectures, a brief introduction to quantum mechanics is given, which is necessary for understanding further. 
We will also touch on quantum algorithms for distributed computing, the advantage of which is the use of quantum nonlocality. The scheme of fermionic quantum computing and its control will be analyzed. We will talk about quantum teleportation - on the example of one qubit. Quite briefly, we will touch on quantum cryptography -  the BB84 protocol.

One lecture will be devoted to adiabatic quantum computing, which clarifies the special status of the most important equation of quantum physics - Schrodinger equation.

Attention will also be paid to the physical implementation of the Feynman model of quantum computing - quantum gates. We will analyze a specific gate technology - on optical cavities, which is based on finite - dimensional models of QED  -quantum electrodynamics.

Finally, we will consider the role of quantum computing in determining the limits of applicability of the quantum theory itself, as well as the possible scheme of a quantum operating system. Quantum algorithms and computations use this operating system as a technical basis, but its limitations directly affect the algorithms themselves. Quantum computing thus becomes an experimental platform for determining the form of quantum laws in the field of complex processes.

In conclusion, we will observe main results and possible ways of developing quantum computing and its applications.

\newpage

\section{Lecture 1. Quantum Mechanics and modeling of Nature}

\bigskip

{\it \ \ \ \ \ \ \ \ \ \ \ Everything true is simple and clear, and where there is fog, there is always 

 \ \ \ \ \ \ \ \ \ \ \ some kind of  mud

\ \ \ \ \ \ \ \ \ \ \ \ \ \ \ \ \ \  \ \ \ \ \ \ \ \ \ \ \ Lev Landau}
\bigskip

Modeling of natural processes is the main task of physics. If until the first half of the 20th century modeling was reduced mainly to algebraic calculations, which were directly compared with the experiment, then at the end of the 20th century a computer stood out as the main device of theoretical physics, into which you can download all the analytical calculations and, moreover, it can go much further than these calculations, presenting us with reality in the form of the result of its work - as a computation.

Let's agree about the basic terminology. Suppose we have a certain device called a {\it computer}, which can be in a certain set of states, the set of which we will denote by ${\cal C}$. {\it Algorithm} we will call the mapping ${\cal F}$ of this set to itself:

$$
{\cal F}: {\cal C}\rightarrow{\cal C}.
$$

The algorithm is set in the form of a certain rule that allows you to get another state of the computer according to a given state. This rule is practically most often made out in the form of a computer program, or in the form of a recipe formulated in ordinary human language "we need to do this and that".

\begin{figure}
\caption{Simulation of a real process on a quantum computer}
\includegraphics[width=1.0\textwidth]{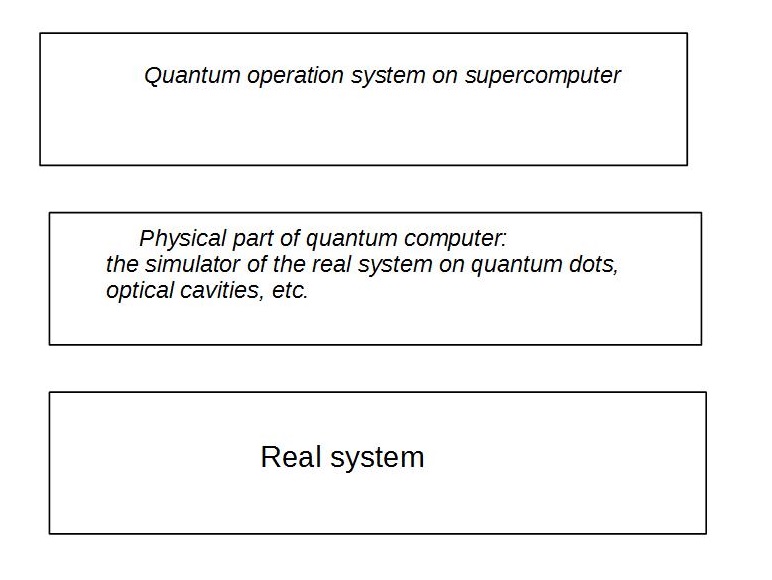} 
\label{fig:QC1}
\end{figure}

\begin{figure}
\caption{The scheme of a quantum computer
}
\includegraphics[width=1.0\textwidth]{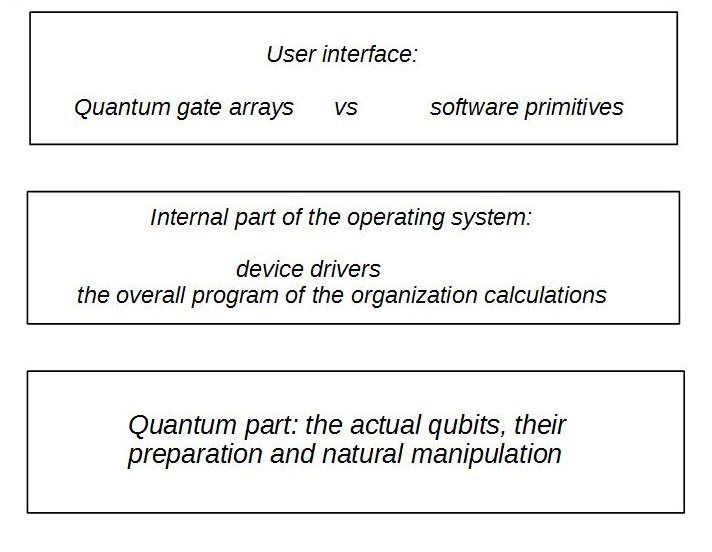} 
\label{fig:QCscheme_}
\end{figure}

In the classical theory of algorithms, the set of computer states ${\cal C}$ is simply a set of all possible Boolean strings of the form $a_0,a_1,...,a_{n-1}$, where $n$ is the size of the computer's memory, $a_j\in\{ 0,1\}$.

The {\it computation} corresponding to this algorithm is the sequence of applying the mapping ${\cal F}$ to the initial state ${\cal C}_0$:

\begin{equation}
{\cal C}_0\ar{\cal F}({\cal C}_0)\ar{\cal F}({\cal F}({\cal C}_0))\ar \ldots\ar \underbrace{{\cal F}(\ldots {\cal F}}_T  ({\cal C}_0)\ldots )=
{\cal F}^{\{T\} } ({\cal C}_0),
\label{iterated}
\end{equation}
where the rule ${\cal F}$ is no longer applicable to the final state. The number $T({\cal C}_0)$ is called the complexity of the algorithm's work on a given initial word ${\cal C}_0$; if $T=\infty$, then the complexity is infinite, that is, the algorithm never terminates. So such defined complexity is actually the running time, expressed in abstract units - the number of applications of this operator.

The operator ${\cal F}$ may depend on the time that we defined above. In order for this situation not to differ from the standard model with the constant operator ${\cal F}$, it is necessary to include time in the states themselves $C\in {\cal C}$. This technique is used in quantum electrodynamics. However, from a practical point of view, the variable operator ${\cal F}$ is a case that requires special approaches - this will be discussed in a lecture on adiabatic computations. In the future, by default, the ${\cal F}$ operator will not depend on time.

We understand a computation in the broad sense of the word: any real process is presented to us in the form of some computation. Therefore

\bigskip

{\bf \large an algorithm is any law of nature that we can formulate in precise terms. }
\bigskip

This algorithmic concept, constructive mathematics, was created by Andrey Markov Jr. (see \cite{Mar},\cite{Ts}) and is the basis for applying computers to modeling real processes.

Any quantum algorithm is ultimately an algorithm that simulates some real process (see Figure \ref{fig:QC1}); and if the computation corresponding to this algorithm does not give the desired result, we must conclude that this algorithm is incorrect. Quantum algorithms are just classical {\it records } of elementary quantum operations that lead to the desired result, provided that we correctly understand the operation of the laws of quantum physics itself in relation to the quantum part of our computer. It is impossible to directly verify the fact of such knowledge - it can only be verified by experiment.

This leads to the unexpected conclusion that the construction of a quantum computer is a test of the quantum theory itself in a field where it has never been tested before - in complex systems. In the problems of physics of the 20th century, we were dealing with simple systems, complexity did not play a special role there, since these problems could be reduced by eliminating the so-called entanglement. For complex systems that are the focus of science today, this cannot be done.

Increasing the number of qubits leads to an exponentially rapid increase in complexity; with qubit memory, we very quickly go beyond simple systems that were easily analyzed by physicists of the past century. Therefore, quantum computing is the physics of modern times, and our ideas about it still need to be tested experimentally. We will follow the standard path of the Copenhagen quantum theory, which has been perfectly tested for simple problems, and see where it leads us to complex problems.

Let us somehow determine the complexity $C({\cal C})$ of the computer state ${\cal C}$. Let's take the maximum complexity of the algorithm beginning with every initial state of complexity no greater than a given natural $n$ :

$$
C({\cal F})(n)=max_{C({\cal C}_0)\leq n}T({\cal C}_0)
$$
then we get a function of the natural argument, called the complexity of the algorithm
 ${\cal F}$. 

For computations, an external device is often used, which is called an {\it oracle}. An oracle is an object that is much more complex than a computer; it can hardly even be called an object, it is rather a subject that cannot be described in terms of algorithms at all. For example, an oracle can be a computer user.

Formally, the oracle is this other function of the form

$$
{\cal O}: {\cal C}\rightarrow {\cal C}
$$

Let there be a subset ${\cal Q} \subseteq {\cal C}$ in the set of computer states ${\cal C}$, whose states are called query states. Pair $({\cal O}, {\cal F})$ is called an algorithm with an oracle. The computation corresponding to a given algorithm with an oracle is a sequence of the form

\begin{equation}
{\cal C}_0\ar{\cal L}({\cal C}_0)\ar{\cal L}({\cal L}({\cal C}_0))\ar \ldots\ar \underbrace{{\cal L}(\ldots {\cal L}}_T  ({\cal C}_0)\ldots )=
{\cal F}^{\{L\} } ({\cal C}_0),
\label{iterated_or}
\end{equation}

where the mapping ${\cal L}$ acts as ${\cal F}$ if its argument does not belong to ${\cal Q}$, and as ${\cal O}$ if its argument belongs to ${\cal Q}$. Here, as above, the mapping ${\cal L}$ is not applicable to the final state of the computer. Such a calculation works as ${\cal F}$ until a query state is encountered. If such a state occurs, the oracle ${\cal O}$ is used instead of the usual function ${\cal F}$.

After a little thought, we will come to the conclusion that the user's interaction with the computer exactly fits into the computation scheme with the oracle, if the latter designates the user.

\bigskip
{\Large\bf Oracle = computer user}
\bigskip

The complexity of the computation with an oracle is the number of applications of the oracle in the chain \eqref{iterated_or}; the complexity of the algorithm with an oracle is determined in the same way as above.

The computation \eqref{iterated} is an abstract model of any natural process that is described by the law ${\cal F}$, that is, any real process. The form of the computation depends on the form of the description of the states of the system under consideration. For example, in classical physics, states from the set ${\cal C}$ are binary strings of length $n$, which total number is $N=2^n$. Note that the application of mathematical analysis requires a limit transition of $n\rightarrow \infty$. However, this requirement of Cartesian mathematics does not exactly correspond to the real world. For example, if we are talking about the air in a given room, we cannot, strictly speaking, consider it continuous: it consists of molecules of finite size. In a computer computation, any representation will be finite, since the computer's memory is always limited. This circumstance means that computer modeling is able to more adequately reflect real physical processes in comparison with analytical formulas.

Is it possible to speed up evolution if you expand the computer's memory? Is it possible to buy time by paying for it with space? You can't! Evolution, in general, cannot be accelerated by involving new resources. Only a narrow range of tasks, called search (or brute force) problems, is parallelized. A quantum computer will not be able to speed up the execution of all tasks as well.

\bigskip
{\it Theorem (\cite{Oz3}).

The probability that an iteration of the length $O(N^{1/7})$ arbitrarily selected from the uniform distribution of the "black boxes" ${\cal F}$ can be spedup by at least one on a quantum computer tends to zero with the dimension of the space tending to infinity.}

\bigskip

The meaning of this theorem is that the proportion of problems that allow quantum speedup by at least one is vanishingly small among all possible problems.

In other words:
 
\bigskip

{\bf Quantum speedup is a rare phenomenon that occurs only for problems of the brute force type
}

\bigskip

This brings quantum parallelism closer to classical parallelism. The advantage of a quantum computer consists in a) the potential possibility of solving iterative problems faster than on a classical computer, and b) in a much more adequate modeling of real processes.

The advantage of point a) is not absolute, since we do not know the physical limitations on the use of the Copenhagen quantum theory in complex systems. The advantage of point b) has huge prospects. In particular, we can use the phenomenon of quantum nonlocality to improve the quality of some computations.

\subsection{Quantum representation of states}

The description of states in the form of binary strings corresponds to classical physics - this is a description through classical algorithms.

For relatively simple processes, the classical computational model \eqref{iterated} is quite adequate. However, this is not the case for complex processes. The fact is that the binary representation of states in the form of binary strings itself is inadequate to reality in the case when very small errors in determining the state are essential. This happens in states of unstable equilibrium, when for accurate modeling we must consider very small segments of time $dt$ and space $dx$, that is,  the microcosm. Here, the value of the elementary action $dS$ (the energy element multiplied by the time element), which is used in modeling, is important: if $dS\gg\hbar\approx 3\cdot 10^{-27}\ erg \cdot\ sec$, the classical description of the dynamics is adequate, if $dS$ becomes comparable to the Planck constant $\hbar$, the description from classical physics becomes inadequate, and it should be replaced by a quantum one.

The main feature of the quantum description of reality is multiplicity. With a quantum description, any object can be in several classical states at once. The quantum state is a vector that changes according to the matrix rule: at the next moment in time, it is multiplied by the evolution matrix. The vector itself has a probabilistic meaning. This means that the quantum description does not refer to one single object, but to a whole huge series of equally prepared objects; only having such a series, we can collect statistics in order to compare theory and experiment.

In the quantum representation of nature, any state from the set ${\cal C}$ is a {\bf linear combination} (superposition) of binary strings of length $N$.

Let ${\cal C}_0, {\cal C}_1,..., {\cal C}_{N-1}$ be the classical states of some system, by which we will always understand the computer's RAM. Then the quantum state of this system will have the form

\begin{equation}
\label{state}
|\Psi\rangle=\sum\limits_{j=0}^{j=N-1}\la_j|{\cal C}_j\rangle
\end{equation}
where the complex numbers $\la_j$ are called {\it amplitudes} of the states ${\cal C}_j$. 
This means that the quantum RAM can be simultaneously in all possible classical states, but in each such state $|{\cal C}_j\rangle$ - with its own amplitude $\la_j$. This is the essence of quantum parallelism and the possibility of quantum spedup of solving iterative problems compared to a classical computer.

For example, if a classical computer has RAM consisting of $n$ bits, then the quantum analogue of such a computer will have memory consisting of $n$ quantum bits (qubits).

In addition to RAM, any computer has a long-term memory. It is in the long-term memory that the description of the algorithm ${\cal F}$ is stored, which does not change when this mapping is applied, like the genetic code in a living cell. Thus, the algorithm can only change the RAM, but not the long-term memory. In the case of a classical computer, these types of memory have an identical mathematical description in the form of binary strings. In a quantum computer, the situation is different. Here, the long - term memory is also a binary string designed for storing the algorithm record, but the operational memory will be in the so-called quantum state, which we also call the vector state. In the future, by memory, we mean, by default, RAM.

The state of the computer is a pair: the state of RAM and long-term memory. For example, the state that determines the end of the algorithm, or the query transformation is always determined in the long-term memory of the computer. Here we restrict the traditional theory of algorithms that allow for "unpredictable" behavior of computations. In the theory of quantum computing, their end is always predictable; "unpredictable" is teh final state of quantum memory.

Let the classical state of the computer's RAM have the form of a binary string $a_0,a_1,..., a_{n-1}$. We will represent such a string as a natural number

$$
a=a_0+2a_1+...+2^{n-1}a_{n-1}
$$
lying in the set $\{ 0,1,..., N-1\}$, where $N=2^n$, and this representation will be one-to-one. Then the quantum state of this memory will have the form
\begin{equation}
|\Psi\rangle=\sum\limits_{j=0}^{N-1}\la_j\ket{j}=\begin{pmatrix}
&\la_0\\ &\la_1\\ &\vdots\\&\la_{N-1}\end{pmatrix};\ \ |0\rangle=\begin{pmatrix}
&1\\ &0\\ &\vdots\\&0\end{pmatrix},\ |1\rangle=\begin{pmatrix}
&0\\ &1\\ &\vdots\\&0\end{pmatrix},\ ..., |N-1\rangle=\begin{pmatrix}
&0\\ &0\\ &\vdots\\&1\end{pmatrix}.
\label{state_}
\end{equation}

So, $| \Psi\rangle$ is a column vector belonging to the space $C^N$, and all the components of $|j\rangle$ also have column vectors, each of which has all zeros, except for the $j$ element, which is equal to one. Thus, \eqref{state_} is a decomposition of an arbitrary vector in an $N$ - dimensional complex space on the basis of $|j \rangle$.

The symbols $ | $ - bra, and $ \rangle$ - ket, were introduced by Dirac; they are very convenient and we will always use them.

A scalar product is defined in this space, so that the vectors $|j\rangle$ form an orthonormal basis in it. The vectors $| \Psi\rangle$ can be added and multiplied by any number - in this case, new quantum memory states will be obtained, which, like $|j\rangle$, are physically realizable quantum states.

We will introduce an object of the form $\langle\Psi|$ as the result of the conjugation of the vector $|\Psi\rangle$, that is, a vector is a string consisting of complex conjugate elements. Then the matrix product of the form $\langle\Psi| \cdot |\Phi\rangle$, which we will briefly write as $\langle\Psi|\Phi\rangle$, will be the scalar product of the vectors $|\Psi\rangle$ and $|\Phi\rangle$. Thus, we can write the normalization condition for one in the form $\langle\Psi|\Psi\rangle=1$. In order to normalize any vector-state $|\Psi\rangle$, you need to divide it by its own norm: $\frac{1}{\sqrt{\langle\Psi|\Psi\rangle}}|\Psi\rangle$. The length of such a normalized vector will be unit.

It is convenient to write matrices in Dirac notation. For example, the matrix product of a column by a row of the form $|i\rangle\langle j|$ is a matrix in which one stands in the place of $(i,j)$, and in other places there are zeros. Then any matrix $A=(a_{i, j})$ can be represented as an expansion

$$
A=\sum\limits_{i,j}a_{i,j}|i\rangle\langle j|
$$
and the conjugate matrix - the result of transposition and complex conjugation of elements, denoted by $A^*$ or $A^+$, will have the form $A^+=\sum\limits_{i, j}\bar a_{i, j}|j\rangle\langle i|=\sum\limits_{i, j}\bar a_{j, i}|i\rangle\langle j|$.

Let the state $|\Psi\rangle$ of the form \eqref{state_} be normalized (by one). Then its {\it measurement} is called a random variable that takes the values $|j\rangle$ with probabilities $p_j=|\la_j|^2$. The total probability will be equal to one due to the normalization of this state.

The unitary operator $C^N\rightarrow C^N$ is a linear operator that preserves the length of any vector. This is equivalent to the fact that it translates one orthonormal basis into another orthonormal basis.

In the space $C^N$, you can consider bases other than $|j\rangle$. This is done by applying some kind of unitary transformation of $U:\ |\tilde j\rangle = U|j\rangle$. Then we can generalize the concept of measurement to a new basis $| \tilde j\rangle$. Namely, we call the measurement of a system in the state $| \Psi\rangle$ in the basis $|\tilde j\rangle$ a random variable taking the values $|\tilde j\rangle$ with the probability $\tilde p_j=|\langle j|\Psi\rangle|^2$.

\begin{figure}[h]
\caption{The coordinate-momentum uncertainty ratio for one qubit. If the coordinate is precisely defined, the momentum is completely indeterminate, and vice versa
.}
\includegraphics[width=1.0\textwidth]{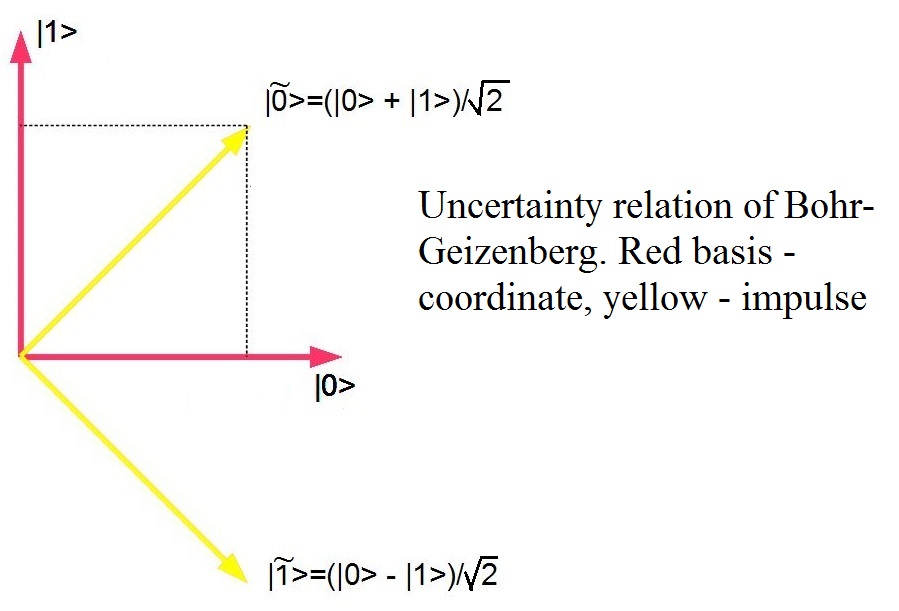} 
\label{fig:uncertainty}
\end{figure}
 \newpage

The linear operator $H:\ C^N\rightarrow C^N$ is called Hermitian (or self-adjoint) if $H=H^+$. From linear algebra, it is known that for any unitary operator $U$ there is such a Hermitian operator $H$ that $U=exp (iH)$, where the matrix exponent is defined, as well as the numeric one, through the series $exp(A)= \sum\limits_{n=0}^\infty A^n/n!$. The opposite is also true, for any Hermitian operator $H$, the operator $e^{iH}$ is unitary.

It is also known that for any unitary or Hermitian operator there is its system of eigenvectors, which is an orthonormal basis of the entire state space. This allows us to introduce the concept of {\it observation}, which is related to the concept of measurement. We will call the Hermite operator $H$ {\it observable} if the basis of the space $C^N$ is allocated, consisting of the eigenvectors $|\phi_j\rangle$ of the operator $H$ with the eigenvalues $a_j$: $H|\phi_j\rangle = a_j|\phi_j\rangle$. {\it Observation} of a system in the state $|\Psi\rangle$ corresponding to {\it observable} $H$ is a random variable taking the values $a_j$ with probabilities $P_j=|\langle\phi_j|\Psi\rangle|^2$.

So, measurement and observation are almost the same thing, the only difference is that when measuring, the result will be the eigenstates of some Hermitian operator - the basis $|\tilde j\rangle$ can be considered exactly as such states, and when observing, the result will be eigenvalues corresponding to these states. If the measurement is carried out by a device called a meter, but the observation is carried out by an {\it observer}.

Measurement is a physical process during which the state $| \Psi\rangle$ of the form \eqref{state_} turns into one of the states $|j\rangle,\ j=0,1,...,N-1$. In this case, all information about the amplitudes of $\la_j$ is lost. This process occurs when the system in question comes into contact with a special device called a meter, or an observer. In quantum mechanics, there is no exact definition of which object can be a meter. However, the term "observer" can mean that such a device can be a subject of any nature, for example, a person. We will not go into this question deeply yet, assuming that a computer user can always initiate a computer memory measurement at will.

Measurement or observation is the only way to learn anything about the quantum state $| \Psi\rangle$ in which this system is located. Quantum mechanics, therefore, describes only probabilities, but not classical states. A paradoxical conclusion follows from this. One single system, generally speaking, does not have any quantum state! The quantum state is a characteristic of not one object, but a huge number of equally prepared objects.

For example, let's consider an electron in a hydrogen atom and wonder where it is located there. You can measure its coordinate by irradiating an atom with hard radiation - a photon of very high frequency. This photon, after interacting with an electron, will fly out of the atom and we will be able to fix it, after which it will be possible to approximate where the electron was located. But as a result of the interaction, the electron will receive such an impulse that it will fly out of the atom completely! And for the next experiment, we will have to take another atom already - the previous one has completely lost its initial state. Thus, in order to talk about the "quantum state of an electron in a hydrogen atom", we need to have a huge number of equally prepared hydrogen atoms, which nature provides us with. If we had only one unique hydrogen atom, we would not be able to say anything about the vector state of the electron inside it.

Quantum theory does not deal with individual particles or individual systems of particles. It deals only with ensembles of independently and equally prepared systems or particles. The vector state characterizes not a separate system of atoms representing the memory of a quantum computer in a given experiment, but the result of a huge number of identical experiments. Quantum computing, therefore, is a probabilistic computation of a special kind. They are radically different from classical probabilistic computations based on classical physics. Quantum computations completely model the quantum dynamics of real complex systems, which is why they apply for not only to describe, but also to control complex processes
.

In the course of quantum computing, we must come to a state of the form $|j\rangle$ - to a basic state, the measurement of which invariably gives the same result. This is the art of quantum computing.

\subsection{Unitary evolution}

What happens to the memory of a quantum computer when it is left to itself, and no one measures it? Then its vector-state satisfies the Schrodinger equation, which has the form

\begin{equation}
\label{shredinger}
i\hbar|\dot{\Psi}\rangle=H|\Psi\rangle
\end{equation}
where $H$ is the Hermitian operator, called the energy operator of the system or {\it is the Hamiltonian}, $\hbar\approx 3\cdot 10^{-27}\ erg\cdot\ sec$ is the Planck constant. 
If the initial vector state is set to $| \Psi(0)\rangle$, then the solution of the Schrodinger equation will have the form
\begin{equation}
\label{shredinger_solution}
|\Psi(t)\rangle = exp(-\frac{i}{\hbar}Ht)|\Psi(0)\rangle
\end{equation}

The formula \eqref{shredinger_solution} expresses the fact that the trajectory of a quantum system is the orbit of a unitary evolution operator $U_t=exp (- \frac{1}{\hbar}Ht)$ acting on the entire state space, in contrast to the dynamics of a classical system that evolves along its trajectory given by the initial value. From this, in particular, it follows that if we made a little mistake in the original state vector $| \Psi(0)\rangle$, then this error will exactly persist for any arbitrarily long period of time, and will not increase, as is possible in the case of classical dynamics.

Suppose we have diagonalized the energy operator $H$, that is, we have found its eigenfunctions-vectors $|\phi_0\rangle,|\phi_1\rangle,...,|\phi_{N-1}\rangle$ and the corresponding eigenvalues $E_0<E_1\leq E_2\leq...\leq E_{N-1}$ (the first inequality is always strict, the others are not strict). Then we can express the solution of the Schrodinger equation with the initial condition $| \Psi(0)\rangle$ in the form of such a decomposition:

\begin{equation}
\label{shredinger_solution_}
|\Psi(t)\rangle=\sum\limits_{j=0}^{N-1}\la_j e^{-\frac{i}{\hbar}E_jt}|\phi_j\rangle,\ \la_j=\langle\phi_j|\Psi(0)\rangle,
\end{equation}
which is checked directly. Thus, the solution of the Schrodinger equation reduces to the solution of the problem on the eigenvalues of the Hamiltonian:

\begin{equation}
\label{eigen}
H|\phi_j\rangle=E_j|\phi_j\rangle.
\end{equation}

We introduce the Pauli matrices:

\begin{equation}
\label{Pauli}
\sigma_x=\begin{pmatrix}&0&1\\ &1&0\end{pmatrix},\ \sigma_y=\begin{pmatrix}&0&-i\\ &i&0\end{pmatrix}\ \sigma_x=\begin{pmatrix}&1&0\\ &0&-1\end{pmatrix}
\end{equation}

The Pauli matrices will be both Hermitian and unitary at the same time, and their eigenvalues will be equal to $\pm 1$ (check it out!).

Consider, as an example, one qubit whose state has the general form $\la_0|0\rangle+\la_1|1\rangle$. We find a solution of the Schrodinger equation for it with the Hamiltonian $H= - \s_x$.
The eigenvectors of the state have the form $|0\rangle+|1\rangle$ - for the eigenvalue $E_0=-1$, and $|0\rangle-|1\rangle$ - for the eigenvalue $E_1=1$, so that the general solution of the Schrodinger equation with the initial condition $| \Psi(0)\rangle=|0\rangle$ by the formula \eqref{shredinger_solution_} after reduction of similar terms will take the form:
\begin{equation}
\label{simple_shred}
|\Psi(t)\rangle = cos(\frac{t}{\hbar})|0\rangle+i\ sin(\frac{t}{\hbar})|1\rangle
\end{equation}
and we see that over time there will be oscillations of the form: $|0\rangle\rightarrow i|1\rangle\rightarrow -|0\rangle\rightarrow ...$, so the qubit will return to the state $|0\rangle$ in time $t=2\pi\hbar$.

The Landau density matrix of the state $|\Psi\rangle$ is defined by the equality $\rho_\Psi=|\Psi\rangle\langle\Psi |$
It is proposed to prove that the Schrodinger equation for the density matrix has the form
$$
i\dot{\rho}=[\rho,H]=\rho H-H\rho .
$$

\subsection{Matrix dynamics
}

The equation \eqref{shredinger_solution} means that the state vector is transformed in time by multiplying by a certain evolution matrix $exp (- \frac{i}{\hbar}Ht)$. We assume that $H$ is a constant matrix, but it can also be time-dependent - in the latter case, the exponent should be interpreted as a chronological exponent; we will not deal with this issue, since such an interpretation will not change anything in essence.

Let the elements of the evolution matrix be denoted as $u_{ij}$, and the initial state is $|\Psi\rangle=|\Psi(0)\rangle$ has the form

\begin{equation}
\label{inistate}
|\Psi(0)\rangle=\sum\limits_i\la_i(0)|i\rangle.
\end{equation}

Then the matrix multiplication rule gives the equality $\la_i (t)=\sum\limits_{j=0}^{N-1}\la_j(0)u_{ij}$. Let's look at it in more detail. It means that the resulting amplitude $\la_i (t)$ of any state $|i \rangle$ is obtained by summing different contributions: from each state $|j\rangle$, from its amplitude $\la_j(0)$, this contribution is obtained by multiplying by $u_{ij}$. Thus, the number $u_{ij}$ is the amplitude of the transition $|j\rangle\rightarrow |i\rangle$.

Now let's consider two consecutive time intervals: $[0,t]$ and $[t, 2t]$. The resulting state vector is $|\Psi(2t)\rangle$ will be obtained by multiplying the initial vector by the second power of the evolution matrix $U_t^2$. By the matrix multiplication rule, we have

\begin{equation}
\label{2evol}
\la_i(2t)=\sum\limits_{j,k=0}^{N-1}\la_ju_{kj}u_{ik}
\end{equation}
that is, the transition is carried out in two stages: first from the state $|j\rangle$ to the state $|k\rangle$, and then from $|k\rangle$ to $|i\rangle$. Generalizing this to the case of finite time $T=nt$, we get a transition from the state $|j\rangle$ to the state $|i\rangle$ along the path

\begin{figure}[h]
\caption{Matrix dynamics}
\includegraphics[width=1.0\textwidth]{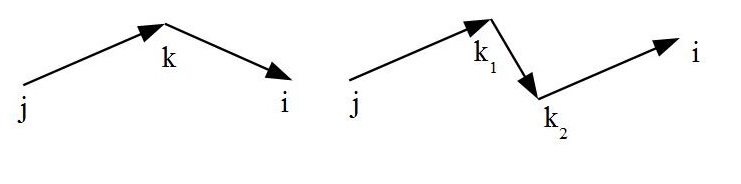} 
\label{fig:matrix_dynamics}
\end{figure}

\begin{equation}
\label{intw}
|j\rangle\rightarrow|k_1\rangle\rightarrow...\rightarrow |k_{n-1}\rangle\rightarrow |i\rangle
\end{equation}
by the polyline with $n-1$s links, so that the resulting amplitude is found by the formula
\begin{equation}
\label{resamp}
\la_i(nt)=\sum\limits_{j,k_1,r_2,...,k_{n-1}}\la_j(0)u_{ik_{n-1}}...u_{k_2k_1}u_{k_1j}
\end{equation}
- see Figure \ref{fig:matrix_dynamics}.

For the evolution matrix, $U_t=exp(-\frac{i}{\hbar}Ht)$ its element $\langle a|U_t|b\rangle$ is the amplitude of the transition from the state $|b\rangle$ to the state $|a\rangle$. In the first approximation of the exponent on $t$ , we get

\begin{equation}
\label{amp_per}
\langle a|U_t|b\rangle\approx\langle a|1-\frac{i}{\hbar}Ht|b\rangle=\delta_{ab}-\frac{i}{\hbar}\langle a|H|b\rangle.
\end{equation}

From this we can draw a simple conclusion.
The rule for finding the amplitude of the resulting transition is that a) it is necessary to add up the transition amplitudes along all paths leading from all starting points to the final one and b) along any of these paths, the transition amplitudes are multiplied. This rule is the basis of the "dial method " proposed by Feynman in the book \cite{Fe1} for a simple explanation of the law of quantum evolution.

So: a single quantum particle behaves like a swarm of independent particles - its duplicates, so

a) to obtain the resulting complex number $ \Psi (x,t)$ at a given point $x$ along each path, the amplitude of the duplicate is multiplied, b along all paths leading to this point $x$, the amplitudes are added, and

b) the probability density of detecting a particle at the point $x$ is its square of the module:
  $|\Psi(x,t)|^2$.

\begin{figure}
\centering
\includegraphics[scale=0.5]{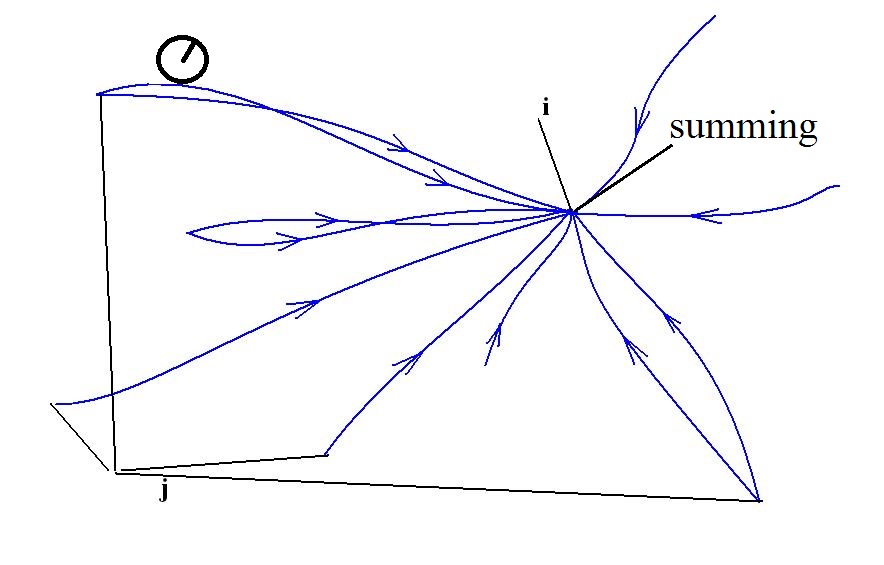}
 \caption{Matrix law: the hands of the dial rotate in proportion to the distance traveled, at the end point all the hands add up}
\label{fig:mat}
\end{figure}

The matrix dynamics can be represented as the hands of the dial plate turning along the traveled path (see the figure
 \ref{fig:mat}).

\begin{figure}
\centering
\includegraphics[scale=0.5]{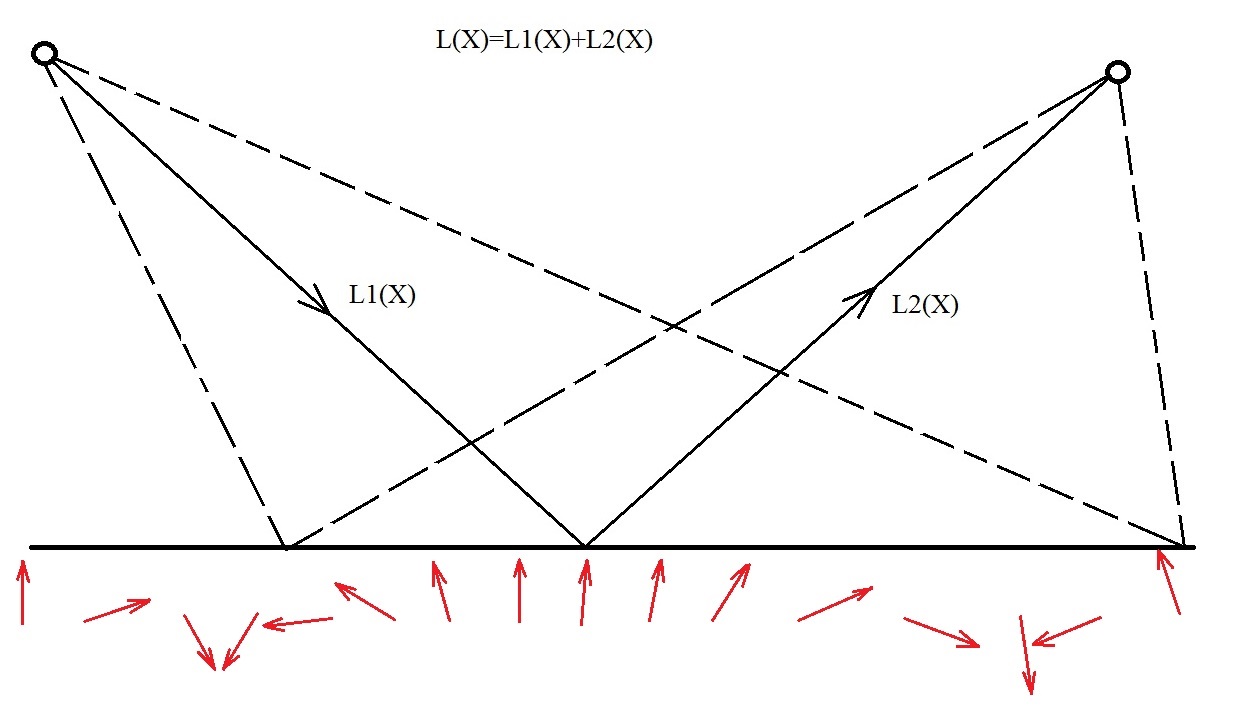}
 \caption{Reflection of light from a mirror
}
\label{fig:reflect}
\end{figure}

\begin{figure}
\centering
\includegraphics[scale=0.5]{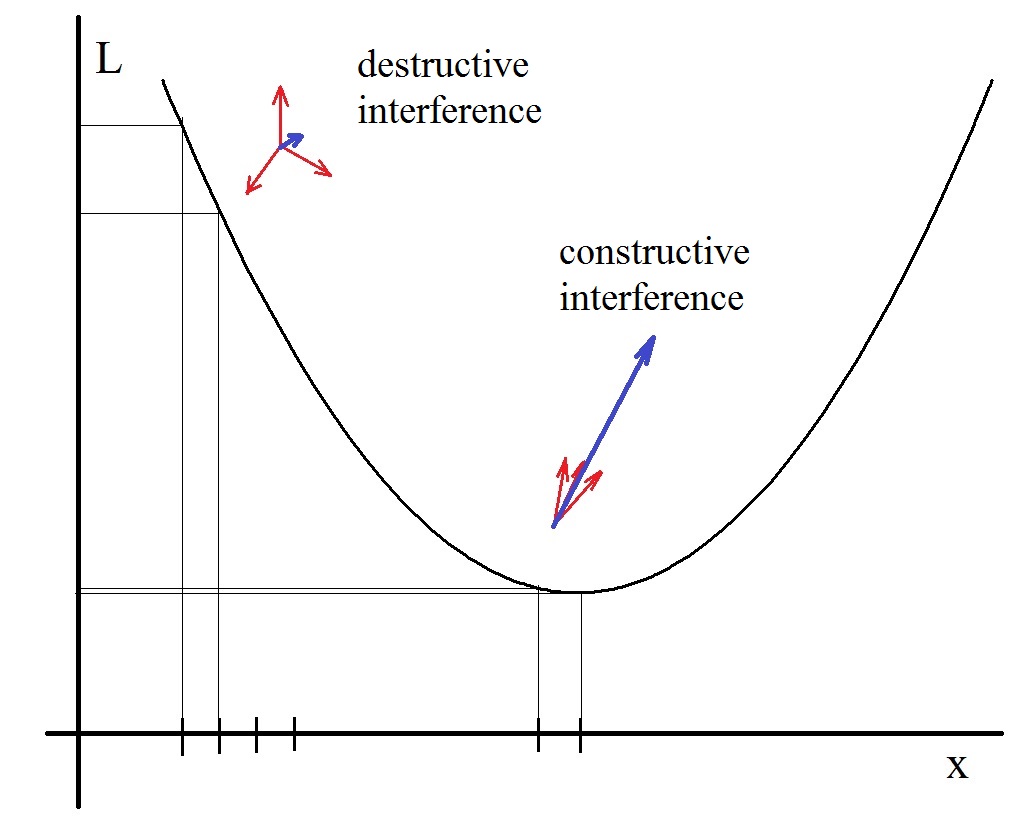}
 \caption{Constructive and destructive interference
}
\label{fig:interf}
\end{figure}

An illustration of constructive and destructive interference is shown in Figures \ref{fig:reflect} and \ref{fig:interf}

\subsection{Path integrals}

What will happen in the limit when we direct the elementary time $t$ to zero, and the number of links $n $ to infinity, so that $T=tn$ will be constant? Polyline paths \eqref{intw} will be replaced by continuous curves of the form $\g:\ x=x(t),\ t\in [0,T]$. For simplicity, let us again consider the case of a one-dimensional particle. The summation in \eqref{resamp} can be divided into two sums: one for $j$, the other for all intermediate points $k_1,k_2,...,k_{n - 1}$. The first sum will give the integral

\begin{equation}
\label{fe}
\Psi(y,t)=\int\limits_R K(y,x,t)\Psi(x,0)dx,
\end{equation}

and the second one will turn into a rule for calculating the matrix of a complex transition in the continuous case:
\begin{equation}
\label{fe1}
K(y,x,t)=\int\limits_{\g:\ x\rightarrow y}exp(\frac{i}{\hbar}S[\g ]){\cal D}\g,
\end{equation}
where $S[\g]$ is the action along the trajectory of $\g$, which is calculated by the formula $S[\g]=\int\limits_0^tL(\dot{x}, x, t)dt$, where $L (\dot{x},x,t)=E_{kin}-V$ is the Lagrangian equal to the difference between the kinetic and potential energy of a particle moving from point $x(0)=x$ to point $x (t)=y$. This function $K (y, x, t)$ is called the Feynman kernel, and the integral \eqref{fe1} is the Feynman interval along the trajectories.

The analogy with the discrete case is simple: $x$ plays the role of $j$, $y$ - the role of $i$, and  $K(y,x,t)$ - the role of evolution matrix $U_t$. The summing then turns to the integration on paths.

If the initial state of the particle $ \Psi (x,t)$ is a delta function concentrated at the point $x_0$, then the Feynman kernel is a wave function at the moment $t$. For the case of a free particle, $V=0$, so that the action will be an integral of the kinetic energy. It can be shown (see \cite{FeH}) that the kernel for a free particle has the form $c\cdot exp (- i m (x-x_0)^2/2\hbar t)$ for a constant $c$ that depends only on time $t$. This determines the spreading of the quantum state of a free particle initially concentrated at the point $x_0$: it will spread over the entire axis $(- \infty,+\infty)$ for any arbitrarily short period of time $t>0$, which illustrates the coordinate-momentum uncertainty relation.

Having considered the integral expression in \eqref{fe1}, we can see some discrepancy with the formula \eqref{shredinger_solution}, namely: there is no minus sign here, and when calculating the exponent, instead of the Hamiltonian, as in \eqref{shredinger_solution}, the Lagrangian is used, whose potential energy has the sign minus. How can you explain this?

Consider the Schrodinger equation for a particle in the potential $V$. If we do not pay attention to the kinetic energy, the signs will be all right: the minus ahead of the exponent compensates for the minus in the Lagrangian. Let's deal with kinetic energy. Its expression in $H=\frac{p^2}{2m}+V_{pot}$ as $p^2/2m $ coincides with the expression in terms of the Lagrangian in \eqref{fe1}: $m\dot{x}^2/2$, but the sign does not fit. However, in the Schrodinger equation \eqref{shredinger}, the impulse enters as a quantum impulse $p=\frac{\hbar}{i}\nabla$ and in \eqref{fe1} - as a classical impulse $m\dot{x}$. In order to move from it to the quantum one, it is necessary to perform the inverse Fourier transform, which will change the sign: $p^2/2m$ will turn into $-p^2/2m$, which is exactly necessary for matching the Schrodinger equation with the path integral. For a small section of the trajectory

$$
 m\dot{x}^2t/2=-p^2t/2m +px.
$$

Of course, this argument is not a proof that Feynman path integrals are equivalent to the Schrodinger equation, the formal proof is given in the book \cite{FeH}, to which we refer the reader for details.

Feynman integrals are thus a continuous analogue of matrix dynamics, which emphasizes the naturalness of the transition from continuous to discrete quantities. These integrals naturally generalize to the case of composite systems of many particles, or charged particles and an electromagnetic field, which allows us to calculate, for example, the amplitude of photon emission by a relaxing atom (see the book \cite{FeH}), as well as to generalize quantum dynamics to the relativistic case when the movements of charges occur at a speed comparable to the speed of light.

The most important advantage of Feynman integrals is a simple explanation of the transition from the quantum description of dynamics to the classical one. 

First, we consider the reflection of light from a mirror, the classical law of which is: the angle of falling is equal to the angle of reflection. Consider the phase $\phi$ of the photon wave function $\psi (x,t)=e^{i\phi(x,t)}$, considering it as a point particle, and its change along various paths, each of which is determined by the reflection point $x$ from the mirror (see figure \ref{fig:reflect}). The classical trajectory has this property. If it is slightly disturbed, the path length will not change too much, and the resulting phases along the original and disturbed paths (the phase is proportional to the path length) will be approximately the same (see figure \ref{fig:interf}). This means that the contribution of trajectories close to the classical one will prevail, and we can assume that the light moves along the classical trajectory, where the angle of falling is equal to the angle of reflection. But if the length of the paths is very small (the light source is very close to the receiver), then this reasoning will not pass, it will be necessary to take into account the contribution of all trajectories, not just the classical one, since then the change in the path length will be comparable to its length.

This reasoning is transferred to the general case and shows the limits of the application of classical physics.

Consider the formula for the core \eqref{fe1}. Here, integration is performed along all paths going from the start point to the end point. But among these paths there is one special path $ \gamma_{class}$ - the classical trajectory. This trajectory stands out from the set of all others in that it satisfies the principle of zero variation of action -  Maupertuis principle :

\begin{equation}
\delta S[\gamma_{class}]/\delta\gamma=0,
\label{mop}
\end{equation}
which is mistakenly called the principle of least action (the action there is not the smallest, its variation with the variation of the trajectory is zero). Indeed, consider the Lagrangian:

$$
L(\dot{x},x,t)=m\ddot{x}/2 - V,\ S[\g]=\int\limits_0^TL(\dot{x},x,t)dt,\ \g:\ x=x(t),\ 0\leq t\leq T
$$
and we will give an increment of $\d x$ to the coordinate. Then $x\rightarrow x+\d x,\ \dot{x}\rightarrow \dot{x}+\d\dot{x}$, and we will have

$$
\begin{array}{ll}
&\d S[\g]/\d \g=\int\limits_{0}^TL(\dot{x}+\d\dot{x},x+\d x,t)dt-\int\limits_0^TL(\dot{x},s,t)dt=\\
&\int\limits_0^T\left(\frac{\partial L}{\partial\dot{x}}\d\dot{x}+\frac{\partial L}{\partial x}\d x\right)dt=\int\limits_0^T\frac{\partial L}{\partial\dot{x}}d\d x+\int\limits_0^T\frac{\partial L}{\partial x}\d x dt=\\
&\frac{\partial L}{\partial \dot{x}}\d x\left|_0^T \right. -\int\limits_0^T\frac{d}{dt}\frac{\partial L}{\partial \dot{x}}\d x dt+\int\limits_0^T\frac{\partial L}{\partial x}\d xdt=0 \equiv \\
&\frac{\partial L}{\partial x}=\frac{d}{dt}\frac{\partial L}{\partial \dot{x}}\equiv -\frac{\partial V}{\partial x}=m\ddot{x}\equiv F=ma.
\end{array}
$$

Let's say we are modeling a process by choosing a time step $dt$. If this process can be adequately represented by choosing such a $dt$, in which the change in the action of $dS$ will be much greater than the Planck constant $\hbar\approx 10^{-27}$ erg sec., then in the formula \eqref{fe1} only those trajectories that are close to $ \gamma_{class}$ will survive, because the action is comparable in order of magnitude to its variation, so that for the environment (the environment is a family of trajectories close to) a non-classical trajectory, it will be very small due to the rapid oscillation of the exponent and the destructive nature of interference resulting from this - the sum will contain the lion's share of reductions and will be much less than the contribution of the environment of the classical trajectory.

If, for an adequate description of the process, it is necessary to take such a small time step $dt$ that the change in the action on it $dS$ will be comparable to $\hbar$, non-classical trajectories will also have to be taken into account. We can describe the flight of a bullet using quantum mechanics, and then at a small $dt$ the bullet will behave like a quantum object; the accuracy of the final result will be the same as with the classical approach, but computational complexity will make such a method unreasonable. Another thing is the movement of an electron in an atom - here it is necessary to make $dt$ very small, so it will be impossible to neglect non-classical trajectories.

So, here we rely on the possibility of simply discarding very small amplitudes - a powerful heuristic approach that will later lead us again to the need for a certain determinism, but no longer reducible to Newtonian mechanics; the post-quantum determinism for complex systems.

\begin{figure}
\centering
\caption{The real part of the Feynman core of a free particle.
}
\includegraphics[width=0.7\textwidth]{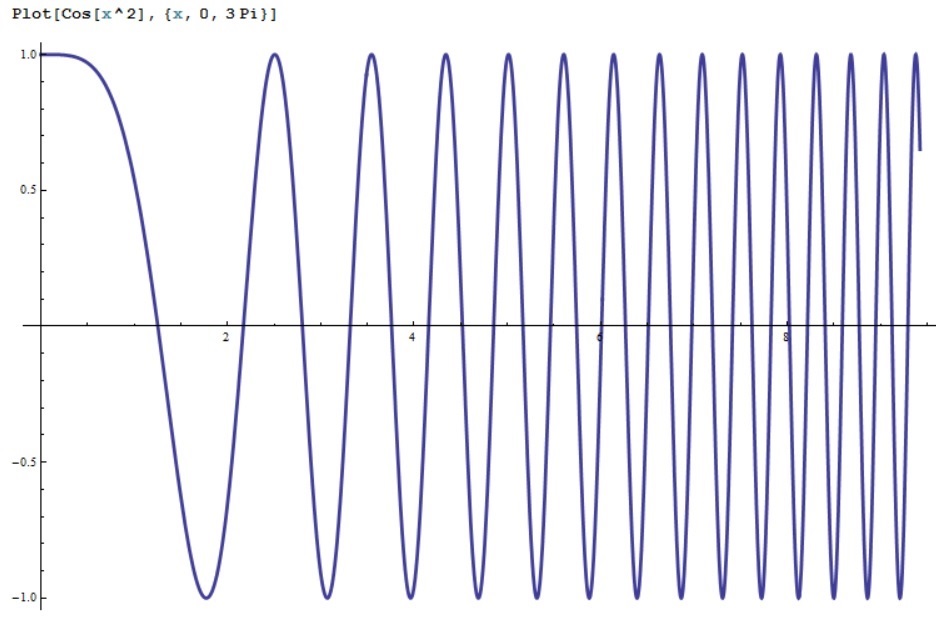}
\label{fig:free_particle} 
\end{figure}

Let's try using Feynman integrals along the trajectories to find out what the state of a free point particle moving along the $OX$ axis will look like at the moment $t>0$, if at the zero moment it was at the origin of coordinates. We assume that the trajectories of duplicates of this particle at a small $t$ are straight line segments, and the speed of its movement along these segments is constant. Then, on a segment of length $x$, the velocity will be equal to $x/t$, and substituting this velocity value into the expression for the Lagrangian, we get the Feynman core in the form of $a\ exp (\frac{imx^2}{2\hbar t})$, where $a$ is a constant. The graph of the real part of this function is shown at the figure \ref{fig:free_particle}.

\newpage

{\it Show that this state is consistent with the de Broglie wave

$$
exp\left(\frac{ipx}{\hbar}-\frac{iEt}{\hbar}\right)
$$
in the following sense:  duplicates of a particle that have reached the point $x$ during the time $t$ will have the same de Broglie oscillation period as Feynman in \ref{fig:free_particle}.}

For a free particle, it is very important to have duplicates with different velocities, and the distribution over all velocities should be uniform, so that any velocity in a virtual swarm of duplicates should be represented by the same number of  duplicates.

The dynamics of a free particle cannot be replaced by simple movements from one point to a neighboring one on a set of points of the form $x=\epsilon, 2\epsilon, 3\epsilon,...$, since a free particle has the ability to " jump" through many points at once. This feature should be taken into account when modeling free motion in terms of photon movements between optical cavities.

\subsection{Exercises}
\begin{figure}
\centering
\caption{Collision of two Gaussians
}
\includegraphics[width=1.0\textwidth]{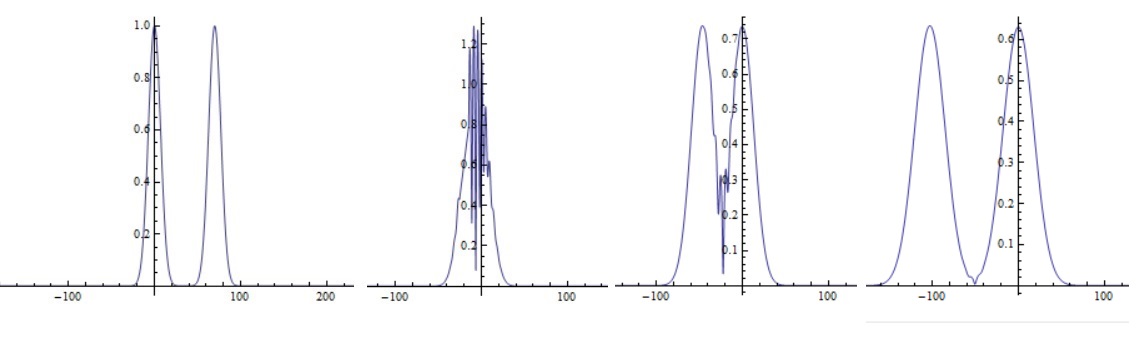}
\label{fig:collision}
\end{figure}

1. Explain the effect of restoring the shape of the parts of the wave function-Gaussians when they "collide" (figure \ref{fig:collision}).

\begin{figure}
\centering
\caption{Gaussian propagation with different dispersion
}
\includegraphics[width=1.0\textwidth]{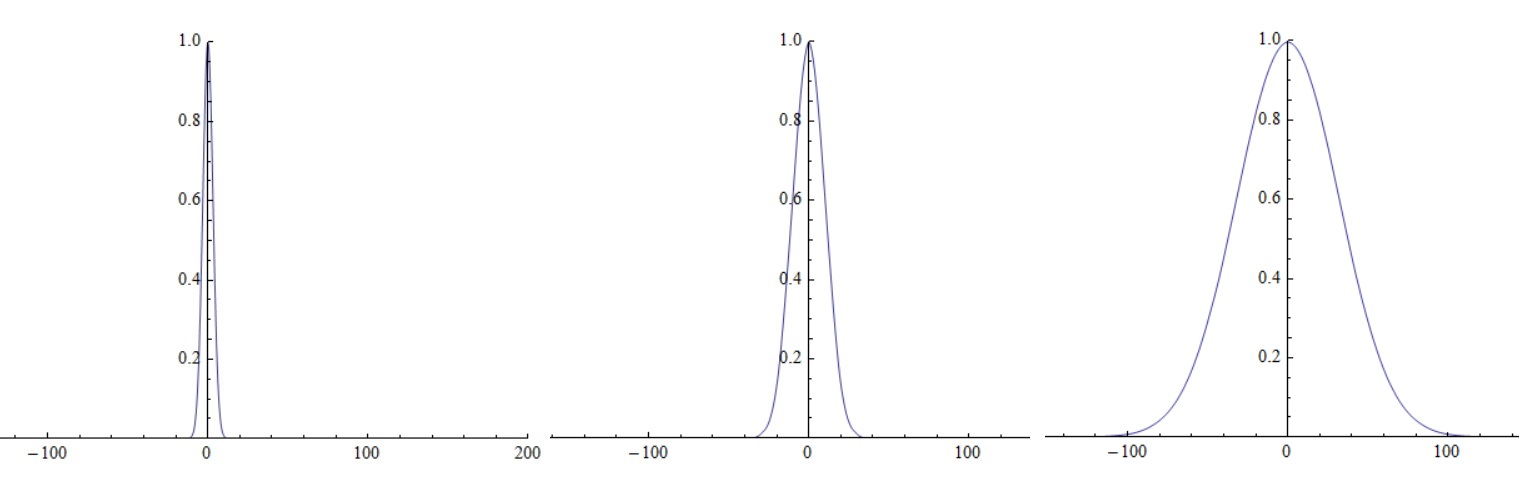}
\label{fig:unc}
\end{figure}

2. To explain the effect of different spreading rates of a wave packet in the form of a Hamiltonian with different degrees of dispersion. Is there a contradiction here with the theorem on the existence and uniqueness of the solution of the Cauchy problem for the Schrodinger equation? (see figure 
\ref{fig:unc}). The speed of spreading much decreases from the left picture to the right.

3. Calculate the approximation of the Dirac delta function in the form of the Fourier transform of the de Broglie wave:
$$
\phi(p)={\cal F}e^{\frac{i}{\hbar}p_0x}=\int_{-A}^Ae^{\frac{i}{\hbar}p_0x}e^{-\frac{i}{\hbar}px}dx
$$

\begin{figure}
\centering
\caption{Approximation of the Dirac delta function by a space segment constraint; argument $p$, $p_0=0$.
}
\includegraphics[width=0.5\textwidth]{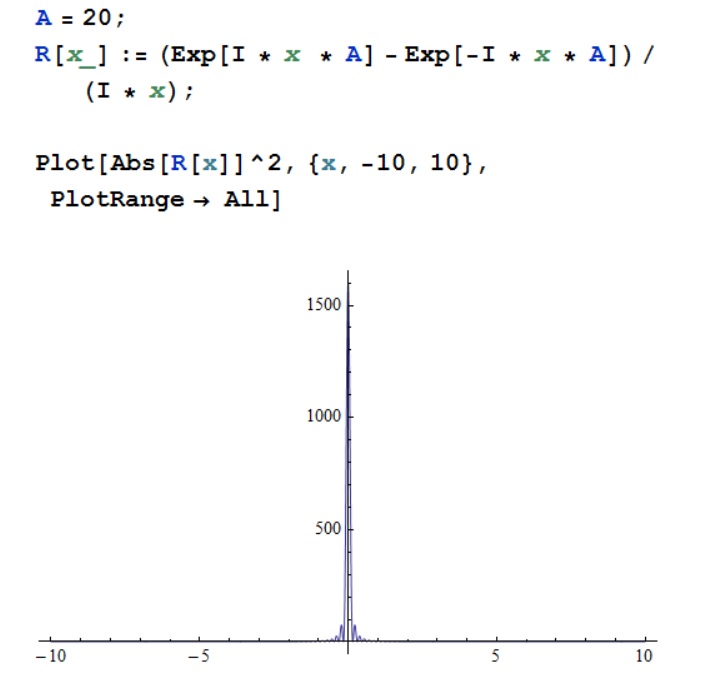}
\label{fig:delta}
\end{figure}

(see figure \ref{fig:delta}).

\newpage

\section{Lecture 2. Composite systems
}

\subsubsection{Tensor products
}

We have already received the simplest gate implementing the unitary operation $\s_x$: we need to create a quantum dot whose evolution obeys the Hamiltonian $\s_x$, and wait for the time $t=2\pi\hbar$. Such a dot is a charge in a two-hole potential.

But for full-fledged quantum computing, we need more complex gates, two-qubit ones. To do this, we need to introduce the concept of a tensor product.

\begin{figure}[h]
\centering
\includegraphics[scale=0.5]{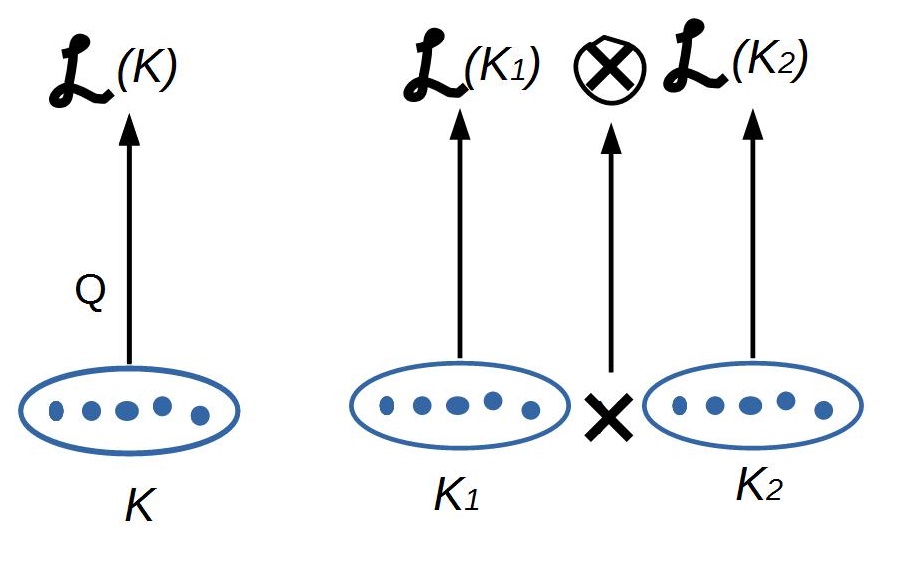}
 \caption{The tensor product is the result of Cartesian lifting using the quantum $Q$ operation-taking a linear span}
\label{fig:quant}
\end{figure}

Let us have two sets (two registers) of qubits: $A$ and $B$ with $n_A$ and $n_B$ qubits in each. The sets of classical states of these registers are: $K_A=\{ 0,1,..., N_A-1\}$ and $K_B=\{ 0,1,..., N_B-1\}$, respectively, where $N_A=2^{n_A}, N_B=2^{n_B}$. The quantum state space for the register $A$ is $L_A=C^{N_A}$, for the register $B$ - $L_B=C^{N_B}$. The tensor product $L_A\otimes L_B$ of the spaces $L_A$ and $L_B$ is the state space of the composite system of qubits $A\cup B$ - $C^N$, where $N=2^n,\ n=n_A+n_B$. Its orthonormal basis will be the Cartesian product $K_A\times K_B$ (see figure  \ref{fig:quant}). 

The general form of a vector from the tensor product of spaces will be:
\begin{equation}
\label{gen_prod}
|\Psi\rangle=\sum\limits_{j\in\{ 0,1,...,N_A-1\},k\in\{ 0,1,...,N_B-1\}}\la_{j,k}|jk\rangle
\end{equation}

Let $|\Psi_A\rangle=\sum\limits_{a=0}^{N_A-1}\la_a|a\rangle$ and $|\Psi_B\rangle=\sum\limits_{b=0}^{N_B-1}\la_b|b\rangle$ be the quantum states of the registers $A$ and $B$. Let's define their tensor product as $|\Psi_A\rangle\otimes|\Psi_B\rangle=|\Psi_A\rangle|\Psi_B\rangle=\sum\limits_{a\in K_A, b\in K_b}\la_a\la_b|a\rangle|b\rangle$. We will omit the ket-bra and just write $|ab\rangle$. The tensor product obeys the same rules as the usual one, for example, we can take out of brackets a common multiplier.

The state of a composite system that cannot be represented as a tensor product of $|\Psi_A\rangle|\Psi_B\rangle$ is called entangled. An example of such a state is the EPR pair: $|00\rangle+|11\rangle$. 

{\it Prove that this state cannot be represented as a tensor product of states of the first and second qubits. 

Try to formalize the question and answer to it: ''Which states are more: entangled or nonentangled?''}

Let $U_A:\ L_A\rightarrow L_A,\ U_B:\ L_B\rightarrow L_B$ be two linear operators on the quantum state spaces of these registers. The tensor product of these operators $U_A\otimes U_B$ is defined for the basis vectors as follows: $U_A\otimes U_B|ab\rangle = U_A|a\rangle\ \otimes U_B|b\rangle$, and we will extend the linearity to the entire space. The rule for finding the tensor product matrix for one-qubit spaces is illustrated by the example:

$$
U_A=\begin{pmatrix}&a_{11}&a_{12}\\&a_{21}&a_{22}\end{pmatrix}, U_B=\begin{pmatrix}&b_{11}&b_{12}\\&b_{21}&b_{22}\end{pmatrix}, U_A\otimes U_B=\begin{pmatrix}&a_{11}U_B&a_{12}U_B\\&a_{21}U_B&a_{22}U_B\end{pmatrix}, 
$$
the generalization of this rule to higher dimensions is obvious. The tensor product matrix will have a dimension equal to the product of the dimensions of the original matrices. 

Prove this by using the standard notation of the basis vectors in the form of columns, and moving on to the Dirac notation. Use the natural ordering on the basis vectors from the tensor product: for one qubit $|0\rangle,\ |1\rangle$, for two qubits $|00\rangle,\ | 01\rangle,\ |10\rangle,\ |11\rangle$.

It is proposed to prove the following formulas:

$$e^{A\otimes I}=e^A\otimes I, (A\otimes I)\cdot (I\otimes B)=A\otimes B, e^{A\otimes I}\cdot e^{I\otimes B}=e^{(A\otimes I)+ (I\otimes B)}, [(A\otimes I), (I\otimes B)]=0,$$
where $ \cdot$ denotes an ordinary matrix product, $I$ denotes an identical matrix.

\subsection{Partial measurements. Mixed states}\label{mix}

\begin{figure}[h]
\centering
\includegraphics[scale=0.7]{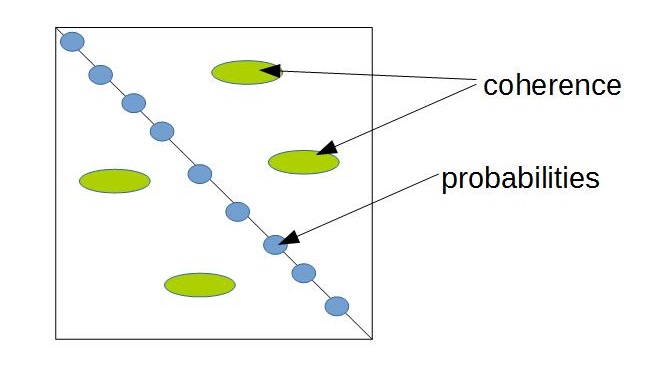} 
\caption{Density matrix
.}
\label{fig:dens_} 
\end{figure}

\begin{figure}[h]
\centering
\includegraphics[scale=0.7]{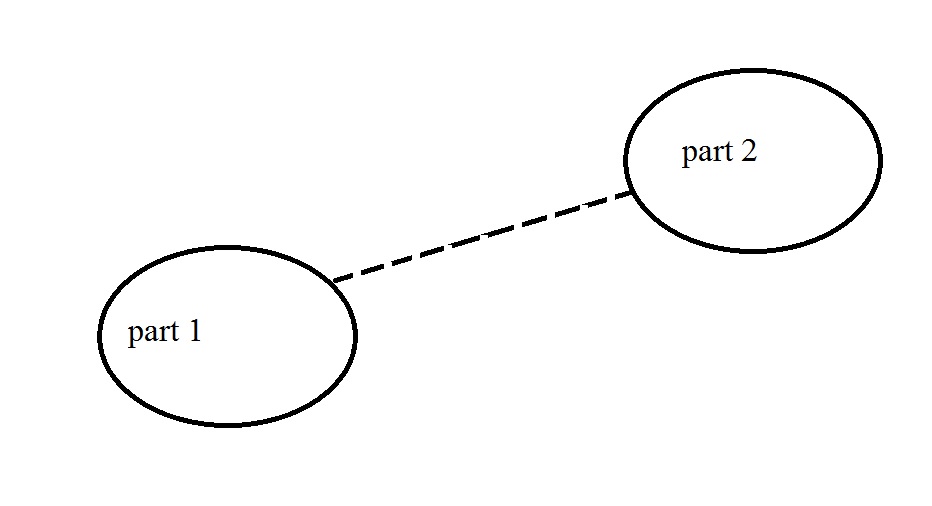} 
\caption{Bipartite entanglement
.}
\label{fig:dens__} 
\end{figure}

For multiparticle systems, a new question about partial measurement arises. If we have two particles, for example, two qubits, we can measure only one of them, leaving the second unaffected. In what state will then be unaffected qubit? To solve this problem, we turn to the algebraic form of the measurement result.

First, let's consider the case of a single qubit and the above-defined measurement of its state $|\Psi\rangle$ as a random variable. The following equality immediately follows from the matrix multiplication rule

$$
p_j=\langle j|\rho_\Psi|j\rangle
$$
The trace of the density matrix will be found by the formula $tr(\rho_\Psi)= \sum_j\langle j|\rho_\Psi|j\rangle$ and for one qubit this sum will be equal to 1. We see that the configuration of the form $\langle a | b \rangle$ always gives a number, and if $a$ and $b$ are vectors from the orthonormal basis of states of one qubit, this number is equal to $\delta_{ab}$ - the Kronecker symbol equal to zero for $a\neq b$ and one for $a=b$. 
This observation allows us to generalize the rule of working with Dirac symbols to tensor products, if we assume that in this case $a$ and $b$ must refer to the same real particle.

Consider the density matrix of the joint system of the form 
\begin{equation}
\rho=\sum\limits_{j',j'',k',k''}\rho_{j',k',j'',k''}|j',j''\rangle\langle k',k''|
\label{dens_comp}
\end{equation}
where one streak denotes the first subsystem, the double streak denotes the second subsystem. 

Let we measure the first qubit and obtain the value $|j\rangle$. Which state the second qubit will be in? We have to obtain the density matrix $\rho_2$ of the second qubit as we found the probability in the case of only one qubit: by enclosing of the density matrix by the same state from the left and right side. But now we must enclose it by the state of the first subsystem only: $\j\rangle$. Taking into account the orthonormality of all basic states of the same subsystem, we have:
$$
\rho_2^j=\sum\limits_{j',j''k',k''}\rho_{j',k',j'',k''}\langle j |j'\rangle|j''\rangle\langle k'|\langle k''| j\rangle=\rho_{j,j,j'',k''}|j''\rangle\langle k''|
$$
We note that the trace of this matrix must not be one, because we separate only one state of the first qubit  $|j\rangle$.

Summing on $j$, we find

\begin{equation}
\label{parttrack}
\rho_2=\sum\limits_{j'',k''}r_{j''k''}|j''\rangle\langle k''|,\ \ r_{j''k''}=\sum\limits_{j}\rho_{jj j''k''}
\end{equation}

The formula \eqref{parttrack} gives a simple mnemonic rule for determining the result of a partial measurement: to obtain the element $\rho_1 (j",k'')$ of the density matrix of the first qubit, it is necessary to sum up all the elements of the density matrix of both qubits with numbers obtained by all possible identical additions of the pair $j",k''$ to a pair of indices of a two-qubit density matrix.

However, the matrix $\rho_2$ obtained by the formula \eqref{parttrack}, generally speaking, will no longer have the form $|\psi\rangle\langle\psi|$ for any state $|\psi\rangle$ of the first qubit. To make sure of this, consider the example: $|\Psi\rangle=| EPR\rangle=\frac{1}{\sqrt 2}(|00\rangle+|11\rangle)$.

The state density matrix $|EPR\rangle$ is 

\begin{equation}
\label{EPRdensity}
\left(
\begin{array}{lllll}
&1/2&0&0&1/2\\
&0&0&0&0\\
&0&0&0&0\\
&1/2&0&0&1/2
\end{array}
\right)
\end{equation}

You are invited to make sure that the measurement of the second qubit in this state gives the matrix

\begin{equation}
\label{decoh2}
\left(
\begin{array}{lll}
&1/2&0\\
&0&1/2\\
\end{array}
\right)
\end{equation}

\begin{figure}[h]
\centering
\includegraphics[scale=0.4]{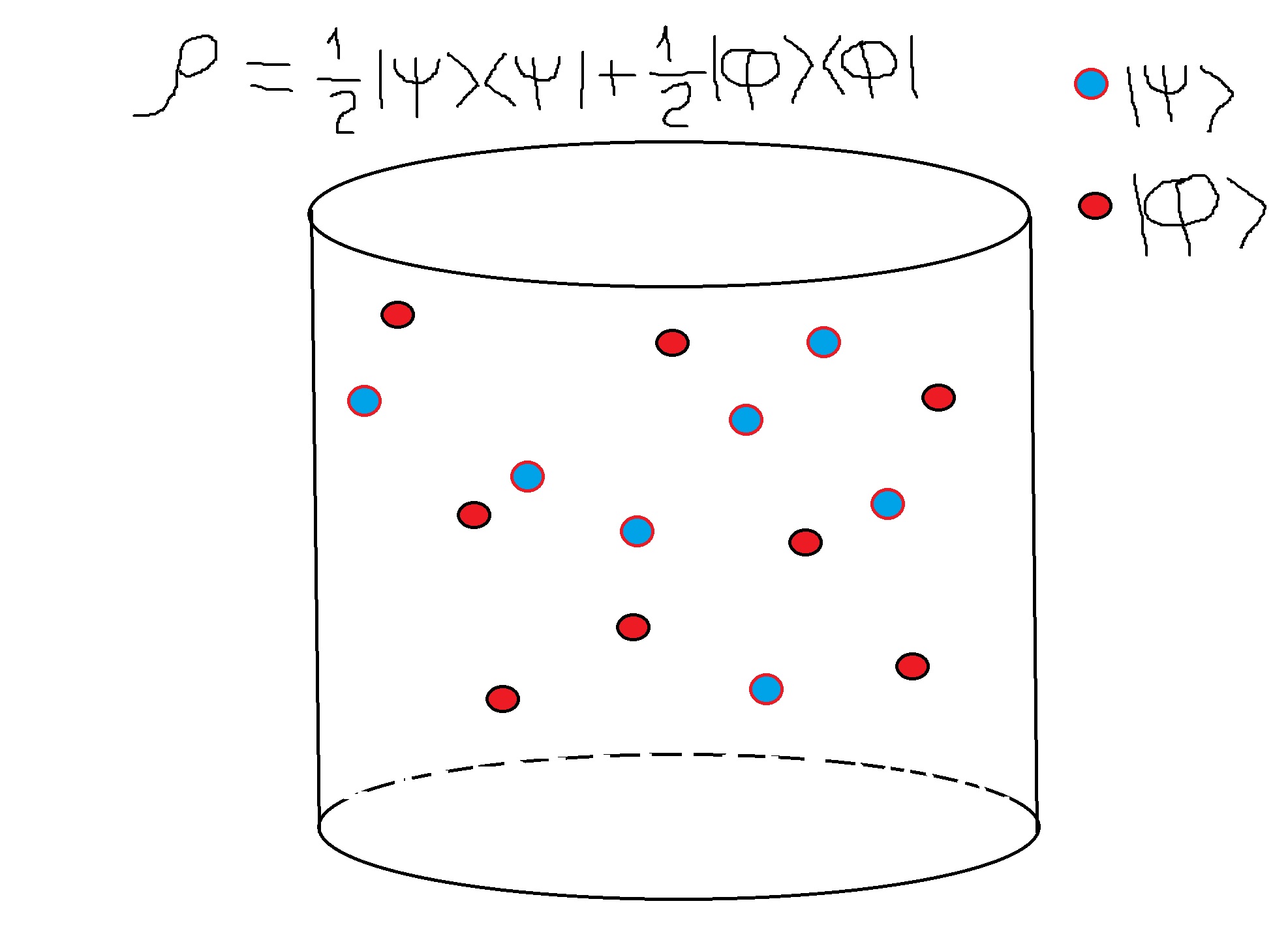} 
\caption{A mixed state is a reservoir with pure states
.}
\label{fig:mixed_state} 
\end{figure}

The matrix \eqref{decoh2} has rank 2, and therefore it cannot be a density matrix of any quantum state vector. Thus, we come to the need to expand the concept of state. We will call the state vectors pure, and we will call the "states" described by matrices similar to \eqref{decoh2} mixed - see Figure \ref{fig:mixed_state}.

So, the mixed state is the result of measuring one part of a vector-the state of a composite system. Let's assume that the vector state of a composite two-qubit system has the form:

$$
|\Psi\rangle=\la_{00}|00\rangle+\la_{01}|01\rangle+\la_{10}|10\rangle+\la_{11}|11\rangle .
$$
If we measure only the second qubit, then the result of such a measurement should be either $|0\rangle$ or $|1\rangle$. How likely is $p_2 (0)$ to get $|0 \rangle$ for the second qubit? This probability, according to the measurement logic, should be equal to $p_2 (0)=\sum\limits_j |\la_{j0}|^2$. Similarly, the probability of getting $|1\rangle$ in the second qubit will be $p_2 (1)= \sum\limits_j |\la_{j1}|^2$. The total probability will, of course, be singular. If we get $|0\rangle$ in the second qubit, then in the first one, which is not directly affected by the measurement, we should get the state $|\psi_0\rangle=a_0\sum\limits_j\la_{j0}|j\rangle$, where the normalization coefficient $a_0=(\sum\limits_j|\la_{j0}|^2)^{-1/2}$. Similarly, if the second qubit has the state $|1\rangle$, then the first qubit will be in the state 
$|\psi_1\rangle=a_1\sum\limits_j\la_{j1}|j\rangle$, where the normalization coefficient $a_1=(\sum\limits_j|\la_{j1}|^2)^{-1/2}$. 

Since in the general case $|\psi_1\rangle$ does not coincide with $|\psi_0\rangle$, we cannot sum these states as vectors in Hilbert spaces. But you can sum up their density matrices by entering a weight coefficient for each of them: $p_2 (0)$ or $p_2 (1)$. as a result, we get a "density" matrix of the form$|\psi_1\rangle=a_1\sum\limits_j\la_{j1}|j\rangle$, where the normalization coefficient

\begin{equation}
\rho_1=p_2(0)|\psi_0\rangle\langle\psi_0|+p_2(1)|\psi_1\rangle\langle\psi_1|.
\end{equation}
You are asked to make sure that this exactly matches the matrix \eqref{parttrack_}.

Thus, in the general case, when the system is divided into two subsystems, the "density" matrix of the measurement result of the second subsystem, found by the formula \eqref{parttrack_}, will have the form

\begin{equation}
\label{density}
\rho_1=\sum\limits_k p_k|\psi_k\rangle\langle\psi_k|,
\end{equation}
where $|\psi_k\rangle$ are the state vectors of the first subsystem obtained as a result of measuring the second subsystem, provided that the result of this measurement for the second subsystem was equal to $|k\rangle$.

The formula \eqref{density} sets the general form of the mixed state, which we will identify with the density matrix $\rho_1$. However, for a given density matrix $\rho_1$, the decomposition of \eqref{density} is not uniquely defined. The fact is that a system that is in a pure state $|\Psi\rangle$ can also be in another pure state $|\Phi\rangle$ with a probability of $|\langle\Psi|\Phi\rangle |^2$.

A mixed state means that the system is in some kind of pure state, but we don't know which one. Therefore, if all the pure states of $|\psi_k\rangle$ in \eqref{density} are mutually orthogonal, there is no coherence between them, and in this case this decomposition is uniquely defined.

A natural question arises: how to physically distinguish the EPR pair $\frac{1}{\sqrt 2}(|00\rangle+|11\rangle)$ from a mixture of the form $\frac{1}{2}|00\rangle\langle 00|+\frac{1}{2}|11\rangle\langle 11|$? Measurements in the standard basis, as we have seen, do not allow this to be done. But if we change the measurement bases, this will affect the diagonal of the density matrix, and we will be able to distinguish the EPR pair from the mixture. You are invited to consider all the details independently.

In what case does a partial measurement of one qubit in a pure state of a two-qubit system result in a pure state, not a mixed one? {\it Prove that this happens if and only if the initial state is not entangled.}

Let $U_1:\ {\cal H}_1\rightarrow {\cal H}_1,\ U_2:\ {\cal H}_2\rightarrow {\cal H}_2$ be two operators in different Hilbert spaces. Their tensor product $U_1\otimes U_2$ acts on the tensor product of the spaces: $U_1\otimes U_2: \ {\cal H}_1\otimes{\cal H}_2\rightarrow {\cal H}_1\otimes{\cal H}_2$. This function is defined naturally as a linear continuation of the action on the basis states: $U_1\otimes U_2: \ | jk\rangle\rightarrow U_1|j\rangle\otimes U_2|j\rangle$.

Let $M$ be the set of qubits of the system under consideration, consisting of $n$ qubits, and $|\Psi\rangle$ is some quantum state of these qubits. It is called non-entangled if there is such a partition of $M=M_1\cup M_2$ into two disjoint non-empty sets and the states $|\Psi_1\rangle,\ |\Psi_2\rangle$ on these sets, such that $|\Psi\rangle=|\Psi_1\rangle\otimes |\Psi_2\rangle$. Otherwise, the state of $|\Psi\rangle$ is called entangled. 

The {\it naive complexity} of the state $| \ Psi\rangle$ on the set $M$ is the size in qubits of the carrier of its maximum entangled tensor divisor. In other words, the naive complexity of a state is the maximum of the natural numbers $s$, such that there is a subset of $M_1\subseteq M$ and the states $|\Psi_1\rangle,\ |\Psi_2\rangle$ on $M_1$ and $M-M_1$, respectively, such that $|\Psi\rangle=|\Psi_1\rangle\otimes |\Psi_2\rangle$, $M_1$ contains $s$ elements and $|\Psi_1\rangle$ is entangled. Such a state $|\Psi_1\rangle$ is called the quantum kernel of the state $|\Psi\rangle$, and the corresponding set $M_1$ is the carrier of the kernel.

There can be several kernels, since the maximum number of $s$ from the definition can correspond to different sets of $M_1$ qubits. Naturally, this definition may depend on very small amplitudes, so that a complex state may be very close to a simple one. However, if we consider only states whose amplitudes $\la_j$ have a "grainy" form

$$
\la_j=\epsilon n_j+i\epsilon m_j,\ \ n_j,m_j\in Z, 
$$
this proximity will be limited by the grain size $ \epsilon$. From what follows, it will be clear that it is impossible to aim $\epsilon$ to zero for complex systems, and therefore the naive complexity is defined correctly in this way.  This definition of complexity depends on the basis in which we consider the states.

\subsection{Schmidt Theorem}

The entangled state in a composite system consisting of two components $S_1$ and $S_2$ has the form \eqref{tens} and is unrepresentable as a tensor product.
\begin{equation}
\label{tens}
|\Psi\rangle=\sum\limits_{j=0,...,N-1,k=0,...,M-1}\la_{jk}|j_1k_2\rangle
\end{equation}

Its storage in the computer memory is very expensive: it is necessary to store the matrix $\la_{jk}$ in contrast to the state \eqref{stateproduct}, 

\begin{equation}
\label{stateproduct}
|\Psi\rangle\otimes|\Phi\rangle=\sum\limits_{j=0}^{N-1}\la_j|j\rangle\otimes\sum\limits_{k=0}^{M-1}\mu_k|k\rangle=\sum\limits_{j=0,...,N-1;k=0,...,M-1}\la_j\mu_k|jk\rangle
\end{equation}

where it is necessary to store only two vectors in memory. It turns out that there is a possibility of a more economical representation of entanglement, but this representation is suitable only for a given fixed state, since this requires changing the basis in both spaces - specifically for this fixed state$|\Psi\rangle\in C^N\otimes C^M$. 

Namely, the following takes place
\bigskip

{\bf Theorem (Schmidt)}.{\it For any state $|\Psi\rangle$ of the form \eqref{tens} of a composite system, there are new orthonormal bases \newline $|J_0\rangle,|J_1\rangle,...,|J_{N-1}\rangle;\ |K_0\rangle,|K_1\rangle,...,|K_{M-1}\rangle$ in the component spaces $C^N,\ C^M$, such that

\begin{equation}
\label{smidt}
|\Psi\rangle=\sum\limits_{q=0}^S\a_q|J_q\rangle|K_q\rangle
\end{equation}
where $S=min(N-1, M-1)$, and $\a_q$ are non-negative real numbers such that $\sum\limits_{q=0}^S|\a_q|^2=1.$ }
\bigskip

The proof of this theorem is carried out by induction on $max (N,M)$. Let $|\Psi\rangle=|\Psi_1\rangle|\Psi_2\rangle$ be an un-entangled state. Then the theorem is fulfilled in an obvious way. Let's analyze the case when $|\Psi\rangle$ is an entangled state.

The set of non-entangled states ${\cal N}$ is closed as a subset of Euclidean space. Indeed, if $|\psi_1^n\rangle|\psi_2^n\rangle\rightarrow |\Psi\rangle$, then the sequences $|\psi_1^n\rangle$ and $|\psi_2^n\rangle$ have limits $|\psi_1\rangle$ and $|\psi_2\rangle$, respectively, and we will have $|\psi_1^n\rangle|\psi_2^n\rangle\rightarrow |\psi_1\rangle|\psi_2\rangle$ with $n\rightarrow \infty$.

So, there is a point in ${\cal N}$, the distance from which to the end of the vector $|\Psi\rangle$ is minimal, let it be the end of the unnormalized vector $|\Phi_0\rangle$: $|\Psi\rangle=|\Phi_0\rangle+ | A\rangle$, so that $\| A\| $ is the distance from $|\Psi\rangle$ to ${\cal N}$. Since $|\Phi_0\rangle\in{\cal N}$, we have $|\Phi_0\rangle=|J_0\rangle|K_0\rangle$ for some vectors $|J_0\rangle\in C^N,\ |K_0\rangle\in C^M$; we will take these vectors as the initial vectors in the \eqref{smidt} decomposition. We need to prove that none of these vectors are present in the decomposition of $|A\rangle$, then an induction step will be taken, since we will then do the same with $|A\rangle$ as with $|\Psi\rangle$. If one of the vectors $|J_0\rangle,\ |K_0\rangle$ were present in the decomposition of $|A\rangle$, we would get a contradiction with the minimality of the vector $|A\rangle$, because it would be possible to "split" a little more from it, which is impossible due to the choice of $|A\rangle$. The details are provided to the reader.

\bigskip
Schmidt theorem gives a numerical characterization of the entanglement measure of the composite state $| \Psi\rangle\in C^N\otimes C^M$ as the entropy of the probability distribution $|\a_q|^2$. Entropy of the distribution $\bar p$ is $E(\bar p)=-\sum\limits_iln(p_i)p_i$.

This Theorem has another useful consequence - the existence of the so - called $SVD$ - decomposition of an arbitrary matrix $A$ in the form $SAV=D$, where $S,\ V$ are unitary matrices, and $D$ is diagonal. This decomposition generalizes the theorem on reducing Hermitian and unitary matrices to a diagonal form; only here the matrix $A$ is arbitrary, not even necessarily square, and $S$ and $V$ are not connected in any way, they may even have different dimensions. This consequence is immediately obtained if we represent the matrix $A$ as a set of coefficients $\la_{jk}$ from the decomposition \eqref{tens} of the state of the composite system; then $S$ and $V$ will be the transition matrices to the bases $|J_i\rangle$ and $|K_j\rangle$ in the condition of the Theorem.

What if we have only one of the two components of the composite system at our disposal, for example, $S_1$, and the other $S_2$ is out of access? In this case, we have, in fact, only the density matrix $\rho_1$ of the first subsystem, so we don't even know about the existence of the second component. Is it possible in this case to "restore" the pure state of $|\Psi\rangle$, such that $\rho_1=tr_2(|\Psi\rangle\langle\Psi|)$? 

Yes, it can be done, and very simply. Let $C^N$ be the space of quantum states of the subsystem $S_1$ Let's take another instance of $S_1$, which we denote by $S_1'$, and the corresponding space of quantum states $C^N$, whose vectors we will denote with the same letters as for $S_1$, which will not cause misunderstandings, since we always write the states $S_1$ first and $S_1'$ second in the tensor product. Taking the eigenvalues $A_i$ of the matrix $\rho_1$ and the corresponding eigenvectors $|\phi_i\rangle$, we put $\a_i=\sqrt A_i$, and define $|\Psi\rangle\in C^N\otimes C^N$ as $ \sum\limits_{i=0}^{N-1}\a_i|\phi_i\rangle|\phi_i\rangle$. Then, from the rule of finding the relative density matrix, we get $\rho_1=tr_2(|\Psi\rangle\langle\Psi|)$. 

This observation also implies the method of finding the matrices $S$ and $V$ in the $ SVD$ decomposition. It is necessary to turn the matrix $A$ into the state $| \Psi\rangle$ of the composite system by taking \eqref{tens}in the decomposition its coefficients, then find its density matrix $\rho_\Psi=|\Psi\rangle\langle\Psi|$, the relative density matrices $\rho_2=tr_1(\rho_\Psi)$ and $\rho_1=tr_2(\rho_\Psi)$, which will have the same sets of eigenvalues that coincide with the numbers $|\a_i|^2$ in the Schmidt decomposition for $|\Psi\rangle$, then search for Schmidt decomposition of the state $|\Psi\rangle$, choosing as $|J_i\rangle,\ |K_i\rangle$ the eigenvectors of the operators $\rho_1$ and $\rho_2$.

\subsection{The paradox of quantum entropy}

What is order in a complex system? Order is an alternative to chaos. If the system is classical, and $ \bar p=(p_0, p_1,..., p_{N-1})$ is a list of probabilities of finding this system in classical states $x_0,x_1,..., x_{N-1}$, then the degree of chaos is the Shannon entropy
$$
Sh(\bar p)=-\sum\limits_{i=0}^{N-1}p_i\ ln(p_i),
$$
When adding new elements to the system, the classical entropy is $Sh(\bar p)$ can only increase, therefore, the order cannot increase.

How to generalize the Shannon entropy to the case of a quantum system? A natural generalization is the von Neumann entropy
$$
N(\rho)=-tr(\rho\ ln(\rho)),
$$
where $ \rho$ is the density matrix, which in the quantum case replaces the probability distribution $\bar p$. 

Consider the state of two qubits $|\Psi\rangle=\frac{1}{\sqrt 2}(|00\rangle+|11\rangle)$. Its entropy is zero. Indeed, the entropy of any pure state in general is zero. {\it Prove this by reducing the matrix $\rho$ to a diagonal form and showing that the entropy of the state of the form $|j\rangle\langle j|$, where $|j\rangle$ is one of the basis vectors, is zero.}

Let's assume that we have removed the second qubit by a large distance, so that only the first qubit remains in our hands. Then this qubit will be in a mixed state $\rho_1=tr_2 (|\Psi\rangle\langle\Psi|)$, and $N(\rho_1)=ln(2)>0$. That is, when adding a second qubit, the entropy of the quantum state will decrease.

The effect of increasing the order during the expansion of the system is a counterintuitive, purely quantum effect. It occurs due to the presence of entanglement, which connects the various physical parts of the system of many bodies.

\newpage

\section{Lecture 3. Quantum gates}

\begin{figure}[h]
\centering
\includegraphics[scale=0.7]{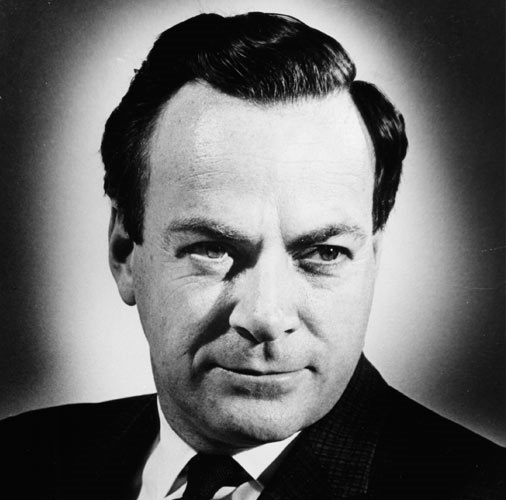} 
\caption{R.Feynman
}
\label{fig:feynman} 
\end{figure}

The user interface of a quantum computer proposed by R. Feynman is based on quantum gates and arrays of them (quantum gate arrays). A quantum gate is a unitary operator operating in the state space of one, two or three qubits, which can be implemented physically. If we take all single-qubit gates and add almost any two-qubit gate to them, for example, the gate \newline CNOT: $|x, y  \rangle\rightarrow |x,y\oplus x\rangle$, we can get a complete system of gates: any unitary transformation can be expressed using gates from this set with any predetermined accuracy (see \cite{universal}). A huge variety of interesting operators can be built on combinations of gates. A reader who loves algebraic exercises can refer to the book \cite{KSV}, which contains many interesting problems on quantum computing.

Thus, the first task of implementing the Feynman scheme of quantum computing is the implementation of single-qubit gates and CNOT. Consider a CiNOT gate that is close to CNOT: $CiNOT|x, y\rangle=e^{i\pi x/2}CNOT|x, y\rangle$. We will show how to implement the one-qubit gate $iNOT: |x\rangle\rightarrow i|x\oplus 1\rangle$ and the quantum gate CiNOT on the charge states of electrons in quantum dots. Having one-qubit gates and CiNOT, it is also possible to implement CNOT, since it is obtained from CiNOT by applying a one-qubit relative rotation of the phase $e^{-i\pi x/2}$to the first qubit. This implementation of CNOT is one of the first proposals for the implementation of entangling gates on charge states (see \cite{Fed}), its scheme is the simplest, although it presents certain technological difficulties.

Shredinger equation for one dimensional quantum particle looks as 
$$
i\hbar\dot{\Psi}=H\Psi,\ H=\frac{p^2}{2m}+V,\ p=\frac{\hbar}{i}\frac{\partial}{\partial x},
$$
where $V=V(x)$ is the potential. 

The potential hole is the simplest form of quantum dot. Cauchy problem for Shredinger equation in the potential hole looks then as 
\begin{equation}
i\hbar\dot{\Psi}=-\frac{\hbar^2}{2m}\frac{\partial^2\Psi}{\partial^2 x}+V\Psi.
\ \Psi(t,0)=\Psi(t,L)=0, \Psi(0,x)=\Psi_0(x),\ x\in [0,L]
\label{c}
\end{equation}

The eigenfunctions of this problem - solutions of the equation $H\phi_n=E_n\phi_n$ with the boundary conditions taken from \eqref{c}, looks as $\phi_n=c_n\ sin{\pi x n/L}$, when $c_n$ is the norming coefficient, which can be found from the condition $\int\limits_0^L|\phi_n(x)|^2dx=1$.

\begin{figure}[h]
\centering
\includegraphics[scale=0.6]{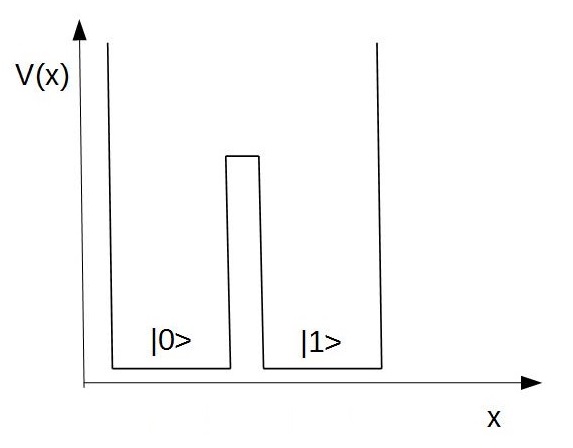} 
\caption{A quantum dot in the form of a two-hole potential. If the particle is in one hole initially, it will oscillate between these two holes.
}
\label{fig:quantum_dot} 
\end{figure}

We introduce the concept of a quantum dot. This is a small region in a solid-state structure in which a potential is created in the form of two wells with a sufficiently high potential barrier between them, and one electron can be in this potential (see figure \ref{fig:quantum_dot}). 

Finding an electron in the right well means the state $|0 \rangle$, in the left - 
 $|1\rangle$. 

\begin{figure}[h]
\centering
\includegraphics[scale=0.7]{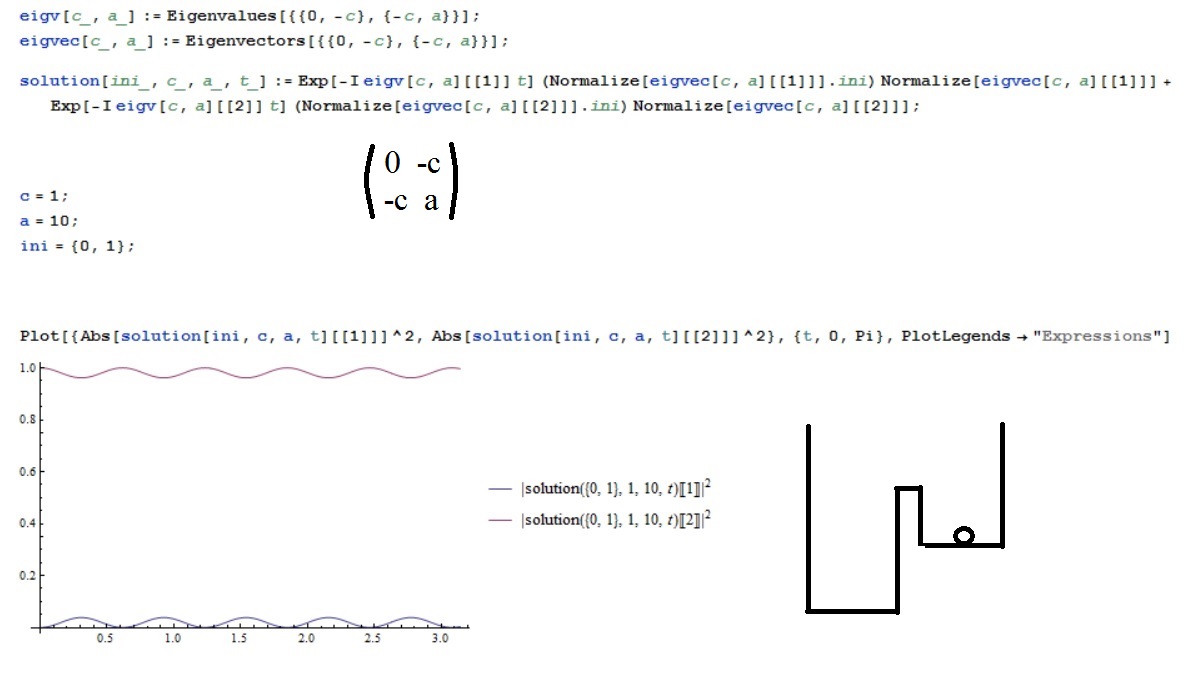} 
\caption{Two assymetric holes. Nonclassical behavior of a particle: given the initial condition in the higher hole it continues to stay in it with the large probability
}
\label{fig:2holes_} 
\end{figure}

The nonlcassical behavior of the particle in the assymetric two hole potential is demonstrated at the figure \ref{fig:2holes_}.

\begin{figure}[h]
\centering
\includegraphics[scale=0.6]{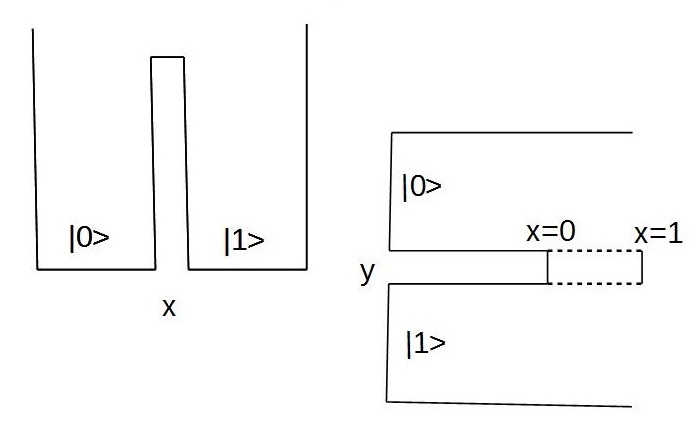} 
\caption{CiNOTon charge states
}
\label{fig:CNOT} 
\end{figure}

\begin{figure}[h]
\centering
\includegraphics[scale=0.6]{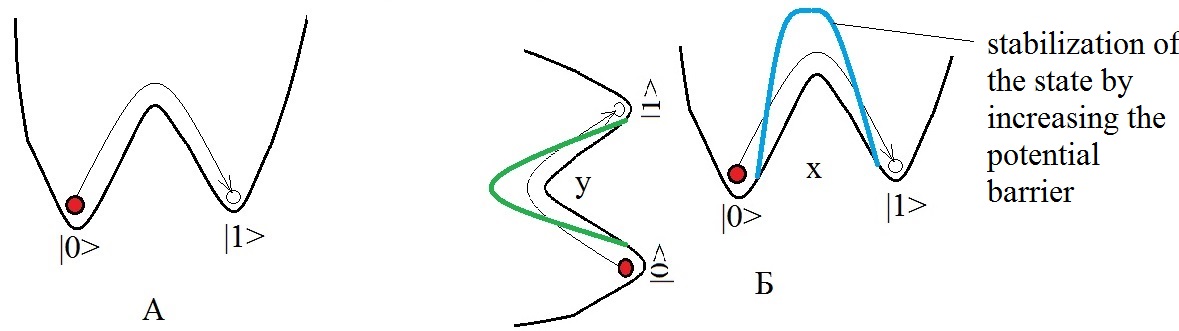} 
\caption{The work of two dots implementing CNOT gate
}
\label{fig:CNOTtwoDots} 
\end{figure}

The Hamiltonian of such a system has the form $H=c_1I-b\sigma_x$, where $\sigma_x$ is the first Pauli matrix defined in \eqref{Pauli}, $b>0$. You can show that the eigenstates of this Hamiltonian will be

\begin{equation}
|\phi_0\rangle=\frac{1}{\sqrt 2}(|0\rangle+|1\rangle,\ \  |\phi_1\rangle=\frac{1}{\sqrt 2}(|0\rangle-|1\rangle, 
\label{eigen_}
\end{equation}
moreover, their eigenvalues are ordered so that $E_0<E_1$, so that $|\phi_0\rangle$ will be the main state, and $|\phi_1\rangle$ will be the excited state. We find a solution to the Cauchy problem for the Schrodinger equation with such a Hamiltonian in the form

\begin{equation}
|\Psi(t)\rangle=A_0e^{-\frac{iE_0t}{\hbar}}|\phi_0\rangle+A_1e^{-\frac{iE_1t}{\hbar}}|\phi_1\rangle=e^{-\frac{iE_0t}{\hbar}}(A_0|\phi_0\rangle+e^{-\frac{i(E_1-E_0)t}{\hbar}}A_1|\phi_1\rangle)
\label{1}
\end{equation}
and now, considering that the states of $e^{i\theta}|\Psi\rangle$ are physically indistinguishable for any vector $|\Psi\rangle$, we come to the conclusion that to implement the gate NOT: $|0\rangle\rightarrow |1\rangle,\ |1\rangle\rightarrow |0\rangle$, it is enough just to wait for a while
 $\frac{1}{2}\tau=\pi\hbar/(E_1-E_0)$. 

It follows from the formula \eqref{1} that the basic states of an electron in a quantum dot oscillate, that is, they pass one into another $ |0\rangle\rightarrow |1\rangle\rightarrow |0\rangle$ and $|1\rangle\rightarrow |0\rangle\rightarrow |1\rangle$ with the period $\tau=2\pi\hbar/(E_1-E_0)$, which we will call the oscillation period.

\bigskip
{\it Here we ignored the phase multiplier $e^{-\frac{iE_0t}{\hbar}}$, which has no physical meaning if the NOT operator is performed for any states. But suppose that NOT is performed conditionally, for example, only if some other qubit has the value $1$, and if its value is $0$, then NOT over $x$ is not performed. In this case, it is necessary to take into account the total run of the phase, and take into account this multiplier. Find $E_0$ and $E_1$ and write an exact expression for the operator implemented by this subroutine at $x=1$ for the time $\tau/2$. Answer: this is the $i\sigma_x$ operator. We will show how to implement an operator close to CNOT on atomic excitations, where the Hamiltonian will have the inverse sign, and the similar operator will have the form $ - i\sigma_x$.}

\bigskip

The implementation of the CiNOT gate requires two quantum dots located perpendicular to each other, as shown in the figure \ref{fig:CNOT}. The Coulomb interaction of two electrons, each of which is located at one of these points, leads to the effect of changing the potential barrier at the point $y$. The potential barrier between the wells at the point $y$ turns out to be higher if the electron of the point $x$ is in the state $|1 \rangle$, compared to the situation when the electron of the point $x$ is in the state $|0\rangle$ due to the fact that the repulsion of electrons is higher at a close distance.

Let's first assume that we managed to fix the position of the electron $x$ in some way, so that it does not tunnel between its wells. Then you can find such a time $\tau_{CiNOT}$ that after this time the conversion of $CiNOT$will occur. Indeed, let the difference of the energy levels of the $ y $ - electron corresponding to the positions of the $x$ - electron $|0 \rangle$ and $|1\rangle$ be equal to $dE^0=E^0_1-E^0_0$ and $dE^1=E^1_1-E^1_0$, respectively. Then the oscillation periods for the $y$ electron when the $x$ electron is at the position $|0 \rangle$ and $|1\rangle$ will be, respectively, $\tau_0=2\pi\hbar/(dE^0)$ and $\tau_1=2\pi\hbar/(dE^1)$. By varying the distance between the points, we can choose these values in such a way that for a certain time value $ \tau_{CiNOT}$, an even number of oscillations with an upper index of $0$ and an odd number with an upper index of $1$ would fit into it, which will give us the required operator $CNOT$ when fixing the position of the$ x $ electron. The details are provided to the listeners.

How to prevent tunneling of the $ x $ - electron? This can be done by increasing the potential barrier between the wells at the $x$ point so that during the tunneling between them, the $x$ electron was significantly less than $ \tau_{CiNOT}$, and then, after making $CiNOT$, again reduce this barrier to the usual level, which is done by the external potential. This is how the $CiNOT$ gate is implemented. The problem is that an electron that is in the excited state $|\phi_1\rangle$ at one point is able to emit a photon, going into the state $|\phi_0\rangle$, which will prevent the implementation of the $CiNOT$ gate according to this scheme. A similar problem always occurs when implementing confusing error gates. For short computations, they may be negligible, but for practically important long computations, they pose a problem. We will return to this topic later, when studying more realistic models of quantum computers.

\begin{figure}[h]
\centering
\includegraphics[scale=0.7]{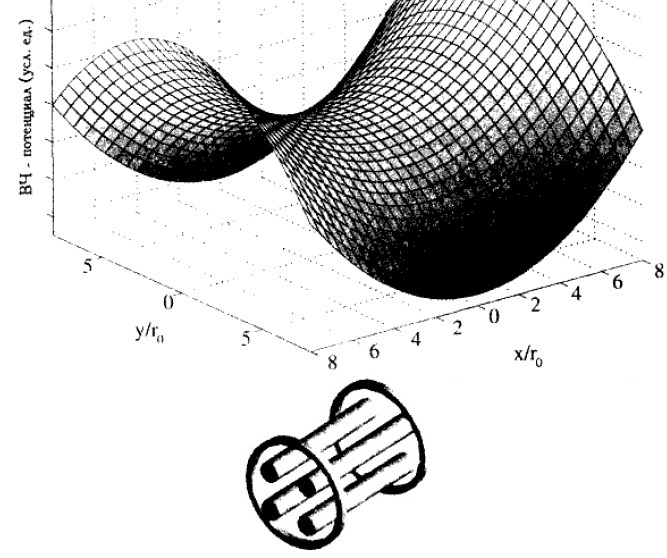} 
\caption{Pseudo-potential of the Paul trap.
}
\label{fig:paul_trap} 
\end{figure}

\begin{figure}[h]
\centering
\includegraphics[scale=0.7]{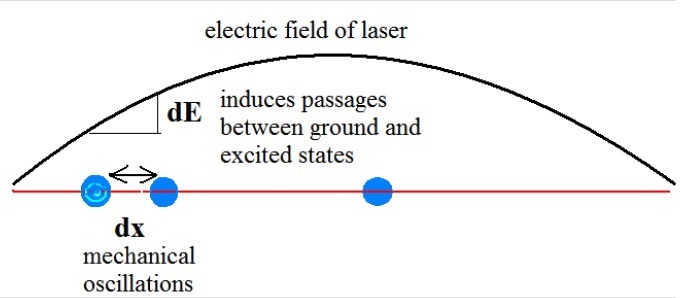} 
\caption{Schematic representation of the realization of CNOT by Paul ion trap. Logical qubits are internal atomic states of positively charged metallic ions; mechanical oscillations plays the role of ancilla. The ions in Paul trap are fixed in their places along the main axes by the oscillating electrical potential on the wires, which with the Coulomb repulsion between ions forms the pseudo potential. 
}
\label{fig:paul_trap2} 
\end{figure}

The other way to realize the CNOT gate is shown at the pictures \ref{fig:paul_trap}, \ref{fig:paul_trap2} 

The scheme of realization of CNOT in Paul trap is shown at the figure \ref{fig:paul_trap2}

\subsection{Single-qubit gates, $CNOT, CSign, \Lambda_\phi$  and $Toffoli$}

Prove that one-qubit gates have the form
$$
e^\alpha\begin{pmatrix}&e^{i(\phi+\xi)}cos(\theta)&e^{i(\phi+\xi)}sin(\theta)\\
&-e^{i\phi}sin(\theta)&cos(\theta)\end{pmatrix}
$$
for some real $\a,\xi,\theta,\phi$. {\it Instruction: find the number of independent real parameters defining the unitary operator.}

How to get the root of the gate $NOT$ - that is, such a gate $V$ that
 $V^2=NOT$?

The gate $CNOT$ is a 1-controlled NOT, it is defined as $CNOT|x, y\rangle=|x, x\oplus y\rangle$, where $\oplus$ is addition modulo 2. Construct the matrix $CNOT$.

2-controlled gate $U$ is defined as

$$
\Lambda_2 U: |x,y,z\rangle =\left\{
\begin{array}{ll}
|x,y,z\rangle\ &if\ xy=0,\\
|x,y\rangle U|z\rangle, \ &if\ xy=1
\end{array}
\right.
$$ 

\begin{figure}
\centering
\includegraphics[scale=0.7]{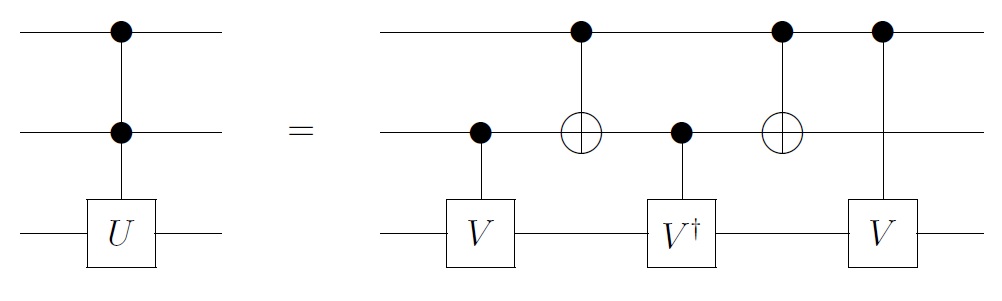}
 \caption{Implementation of a 2-controlled gate using CNOT and a one-bit $V$, where
 $V^2=U$}
\label{fig:1__}
\end{figure}

Toffoli gate is $\Lambda_2NOT$. 
Show that the gate $\Lambda_2 U$ can be implemented using the quantum gate system shown in Figure \ref{fig:1__}.

{\it Instructions. Consider only the actions of gates on the basic states. }

\subsection{The concept of quantum cryptography
}

The practical application of quantum one-qubit gates is provided by quantum cryptography. We will show its advantage over classical cryptography using the example of the quantum cryptographic protocol BB84, the first of a large series of similar protocols.

\begin{figure}
\centering
\includegraphics[scale=0.6]{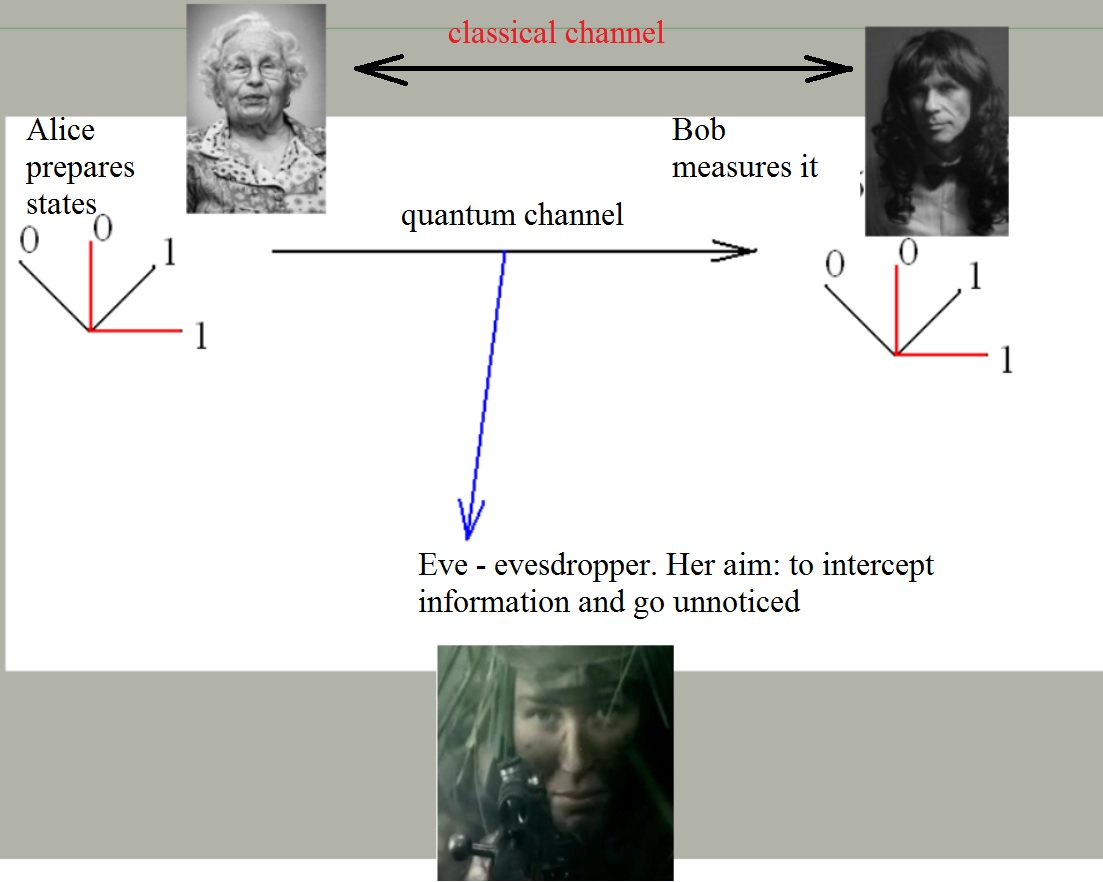}
 \caption{Cryptography scheme: quantum key distribution between Alice and Bob}
\label{fig:crypto}
\end{figure}

We will give a short introduction to the problem of cryptography - see figure \ref{fig:crypto}. It consists in providing a secure connection between the transmitting subject (Alice) and the receiving one (Bob). If Alice sends Bob a binary string $e_1,e_2,...$, then Eve (the eavesdropper) can intercept it on the communication line, copy it and forward it to Bob without changes, thus finding out the information without revealing herself. To prevent such a scenario, Alice and Bob must have a common key - the binary string $k_1, k_2,...$. Alice encodes her message by sending Bob not $e_1,e_2,...$, but $e_1\oplus k_1,e_2\oplus k_2,...$ and Eve, without knowing the key, will not be able to decrypt anything. Bob can easily do this by adding $k_i$ to the received message $e_i\oplus k_i$, and getting the original $e_i$. The task of cryptography, therefore, is reduced to the distribution of the secret key $\bar k$.

No classical method can ensure the safe distribution of the key for the following reason: any classical message can be copied. Copying is a mapping of the form

\begin{equation}
\label{noclon}
U_{clon}:
|\Psi\rangle|0\rangle \rightarrow |\Psi\rangle|\Psi\rangle
\end{equation}
and in the case of the classical - base state, this is done by simply using the $CNOT$ operator. However, if $| \Psi\rangle$ is a superposition of classical states, this technique will not work. Indeed, suppose that there exists such a unitary operator $U$ that the equality \eqref{noclon} is satisfied for any quantum state $|\Psi\rangle$. Let's take as $|\Psi\rangle$ the state $\frac{1}{\sqrt 2}(|0\rangle+|1\rangle)$. Then we have:

$$
\begin{array}{ll}
&U|\Psi\rangle|0\rangle=\frac{1}{2}(|00\rangle+01\rangle+|10\rangle+|11\rangle),\\
&U|\Psi\rangle|0\rangle= \frac{1}{2}U||00\rangle+U|10\rangle=\frac{1}{2}(|00\rangle+|11\rangle)
\end{array}
$$
that is contradiction. 

So, copying quantum states is impossible, and precisely because we can use the superposition of the basic states. It can be practically used like this.

Let $A=\frac{1}{\sqrt 2}(\s_x+\s_z)$ be the Hadamard gate. 
Alice, in addition to the main "blank" key $k_1, k_2,...$, also creates a random sequence of binary characters $bas_1, bas_2,...$, and encodes any $k_i$ in the form of $k_i$ if $bas_i=0$ and in the form of $A|k_i$ if $bas_i=1$. So she sends the encoded string $bas(k_1),bas(k_2),...$ to Bob. If Bob knew the sequence $bas_1,bas_2,...$, he would quickly decipher Alice's message, but he does not know this sequence. Then Alice acts in an unexpected way - she sends Bob a second message - this sequence $bas_1, bas_2, ... $ - over an open classical communication channel, which Eve listens to, but cannot distort! Bob, of course, immediately restores the ''blank" key
 $k_1,k_2,...$. 

If Eve is in the channel, she must distort the states of $bas(k_i))$ by her intervention, otherwise she will not know anything, that is, in fact, she is not in the channel. How do I find out if Eva is in the channel? Very simple. Alice sends Bob the values of $ k_i$ from a randomly selected sequence $i=i_1, i_2,...$ and Bob, having received this message, and comparing it with his decryption of $k_i$, determines the presence of Eve by a mismatch. Thus, the presence of Eve in the communication channel can be reliably detected, and if it is not present, it can be transmitted safely. This is the idea of the BB84 protocol. Then the so-called increased secrecy is applied, when this procedure is done not on the entire billet $k_i$ but only on its part, etc. We will not go into details here.

Quantum cryptography has now become a huge field, many quantum cryptographic protocols are known, and all of them are based on the prohibition of quantum cloning of arbitrary states. We see that even without the use of entanglement, the quantum mechanics of a single qubit gives us a tangible advantage over classical methods in the field of information protection.

Let's pay attention to the fact that the concept of "someone knows" is significantly used in this area. This is not a mathematical concept, but rather a humanitarian one, and its use in mathematics cannot be considered correct. Therefore, to ensure the secrecy of quantum cryptographic protocols, it is necessary to formalize the concept of "someone knows something".

The von Neumann quantum entropy for the mixed state $ \rho $ is defined as

$$
S(\rho)=-tr(\rho\ log(\rho)).
$$
We define the concept of quantum relative entropy for a pair of density matrices $ \rho, \s $ as

$$
S(\rho\|\s)=tr(\rho(log(\rho) - log(\s))
$$. 

Mutual information between two entities $ A $ and $ B $ can now be defined as
$$
I(A:B)=S(\rho_A)+S(\rho_B)-S(\rho_{AB})
$$
where $ A $ and $ B $ are two parts of the same $ AB $ system, and the subscript denotes the relative density matrix.

It can be shown that  
$$
I(A:B)=S(\rho_{AB}\|\rho_A\otimes\rho_B).
$$
This is the standard definition of the concept of the degree of" knowledge " of the subject $A$ about the subject $B$. Thus, it completely fits into the mathematical formalism of quantum theory. A possible attack on a quantum cryptographic channel may consist either in using noise as a cover, or using restrictions on the quantum description of states as such. The first method works only if the noise level in the channel significantly exceeds 12\% . The second way assumes the existence of restrictions on the quantum formalism as such, which is impossible in simple systems: these restrictions occur only in complex systems. 
 Thus, quantum cryptography provides a fundamentally higher reliability of key distribution than classical cryptography, and is used for transmitting data of special importance.

\subsection{Quantum teleportation}

The anti-cloning theorem plays an important role in quantum cryptography, making it an absolutely reliable way of transmitting information. However, it is possible to move a quantum state over a distance without moving its carrier. To do this, you need a kind of information channel formed by the EPR pair $|epr\rangle=\frac{1}{\sqrt 2}(|00\rangle+|11\rangle)$. This protocol is called quantum teleportation.

\begin{figure}
\centering
\includegraphics[scale=0.60]{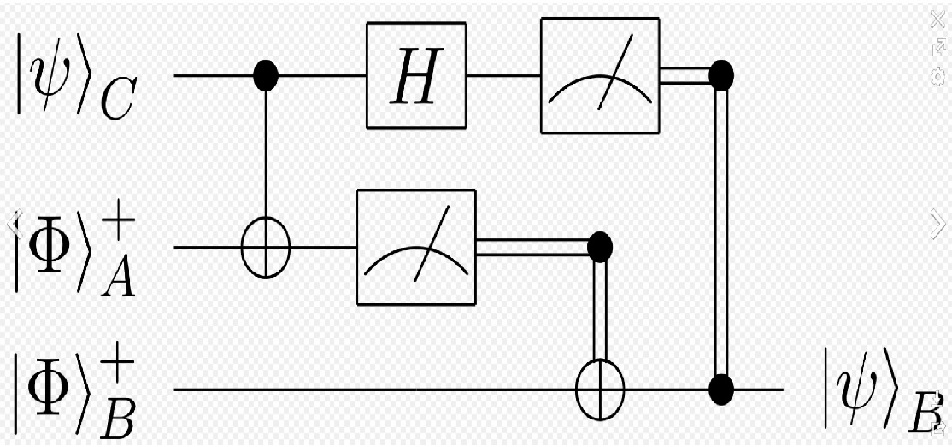}
\caption{Teleportitionn scheme}
\label{fig:tele1}
\end{figure}

It consists of the following. Alice and Bob have an EPR pair, whose qubits we denote by the indices $A$ and $B$, respectively. Alice also has another additional qubit, which we denote by $C$ in an unknown state $\la|0\rangle+\mu|1\rangle$, the state of which Alice wants to transfer to Bob. To this end, Alice performs the $CNOT$ operator on the qubits $C$ and $A$, then she performs the Hadamard transformation on $C$:

$$
|0\rangle\rightarrow \frac{1}{\sqrt 2}(|0\rangle+|1\rangle),\ |1\rangle\rightarrow \frac{1}{\sqrt 2}(|0\rangle-|1\rangle)
$$
after that, she measures both of its qubits and sends the measurement result - the classical state of two qubits - to Bob. Bob is able to restore an unknown state $|\Psi\rangle$in his qubit based on the information received. The teleportation scheme is shown in the figure  \ref{fig:tele1}.

\begin{figure}
\centering
\includegraphics[scale=0.70]{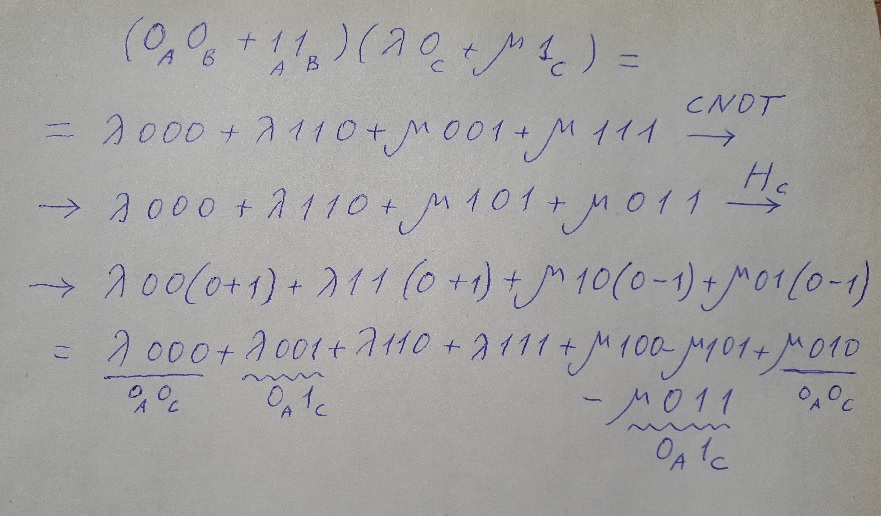}
\caption{Calculating the teleportation result
}
\label{fig:tele2}
\end{figure}

The correctness of the operation of such a scheme is checked by the calculation shown in the figure \ref{fig:tele2}. Here the order of the qubits is accepted: $A, B, C$, and the normalization coefficients are omitted.

\begin{equation}
\label{tele}
\begin{array}{ll}
&\ (|0_A0_B\rangle+|1_A1_B\rangle)(\la|0_C\rangle+\mu|1_C\rangle)=\\
&\la|000\rangle+\la|110\rangle+\mu|001\rangle+\mu|111\rangle\rightarrow\\
&\la|000\rangle+\la|001\rangle+\mu|101\rangle+\mu|011\rangle\rightarrow\\
&\la|00\rangle(|0\rangle+|1\rangle)+\la|11\rangle(|0\rangle+|1\rangle)+\mu|10\rangle(|0\rangle-|1\rangle)+\mu|01\rangle(|0\rangle-|1\rangle)=\\
&\la|000\rangle+\la|001\rangle+\la|110\rangle+\la|111\rangle+\mu|100\rangle-\mu|101\rangle+\mu|010\rangle-\mu|011\rangle
\end{array}
\end{equation}

\begin{figure}[h]
\centering
\includegraphics[scale=0.7]{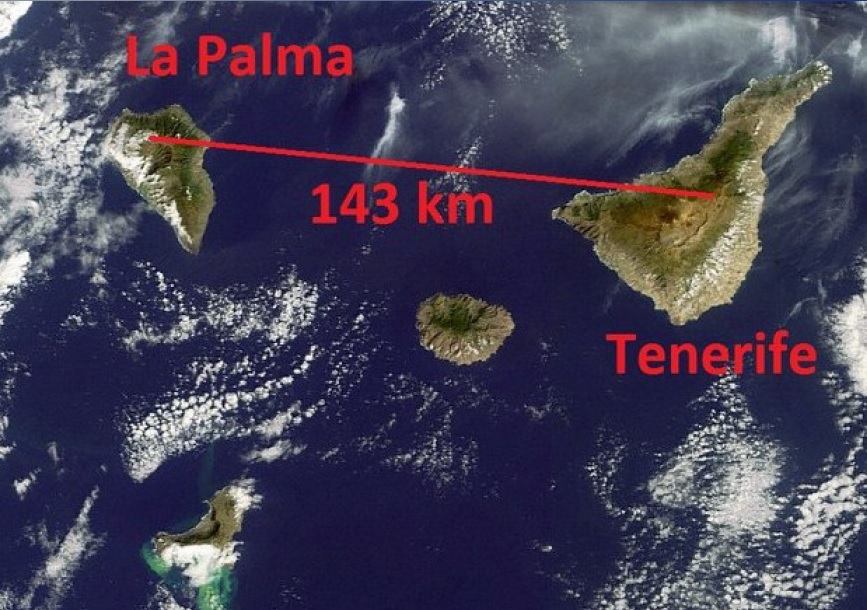} 
\caption{Quantum teleportation between two Canary Islands  
}
\label{fig:canar} 
\end{figure}

The experimental quantum teleportation has been realized between two anary Islands (see figure \ref{fig:canar}).

\newpage

\section{Lecture 4. Grover search algorithm
}

For the practical construction of quantum algorithms, we must now specify the form of the function ${\cal F}$, which we previously called an algorithm. {\it The quantum algorithm} is a drawing consisting of $n$ parallel wires located on top of each other, with highlighted beginnings and ends. These wires are connected by perpendicular jumpers, each of which corresponds to a specific quantum gate. An example of the algorithm is shown in the figure \ref{fig:1}. 

{\it Quantum computation} corresponding to a given algorithm is a sequence of the form

\begin{equation}
{\cal C}_0\ar{\cal F}_0({\cal C}_0)\ar{\cal F}_1({\cal F}_0({\cal C}_0))\ar \ldots\ar {\cal F}_{T-1}(\ldots {\cal F}_0  ({\cal C}_0)\ldots ),
\label{iterated_}
\end{equation}

consisting of the results of successive applications of the algorithm gates from left to right, where the initial state $|{\cal C}_0\rangle$ of the memory is placed in the beginnings of the wires along the qubits, from bottom to top. Thus, the wires actually set the direction of the algorithm's running time. The final state is obtained as the state of the final vertices of the wires. After the end of the work, it can be measured, then the result of the algorithm will be a binary string, or not measured - in this case, the algorithm can be used as a subroutine, embedding it in other algorithms. Often the drawings of the algorithms are similar, which allows you to parameterize a set of algorithms using the number $n$ - the number of wires that matches the amount of RAM, and call such a set a single algorithm. In this case, the complexity is determined as above.

Instead of depicting the algorithm with a drawing, you can solve it with the words: "first we do this and that with the first and second qubits, then we do this and that with the third, etc. "Note that some of the gates can be an oracle-a gate with many variables that implements some fixed unitary operator, so we have actually defined quantum computing with an oracle. Its complexity is determined verbatim in the same way as above.

We will analyze only one fast quantum algorithm found by Lov Grover in 1996 (see \cite{Gr}) - the GSA algorithm (Grover search algorithm). This algorithm contains a minimal number of details, and therefore it can most clearly show the most important property of quantum dynamics - the ability to concentrate the amplitude on individual states, and those that are not known in advance. The speed of such concentration is extremely high, so that this process cannot be reproduced on a classic computer.

GSA is a fundamental quantum algorithm. It can serve as a model of complex processes at the quantum level, which will be discussed in more detail. The transformations of the amplitude of quantum states when calculated using this algorithm will also be considered there. Here we will describe the GSA from the "external" side, in terms of the Hilbert formalism. This description is short and beautiful, and therefore we will start with it.

Let be a Boolean function $f$ of $n$ variables, and the equation

\begin{equation}
f(x)=1
\label{search}
\end{equation}
it has exactly one root $x_{tar}$, which we need to find by referring to the function $f$ the least number of times. If we had a classical computer, the number of such calls would be at least $N=2^n$ in order of magnitude, since this is a classic iterative problem in which there is no better way to find the answer than by directly iterating through all possible options - all Boolean $n$ - ok. This is obvious if $f$ is given to us in the form of a "black box"; if we have an explicit scheme of functional elements that calculates $f$, the need for iteration is not strictly proven, just no faster method for finding a solution \eqref{search} has yet been found.

On a quantum computer, you can find $x_{tar}$ for $[\pi\sqrt{N}/4]$ calls to the function $f$. If we have a classical device that calculates $f (x)$ for any $x\in\{ 0,1\}^n$, we can make a quantum algorithm from it that computes a function of the form

\begin{equation}
\label{quantor}
f_{quant}:\ |x,y\rangle\rightarrow |x,f(x)\oplus y\rangle
\end{equation}
We will demonstrate the idea of such a construction using the example of the simplest identical function $I:\ |x\rangle\rightarrow |x\rangle$. Then $I_{quant}$ is called $CNOT$ and acts as $CNOT|x, y\rangle=|x, x\oplus y\rangle$. It can be shown independently for various quantum computer technologies that such a unitary operator can theoretically be implemented in any of the technologies; for details, the listener can refer to the archive of preprints.

The reflection of the space of quantum states along the vector $|a \rangle$ is called a mirror reflection with respect to a subspace orthogonal to
 $|a\rangle$: 
\begin{equation}
\label{refl}
I_a|b\rangle=\left\{ \begin{array}{lll}&|b\rangle,\ \ \text{если} \ &\langle a|b\rangle=0,\\
-&|a\rangle,\ \ \text{если} \ &|a\rangle=|b\rangle\end{array}
\right.
\end{equation}
The map defined in this way continues linearly over the entire space; we will denote this continuation with the same symbol $I_a$.

\begin{figure}
\centering
\caption{Reflection of the space along the vector
 $|a\rangle$.}
\includegraphics[width=0.5\textwidth]{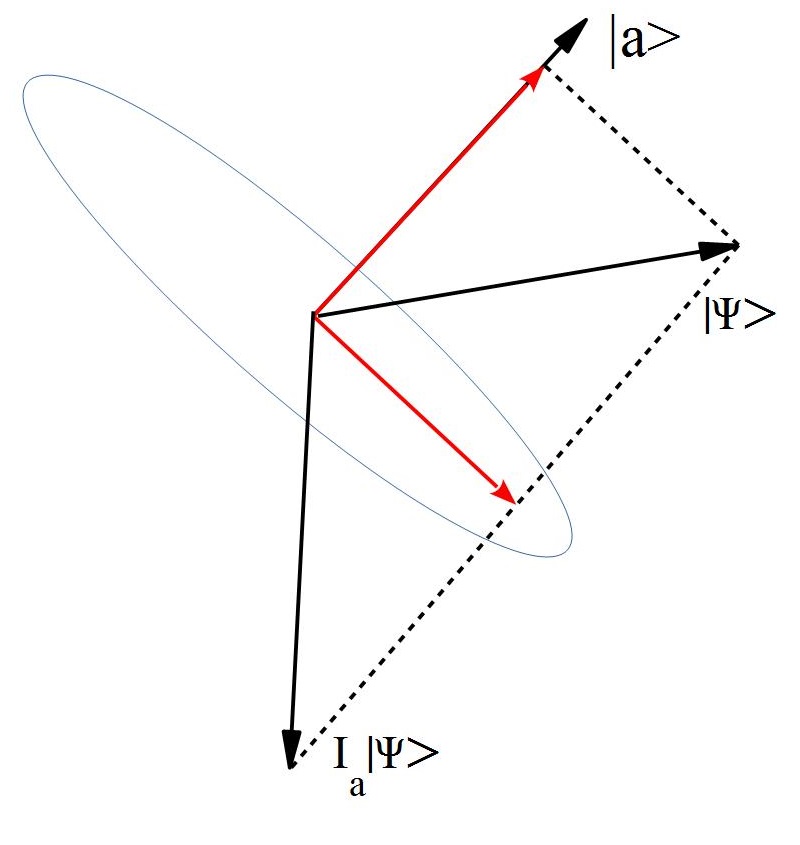} 
\label{fig:reflection}
\end{figure}

The reflection is graphically depicted in the figure  \ref{fig:reflection}.

Having the operator $f_{quant}$, which acts on all linear combinations of basis states, and not only on one basis state, as in the classical case, we can construct the reflection operator $I_{x_{tar}}$ along the vector $|x_{tar}\rangle$, although this vector itself is unknown to us. To do this, we introduce an ancilla (auxiliary qubit), initializing it with the state $|anc\rangle=\frac{1}{\sqrt 2} (|0\rangle-|1\rangle)$, and apply the operator $f_{quant}$ to the state of the form $|\Psi\rangle|anc\rangle$. It follows from the definitions that the state $I_{x_{tar}}|\Psi\rangle|anc\rangle$ will be obtained and the ancilla can be thrown out without fear of spoiling the current state, since the ancilla, which played its role in introducing a minus at the base state $|x_{tar}\rangle$ in the superposition $|\Psi\rangle$, is again not entangled with the main array of qubits.

An important remark should be made here. If we initialized the ancilla with the state $|0\rangle$, and then performed the transformation $f_{quant}$, in order to then change the sign at $x_{tar}$ with the operator $\sigma_z$ applied to the ancilla (which would be natural in a classical computer), this would create, generally speaking, an entangled state between the main array of qubits and the ancilla, and it would be impossible to simply throw out the ancilla: its measurement would lead to irreversible damage to the ground state, and we would not get would be $I_{x_{tar}}|\Psi\rangle$ as a result. Here it would be necessary to apply $f_{quant}$ again, so that the ancilla would again switch to a separate state $|0\rangle$, that is, for one inversion along $|x_{tar}\rangle$, we would spend two calls to the function $f_{quant}$ instead of one with non-trivial initialization of the ancilla; with such initialization, the change of the desired sign in the linear combination at the input occurs with simultaneous cleaning of the ancilla.

Let's construct a state of the form $|\tilde 0\rangle=\frac{1}{\sqrt N}\sum\limits_{j=0}^{N-1}|j\rangle$ - this can be done by performing the Hadamard transformation

\begin{equation}
\label{Hadam}
H=\left(\begin{array}{lll}&1/\sqrt{2}&1/\sqrt{2}\\
&1/\sqrt{2}-&1/\sqrt{2}\end{array}\right)
\end{equation}
on each qubit in the state of the main array $n$ is a qubit $| \bar 0\rangle=|00...0\rangle$, where all qubits have the value $|0\rangle$ (prove it!). Otherwise, such an operator can be written as the tensor $ n $ - th degree of the operator $H$; it is also called the Walsh-Hadamard operator:
 $WH=H^{\otimes n}$.

{\it The listener can try (this is optional) find out the general form of the matrix element of the operator $WH$: $w_{i, j}$. Note: it is necessary to use the qubit representation of the natural numbers $i$ and $j$.
}

Next, we have already seen how to implement the Toffoli gate $T:\ |x, y, z\rangle\rightarrow |x, y, xy\oplus z\rangle$ on any quantum computer technology on which $CNOT$ can be implemented (it can be shown that $T$ is expressed in terms of $CNOT$ and single-qubit gates).

Let's show how to perform the transformation
 $I_{\bar 0}$.

\begin{figure}
\centering
\caption{Sequential application of $R$ - each application with the simultaneous shift of arrows}
\includegraphics[width=0.8\textwidth]{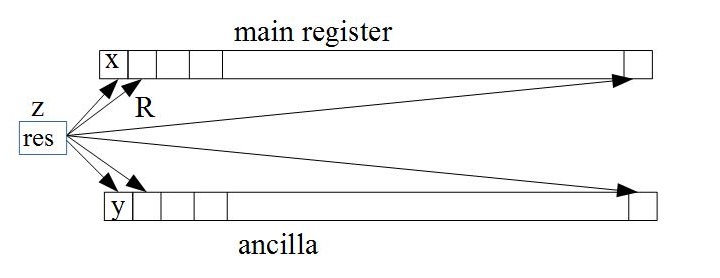} 
\label{fig:R}
\end{figure}

\begin{figure}
\centering
\caption{A scheme that implements the $R$ operator. Its main property is: mapping
 $|000\rangle\rightarrow|000\rangle, |100\rangle\rightarrow|111\rangle,|101\rangle\rightarrow|101\rangle,|001\rangle\rightarrow|011\rangle$.}
\includegraphics[width=0.6\textwidth]{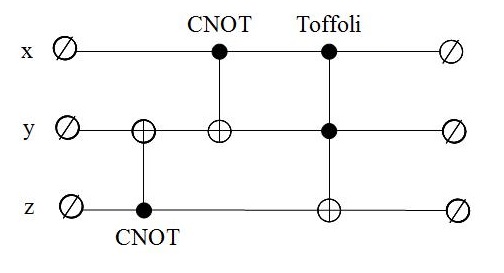} 
\label{fig:gatesG}
\end{figure}

Consider the operator $R$, which is implemented by an array of gates, shown in the figure \ref{fig:gatesG}. Let's create an additional$ n $ ancilli initialized with zeros, and number them with natural numbers $1,2,..., n$. and another additional qubit, which we will call the result $res$. We will perform the transformation of $R$ sequentially over $x,y, z$, which are: $i$ - th qubit of the main array,$i$ - th qubit of the ancilla and $res$, respectively (see figure \ref{fig:R}). Then we will commit $ - \sigma_z(res)$. In a qubit, $res$ will be 1 if and only if $x\neq is 1$. For mandatory cleaning of the ancilla, we will perform all the described transformations in reverse order.

Now we can also realize $I_{\tilde 0}$, noting that
$I_{\tilde 0}=H^{\otimes n}\ I_{\bar 0}\ H^{\otimes n}$. 

After that, we will make consecutive applications of the operator $G=-I_{\tilde 0}I_{x_{tar}}$, starting with $|\tilde 0\rangle$ $[\pi\sqrt{N}/4]$ times. We will show that the result will coincide with $|x_{tar}\rangle$ with high accuracy. Indeed, the entire evolution of the state vector of the $n$ - qubit system will occur in the real linear shell of two almost orthogonal vectors $| \tilde 0\rangle$ and $|x_{tar}\rangle$, and $G$ will invert the orientation of this two-dimensional real plane twice. So, $G$ is its rotation by a certain angle $ \beta$, which can be found by following a single point, for example, the end of the vector $|\tilde 0\rangle$. It is easy to show (do it!) that $ \beta=2\ arcsin(1/\sqrt{N})$. Now from the high-precision equality $\alpha\approx\arcsin(\alpha)$ follows the desired equality
$$
|x_{tar}\rangle\approx G^\tau |\tilde 0\rangle,
$$ 
which is exactly what was required.

\begin{figure}
\centering
\caption{The work of the GSA is consecutive turns at an angle approximately equal to
 $2\ arcsin(\sqrt{l/N}$.}
\includegraphics[width=0.5\textwidth]{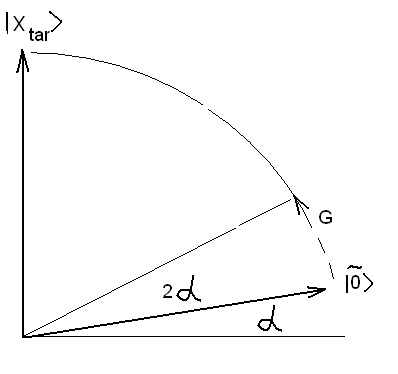} 
\label{fig:GSAwork}
\end{figure}

The work of the GSA is shown in the figure
 \ref{fig:GSAwork}

So, Grover's quantum algorithm requires about $ \sqrt{N}$ calls to the oracle, that is, it accelerates the calculation of an unknown solution \eqref{search} at a level inaccessible to any classical computer. It can be shown (see \cite{Oz2}) that this algorithm is optimal in the following exact sense. Any other algorithm that works significantly faster will give an incorrect answer for the iterative task \eqref{search} for the vast majority of functions (See also \cite{BBBV},\ \cite{Za2}).

\bigskip

\begin{figure}[h]
\centering
\caption{It is sufficient to get to the good area on the circle}
\includegraphics[width=0.7\textwidth]{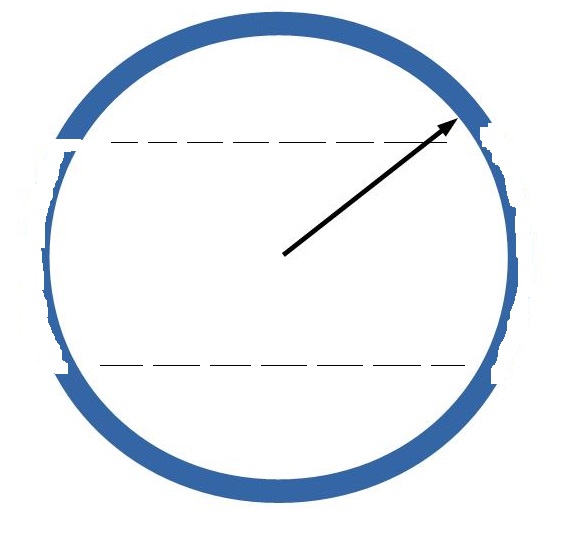} 
\label{fig:circle}
\end{figure}

{\it If the equation \eqref{search} has several solutions: $x_1,x_2,...,x_l$, then exactly repeating the GSA scheme, only taking $\tau=[4\pi\sqrt{N/l}]$, we will get a good approximation of the state $|X_{tar}\rangle=\frac{1}{\sqrt l}\sum\limits_{j=1}^l |x_j\rangle$, after which the measurement will allow us to find one of the $x_j$. Check this fact, making sure that all the arguments are preserved, only $x_{tar}$ should be replaced with $X_{tar}$ with the appropriate time correction
$\tau$.

If $l$ is unknown to us (practically an important case), we can iterate the GSA scheme by performing $ \tau_s$ GSA operations for $\tau_s = 2^s,$ sequentially, for $s=1,2,...$ (see figure \ref{fig:circle}). Show that the number of steps of such an iterative application of GSA will have the order of $O (\sqrt{N/l})$, that is, the root of classical time. This is the maximum possible quantum acceleration for most classical algorithms with an unlimited calculation length - we will show this below; if we consider short classical algorithms, in most cases they cannot be accelerated even by one step on a quantum computer (see \cite{Oz3}).}

\subsection{A continuous version of Grover algorithm
}

Grover algorithm has a continuous version, when the final state $|x_{tar}\rangle=|w\rangle$ is obtained as a unitary evolution of the state vector in a two-dimensional subspace generated by the vectors $|w\rangle$ and $| \tilde 0\rangle$ under the action of the Hamiltonian

$$
H+\frac{1}{\sqrt 2}(|\tilde 0\rangle\langle\tilde 0|+|w\rangle\langle w|)
$$

where $|\tilde 0\rangle=\frac{1}{N}\sum\limits_{j=0}^{N-1}|j\rangle$. We have: $\langle\tilde 0|w\rangle=\a=\frac{1}{\sqrt N}$. The matrix of the Hamiltonian $H$ in the standard basis has the form

$$
|w\rangle\langle w|=
\begin{pmatrix}
&0 &0\\
&0 &1
\end{pmatrix}, |\tilde 0\rangle\langle\tilde 0|=
\begin{pmatrix}
&cos\ \a\\
&-sin\ \a
\end{pmatrix}
\begin{pmatrix}&cos\ \a\ &-sin\ \a
\end{pmatrix}
=\begin{pmatrix}
&cos^2\ \a &-cos\ \a\ sin\ \a\\
&-cos\ \a\ sin\ \a &sin^2\ \a
\end{pmatrix}.
$$

Finding the eigenvalues of the energy $H$, we get $E_{1,2}=\frac{1}{\sqrt 2}\pm sin (\a /2)$, so that the difference between the main and excited levels is $2\ sin (\a /2)$, which is equal to $\a$with great accuracy. From here, solving the Schrodinger equation for $H$, we get that in the time of the order of $\sqrt N$ from the state $|\tilde 0\rangle$, evolution will lead us to the state $|w\rangle$, that is, the continuous version of the Grover algorithm gives the same spedup of the classical calculation as the standard version.

\subsection{Quantum spedup of classical computations and its limits
}

\begin{figure}
\centering
\caption{The quantum spedup of the classical computation of $x_{tar}$ is obtained if the quantum search time is less than the classical one: $t_{quant}<t_{class}$. }
\includegraphics[width=0.5\textwidth]{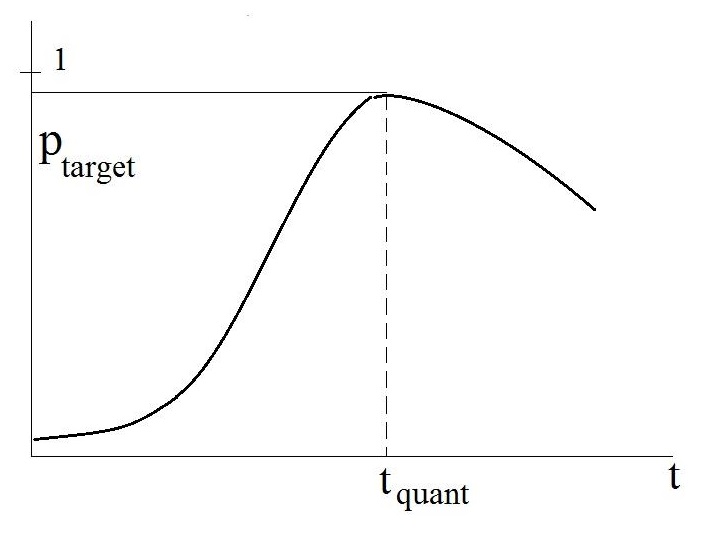} 
\label{fig:qspeedup}
\end{figure}

Quantum spedup of classical computatins is obtained if the search time for solving a problem on a quantum computer is less than the same time on any of the classical algorithms (see figure
 \ref{fig:qspeedup}). 

In short, the quantum spedup of computations consists of this. Let's say we have some abstract quantum computer in which the evolution of $exp(-\frac{i}{h}Ht)$ is implemented under our control over the Hamiltonian $ H $. Can we use such a device to predict the future of an arbitrary classical system? And if so, how fast? This question, in essence, reduces the most important problem of verifying the fact that someone built a quantum computer, and did not fake its work with the help of a supercomputer hidden in the basement.

This is what is called the quantum acceleration of classical computing. We will show that if we understand by a classical system a specific function $f:\ {\cal K}\ar{\cal K}$ from the configuration space to itself (the law of classical evolution), then the answer to this question will depend on the time interval $t$ at which we consider the prediction. If we do not impose any restriction on $t$, then the quantum time will have an order of not less than the square root of the classical one, that is, the quantum spedup for most classical problems will not exceed Grover's - for brute force.

If we demand that the time $t$ be sufficiently small (compared to the number of all possible computer configurations), we will get a completely surprising fact: a quantum computer will spend the same time on modeling as classical evolution itself (see  \cite{Oz3}). 

\nnn
{\bf Quantum Achilles may not catch up with the classic turtle
!}
\nnn

We will show how the lower bound is set to the square root of classical time for the quantum complexity of finding the result of iterations of the classical oracle. Note that such results show the fundamental limits of the speed of the "quantum Achilles"; they cannot be overcome by any improvement of its structure.

So, the classical evolution is represented as an iteration of some function $f$, so it has the form

$$
x_0\ar f(x_0)\ar f(f(x_0))\ar\ldots\ar f^k(x_0)\ar\ldots\ar f^T(x_0),
$$
where in $f^k (x_0)$ is denoted by $k$ - a multiple iteration of $f$. The value of $x_0$ does not play any role in this case, so we just write $f^k$.

\begin{figure}
\centering
\caption{Quantum Achilles and the classical Turtle
}
\includegraphics[width=0.6\textwidth]{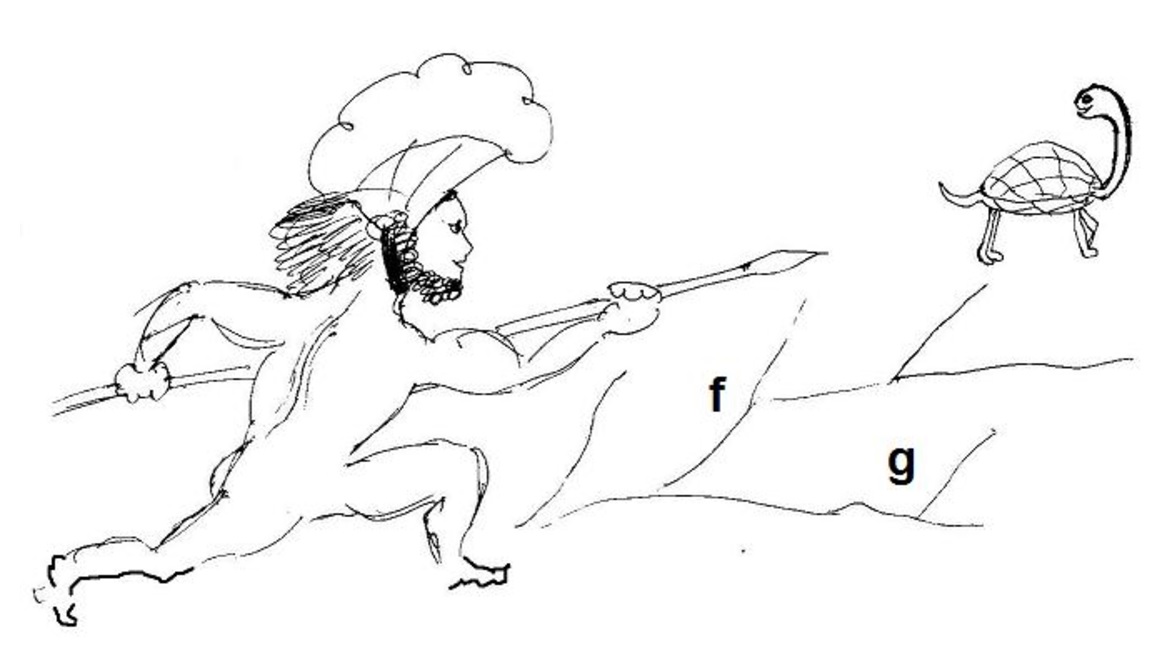} 
\label{fig:ahill}
\end{figure}

A quantum computer, our Achilles - see figure \ref{fig:ahill}, has the function $f$ at its disposal, and can use it as a quantum oracle $Qu_f:|x, y \rangle\ar |x, y\oplus f(x)\rangle$. All the words $f^k$ belong to the basis states of the quantum Hilbert space, so any of these words can be substituted for $x$ or $y$. So, we can assume that $f^k$ belong to the basic states of our computer. Then, after properly grouping several consecutive operations in the calculation, we can assume that each such state $e$ calls the oracle $f$ on some word $q (e)$ from the same set (grouping is necessary so that there is exactly one query state in each group).

We can group elementary operations as we like; we only need the unitarity of all quantum transitions, which is used in further inequalities. Then the probability that the quantum state

$$
|\Psi\rangle = \sum\limits_j\la_j|j\rangle
$$
our Achilles will be called by the oracle $f$ on the word $a$, it will be calculated by the formula
$$
\delta_a(\Psi )=\sum\limits_{j:\ q(j)=a}|\la_j|^2,
$$
resulting from the Born rule. Let's put
$d_a(\Psi )=\sqrt{\delta_a(\Psi )}$. 

The whole speed of Achilles is in this parallelism! He can catch up with the turtle - a classic computation - only due to the fact that the oracle asks for all the words at once, and not just one, like Turtle. But let's see what he can do?

\begin{figure}
\centering
\caption{Computations with two oracles
}
\includegraphics[width=0.6\textwidth]{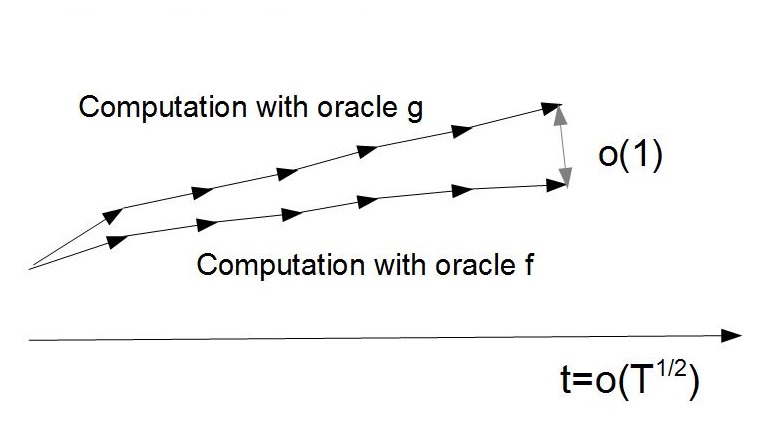} 
\label{fig:lower}
\end{figure}

How to determine the difference between the two strategies of the classical turtle: the functions $f$ and $g$, which determine the classical dynamics? The most natural thing is to generalize the definition of $d_a (\Psi )$, and define the distance between the turtle strategies as

$$
d_\Psi(f,g)=[\sum\limits_{a:\ f(a)\neq g(a)}\delta_a(\Psi )]^{1/2}.
$$
It immediately follows from this definition that

\begin{equation}
\| Qu_f(\Psi )-Qu_g(\Psi )\|\leq 2d_\Psi (f,g).
\label{Ah}
\end{equation}
In this assessment, there is a weak side of Achilles. The oracle query operator $Qu_f$ is unitary, and this relates the quantum velocity. Since we are analyzing the capabilities of a quantum computer in relation to all classical ones, our turtle can use, so to speak, a deceptive move. What happens if you change the value of $f$ on only one word? It is clear that this, in most cases, will also change the value of the final state of $f^T$. But will our Achilles be able to catch the substitution? If his quantum states differ a little when working with these two oracles, he will not be able to distinguish them when measuring, and will be deceived! The illustration is shown in the figure \ref{fig:lower}.

Let's consider two ways of Achilles: with the oracle $f$ and with the oracle $g$, respectively:

\begin{equation}
\begin{array}{ll}
&\Psi_0\ar\Psi_1\ar\ldots\ar\Psi_t,\\
\Psi '_0=&\Psi_0\ar\Psi '_1\ar\ldots\ar\Psi '_t,
\end{array}
\label{Ahi}
\end{equation}

Let $f$ act the same everywhere, except for one word $a$, on which $f(a)\neq g (a)$.
From the inequality (\ref{Ah}), a simple induction on $t$ directly establishes that

\begin{equation}
\| \Psi_t-\Psi '_t\|\leq 2\sum\limits_{i=0}^{t-1}d_a(\Psi_i).
\label{Ahil}
\end{equation}

Now we need to choose $a$ so as to give Achilles the maximum inconvenience. Let's define the matrix

$$
a_{ij}=\delta_{f^j}(\Psi_i),\ i=1,2,\ldots,t;\ j=1,2,\ldots,T.
$$
Here $t$ is the Achilles time, and $T$ is the turtle time; $T/t$ is the quantum spedup coefficient. How are these times related? The maximum inconvenience for our Achilles will be delivered by this choice $a=f^\tau$, where $\tau$ is chosen so that 
$\sum\limits_{i=1}^t\leq t/T$. This is possible because the sum of matrix elements for any row is equal to 1 - this is the full probability; therefore, the sum of all elements in general will not be greater than $t$. Now we have
$$
\|\Psi_t-\Psi '_t\|\leq 2\sum_i\sqrt{a_{i\tau}}\leq 2\sqrt{t\sum_ia_{i\tau}}\leq 2t/\sqrt{T}.
$$
The second transition here follows from the inequality between norms in the spaces $l_1$ and $l_2$, the proof of which we provide to the reader as an exercise.
We see that if $t=o (\sqrt{T})$, then Achilles has lost, because he will not be able to distinguish the position of the classical turtle $f^T$ from others. That is, it is impossible to obtain more than a quadratic quantum spedup for most classical algorithms.

There are also lower bounds of a different type of quantum complexity. For example, they establish that a quantum computer cannot solve an iterative problem significantly faster than using Grover's algorithm (see \cite{BBBV}, \cite{Za2}), and also that any quantum algorithm faster than Grover's must give an erroneous answer for almost all iterative problems
 (см. \cite{Oz2}). 

A more detailed consideration of the quantum computation \cite{Oz3} shows that Achilles in most cases is not able to catch up with the turtle of a classical computer at all.

\bigskip

{\it Theorem (\cite{Oz3}).

The probability that an iteration of the length $O(N^{1/7})$ arbitrarily selected from the uniform distribution of the "black box" ${\cal F}$ can be spedup by at least one on a quantum computer tends to zero with the dimension of the space tending to infinity
.}

Thus, the quantum spedup of classical calculations is a rare phenomenon. It is the case for algorithms of the brute force type, the ones that allow spedup by parallelization (see \cite{VV}). The iteration-type problem we have considered belongs to the GMSP type (see \cite{Ruzzo}) and in general does not allow this type of acceleration; for it, a quantum computer is, in the vast majority of cases, no better than a classical one.

This is an indirect evidence that quantum and classical parallelism are close to each other. This reinforces the confidence that the attempts to find some forms of deterministic description of quantum evolutions, which we undertook in the third chapter, are not just mathematical exercises.

\section{Lecture 5. Discretization of functions and operators
}

To realize the main purpose of a quantum computer-modeling of real micro processes, we need to learn how to move from their standard, analytical description to a discrete one. In Copenhagen quantum mechanics, the state is described by the wave function $ \Psi (x,t)$, for which the Schrodinger equation $i\hbar\dot{\Psi}=H\Psi$ is valid, where $H$ is a continuous energy operator,which for a single particle in ordinary space $(x,y, z)$ has the form

\begin{equation}
\label{sh}
H=\frac{p^2}{2m}+V(x), \ p=\frac{\hbar}{i}\nabla,\ \nabla = grad = \left(\frac{\partial}{\partial x},\frac{\partial}{\partial y},\frac{\partial}{\partial z}\right)
\end{equation}
where $p$ is the momentum operator of the particle, $V$ is the potential in which it moves. 

The main stage is the representation of the so-called wave function to the state vector.

\begin{figure}
\caption{Discretization of a continuous function
.}
\includegraphics[width=1.0\textwidth]{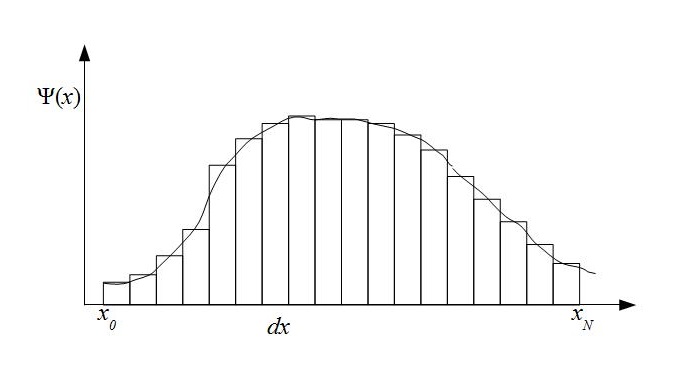} 
\label{fig:discr}
\end{figure}

If there is an abstraction - a wave function $\Psi (x)$ from a continuous variable $x$, it can be made realistic if you enter a discrete set of possible values of the variable $x=x_0,x_1=x_0+dx, x_2=x_0+2dx,..., x_{N}=x_0+Ndx$, and then represent an approximately continuous function $\Psi (x)$ how

\begin{equation}
\Psi(x)\approx \sum\limits_{j=0}^{N-1}\Psi(x_j)d_j(x),
\label{discr}
\end{equation}
where $d_j (x)$ is the characteristic function of the $j$ - th segment $[x_j, x_{j+1}],\ j=0,1,..., N-1$ (see figure \ref{fig:discr}). Given the scalar product of continuous functions $\langle f|g\rangle=\int\limits_R\bar f g\ dx$, we can pronormalize the orthogonal vectors $d_j$ by obtaining an orthonormal basis $|j\rangle=d_j/\sqrt{dx}$, and defining $\la_j=\Psi(x_j)\sqrt{dx}$, we come to the representation of our function as a state vector \eqref{state}. For a wave function defined on the space $R^2$ or $R^3$, instead of $\sqrt{dx}$, there will be $\sqrt{dx^2}$ or $\sqrt{dx^3}$, respectively.

The transition from discrete to continuous recording consists in the fact that all sums are replaced by integrals, and summation variables are replaced by integration variables. For example, the formula \eqref{discr} will turn into $\Psi(x)=\int\limits_R \Psi(y)\delta_y(x)dy$ where $\delta_y (x)$ is the limit of the functions $d_j (x)$ for $dx\rightarrow 0$, so that $x_j\rightarrow y$. Such a limit, of course, does not exist in mathematical analysis - among ordinary functions, since at $dx\rightarrow 0$, the function $d_j (x)$ will turn into a needle infinitely high and infinitely thin. This is the so-called generalized Dirac function.

This procedure of discretization will always be kept in mind by default. Moreover, performing a linear transformation of the ${\cal D}$ coordinates of $x$, which is equivalent to choosing new units of length measurement, we can assume that the segment of the definition of the wave function $[x_0, x_N]$ coincides with the segment $[0,1]$, and $N=2^n$, so that $x_j=j/N,\ j=0,1,..., N-1$ and we will write the approximate,with an accuracy of $1/N$, the value of the coordinate $x$ in the form of a sequence of binary signs of the binary expansion $j=\sum\limits_{k=1}^{n}2^{n-k}e_k$, that is, as a binary string $|e_1e_2...e_n\rangle,\ e_k\in\{ 0,1\}$. 

Any such string is the basic state of a system of $n$ qubits (quantum bits), so we will call this discrete representation of wave functions a qubit. In the qubit representation, the wave function will have no physical dimension. Only the ${\cal D}$ operator of the transition to the qubit representation from the physical continuous function $\Psi(x)$ will be dimensional, and the basis states $|j\rangle$.

{\it The listener is invited to practice using natural numbers in binary notation: enumerate, add, multiply and divide. If it is true that Nature speaks to us in the language of mathematics, then the basis of this language is precisely operations with integers in binary notation
.}

\subsection{Physical quantities as observables}

Any physical quantity, except time, corresponds to a certain observable in quantum theory. In this case, the eigenvalues of this observable will be the possible values of this value, and the eigenstate in which it has this value is the state in which this value is uniquely determined - precisely by this value.

Consider three examples: observation of the coordinate, impulse and energy.

\subsection{Observation of the coordinate} We will consider only the case of a one-dimensional particle, generalization to the three-dimensional case is not particularly difficult. The observable will be the operator of multiplying the wave function by its argument-coordinate, which in the continuous representation has the form: $x:\ f (x) \rightarrow xf (x)$. Remembering the transition \eqref{discr} from a continuous representation of the state vector to a discrete one, we can directly check that in the qubit representation the matrix of the coordinate operator is diagonal, and the arithmetic progression $0,1/N,...,(N-1)/N$ is diagonal. We will denote this matrix by $x_{discr}$. Thus, the eigenstates of this operator will be the basic states of the $n$ - qubit system, which we have agreed to denote by binary expansions of the natural numbers $|0\rangle,|1\rangle,...,|N-1\rangle$, and we will identify them with these numbers themselves when writing. Their eigenvalues will be the numbers themselves $0,1/N,..., (N-1)/N$.
These basic states, by definition, constitute the orthonormal basis of the space $C^N$ of the system of $ n $ qubits.

\begin{figure}
\centering
\includegraphics[scale=0.8]{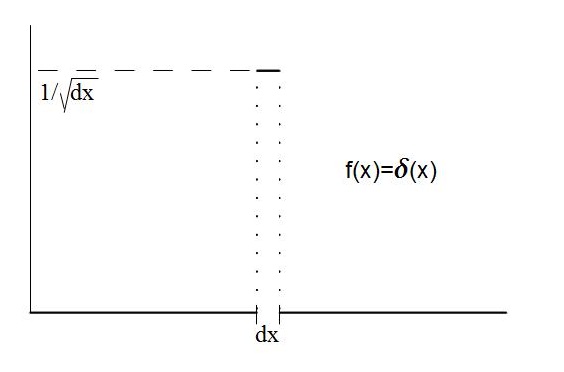} 
\caption{The eigenfunction of the coordinate operator. Discrete form}
\label{fig:dx} 
\end{figure}

In continuous form, they correspond to the so-called Dirac delta functions $\delta_{\lambda}$, which are defined as linear functionals of the form $\delta_\lambda :\ f\rightarrow f(\lambda)$. These functionals are not ordinary functions, their geometric representation is infinitely high needles growing from points $\lambda$. They cannot be normalized, they cannot be multiplied by each other. This is another example of how mathematical analysis comes into conflict with quantum physics. The contradiction arises due to the continuous nature of the variable $x$; as soon as we carry out the discretization, this contradiction will disappear, and the "needles' 'will turn into high steps $\delta_j (x)$ of a finite value $1/\sqrt{dx}$ where $dx$ is the selected grain of spatial resolution (see Figure \ref{fig:dx}).

The element of the discrete space $C^N$ will be the state vectors $|\Psi\rangle$ of the form \eqref{state}, and the conjugate space of linear functionals will consist of strings of the form $\langle\Psi|$ acting on states in a natural way: $\langle\Psi|:\ |\Phi\rangle\rightarrow \langle\Psi|\Phi\rangle$, which makes the state space $C^N$ isomorphic to the conjugate (which is violated in the case of continuous formalism).

Discretization removes all contradictions between physics and the mathematical apparatus, and therefore we will always talk about finite-dimensional spaces, even using integration and differentiation as approximate calculation techniques; we will always control such techniques for the absolutely necessary possibility of discretization.

\subsection{The quantum Fourier operator and the observation of impulse  \label{QFT_}}

The quantum impulse operator in the one-dimensional case has the form in the continuous formalism
\begin{equation}
\label{imp}
p:\ f(x)\rightarrow \frac{\hbar}{i}\nabla f
\end{equation}

Its eigenfunctions are complex exponents $exp (ipx/\hbar)$, with eigenvalues of $p$.

{\it Prove that this operator is Hermitian using the equivalent definition of the Hermitian matrix $A$: $\langle i|A|j\rangle = \bar{\langle j|A|i\rangle}$ (the dash denotes the complex conjugation); apply the formula for calculating the scalar product through the integral.}

To construct a correct discrete form of the impulse operator, you should use the Fourier transform,

\begin{equation}
\label{class_fou}
f(x) \rightarrow  \frac{1}{\sqrt{2\pi\hbar}}\int\limits_R exp(-ipx/\hbar)f(x)dx=\phi(p)
\end{equation}
by converting the \eqref{imp} functions to Dirac delta functions, as well as by the inverse Fourier 

\begin{equation}
\label{class_fou-1}
\phi(p) \rightarrow \frac{1}{\sqrt{2\pi\hbar}}\int\limits_R exp(ipx/\hbar)\phi(p)dp=f(x) 
\end{equation}
making the reverse transition.
\bigskip

{\it The listener is invited to verify this by accepting the simplifying equation $\hbar =1$, which is achieved by switching to a suitable system of physical units. Substitute the proper function of the impulse operator $f_{p_0} (x)=e^{ip_0x/\hbar}$ into the formula\eqref{class_fou}, and perform integration over a finite interval of the form $(- A, A)$. The integral is taken in a finite form, and the result for $A\rightarrow +\infty$ will more and more resemble a needle resting on the point $p=p_0$, and going indefinitely to infinity. Thus, the eigenfunction of the impulse operator will be translated into the eigenfunction of the coordinate operator; the name of the arguments $x$ or $p$ does not play any role. Do the same with the reverse transformation. But when integrating along the entire straight line, divergence will result; moreover, neither the delta function nor the complex oscillation $exp(ipx)$ can be normalized. Everything is corrected only by switching to a discrete representation.}
\bigskip

The discrete form of the Fourier transform and its inverse is represented by operators acting on the basis states of an $n$- qubit system as follows:
\begin{equation}
\begin{array}{ll}
&QFT: |c\rangle\rightarrow \frac{1}{\sqrt N}\sum\limits_{a=0}^{N-1}exp(-2\pi i ac/N)|a\rangle\\
&QFT^{-1}: |a\rangle\rightarrow \frac{1}{\sqrt N}\sum\limits_{c=0}^{N-1}exp(2\pi i ac/N)|c\rangle\end{array}
\label{Fourier}
\end{equation}

Both of these are mutually inverse operators (prove!) with a linear extension to the entire space of quantum states, $C^N$ will give unitary operators - Fourier and inverse to it.

For applications, it is convenient to assume that for the variable $a$, the number $a/\sqrt{N}$ is the coordinate belonging to the segment $[0,\sqrt{N}]$ (Planck's constant can be considered a unit in the proper system of units). Then $c/\sqrt{N}$ must be associated with the impulse. It is natural to assume that the impulse belongs to the segment $[-\sqrt{N}/2,\sqrt{N}/2]$, since a particle located on the segment $[0,\sqrt{N}]$ can move in both directions. Therefore, the impulse should be equal to $\sqrt{N}(c/N-1/2)$.

Accordingly, the discrete form of the impulse operator will be the $N$- dimensional Hermitian operator $p_{discr}=QFT^{-1}\sqrt{N}(x_{discr}-I/2)QFT=A^{-1}QFT^{-1}\sqrt{N}x_{discr}QFT\ A$, where the diagonal operator $A=diag(exp(\pi i a))_{a=0,1,...,N-1}$. Its eigenvectors will have the form $A^{-1}FT^{-1}|a\rangle$ of the form \eqref{Fourier} and their eigenvalues will be numbers $\sqrt{N}(a-1/2);\ a=0,1/N,...,(N-1)/N$.

So, in the discrete representation, all the eigenstates of the main operators are normalized by one, and there are no contradictions with mathematical analysis. Here we used the analytical technique of continuous Fourier transforms to correctly write its discrete analog. It is not difficult to show that all the useful properties of the Fourier transform: the transition from differentiation (application of the impulse operator) to multiplication by a constant, as well as the identification of the hidden period of the complex exponent will be preserved during the transition from a continuous form to a discrete one, so that we can use discrete operators in finite-dimensional spaces in all physical problems related to quantum theory.

The operators \eqref{Fourier} are called direct and inverse quantum Fourier transforms. With their help, you can build a polynomial quantum algorithm that finds the decomposition of a number into non-trivial factors (\cite{Sh}).

\subsection{Implementation of the quantum Fourier transform on a quantum computer}

Let's agree to represent an integer of the form $a=a_0 +a_0 2+\ldots +a_{l-1} 2^{l-1}$ with the base state $|a_0 \ a_1 \ \ldots\ a_{l-1} \ \rangle$ and place all $a_j$ from top to bottom. We will accept the same agreement for the output, only the binary signs $b_j$ of the number $b=b_0 +b_0 2+\ldots +b_{l-1} 2^{l-1}$ will be written in reverse order-from bottom to top.

\begin{figure}
\caption{Implementation of $QFT^{-1}$ in the form of an array of quantum gates.}
\includegraphics[width=0.9\textwidth]{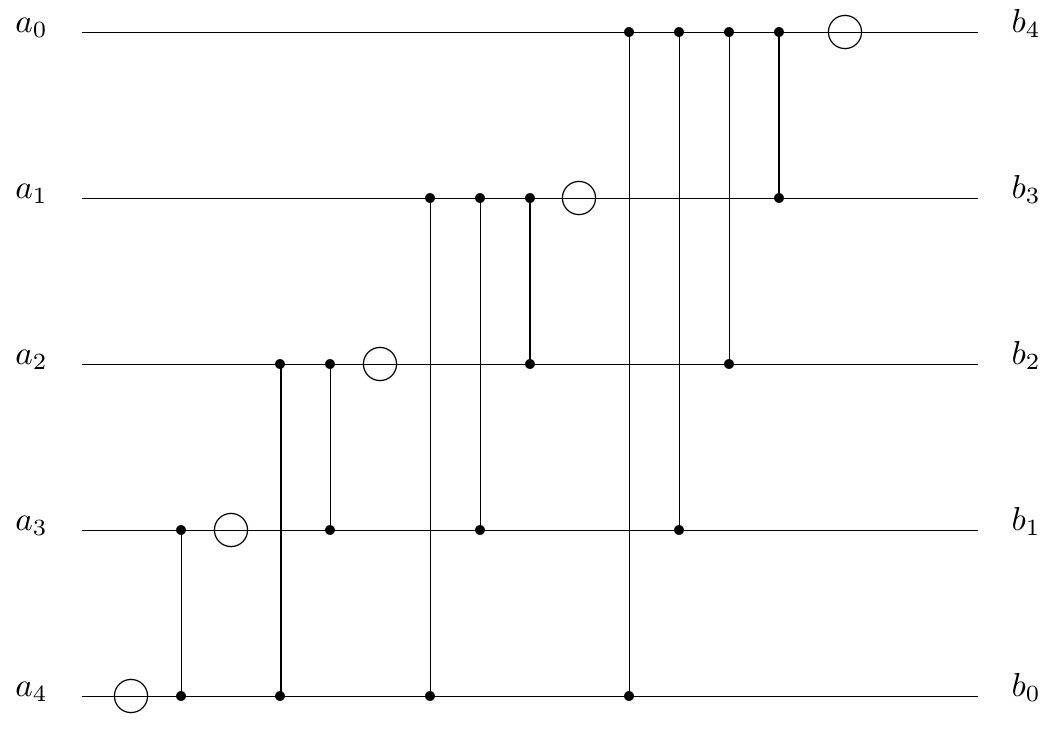} 
\label{fig:QFT-1}
\end{figure}

The circles here denote the transformation $W^1$, two-qubit operations have the form:
\begin{equation}
U_{k,j}=\left(
\begin{array}{ccccc}
&1 &0 &0 &0\\
&0 &1 &0 &0\\
&0 &0 &1 &0\\
&0 &0 &0 &e^{i\pi/2^{k-j}}
\end{array}
\right) ,
\ k>j.
\label{shift}
\end{equation}

To make sure of this, we will consider the amplitude of the transition from the base state $a$ to the base state $b$. This concept is legitimate, this is the name of the corresponding element of the matrix of the operator under consideration. Here we will have to be patient - the calculation is ideologically simple, but it requires some tedium. First, we note that the modules of all such amplitudes are the same and, as in the inverse Fourier transform, are equal to $ 1/2^{l/2}$, so we only need to monitor the phase shift, i.e., the argument $\phi$ of the complex amplitude $e^{i\phi}$. We will account for this phase incursion by summing the contributions from Walsh transformations with the contributions from two-qubit phase shifts.

To simplify the calculation, we will introduce the following abbreviated notation: $b'_j =b_{l-1-j},\ j=0,1,\ldots, l-1$ - this will be necessary in order to take into account the reverse order of the binary digits in $a$ and $b$at the right time. Let's imagine how the states change when moving from left to right along the wires of our circuit. Actually, the transition from $a$ to $b$ occurs only when performing the Hadamard operation, two-qubit operations do not change the diagonal and basic states, adding only the terms to the phase.

The contribution from the Hadamard operation will be as follows: $\pi a_j b'_j$. This number is not equal to zero only if both the $j$ digits of our input and output numbers are equal to 1, which exactly corresponds to the definition of the Hadamard transform. The contribution from a two-qubit operation for $j<k$ will be $\pi a_j b'_k /2^{k-j}$, because the state of $a$ changes to $b$ only when the Hadamard device passes, and as can be seen from Figure \ref{fig:QFT-1}, such a two-qubit operation is performed at the moment when the $j$-th qubit is still in the state of $a_j$, and $k$ - th is already in the state of $b'_k$. Summing up all these terms of the phase shift, and taking into account that an integer multiple of $\pi$ can not be taken into account at all, we get this:

\begin{equation}
\begin{array}{ll}
&\pi\sum\limits_{l>k>j\geq 0}\frac{a_j b'_k}{2^{k-j}} +
\pi\sum\limits_{l>j\geq 0} a_j b'_j=\\
&2\pi\sum\limits_{l>j+k\geq 0}\frac{a_j b_{k}2^{j+k}}{2^l}=\\
&2\pi\sum\limits_{l>j,k\geq 0}\frac{a_j b_{k}2^{j+k}}{2^l}=\\
&\frac{2\pi}{2^l}\sum\limits_{l>j\geq 0}a_j 2^j \sum\limits_{l>k\geq 0}b_k 2^j =\frac{2\pi}{2^l}.
\end{array}
\end{equation}
This is exactly what is required in the definition of the inverse Fourier transform. If we need to perform a direct transformation, it is enough to reverse the order of the functional elements in the scheme under consideration and put a minus sign before the phase shift in the definition of two-qubit operations.

Now let's look at what we just did. The scheme we have constructed that implements the Fourier transform contains about $l^2 $ functional elements. Note that if we do not chase the accuracy of this transformation, it will be possible to discard all two-qubit operations that involve qubits that are too far apart from each other. Indeed, the denominator in $\pi/2^{k-j}$ for them makes the entire fraction negligible, the exponent will be almost one, i.e. such transformations are almost singular and they can be discarded. The scheme will then be much simpler - its size will generally be linear-of the order of $C\ l$, where the constant $C$ will, of course, depend on the accuracy we have chosen.

\subsection{The Zalka-Wiesner algorithm \label{Za}}

The RSA algorithm, which we have studied, operates with qubits, converting their classical (basic) states into quantum ones using the Hadamard operator. This technique illustrates the most important features of the description of evolution at the quantum level, but with a great loss of accuracy. A real particle can occupy several classical positions, not just two, like a qubit.

We will consider the algorithm Z for modeling quantum unitary evolution proposed in \cite{Za} (see also \cite{Wie}), which actually generalizes the GSA to the case of many classical states of each particle. In it, instead of the Hadamard operator "smearing" the amplitude over two possible qubit states, the wave function of a particle capable of being in many classical spatial states is calculated at each step.

The Z algorithm differs from the direct solution of the Schrodinger equation on a classical computer only in that the amplitudes $\la_j$ of the current quantum state $|\Psi(t)\rangle$ are not calculated directly, but are modeled by the quantum dynamics of qubits in a discrete representation $|j\rangle=|0\rangle,|1\rangle,...,|N-1\rangle$ of the space of classical states in the computational memory $n$ qubit, $N=2^n$, in which the wave function is represented as $|\Psi(t)\rangle=\sum\limits_{j=0}^{N-1}\la_j|j\rangle$. 

Recall that the real one-dimensional space of classical states is first translated by a linear transformation ${\cal D}$ into a segment $[0,\sqrt{N}]$, which is then discretized by a qubit representation of numbers with an approximation accuracy of $1/N$: $x_k\approx k/N,\ k=0,1,..., N-1$. Such a representation of the wave vector requires an appropriate discretization of the operators. The discrete form of the coordinate operator $x_{descr}$ and the momentum operator $p_{discr}$ was considered in the section \ref{QFT_}.

In this case, the potential energy operator $V$ becomes a diagonal matrix 
$$
diag (V (X_0), V (X_1), V (X_2),..., V(X_{N-1})),\ X_k=\sqrt{N}x_k
$$
 with potential energy values on the main diagonal, a diagonal representation of the kinetic energy operator (in the space of its own eigenvectors of the momentum operator) also diagonally: $K_{diag}=diag(-\hbar^2 p_0^2/2m),-\hbar^2 p_1^2/2m),- \hbar^2 p_2^2/2m),...,- \hbar^2 (p_{N-1})^2/2m))$, where $p_k=\sqrt{N}(x_k-1/2)$, so that in the coordinate basis the kinetic energy is represented by the operator

\begin{equation}
K=A^{-1}QFT^{-1}\ K_{diag}\ QFT\ A,
\label{Za_}
\end{equation}
where $A=diag(exp(\pi i a))_{a=0,1,...,N-1}$.

Then the part of the evolution corresponding to the potential energy operator \newline $exp (- iVt/\hbar)$ in the simple form of the potential will be implemented as a quantum subroutine, the quantum Fourier transform can also be implemented according to the Shor scheme, that we have considered above, and the operator corresponding to the kinetic energy and time $t$ can also be implemented as a quantum subroutine. Applying the Trotter approximation

$$
exp(A+B)\approx [exp(A\ dt)\ exp(B\ dt)]^{t/dt},
$$
we will get an algorithm for computation the evolution Z in the form:

\begin{equation}
U_t=exp(-\frac{i}{\hbar}Ht)\approx [exp(-\frac{i}{\hbar}K\ dt)\ exp(-\frac{i}{\hbar}V\ dt)]^{t/dt}
\label{Z}
\end{equation}

We obtain a model of unitary dynamics with quadratic deceleration compared to the real process. {\it Prove this by using the exponential expansion to the first order by $dt$. Fix the order of the error $ \epsilon=const$ and, using the accuracy of the Taylor approximation for the exponent, set the number of operations necessary to find the approximation of the resulting state. This number will be equal to $t/dt$, which will result in a quadratic time dilation compared to the time $t$ of the real process.}

The Z algorithm can be generalized to the case of several particles. In this case, the Fourier transform must be applied for each coordinate of each particle separately. This algorithm requires memory that grows proportionally to the first power of the number of real particles, but cannot be used to control a complex system, since it assumes a priori modeling of the process with the transfer of the result to a new similar process, whereas in reality any complex process is not exactly reproducible, and therefore its control requires modeling in real time.

Comparing this calculation with the calculation using the GSA algorithm, which has the form $G^tau=(- I_ {\tilde 0}I_{x_{tar}})^\tau$, we see a complete analogy with the formula \eqref{Z}. In this case, the role of the Walsh-Hadamard operator in the representation of $I_{\tilde 0}=WH\cdot I_{\bar 0}\cdot WH$ is played by the quantum Fouret operator in \eqref{Za_}. For a single qubit, the Fourier operator just coincides with the Hadamard operator (see the implementation of the Fourier operator in the Appendix), so that the Z algorithm can be considered a generalization of the GSA for the case of many classical states of each of the particles.

So, we see that there are two methods of ultra-fast, inaccessible to a classical computer, concentration of the amplitude on the target unknown state. The first is the GSA algorithm, the second is the quantum Fourier transform. It can be shown (prove it!) that the roughest approximation of the Fourier transform is just the Walsh-Hadamard operator, which brings these two techniques together. The fast Shor integer factorization algorithm actually uses the same fundamental features of quantum dynamics as GSA. The arsenal of quantum methods for spedup classical calculations is thus limited to these general methods of ampitude concentration for problems of the search type, in accordance with the general result \cite{Oz3}. In problems that are not spedup by parallelization, the quantum computer does not show any advantages over the classical one, except only for its amazing property of non-locality.

\subsection{Revealing hidden periods with QFT}

We recollect the definnition of QFT and its conversion: 
\begin{equation}
\QFT :\ |a\rangle\ar\frac{1}{\sqrt{N}}\sum\limits_{b=0}^{N-1}e^{-\frac{2\pi i\ ab}{N}}|b\rangle ,
\ \ \ \ \ \ \ \ 
\QFT ^{-1} :\ |a\rangle\ar\frac{1}{\sqrt{N}}\sum\limits_{b=0}^{N-1}e^{\frac{2\pi i\ ab}{N}}|b\rangle .
\label{qft}
\end{equation}

We now can demonstrate how this transform can reveal hidden periods. We introduce one notion playing the important role in the quantum computing. This is the conditional application of an operator. The idea consists in the following. Let we are given a unitary operator $U$. For the obviousness, we can assume that it is given in the form of scheme of functional elements, though it is not necessary. Further, let we have some auxiliary quantum register consisting of several qubits, which we call controlling. The aim is to apply the operator $U$ sequentially so many times as is written in the controlling register that is the natural generalization of the operator ''conditional U", defined earlier. We wrote the conditional U formally in the form:
$$
U_{cond}:\  |x,\a \rangle \ar\left\{
\begin{array}{cc}
|U \ x,\a\rangle,\ &\mbox{if} \ \a =1,\\
|x,\a\rangle\ &\mbox
{if} \ \a=0.
\end{array}
\right.
$$
We will not discuss the question of the realization of this transform now. We only note that if the operator $U$ is determined as the scheme of functional elements, we can easily build the scheme of the same type realizing $U_{cond}$. For this it is sufficient to make conditional each operator contained in this scheme. We leave the details for the listener.

We then can introdue the further generalization of $U_{cond}$ - the operator $U_{seq}$:

$$
U_{seq} |x,\a \rangle=
|U^\a \ x,\a\rangle
$$
Its implementation can be fulfilled by the simple algorithm, which takes the digit figures of $\a$ and applies conditionally the sequence of $U$ corresponding to the degree of the digit - for every digit. We propose the detailes for the listeners.a

We now take up the important problem of the finding of eigen frequency of the operator $U$. 

Eigenvalues o $U$ have the form $exp(2i\pi w_k)$, where $w_k\in [0,1)$, we call $w_k$ the frequency corresponding to the $k$th eigenvalue.  This frequency will result from the measurement of some special register from $n$ qubits denoted by $\a$, in which we store the sequential binary figures of this frequency, with the limited accuracy. We suppose that the real frequency can be written in this register with the absolute accuracy, it is not important for the general scheme we describe. That is we suppose that for every $k$ $w_k$ has the form 
\begin{equation}
\label{freq}
w_k=\frac{c_k}{N}
\end{equation}
for some natural $c_k\in\{ 0,1,...,N-1\}$. The listener who is interested in the general case, we address to the book \cite{Oz5}. Our computer thus works with two registers: the register of the argument of the operator $U$, and the register of the value of its eigen frequency. The initial state we choose $|\xi ,\bar 0\rangle$, where $\xi=\sum\limits_k x_k \psi_k$, and $\psi_k$ are eigen states of our operator $U$, corresponding to the eigen frequencies $w_k$. 

The central trick for the revealing of the eigen frequencies is the operator, which was introduced by Shor for the particular case where $U$ is the numerical multiplication, and was generalized to the case of arbitrary unitary operators by Abrams and Lloyd. Its definition is the following:
\begin{equation}
Rev=\QFT_2\ U_{seq}\ QFT_2.
\end{equation}
Here Fourier transform is applied to the second register - to $\a$. We find what gives this procedure $Rev$ as applied to our initial state $|\psi\rangle|0\rangle$. The first Fourier transform gives the uniform amplitude distribution in the second register: $=\frac{1}{\sqrt{N}}\sum\limits_k \sum\limits_{\a=0}^{N-1} x_k |\psi_k ,\a\rangle$. The operator of condition application of $U$, by virtue of that $\psi_k$ are eigenvectors of $U$ gives $U_{seq}|\psi_k,\a\rangle=|U^\a \psi_k,\a\rangle= e^{2i\pi w_k \a}|\psi_k, \a\rangle$, therefore, all the state after the application of the conditional operator transforms to $\frac{1}{\sqrt{N}} \sum\limits_k \sum\limits_\a e^{2i\pi w_k \a}|\psi_k, \a\rangle$. At last, the final application of Fourier transform gives the state:
\begin{equation}
\frac{1}{N} \sum\limits_k \sum\limits_c\sum\limits_{\a=0}^{N-1} e^{2i\pi \a (w_k -\frac{c}{N})}|\psi_k, c\rangle 
\end{equation}
where $w_k=c_k/N$ due to our supposition \eqref{freq}.
If $c$ is just the list of binary figures of $c_k=w_kN$, then the exponential degree is zero and we obtain after the summing on $\a$ the sun of units of the total number $N$ so that the coefficient at the state with this $c$ will be $x_k$. It follows from this, due to the normalizing - the sum of squared modules of all $x_k$ is 1, that the amplitude of the basic vectors with the others $c$ equals zero. We can check it straightforwardly: 
$\sum\limits_{\a=0}^{N-1} e^{2i\pi \a \b} =0$ when $\b=(c_k-c)/N\neq 0$. Indeed, this is the sum of the geometrical progression with the ratio not equal 1, which sum is $(1-exp(2i\pi)/(1-exp(2i\pi(c_k-c)/N))=0$. 

Our procedure thus results in the state 
$$
\sum\limits_k |\psi_k ,c_k\rangle ,
$$
where by $c_k$ we mean its binary notation. If we thus observe this resulting state in the basis consisting of the eigenvectors of the operator $U$, we obtain as the addition to the eigenvector the binary notation of the corresponding eigen frequency. In particular, if the initial state $\xi$ was eigenvector itself, we simply obtain its frequency.


\subsection{Factoring of integers}

The general method of finding eigenvalues, which we represented in the previous section, was invented by P.Shor for the particular case arising in the problem of factoring integers. The factoring problem or the problem of decomposition of an integer number to the integer multipliers is the famous computational problem. The known classical algorithms for its solution require of the order $e^{a\ n^{1/3}}$ steps. This problem therefore belongs to the class of (supposedly) difficult problems. Here we show the quantum algorithm giving the solution of his problem. Shor algorithm was the first fast quantum algorithm solving a problem of the practical significance. The point is that the cryptography protocol RSA rests on the difficulty of the factoring integers, namely the security of this protocol depends on the complexity of factoring problem. This protocol is used in the numerous commercial applications, for example in the defense of the Windows operational system. There to overcome the highest level of defense one must be able to factorize integers with 200 decimal signs. This problem is out of the capacity even for the modern supercomputers. The quantum computer with only 1000 qubits with the frequency about 1 GHz is able to cope with this problem in a few minutes. The practical construction of the quantum computer would mean the final of the modern cryptography.

Shor algorithm has also the theoretical value. It illustrates the significance of the method how Fourier transform is applied, namely the significance of the auxiliary transformations. The point is that the main time is spent here not to Fourier transform, complex from the classical viewpoint, but to the multiplication of integers. 

We take up the factoring problem. Let we have to find the nontrivial decomposition $q=q_1 q_2$ of the known natural number $q$ to the multipliers. This task can be reduced to the problem of the finding of the minimal multiplicative period $r$ of the arbitrary natural number $y$ modulo $q$: $y^r \equiv 1\ (mod\ q)$. In a few words, this reduction looks as follows. Let we have the method of finding of $r$. We will choose $y$ randomly and find $r$. Then with the non-vanishing probability $r$ turns to be even. We then have $y^r-1=(y^{r/2} -1)(y^{r/2}+1)\equiv 1\ (mod\ q)$, and one of the multipliers is the divisor of multiple $q$ number, we thus obtain with the non-vanishing probability the factoring of the number $q$ itself. We thus have only to learn how to find $r$ quickly given $q$ and $y$. This is analogous to the finding of the unknown period when in place of the operator $U$ stands the operator of multiplication on the number $y$.

We take $n$ such that $2^{n-1}\leq q<2^n$ and will work with the quantum memory of $n$ qubits. 
We consider the following operator $U$: $U |x\rangle\ar |y x\ 
({\rm mod}\ q)\rangle$, where $y x$ is the numerical multiplication. To make this operator acting on all our basic vectors we agree that this equality defines it on the numbers less than $q$, whereas on the rest: $q,q+1,\ldots , 2^n -1$ it acts as the identical operator. It brings the little difficulty: this operator can be not unitary. If $y$ and $q$ have the common divisors, some elements will "stick" together. To exclude this trouble we assume that these numbers are mutually disjoint: $(y,q)=1$. Since we choose $y$ at random, then this is the case with the non-vanishing probability. All is ready now. We can apply the powerful technique of the quantum computing we developed earlier. Eigenvectors of $U$ have the form $\frac{1}{\sqrt{r}}\sum\limits_{j=0}^{r-1} \exp (-2\pi i k j/r) |y^j \
({\rm mod}\ q)\rangle$ and the corresponding eigenvalues are $\exp (2\pi i j/r)$. If we apply the procedure of the revealing of eigen frequencies from the previous section, the measurement results in the number $j/r$\footnote{in the reality we obtain the approximation of this number within $O(1/N)$ with the high probability. The details of the proof can be found (in some equivalent form) in the paper \cite{Sh}}, or in \cite{Oz5}. If we know with the high probability the binary approximation of the fraction, it is easy to find its denominator, if we assume that this fraction irreducible. The formal algorithms for this search are based on the method of continued fractions, it can be found, for example, in \cite{Kn}. This fraction will be irreducible with the non-vanishing probability, since the all possible $j$ appear uniformly if we guarantee that the initial state $\xi$ for the procedure of revealing (see the previous section) is chosen arbitrary. We then repeat this procedure many times, which results in the frequent appearance of values $j$ mutually disjoint with $r$ and thus can find $r$ itself. This is Shor algorithm. 

We now come to the main thing: we estimate how good this algorithm is. As for Fourier transform all is clear here, it is very fast, generally speaking in the linear time relatively to the length of the notation of the number $q$, which we have to factorize. However, there is the other routine operation, which threatens to eliminate all the advantages of quantum Fourier transform. This is the operation of multiplication on the number $y$ containing in the operator of the conditional application $U_{cond}$. To find $U^\a$ we have to multiply to $y$ $\a$ times, which is about $q$ actions. This difficulty in the general case of the quantum Fourier transform application bears the principal character. It is irremovable for the arbitrary operator $U$. However, in the case of factoring we are lucky: we can fulfill the conditional application in the time of the order $\log^2 q$. To multiply to the number $y^\a$ we will obtain the number $y^\a$ by the sequential involution to the second power, beginning with $y$: $y,y^2, y^4, \ldots$. Of course, at each step we take the remainder from the division to $q$. We thus reach the closest to $y$ degree of two: $2^{l_1}$. We then take the quotient $q/2^{l_1}$ and do the same with it, etc. We then reach $y$ in the time of the order logarithm of $q$, e.g., in the number of steps of the order of the length of the $q$ notation. At each step we use about $log^2 q$ actions for the computation of the multiplication of numbers by the direct method, which results in the realization of the operator of the conditional application $U$ in the time $O(\log^3 q)$.

It is the complexity of Shor algorithm. We see, that the most difficult part of this algorithm is the routine operation contained in the preparation of the input state for Fourier transform - the sequential multiplication of natural numbers. 

\subsection{Solution of the problem of discrete optimization}

We continue to consider the examples of problems for which fast quantum algorithms can be obtained by some successful modification or combination of the main quantum tricks: GSA and QFT. At first we take up the natural generalization of the search problem: the search of the extreme point of an integer function. Let a function $f:\ \{ 0,1\}^n \ar\{ 0,1\}^n$ be defined by its oracle (or the scheme of the functional elements). We treat it, as usual, as the integer function. The problem is to find its extreme point: maximum or minimum. We note that in this most general formulation we cannot apply any trick essentially simplifying the search, like simplex method or the differentiation. The classical solution of this problem thus requires of the order of $N=2^n$ actions. 

The idea of its quantum solution rests on the GSA algorithm. We try to find the poin of maximum by the sequential approximations. Namely, we place all argument into the order of the growth of the function $f$ on them: $f(x_0 )\leq f(x_1)\leq\ldots \leq f(x_{N-1})$. On each step $j$ the input value will be some $x_{j_k}$. We apply G-BBHT algorithm with the oracle taking the value 1 exactly on the arguments $x_j$, for which $f(x_{j_k})<f(x_j)$, e.g., on $x_{j'},\ j'>j_k$. After the regular observation and the check of correctness we obtain the following value $x_{j_{k+1}}$ etc., up to the step when we reach $x_{N-1}$. The detailed analysis (see \cite{Ho}) shows that the complexity of such algorithm has the order $\sqrt{N}$, that gives us the same acceleration as GSA.

\section{Lecture 6. Adiabatic quantum computations}

So far, we have considered operational-type computations consisting of sequential applications of quantum gates. There are also continuous quantum computations called adiabatic. They are based on a slow change in the control Hamiltonian of the qubit system. In this case, the ground state of the original system will pass into the ground state of the modified system - this is the essence of the adiabatic theorem.

Consider the Schrodinger equation with a changing Hamiltonian $H(t)$ 
\begin{equation}
\label{Shr_}
i\hbar|\dot{\Psi}\rangle=H(t)|\Psi\rangle.
\end{equation}
Its solution will be the same as in the stationary case, given by the formula
\begin{equation}
\label{sol_}
|\Psi(t)\rangle=exp(-\frac{i}{\hbar}H(t)t)|\Psi(0)\rangle,
\end{equation}
but only now the exponent should be understood as a chronological exponent. How will its eigenstates change with a smooth change of the Hamiltonian $H$? If $H$ did not change at all, they would remain unchanged (except for the phase, of course). But if $H$ changes, will the eigenstates pass under the action of unitary evolution into the eigenstates of the new Hamiltonian?

\begin{figure}
\caption{The deviation of the ground state from the result of a smoothly changing unitary evolution. Direct calculation according to this scheme leads to very cumbersome expressions. To prove the adiabatic theorem, it is necessary to cover the entire process at once over a large time interval. Interference effects play a key role here - a change in the phase of the initial state with a period of $2\pi/E_0$ and the first excited state with a period of $2\pi/E_0$, and the periods also change with time! Straight lines represent the unitary evolution induced by $H(t)$.}\includegraphics[width=0.9\textwidth]{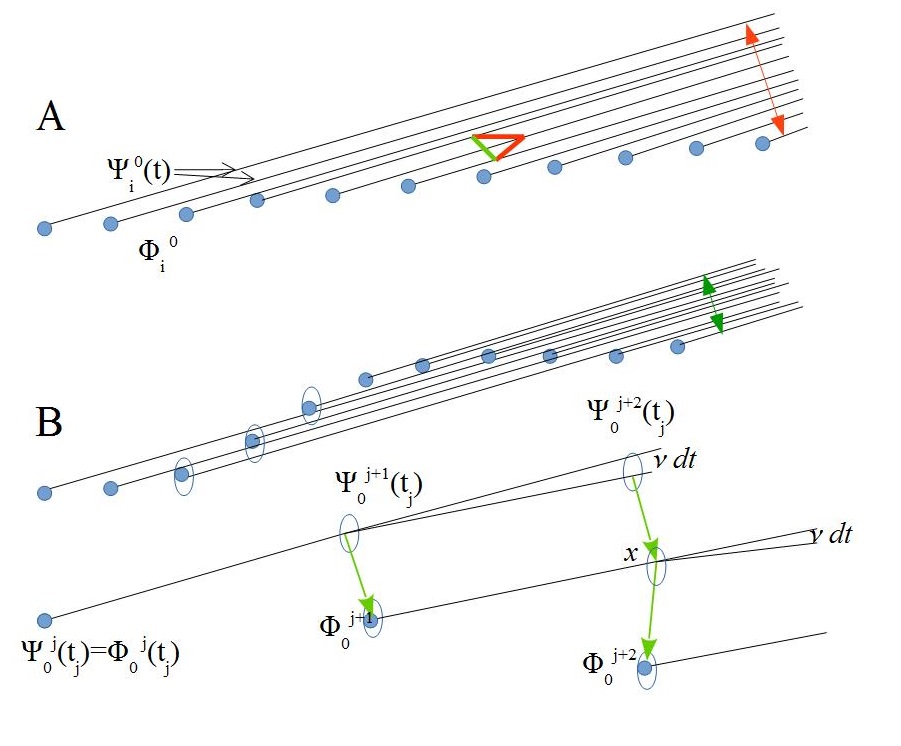} 
\label{fig:adia_int}
\end{figure}

\subsection{The Adiabatic theorem}

First, we consider the idea of quantum adiabatic processes, and prove a weakened version of the adiabatic theorem.

Let us have the main Hamiltonian $H_0$ and the target Hamiltonian $H_1$. We want to consider the slow change of the first Hamiltonian, which eventually leads to the second one. This change is a homotopy given by the real function $s (t):\ s(0)=0, s(T)=1$, where $T$ is a large number, so that $H(t)=(1-s(t))H_0+s(t)H_1$ is the value of the modified Hamiltonian at time $t$.

The adiabatic theorem is that for a very small value of $ \partial H/\partial t$, any eigenstate $|\Phi^0_k\rangle$ of the Hamiltonian $H_0$ as a result of the quantum evolution induced by the Hamiltonian $H(t)$ will move with great accuracy to the corresponding eigenstate $|\Phi^1_k\rangle$ of the Hamiltonian $H_1$; $k=0,1,...,N-1$. At the same time, the slowness of the change of $H (t)$ means that the maximum value of $ \partial H/\partial t=\partial s/\partial t$ is very small compared to the minimum energy gap $g$ - that is, the minimum value of the eigenergy difference $E_k(t)-E_s(t)$ for all $k> s$ and for all values of $t:\ 0\leq t\leq T$. We assume that the eigenstates are non-degenerate, so that the equality $E_k(t)-E_s (t)=0$ is achieved only for $s=k$.

A more precise analysis shows that in order to accurately approximate the eigenstates of the target Hamiltonian by images of the eigenstates of the main Hamiltonian, it is still necessary to learn the requirement, namely, it is necessary to require that $max|\partial s/\partial t|/g^2$ be a very small value.

We will not prove the adiabatic theorem in this strong form, but only show the idea of the adiabatic theorem, and proceed to its use in quantum computations.

For convenience, we will use $|n\rangle=|n(t)\rangle$ to denote the eigenstate of the Hamiltonian $H(t)$ with the number $n$, and $E_n$ to denote the corresponding eigenvalue of energy, and omit the explicit mention of time.

Let some $n_0$ be fixed, so that $|\Psi(0)\rangle=|n_0(0)\rangle$. 

The state of the system $| \Psi\rangle$ at the moment $t$ obeys the non-stationary Schrodinger equation
\begin{equation}
\label{shift10}
i\hbar|\dot{\Psi}\rangle=H(t)|\Psi\rangle
\end{equation}

and we can decompose it by the eigenstates of the current Hamiltonian: $| \Psi\rangle=\sum\limits_n a_n|n\rangle$, where all states and coefficients will depend on the time $t$.

We have:
\begin{equation}
\label{shif}
a_{n}(0)=\delta_{n,n_0}.
\end{equation}
Substituting this decomposition into \eqref{shift10}, we get:
\begin{equation}
\label{shift11}
\sum\limits_n(\dot{a}_n|n\rangle+a_n|\dot{n}\rangle)=
H\sum\limits_n a_n|n\rangle=\sum\limits_n E_n|n\rangle.
\end{equation}

Now we multiply this equality on the left by $\langle m|$ and use the orthonormality of the basis vectors $|n\rangle$: $\langle n|m\rangle=\delta_{nm}$.
The value of the summation index will be important to us, and we will divide the sum into two parts like this:

\begin{equation}
\label{shift12}
i\hbar \dot{a}_m+i\hbar\sum\limits_{n\neq m}a_n\langle m|\dot{n}\rangle+i\hbar a_m\langle m|\dot{m}\rangle=a_mE_m.
\end{equation}

Now we transform the equality $H|n\rangle=E_n|n\rangle$ by differentiating it in time $t$ and multiplying the resulting equality by $\langle m|$:

\begin{equation}
\label{shift13}
\dot{H}|n\rangle+H|\dot{n}\rangle=\dot{E}_n|n\rangle+E_n|\dot{n}\rangle, \langle m|\dot{H}|n\rangle+\langle m|H|\dot{n}\rangle=\dot{E}_n\delta_{nm}+E_n\langle m|\dot{n}\rangle.
\end{equation}

Next, we assume that $m\n n$, and given that $\langle m|H=E_m\langle m|$ we get:
\begin{equation}
\label{shift14}
\langle m|\dot{n}\rangle=\frac{\langle m|\dot{H}|n\rangle}{E_n-E_m}.
\end{equation}

Taking into account \eqref{shift14}, the equality \eqref{shift12} takes the form

\begin{equation}
\label{shift15}
i\hbar\dot{a}_m+i\hbar\sum\limits_{n\neq m}a_n\frac{\langle m|\dot{H}|n\rangle}{E_n-E_m}+i\hbar a_m\langle m|\dot{m}\rangle=a_mE_m.
\end{equation}

From here we immediately get the differential equation for $a_m$:

\begin{equation}
\label{shift16}
\dot{a}_m=-a_m(\frac{i}{\hbar}E_m+\langle m|\dot{m}\rangle)+\sum\limits_{n\neq m}a_n\frac{\langle m|\dot{H}|n\rangle}{E_m-E_n}
\end{equation}

We see that if the quotient $\frac{\langle m|\dot{H}|n\rangle}{E_n-E_m}$ is very small for any $m\neq n$ in absolute value, then there integrals
\begin{equation}
\label{cond_adia}
\Delta_{n,m}=\int\limits_0^Ta_n\frac{\langle m|\dot{H}|n\rangle}{E_n-E_m}
\end{equation}
for $n\neq m$ are very small, then the last term in \eqref{shift16} can be discarded. and we get the Cauchy problem for the coefficient $a_m$ :
$$
\dot{a}_m=A(t)c_m,\ a_m(0)=0
$$

Given the initial condition \eqref{shift} and the uniqueness theorem of the solution of the Cauchy problem, we get $a_m=0$ for any $m\neq n_0$.

So, if the condition $\Delta_{n, m} = o (1)$ is met, the adiabatic approximation works well.

However, for quantum adiabatic calculations, we will need a more precise formulation of the adiabatic theorem, which can be found in the book \cite{mess}:
\bigskip

{\it Adiabatic theorem (refined version).

If $|0(t)\rangle$ is the ground state, and $|1(t)\ rangle$ is excited state of the Hamiltonian $H (t)$,
such that $g=min_{\ 0\leq t\leq T,\ k=1,2,..., N-1} |E_k-E_0|$, where the minimum is reached at $k=1$, and for any $t:\ 0\leq t\leq T$, the following inequality is fulfilled:

\begin{equation}
\label{ad}
\left|\frac{\langle 1|\dot{H}|0\rangle}{g^2}\right|\leq \varepsilon
\end{equation}
and $| \Psi\rangle$ is the solution of the Cauchy problem for the Schrodinger equation with the Hamiltonian $H$, and the initial condition $|\Psi(0)\rangle=|0(0)\rangle$, then $|\langle 0(T)|\Psi(T)\rangle|^2\leq 1-\varepsilon^2$.}

Thus, in comparison with our previous arguments, it is necessary to require that the rate of change of the Hamiltonian be less than the square of the minimum gap between the energies.

We will not strictly prove this version of the adiabatic theorem, referring the listener to the book \cite{mess}; we will only show why it is necessary to require a small rate of change of $H$ greater than $g^2$. Such a requirement does not directly follow from the form of the integral \eqref{cond_adia}. However, let's consider a small time interval $dt$, at which the Hamiltonian changes little, so that we can consider it a constant. Then the solution of the equation \eqref{Shr_} will be written as an eigenstate expansion of the current Hamiltonian $|\phi_i\rangle$:

\begin{equation}
\label{expa}
|\psi\rangle= \sum\limits_i e^{\frac{1}{\hbar}E_i dt}\la_i|\phi_i\rangle
\end{equation}

If we take the ground state, we can assume that $E_0=0$ by shifting the spectrum by $E_0$, and then we will see that all excited states will have a coefficient in the form of an oscillating multiplier $exp(-\frac{i}{\hbar}(E_i-E_0)t)$, so that in the case of a smooth change in the Hamiltonian, these oscillations will result in an alternating Leibniz series, in which the terms will almost completely reduce each other, so that the integral \eqref{cond_adia} will be reduced to an integral over one period of such an oscillation, which is approximately $O(1/g)$, and the value of it will remain, in order equal to $O(\langle 1|\dot{H}|0\rangle/g^2$, that is, proportional to the rate of change of the Hamiltonian itself $H$.

Consider the main integral \eqref{cond_adia}, and evaluate it roughly. Let the Hamiltonian change slowly, so that in the interval $[0, \Delta t]$ it changes little, but the ground states change the phase significantly. We assume that $\hbar =1$. Then the states $|0\rangle$ and $|1\rangle$ - the main and excited, on this interval - evolve in this way:

\begin{equation}
\label{evol_}
|0(dt)\rangle = e^{-iE_0t}|0(0)\rangle,\ |1(t)\rangle = e^{-iE_1dt}|1(0)\rangle,
\end{equation}

Then the numerator in the expression \eqref{cond_adia} will take the form
\begin{equation}
\label{interf}
e^{i(E_1-E_0)t}\langle 1(0)|\dot{H(t_{average})}|0(0)\rangle
\end{equation}
and it will oscillate with a period of the order of $O (1/(E_1-E_0)$. Suppose that the Hamiltonian changes smoothly, so that $\langle 1|\dot{H}|0\rangle$ has a limited number of monotone sections. We estimate the value of \eqref{cond_adia}, taking into account the reduction in interference generated by the oscillations of the complex exponent in \eqref{interf} in the monotonicity section of the numerator. Here we will have a Leibniz series, the sum of which is equal in order to the highest member of the series. So the integral will be reduced to an integral over a section of length $O(1/E_1-E_0)$ from the value of the same order muliplied to coefficient $\langle 1|\dot{H}|0\rangle$, which will result in the formula \eqref{ad}. A graphic illustration is shown in the figure \ref{fig:adia_int}.

\begin{figure}
\caption{Interference in the deviation of the ground state from the result of a smoothly changing unitary evolution.}
\includegraphics[width=0.9\textwidth]{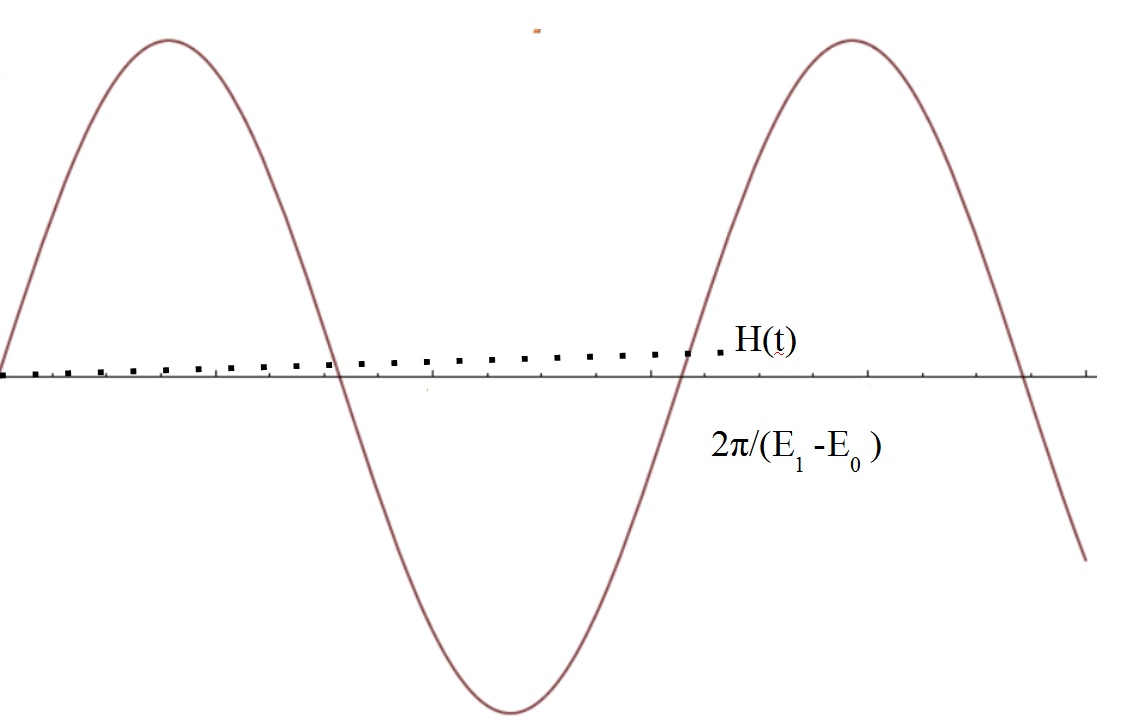} 
\label{fig:adia_int}
\end{figure}

\subsection{Adiabatic form of Grover algorithm}

Grover's algorithm, originally formulated in terms of quantum operations (a sequence of gates), can also have a continuous form.

For the first time, this form was proposed by Farhi and Gutman. It consists of the following. Let's say we need to find an unknown basic state $|m\rangle$, which we will call the target state. We will create an initial state $|\tilde 0\rangle$ - an arbitrary initial state that is convenient for us to build. If we choose some functions $f(t)$ and $r(t)$, then the Shredinger equation induced by Hamiltonian$H(t)=I+f(t)|m\rangle\langle m|+r(t)|\tilde 0\rangle\langle \tilde 0|$ will lead us from the initial state to the target state in time $\pi/(2E\langle m|\tilde 0\rangle)$. As one can see, the continuous version of Grover's algorithm delivers the same quantum spedup of the solution of the iterative problem as GSA.

This statement is not trivial. Indeed, the canonical version of Grover's algorithm uses the rotation transformation $U=I_ {\tilde 0}I_{m}$, which does not reduce to the application of the evolution induced by the Hamiltonian $H$, since the Hamiltonians $|m\rangle\langle m|$ and $|\tilde 0\rangle\langle\tilde 0|$ do not commute.

Thus, with the adiabatic form of Grover's algorithm, everything is not so simple.
We will create the initial state $|\Psi(0)\rangle=|\tilde 0\rangle=\frac{1}{\sqrt N}\sum\limits_{a=0}^{N-1}|a\rangle$, as in the GSA operating form, and let's assume that we can create an evolution induced by the Hamiltonian $\tilde H (s)=(1-s)H_0+sH_m$, where
$$
H_0=I-|\tilde 0\rangle\langle \tilde 0|, \ H_m=I-|m\rangle\langle m|
$$

The adiabatic algorithm consists in applying to the initial state $|\tilde 0\rangle$ a variable Hamiltonian $H (s)$, where the function $s (t)$ is such that $s (0)=0,\ s (T)=1$ for a large $T$. The art of adiabatic computation consists in choosing the deceleration function $s (t)$.

Note that $|\tilde 0\rangle$ and $|m\rangle$ are the ground states of the Hamiltonians $H_0$ and $H_m$, respectively, with zero eigenvalues.

We will use $H(t)$ to denote the dependence of the Hamiltonian on the real time $t$, to distinguish it from $\tilde H(s)$ - the dependence on the abstract parameter $s$. The dependence of $s$ on $t$ means a slowdown in evolution.

Under the conditions of the adiabatic theorem for the Hamiltonian $H (t)$, we will again consider $|0 \rangle$ to be the ground state, and $1\rangle$ to be the excited state for which an energy gap of size $g$ is realized. Then we have:
\begin{equation}
\label{sh1}
\langle 1|\dot{H}|0\rangle=\frac{ds}{dt}\langle 1|\frac{d\tilde H}{ds}|0\rangle=\frac{1}{T}\langle 1\frac{\tilde H}{ds}|0\rangle.
\end{equation}

First, we will try to follow a simple path, and choose $s (t)$ as a linear function:
$s=t/T$ for a sufficiently large $T$; then the Hamiltonian will change slowly.
Solving the eigenvalue problem for the Hamiltonian $H$, we can also find the energy gap. The result of the computation is as follows:

\begin{equation}
\label{sh2}
g=\sqrt{1-4\frac{N-1}{N}s(1-s)},
\end{equation}
here $|\langle 1|\frac{d\tilde H}{ds}|0\rangle|\leq 1$. 

\begin{figure}
\caption{The eigenvalues of the Hamiltonian $\tilde H$ for $N=64$. The picture is taken from the article Roland, Cerf, Quantum Search by Local Adiabatic Evolution (arxive.org, quant-ph/0107015).}\includegraphics[width=0.9\textwidth]{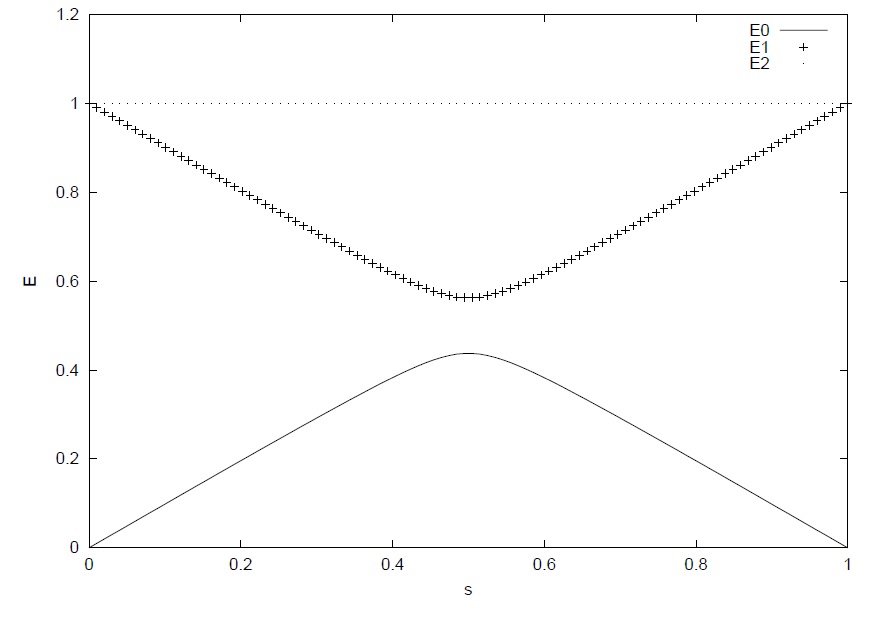} 
\label{fig:gap}
\end{figure}

The formula \eqref{sh2} is graphically illustrated in the figure \ref{fig:gap}. Now we will find the minimum gap, which is equal to $ g_{min}=1/\sqrt{N}$ (obtained for $s=1/2$).
Then the condition of the adiabatic theorem gives us the inequality $T\geq N/\varepsilon$. Thus, we get the same  time of the  computation of $|m \rangle$ as on a classical computer, and the adiabatic algorithm does not give quantum spedup.

To obtain the quantum spedup, we will set a more complex, nonlinear time dilation $s (t)$. Since the critical value of the gap $g$ is not reached at any time, but only at $s=1/2$, we can improve the estimate of the total running time of the algorithm by selecting a nonlinear deceleration. To do this, we will apply the \eqref{ad} relation locally, obtaining the formula

\begin{equation}
\label{ad1}
|\dot{s}|\leq g^2(t)/|\langle 1|\frac{d\tilde H}{dt}|0\rangle |.
\end{equation}

\begin{figure}
\caption{Optimal time dilation for the adiabatic version of GSA. The picture is taken from the article Roland, Cerf, Quantum Search by Local Adiabatic Evolution (arxive.org, quant-ph/0107015).}\includegraphics[width=0.9\textwidth]{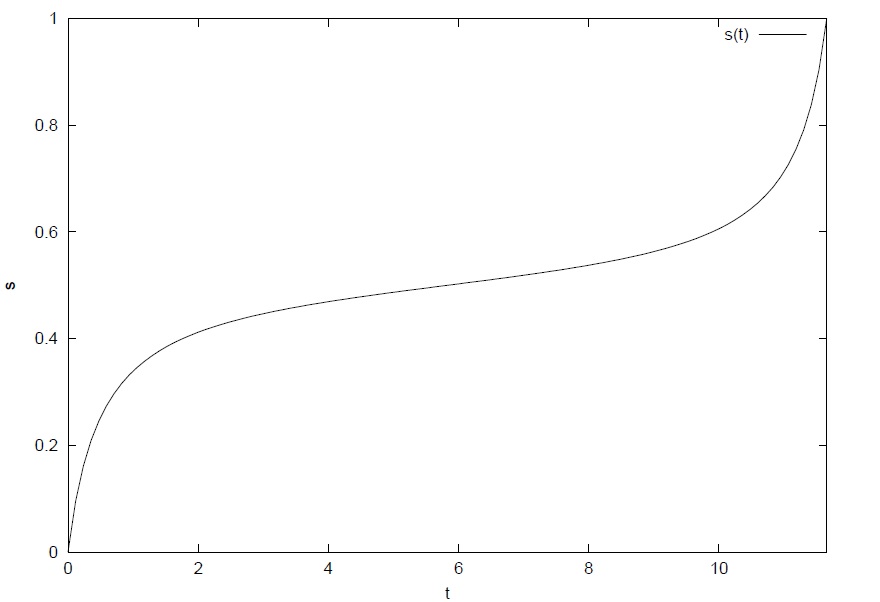} 
\label{fig:slow}
\end{figure}

Using the ratio \eqref{sh2} under the condition $|\langle 1|\frac{d\tilde H}{ds}|0\rangle|\leq 1$, we obtain a time dilation equation of the form:	
\begin{equation}
\label{sh3}
\dot{s}=\varepsilon g^2(t)=\varepsilon (1-4(1-1/N)s(1-s))
\end{equation}
for the small $\varepsilon$. Integration of  \eqref{sh3} gives 
\begin{equation}
\label{sh4}
t=\frac{N}{2\varepsilon \sqrt{N-1}}(arctg((2s-1)\sqrt{N-1}+arctg\sqrt{N-1}),
\end{equation}

inverting, we find the final expression for time dilation in the form of a graph shown in the Figure \ref{fig:slow}. Here it can be seen that time flows most quickly in those areas where the gap is large, and slowly - where the gap is small.

We will find the full running time of the algorithm by substituting $s=1$ in \eqref{sh3}: $T=\pi\sqrt N/2\varepsilon$ - this is exactly the acceleration that the standard Grover algorithm gives.

Now we show that the quantum acceleration found by the adiabatic method is optimal. To do this, we consider two competing basic states $|m \rangle$ and $|m'\rangle$, each of which can be a target for Grover's algorithm. Let $\psi_m\rangle$ and $|\phi_{m'}\rangle$ be the states of a quantum computer under adiabatic computation with some time dilation $s (t)$. A properly working algorithm must reliably distinguish them in time $T$, that is, the following inequality must be fulfilled

\begin{equation}
\label{sh6}
1-|\langle\psi_m(T)|\psi_{m'}(T)\rangle|^2\leq\varepsilon
\end{equation}

In the notation we have introduced, we will divide the Hamiltonian into two terms:
$\tilde H(s)=\tilde H_1(s)+\tilde H_{2m}(s)$, где 
$$
\tilde H_1(s)=I-(1-s)|\psi_0\rangle\langle\psi_0|,\ \tilde H_{2m}(s)=-s|m\rangle\langle m|
$$

For the time dependence $t$, we, as before, use the notation $H$ without tildes. Then $|\psi_m\rangle$ and $|\psi_{m'}\rangle$ will be solutions of the equations

$$
i|\dot{\psi}_m\rangle=(H_1+H_{2m})|\psi\rangle,\ i|\dot{\psi}_{m'}\rangle=(H_1+H_{2m'})|\psi\rangle
$$
with the common initial condition $|\psi_m(0)\rangle=|\psi_{m'}(0)\rangle=|\psi_0\rangle$. 

We have:
\begin{equation}
\label{sh7}
\begin{array}{ll}
&\frac{d}{dt}(1-|\langle\psi_m|\psi_{m'}\rangle|^2)=\\
&2Im(\langle\psi_m|H_{2m}-H_{2m'}|\psi_{m'}\rangle\langle\psi_{m'}|\psi_m\rangle)\\
&\leq 2|\langle\psi_m|H_{2m}-H_{2m'}|\psi_{m'}\rangle| \ |\langle\psi_{m'}|\psi_m\rangle|\\
&\leq 2(|\langle\psi_{m}|H_{2m}|\psi_{m'}\rangle|+|\langle\psi_{m}|H_{2m'}|\psi_{m'}\rangle|).
\end{array}
\end{equation}

 Now let's take the sum of $m, m'$ and get:
\begin{equation}
\label{sh8}
\begin{array}{ll}
&\frac{d}{dt}\sum\limits_{m,m'}(1-|\langle\psi_m|\psi_{m'}\rangle|^2)\leq 4\sum\limits_{m,m'}|\langle\psi_m|H_{2m}|\psi_{m'}\rangle|\\
&\leq 4\sum\limits_{m,m'}\| H_{2m}|\psi_m\rangle\| \ \| |psi_{m'}\rangle\|\leq 4N\sum\limits_m \| H_{2m}|\psi_m\rangle\| .
\end{array}
\end{equation}
In the last transition, the Cauchy - Bunyakovsky - Schwarz inequality was used. Note also that for the normalized state $|\psi\rangle$ from $ \sum\limits_m\| H_{2m}|\psi\rangle\|^2=s^2$ follows $\sum\limits_m \| H_{2m}|\psi\rangle\|\leq\sqrt{N}s$ (the inequality between norms in the spaces $l_2$ and $l_1$).

As a result, we have:
\begin{equation}
\label{sh9}
\frac{d}{dt}\sum\limits_{m,m'}(1-|\langle\psi_m|\psi_{m'}\rangle|^2)\leq 4N\sqrt{N} s.
\end{equation}
Integrating this inequality, we have:

$$
\sum\limits_{m,m'}(1-|\langle\psi_m|\psi_{m'}\rangle|^2)\leq 2N\sqrt{N}\int\limits_0^Ts(t)dt
$$
and, given \eqref{sh6}, we find 
$$
T\geq \varepsilon (N-1)/(4\sqrt{N})
$$

which proves that a properly working adiabatic algorithm for the brute force problem cannot work faster than the root of classical time.

The adiabatic theorem has many forms, and error estimates. The simplest one belongs to Landau and Zener; it is valid only for a two-level system, that is, for a single qubit. It has the form
$$
err = O(e^{-C\Delta^2 t_f})
$$
where $t_f$ is the total time of the process.

\subsection{Construction of Hamiltonians for adiabatic computations}

To implement the adiabatic method, it is necessary to practically construct Hamiltonians with the necessary property: for the initial $H_0$, its ground state should be quite simple, and for the target $H_1$, the ground state should give a solution to the desired problem.

As a rule, the ground state of the initial Hamiltonian is chosen, as in the operator version of Grover algorithm, in the form $|\tilde 0\rangle=\frac{1}{\sqrt N}\sum\limits_{j=0}^{N-1}|j\rangle$. Such a Hamiltonian, with such a ground state, has the form
$$
H_I=\sum\limits_{i-1}^m\frac{1-\s_i^x}{2}.
$$
Indeed, for $m=1$, this is checked directly, and for large $m$, the Hamiltonian decomposes into the sum of terms-operators, for each of which the eigenstates will be states having in the notation from linear algebra of the form $(...a, a...)^*$, where the positions in the vector - column - of the basis state corresponding to the values $0$ and $1$ of a fixed qubit are highlighted, and the dots denote arbitrary states. But this also means that the sum of such operators has its eigenstate $|\tilde 0\rangle$.

Now we will deal with the target Hamiltonian $H_1$. Let's first consider a simple example. We have two qubits and the task is to determine whether their values are equal. For such a search problem, the ground state of the target Hamiltonian will be
$$
\frac{1}{\sqrt 2}(|01\rangle+|10\rangle.
$$
and the target Hamiltonian will have the form
$$
H_1=\frac{1}{2}(I+\s_1^z\s_2^z).
$$

\begin{figure}
\centering
\includegraphics[scale=0.7]{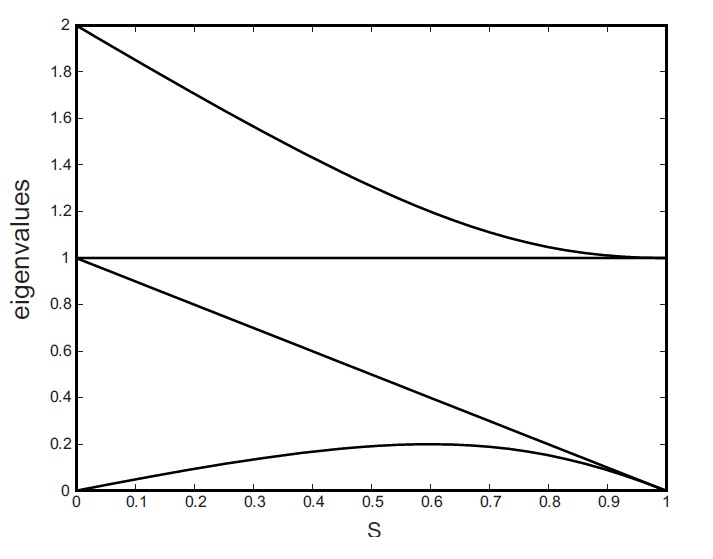}
 \caption{The behavior of 4 eigenvalues for the 2-bit disagree algorithm. The picture is taken from the article Adiabatic quantum computation and quantum annealing (Catherine McGeoch).}
\label{fig:2bitdis}
\end{figure}

This is the so-called 2-bit disagree algorithm. The figure \ref{fig:2bitdis} shows the behavior of its spectrum depending on the mode $s (t)$. Gluing two eigenvalues at the end of the process is not fatal, since we do not care which of the ground states will be output when measuring the final state.

Let's consider a slightly more complex problem - the exact coverage. This is the problem of determining the truth of the conjunction $\&_cC_c$ where each factor $C_c$ has the form $(x^c_1, x^c_2, x^c_3)$, where $x^c_j$ is either $x_i$ or $not \ x_i$ for some $i$. Here $C_c$ is true if and only if exactly one term of $x^c_1, x^c_2,x^c_3$ is true.

We will introduce a covering function for each $c$
$$
f_c=(1-x^c_1-x^c_2-x^c_3)^2
$$

abd put $f(\bar x)=\sum\limits_cf_c(\bar x).$ Now we have

$$
f(\bar x)=-2m-\sum\limits_i(B_ix_i+B_ix_ix_i)+\sum\limits_{i<j}C_{ij}x_ix_j=-2m+\sum\limits_{i<j}C_{ij}(1-s_i)(1-s_j)
$$

where $x_i=(1-s_i)/2$ is the formula for the transition between Boolean variables $x_i=0,1$ and spin variables $s_i=\pm 1$.

The target Hamiltonian for this problem has the form
$$
H_1=\sum\limits_{i<j}C_{ij}(1-\s_i^z)(1-\s_j^z).
$$

Now let's consider the general form of the $ SAT$ - problem-satisfiability of the formula of the logic of statements. It is enough to assume that this formula is given in conjunctive normal form, moreover, that each conjunctive term has the form $f_c$, only now the truth is determined by the rules of the logic of statements, where $f_c$ is the disjunction of three variables or their negations (the so called $SAT-3$ problem).

In this case, the definition of, for example, the elementary function $f_c$ for the elementary disjunction $x_1\ or\ x_2\ or\ x_3$ will have the form
$$
f_c(s_1,s_2,s_3)=(5-s_1-s_2-s_3+s_1s_2+s_2s_3+s_1s_3+3s_1s_2s_3)/8.
$$

The {\bf Ising model}. The energy function for the basis state of $n$ spins connected in pairs has the form
$$
H(\bar s)=\sum\limits_ih_is_i+\sum\limits_{i<j}J_{ij}s_is_j.
$$
The problem of finding $\bar s$ that minimizes this function is NP-complete if the spins are connected in the form of a three-dimensional structure (Istral). Fu and Anderson showed that it is NP-complete even for 2D connections, provided that nonzero $h_i$are present. The problem of finding the minimum of the function $H(\bar s)$ is equivalent to finding the ground state of the Hamiltonian
$$
H=\sum\limits_ih_i\s^z_i+\sum\limits_{i<j}J_{ij}\s^z_i\s^z_j.
$$

To solve this problem, we can use the method of quantum annealing.

It consists of the following. We have a target Hamiltonian $H_{tar}$, whose ground state we need to find. We choose a perturbing Hamiltonian $H_d$ and a schedule $G(t)$ expressing the intensity of the perturbation. This schedule is arranged as follows: $G (0)\gg 1,\ G(t)\rightarrow 0\ (t\rightarrow \infty)$. The current Hamiltonian is selected in the form
$$
H(t)=H_{tar}+G(t)H_d.
$$

\begin{figure}
\centering
\includegraphics[scale=0.7]{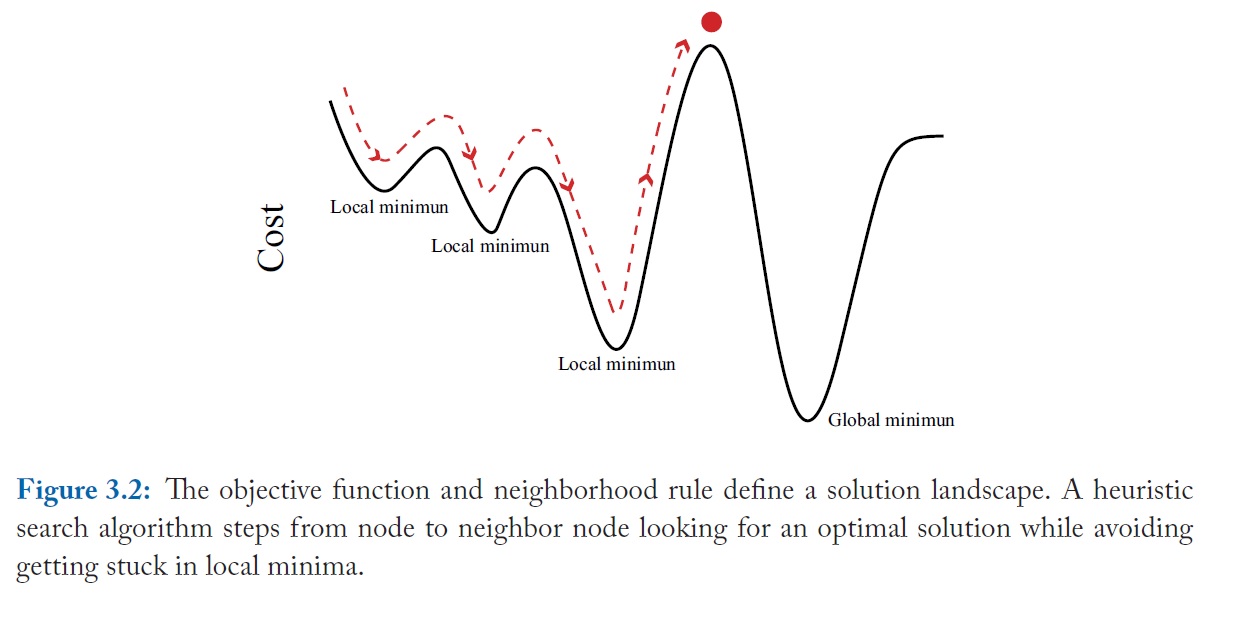}
 \caption{Schematic representation of quantum annealing. The picture is taken from the article Adiabatic quantum computation and quantum annealing (Catherine McGeoch).}
\label{fig:annea}
\end{figure}

An illustration of quantum annealing is shown in the figure \ref{fig:annea}
\newpage

\section{Lecture 7. Simplified control and fermionic identity in quantum computations}

Quantum computer is the unexampled testing of quantum physics because it requires such level of control over nano-sized objects, which has been never reached artificially. Whereas the mathematical theory of quantum computing in the framework of standard quantum formalism is well developed its physical realization represents the serious challenge to our understanding of the Nature. This is why it is important to look for its simplest possible realization, so that it rests on the basic principles of quantum theory and includes the minimal technological difficulties. Two requirements for such schemes we can formulate: the adequate description of states forming the computational Hilbert space, and the realistic method of control over computations. 

Typically one computational element - a qubit we represent as some characteristic, like a spin , charge or a position of some elementary particle. This approach works well for the isolated qubit. For a system of several qubits, this approach meets serious difficulties. These difficulties come from the fundamental principle of the non-distinguishability (or identity) of elementary particles of the same type. To control a computation we must be able to address to a separate qubit, whereas the different particles are identical. 

Of course, we can distinguish particles placing them on the big distance one from another, but in this case, it will be difficult to keep them in entangled states what is necessary for quantum computations. On of the solutions of this dilemma is to use Fock space of the occupation numbers for the description of quantum computations. Here the natural identification of qubits with the energetic levels in Fock space is used, so that the unit is treated as the occupied level, and zero - as the free level. 

This approach gives the universal quantum computations by the high cost. It requires to control not only the external field and the tunneling, but also the diagonal interactions between qubits, and the contact with the superconductor, e.g., we have to control the coefficients $\a ,\b , \g$ in (\ref{ham}) and to control on the additional summand $\d a_k^+ a_j^+ +\d^* a_k a_j. $

In this section, we will see how to decrease this cost by means of the idea of the continuous and non-controlled interaction. To this, we need two things: The supposition that the initial Hamiltonian of interaction contains only the external field, the diagonal and the tunneling summands, and the modified correspondence between states in the occupation numbers space and the computational Hilbert space. To control quantum computation we then need only to switch the external field and the tunneling. Lasers can fulfill this type of the control. The main scheme we give further. It rests on the idea of the continuous interaction, proposed in the work \cite{OF} and adapted to the language of Fock space of occupation numbers.

\subsection{One qubit control in quantum computations}

 The main difficulty in the practical realization of quantum computations is that it is technically difficult to fulfill two qubit operators playing the necessary role in such computations. To fulfill these operators we must control the degree of entanglement of particles that is determined by the overlapping of the spatial parts of their wave functions. However, to fulfill computations we need to distinguish particles with certainty, which is possible only if the overlapping is sufficiently small. 

This is the evident contradiction in requirements for the physical realization of quantum computations. We see that it is much more difficult to realize two qubit gates then one qubit. The following approach would be appropriate here. Since the interactions between particles with the varying degree of entanglement follows from the wave equation and is confirmed in experiments, the two-qubit transformations go permanently in course of natural time evolution of a quantum system. As for the control over such evolution, we can fulfill it by one-qubit impulses, whereas two qubit gate will go in the non controlled background regime. This is the essence of the proposed quantum computations with one qubit control. 

This model is much more realistic than the abstract model of quantum computer, which supposes the control on two-qubit interaction. We temporarily leave the question about the general possibilities of this approach and demonstrate how the concrete problem about the simulation of the behavior of many bodies system with quadratic interaction of the diagonal form can be solved in the frameworks of the proposed model. The main difficulty of the proposed model is that two-qubit interaction is out of the control, in particular, it goes with the outside qubits that seriously distorts the picture of quantum computations.
To perform computations in such model we have to create the method of the correction of "undesirable" transformations by means of one qubit impulses. 

For the demonstration of abilities of this approach, we first show how to realize quantum Fourier transform in the framework of this model. The main supposition will be that the Hamiltonian matrix of two qubit interaction has the diagonal form. For the convenience, we at first impose some limitations on the speed of decreasing of this interaction with the distance. Namely, we suppose that the potential falls with the distance as Yukawa potential.. This method then can be applied to the more wide class of the diagonal interaction. Moreover, this method can be generalized to the case when the different qubit pairs interact differently. At last, we apply this approach to the system of many particles with the potential of quadratic type. 
\nn

\subsubsection{Realization of quantum Fourier transform on one qubit control} 
\nn

Quantum Fourier transform is the key subroutine in quantum computing. It is used in the big number of other algorithms. The quantum gate array realizing this transformation is represented at the picture 6. It was proposed and used for the fast quantum factoring by P.Shor (see \cite{Sh}). We agree to represent an integer of the form $a=a_0 +a_
0 2+\ldots +a_{l-1} 2^{l-1}$ by th ebasic state $|a_0 \ a_1 \ \ldots\ a_{l-1} \ \rangle =|a\rangle$ . These states form orthonormal basis for the input states of the quantum gate array. We place them from the top to the bottom. The analogous agreement we accept for the output state but the binar signs $b_j$ of the number $b=b_0 +b_0 2+\ldots +b_
{l-1} 2^{l-1}$ we place in the opposite order. 

This scheme fulfills the reverse transformation $QFT^{-1}$ in $O(l^2 )$ steps, whereas its matrix $N=l^2$ - dimensional. However, in this scheme, the two qubit control is required, it cannot be directly applied in terms of our model. We show how to do that. We treat the interactions of the form
 \begin{equation} {\rm A)}\ H=\left( \begin{array}
{ccccc} &0 &0 &0 &0\\ &0 &0 &0 &0\\ &0 &0 &0 &0\\ &0 &0 &0 &\rho \end{array} \right) ,\ \ \rho >0, \ \ \ \ \ \ \ \ \ \ \ 
{\rm B)}\ H=\left( \begin{array}{ccccc} &\rho_1 &0 &0 &0\\ &0 &\rho_2 &0 &0\\ &0 &0 &\rho_3 &0\\ &0 &0 &0 &\rho_4 \end
{array} \right) , \label{Ham} \end{equation} where all $\rho =\rho_0 \frac{e^{-br}}{r}$; $b=const$; $r$ is the distance between qubits-particles, and $\rho_1 +\rho_4 \neq \rho_2 +\rho_3$.

 We place $l$ qubits on one line with the equal intervals. Let the interaction between $j$ - the and $k$ - th qubits have Hamiltonian $H_{j,k}$ of the form (\ref{Ham}). This type of Hamiltonian arises, for example, in Izing model for particles with the spin 1/2. The required decreasing on the interaction with the distance we can reach placing qubits to the appropriate potential hole. Choosing the proper unit of the length we can ensure that $b=1$. 
At first we study the case of interaction of the form (\ref{Ham}, A) and then extend the results to the case (\ref{Ham}, B). 
\nn

{\bf Realization of QFT within phase shift} 
\nn

We remind that QFT and its reverse have the form: \begin{equation} QFT :\ 
|a\rangle\ar\frac{1}{\sqrt{N}}\sum\limits_{b=0}^{N-1}e^{-\frac{2\pi i\ ab}{N}}|b\rangle , \ \ \ \ \ \ \ \ QFT ^{-1} :\ |a
\rangle\ar\frac{1}{\sqrt{N}}\sum\limits_{b=0}^{N-1}e^{\frac{2\pi i\ ab}{N}}|b\rangle . \label{qft} \end{equation} 
The reverse transform can be then fulfilled by the following scheme. 
\newpage
\begin{picture}(500,310)(0,-30) \multiput(80,50)(0,50){5}
{\line (30,0){30}} \multiput(140,50)(0,50){5}{\line (30,0){30}} \multiput(200,50)(0,50){5}{\line (30,0){30}} \multiput
(260,50)(0,50){5}{\line (30,0){30}} \multiput(320,50)(0,50){5}{\line (30,0){30}} \multiput(110,40)(60,0){4}{\framebox 
(30,220)} \multiput(95,50)(60,50){5}{\circle{10}} \put(30,200){j} \put(30,100){k} \put(60,50){$a_4$} \put(60,100){$a_3$} 
\put(60,150){$a_2$} \put(60,200){$a_1$} \put(60,250){$a_0$} \put(360,50){$b_0$} \put(360,100){$b_1$} \put(360,150){$b_2$} 
\put(360,200){$b_3$} \put(360,250){$b_4$} \put(60,10){Picture 2. \ Rectangles denote continuous interaction} 
\put(120,-5){circles - Hadamard operators} \end{picture} Here rectangles denote unitary transforms of the form $U=e^{-i\tilde H}$, where $\tilde H=\sum\limits_{l>j>k\geq 0}\tilde H_{j,k}$, and each from $\tilde H_{j,k}
$ has the form (\ref{Ham} , A) with $\rho_0 =\pi$, $r=j-k$. If we choose the system of physical units such that Planck constant multiplied to 
 $\rho_0$ equals to $\pi$ when the unit of the length such that $r=j-k$, then $U$ will be exactly the transformation of the state vector induced by the considered Hamiltonian in the unit time. Here we suppose that the time of one-qubit operations is negligible, and the interaction between qubits cannot substantially change the phase in this time. 

This scheme can be obtained from the previous one by the insertion of the missing elements corresponding to the interaction going in the system with this Hamiltonian. To prove that this scheme really fulfills $QFT^{-1}$, we apply the method of the amplitude counting proposed in the paper \cite{Sh}. Let the basic input state be given: $|a\rangle$, we consider the corresponding output state. 
This output state is the linear combination of basic states $|b\rangle$ with some amplitudes. All modules of these amplitudes are the same and equal 
$1/\sqrt{L}$, and we have to look after their phases only. For the simplicity we introduce the notation $a'_j =a_{l-1-j},\ j=0,1,\ldots , l-1$. 

In the course of application of our scheme the value of qubits number $j$ and $k\leq j$ pass through elements form the picture 2 from the left to the right. Following this direction, we separate four types of interactions: The interaction of $a'_j$ with itself and $a'_k$ with itself in Hadamard gate, the interaction of $a'_j$ with $a'_k$ ($j>k$), the interaction of $a'_j$ with $b_k$ for $j>k$, and the interaction of 
$b_j$ with $b_k$ ($j>k$). The times of the second and third interactions will be: zero and $j-k$  correspondingly. Summing the deposits of these interactions to the phase we obtain the resulting phase of the form 
\begin{equation} 
\pi\sum\limits_{l>j>k\geq 0}\frac{a'_j a_k 
k}{2^{j-k}}C_1 + \pi\sum\limits_{l>j>k\geq 0}\frac{a'_j b_k (j-k)}{2^{j-k} (j-k)} + \pi\sum\limits_{l>j\geq 0} a'_j b_j 
+ \pi\sum\limits_{l>j>k\geq 0}\frac{b_j b_k }{2^{j-k}}C_2, 
\end{equation} 
for some $C_1,C_2$.

We denote the first and the fourth summands by $A$ and $B$ correspondingly. Their deposit corresponds to the action of the diagonal Hamiltonians to $|a\rangle$ and $|b\rangle$ correspondingly. We temporarily leave these deposits. We take up the second and the third summands of this sum. After the replacement of $j$ by $l-1-j$ this part acquires the form 
\begin{equation} \begin{array}{cc} \pi\sum\limits_{l-1>k+j\geq 0}\frac{a_j b_k 2^{j+k}}{2^{l-1}}+
\pi\sum\limits_{l- 1\geq j\geq 0} a_{l-1-j} b_k &=2\pi\sum\limits_{l>k+j\geq 0}\frac{a_j b_k 2^{j+k}}{2^l} =2\pi S+ 2\pi
\sum\limits_{l>k,j\geq 0}\frac{a_j b_k 2^{j+k}}{2^l}= \\ &2\pi S+ 2\pi\frac{ab}{2^l} \end{array} 
\end{equation} 
for some integer $S$. The first summand does not change the phase and we obtain that is required within the deposit of $A$ and
$B$. 

\nn
{\bf Correction of phase shift}
\nn

 To account the deposit of diagonal summands $A$ and $B$ to the phase we apply one trick. At first we consider the summand $A$. It consists of the members of the form $A_{j,k}=c_{j,k} a'_j a'_k$, where $c_{j,k}$ depends only on $j$ and $k$, but not on $a$. We call $j$-th and $k$-th qubits separated. 

We will apply one-qubit operator NOT several times to each qubit but separated in order to suppress all interactions but the interaction going between the separated qubits. At first we consider the pair of not separated qubits with the numbers $p,q,\ q>p$. Their continuous interaction in time $\D t$ gives the summand $d_{p,q}\D t\ a'_p a'_q$ to the phase, where the real number $d_{p,q}$ depends only on how fast the interaction falls with the distance, but not on  $a'_p , a'_q$. For example, for the decreasing of Yukawa type we have $d_{p,q}=e^{-|q-p|}/|q-p|$. 

Now we invert one of these two qubits, no difference which exactly, let it be $q$-th, by means of the NOT gate. It state will be $1-a'_q$. Now the second period of the longitude $\D t$ 
of the continuous interaction gives the summand $d_{p,q}\D t\ a'_p (1-a'_q)$ to the phase. At last, we restore the contents of $q$-th 
qubit by the second application of NOT. The resulting phase shift in these four actions will be $d_{p,q}\D t\ a'_p$ and it depends on the contents of $p$-th qubit only. 

Now we can compensate this phase shift by means of single one qubit transformation. If we consider the pair of qubits with the numbers
 $p,q$, where one, say, $p$-th is the separated, the other is not separated, we then can compensate its interaction using only one qubit operations: two NOTs for $q$-th and some phase shift for $p$-th. Now we have to modify this method so that to compensate all influences of not separated qubits simultaneously. 

For this, we will fulfill NOT operations on each such qubit with the sufficiently small time intervals so that the deposits to the phase of non-separated qubits will cancel each other. There are two ways to do it: to use the random process for the generation of the moments for one-qubit operations, or to realize them periodically with the different periods for the different qubits. At first, we study the first approach. 
\nn

 {\bf Method of random processes} 

\nn
 For each not separated qubit number $p$ we consider the Poisson process ${\cal A}_p$, generating time instances $0<t_1^p <t_2^p <\ldots <t_{m_p}^p <1$ with some fixed density $\la\gg 1$. Let all ${\cal A}_p$ be independent. We fulfill NOT operators on each qubit with number $p$ in time instances $t^p_m$ sequentially. In the moment 1 we fulfill NOT on $p$-th qubit if and only if $m_p$ is odd. Therefore, after this procedure each qubit restores its initial value. 

We count the phase shift generated by this procedure. Interactions between the separated qubits remain untouched. We fix some non-separated qubit and count its deposit to the phase. It consists of two summands: the first comes from the interaction with the separated, the second - from the interaction with non-separated qubits. We find them sequentially. 

1. In view of high density $\la$ of Poisson process ${\cal A}_p$ about the half of all time $p$-th qubit will be in the state $a'_p$, and the remaining half - in the state $1-a'_p$. Its interaction with the separated qubit, say, with $j$-th, gives the deposit $\frac{1}{2} d_{p,j}a'_p a'_j +\frac{1}{2} d_{p,j} (1-a'_p)a'_j$ e.g., $\frac{1}{2} d_{p,j}a'_j$. 

2. We consider the different non-separated qubits with the numbers $q\neq p$. In view of independency of the time instants on the fulfillment of NOT- operators on $p$м and $q$-th qubits and the high density $\la$, these qubits will be in each state ($a'_p ,\ a'_q$), ($a'_p ,\ 1-a'_q$), ($1-a'_p , 
\ a'_q$), ($1-a'_p ,\ 1-a'_q$) approximately the quarter of the all time. The resulting deposit will be $\frac{1}{4}d_{p,q} [a'_p 
a'_q +a'_p (1-a'_q )+(1-a'_p )a'_q +(1-a'_p )(1-a'_q )]$ $=\frac{1}{4}d_{p,q}$. The common phase shift coming from the presence of non separated qubits is found by the summing of the values from the points 1 and 2 for all $p\notin\{j,k\}$. It will be 
$$
\frac{1}{2}[\sum\limits_{p\notin\{j,k\}}d_{p,j}a'_j +\sum\limits_{p\notin\{j,k\}}d_{p,k}a'_k ]+\frac{1}{4}\sum
\limits_{p,q\notin\{j,k\}} d_{p,q}.
 $$

 This shift can be compensated by only one qubit gates because the first two summands depend on the values of qubits only, and the other are constants. We thus obtain the scheme with the continuous two qubit interaction and one qubit operations which fulfills the appropriate phase shift to $d_{j,k} a'_j a'_k$. 

If we take the time segment $\D t$ instead of the unit time in this procedure, we obtain the phase shift to $\D t\ d_{j,k} 
a'_j a'_k$. If we want to obtain the phase shift to $-\D t\ d_{j,k}a'_j a'_k$, we must at first apply NOT to $j$-th qubit, then the previous procedure, then again NOT to $j$-th qubit, and at last, add $-\D t\ d_{j,k}a'_k$ by the one qubit operation. 

Therefore, we can do any addition to the phase of the form $c\cdot a'_j a'_k$ for a real $c$ independently of its sign. The appropriate combination of these schemes gives the phase shift 
\begin{equation} \sum_{j,k} c_{j,k} a'_j a'_k \label{Phase} 
\end{equation} 
for any $c_{j,k}$. Disposing these operations before and after $QFT^{-1}$ in the procedure of the previous point, we compensate the summands $A$ and $B$ in the phase and obtain the scheme realizing $QFT^{-1}$. The errors appearing in this scheme come from the possible low quality of Poisson processes generating the moments of the fulfillment of  NOT operations, and from the interaction in course of these operations. They can be minimized by the increasing of the density $\la$ and the decreasing of the time of NOT operations comparatively with the typical time of two qubit transformations determined by $d_{j,k}$. 

We evaluate the slowdown induced by the insertions of NOTs with the high density in comparison with the abstract realization of quantum computations of quantum gate arrays. We fix the unit of time such that the application of one operation in the scheme requires the unit time. 
Let the time axes be divided to the equal short intervals of the length $\d t$ units, NOT-th can be applied in the moments of the form $k \d t$ only, for any integer $k$ with the probability $p=1/\la$, where $\la$ in the density of process.

 Let the time of the whole computation equal $T$, and $M=T/\d t$. The error in the phase shift coming from the low quality of this model of random process, will be $\d t\ D$ where $D$ is the dispersion of the sum of random variables taking values 1 and 0 with the probabilities $p$ and $1-p$ that is $O(\sqrt{M})$. Consequently, the resulting error will be of the order $T/\sqrt{M}$ and must be negligible. For QFT we have $T=O(\log N)$ and we obtain that $M=O(\frac{\log^2 N}{\e} )$ is sufficient for the negligible $\e$. We see that the method of random processes gives a bit more than the quadratic slowdown comparatively to the standard abstract model that is sufficient for so fast quantum algorithms as Shor factoring.

\n 
We now prove the universality of the proposed model of quantum computations. We suppose that the interaction between qubits depends on their spatial positions only that we set fixed. The single condition we impose to the interaction is that it must be diagonal. Thus if $j$ and $k$ denote the number of two qubits, Hamiltonian of their interaction has one of the forms
 \begin{equation} {\rm A)}\ H_{j,k}=\left( \begin{array}
{ccccc} &E^{j,k}_1 &0 &0 &0\\ &0 &E^{j,k}_2 &0 &0\\ &0 &0 &E^{j,k}_3 &0\\ &0 &0 &0 &E^{j,k}_4 \end{array} \right) , \ \ \ 
\ \ \ \ \ \ \ \ \ \ {\rm B)}\ H_{j,k}=\left( \begin{array}{ccccc} &0 &0 &0 &0\\ &0 &0 &0 &0\\ &0 &0 &0 &0\\ &0 &0 &0 &E_
{j,k} \end{array} \right) ,\ \ E_{j,k} >0. \label{Hamiltonian} 
\end{equation} 

At first we note that any interaction on the general form (\ref{Ham}, A) can be reduced to the form (\ref{Ham}, B) by the addition of the proper one qubit Hamiltonians $H'_{j,k}$, which matrices have the forms
 $$ \left( \begin{array}{ccccc} &a &0 &0 &0\\ &0 &a &0 &0\\ &0 &0 &b &0\\ &0 &0 &0 &b
\\ \end{array} \right), \ \ \ \ \ \ \ \ \ \left( \begin{array}{ccccc} &\a &0 &0 &0\\ &0 &\b &0 &0\\ &0 &0 &\a &0\\ &0 &0 &0 
&\b\\ \end{array} \right) . 
$$ 

This addition reduces Hamiltonian of the form (\ref{Hamiltonian}, A) to  (\ref{Hamiltonian}, B) and it can be realized by the one qubit gates, since all diagonal matrices commutes. We note that the different pairs of qubits can interact differently, they can be placed on the different distances, not necessary on one line, etc. 

To prove the universality of the computational model with the continuous interaction we have to show how to fulfill an arbitrary two-qubit operation. Let we be given a unitary operation induced by Hamiltonian (\ref
{Hamiltonian}, B) in the unit time: $U_{j,k}=\exp (-iH_{j,k})$ (Plank constant we set equal unit, as usual). We show how to make this operation on two qubits:  $j$-th and $k$-th, preserving all the rest. If we can do it, we will be able to realize any two qubit operation on any pair of qubits. Then for the far interaction, we have at most the linear slowdown comparatively to the standard model, and for the short interaction, we have to fulfill SWAP operation to bring the required pair of qubits together. We thus obtain the multiplier to the time of computation proportional to the size of memory.

To make transformations $U_{j,k}$ it is needed to apply the method of NOT operations on non separated qubits described in the previous point, in moments of time generated by the independent Poisson processes of high density. However, now the advantage of this method is not as evident as in the case of QFT, because, for example, the fast quantum search requires more than logarithmic time: the square root of classical time. For such cases, one can apply the following modification of our trick.
\nn

{\bf Method of periodic NOTs} 
\nn

 We will make NOT operations on each of $j$ qubits in the time instants of the form $jk\d t$ for integer $k$, where $\d t$ is again the small period. We then can repeat the construction described above, and get rid of undesirable phase shifts by means of appropriate choice of $\d t$. This method gives the slowdown as the multiplier of the order $n^2$ comparatively to the complexity of the abstract model of quantum gate arrays. 

Now it is sufficient to show how by means of transformations $U_{j,k}$ we can make any two gubit gate. For example, we demonstrate how to realize CNOT operator on this pairs of qubits. Let $j,k$ be fixed and we omit indexes. We denote $\Delta E=E_1-E_2-E_3+E_4$. If $\frac{\Delta E}{\pi} \notin Q$ ($\frac{\Delta E}{\pi}$ not rational, then (because the physical parameters of our system affecting on phases, for example, cycle periods, can be slighly changed to avoid the rational parameter, we can treat it irrational without loss of generality) we can fulfil CNOT operation 
$$
 CNOT=\begin{pmatrix}1&0&0&0\cr0&1&0&0\cr0&0&0&1\cr0&0&1&0 \end{pmatrix}
$$ 
on the chosen pair of close qubit using only one qubit operators and the fixed diagonal operation $E$, where 
$ E=\begin{pmatrix}\exp{\left(i E_1\right)}&0&0&0\cr0&\exp{\left(i E_2\right)}&0&0\cr 0&0&\exp{\left(i E_3\right)}&0\cr0&0&0&
\exp{\left(i E_4\right)} \end{pmatrix}$ by the following way.  

I. We denote the sequence of rotations of the phase of the first qubit by 

$ A=
\begin{pmatrix}1&0\cr0&\exp{\left(i \left(E_1 -E_3\right)\right)}\end{pmatrix}$, of the second qubit by $ B=\begin{pmatrix}\exp\left(- i E_1
\right)&0\cr 0&\exp\left(- i E_2\right) ,\end{pmatrix}$ and the operation  $ U=E \, (A \bigotimes B) = \begin{pmatrix}1&0&0&0\cr0
&1&0&0\cr 0&0&1&0\cr0&0&0&\exp\left( i \Delta E\right) . \end{pmatrix}$ 		

II. Using the irrationality of $\frac{\Delta E}{\pi}$ it is possible to show that $ \forall \varepsilon > 0 \exists m \in N \exists n \in N : |\Delta E n - \pi (2 m + 1)| < \varepsilon , $ 
e.g., for any chosen accuracy $\varepsilon$ there exists $n=n(\varepsilon)$ such that $U^n$ approximates the operator $\Pi
$ where $ \Pi = \begin{pmatrix}1&0&0&0\cr0&1&0&0\cr 0&0&1&0\cr0&0&0&-1 \end{pmatrix}$ within the given accuracy. 

 III. Using the equality $ (I 
\bigotimes H) \Pi (I \bigotimes H) = CNOT, $ where $I$ is the identity matrix and $H$ - 
is Hadamard operation $ H=\frac{1}{\sqrt{2}}\begin{pmatrix}1&1\cr1&-1 \end{pmatrix}$ we see that CNOT is obtained as the sequence 
$ (I \bigotimes H) \left(E \, (A \bigotimes B)\right)^n (I \bigotimes H) $ of one qubit rotations and operation E. 

\subsection{Formalism of occupation numbers}

We consider the system consisting of $n$ identical particles. At first, we make the not physical supposition that they can be certainly distinguished. Then the state of such system belongs to Hilbert space with the basis  $\psi (r_1 , r_2,\ldots , r_n )$ $=\psi_{j_1} (r_1 )\psi_{j_2} (r_2 )\ldots\psi_{j_n} (r_n )$ where $\{\psi_j \}$ is some basis for the one particle states, $j_s$ belongs to some fixed set of indexes $1,2,\ldots ,J$, so that $r_j$ includes spatial and the so-called spin coordinates as well. 
The choice of basis means simply that the system can be found only in some of basic states after the observation. 

However, in the real system of identical particles they cannot be distinguished. Therefore, each basic state must contain all summands of the form $\psi_{j_1} (r_1 )\psi_{j_2} (r_2) )\ldots\psi_{j_n} (r_n )$ with some coefficients. Now we need some information about the nature of the considered particles. They can be fermions, like electrons or protons, or bosons (as photons). The difference between these two types of particles is that the maximal value of the fermionic spin is half integer (1/2, 3/2, ...) and for bosons, it is integer (0,1,2,...). 

For us it is significant that the wave function of the system of fermions must change its sign in the permutation of two particles, for bosons, the sign is preserved. Algebraic correspondence we established between functional notations and qubit formalism dictates the representation of the wave function for the system of $n$ fermions in the form of the determinant:  
\begin{equation}
\Psi=\frac{1}{\sqrt{N!}}\left|
\begin{array}{ccccc}
&\psi_{j_1} (r_1 ) &\psi_{j_1} (r_2)&\ldots &\psi_{j_1} (r_n )\\
&\vdots &\vdots &\vdots &\vdots\\
&\psi_{j_n }(r_1 ) &\psi_{j_n} (r_2 ) &\ldots &\psi_{j_n} (r_n )\\
\end{array}
\right| ,
\label{state}
\end{equation}
and for the system of bosons in the form of the corresponding permanent.

The permanent of matrix differs from its determinant only in that there are pluses instead of minuses in its computation, so that it remains unchanged in the permutations of rows of columns.

 Such state we can treat as the situation when only the states $\psi_{j_s}$ for $s=1,2,\ldots ,n$ are occupied by particles of our system, whereas the rest   $\psi_k$ for $k\in\{ 1,2,\ldots , J\}$, which have not the form $j_s$ are free. If $\psi$ with indices denotes an eigenvector of the one particle Hamiltonian we speak about the occupied or free energetic levels, but in general, $\psi_k$ can form the arbitrary orthonormal basis in the space of one particle states. 

The state of the form (\ref{state}) can be represented as the symbol $|\bar n_\Psi\rangle =|n_1 ,n_2 ,\ldots ,n_J \rangle$ where 
$n_k$ is the unit, if $k$-th energetic level is occupied and zero, if it is free. It is the natural representation of the fermionic ensemble state in terms of occupation numbers. Such vectors $\bar n$ form the basis of Fock space and the general form of a state of our system in this basis is $\sum\limits_{\bar n}\la_{\bar n}|\bar n\rangle$ with amplitudes $\la$. 

The operator of annihilation $a_j$ of the particle in the state $j$ ($j$-th energy level) and its conjugate operator $a_j^+$ 
(creation of the particle in this state), is defined as  
$a_j |n_1 ,\ldots ,n_J \rangle =\d_{1,n_j}(-1)^{\s_j}| n_1 ,\ldots
 ,n_{j-1},n_j -1,n_{j+1},\ldots ,n_J \rangle$ 
where $\s_j =n_1 +\ldots +n_j$. They possess the known commutative relations: 
$a_j^+ a_k +a_k a_j^+ =\d_{j,k}$, $a_j a_k +a_k a_j=a_j^+a_k^++a_k^+a_j^+=0$.

Let us suppose that any interaction in Nature touches no more than two particles. Each interaction in many body ensemble then can be decomposed to the sum of one - two particle interactions of the form $H=H_{one}+H_{two}$ with the corresponding potentials $V_1 (r)$ и $V_2 (r,r')$. Each of them can be represented through the operators of creation and annihilation in the form  
$H_{one}=\sum\limits_{k,l}H_{k,l} a^+_k a_l$, $\ \ H_{two}=
\sum\limits_{k,l,m,n}H_{k,l,m,n}a^+_l a^+_k a_m a_n$ где 
$$
\begin{array}{ll}
&H_{k,l}=\langle\psi_k |\ H_{one}\ |\psi_l \rangle =\int \psi^*_k (r) V_1 (r)\psi_l (r)dr,
\\
&H_{k,l,m,n}=\langle\psi_l ,\psi_k\ |H_{two}\ |\ \psi_m \psi_n \rangle
 =\int\psi^*_k (r)\psi^*_l (r') V_2 (r,r')\psi_m (r)\psi_n (r') drdr'.
\end{array}
$$
Hence, given potentials of all interactions and all basic states $\psi_i$, we can in principle find their representation in terms of operators of creation and annihilation, that is in the language of occupation numbers. 

We consider the ensemble with Hamiltonian of the form $H=\sum_i H^i_{ext. f.}+\sum_{i,j}
(H^{i,j}_{diag.}+H^{i,j}_{tun.})$, where Hamiltonians of external fields, diagonal interaction and tunneling are represented in terms of creation and annihilation operators as 
\begin{equation}
\begin{array}{lll}
&H^i_{ext.f.} &= \a_i a^+_i a_i ,\ \ \ \ \a_i\in{\rm R},\\
&H^{i,j}_{diag.} &= \b_{i,j} a^+_i a_i a^+_j a_j ,\ \ \ \ \b_{i,j}\in{\rm R},\\
&H^{i,j}_{tun.} &= \g_{i,j} a^+_i a_j +\g_{i,j}^* a^+_j a_i .
\end{array}
\label{ham}
\end{equation}

We note that to realize the control on the diagonal Hamiltonian would be difficult, because this interaction touches two arbitrary particles in the considered ensemble, which are non-distinguishable by the identity principle. It is thus natural to treat this interaction as constant and independent from our control, whereas we can effectively control the tunneling interaction. This form of control makes possible to realize any quantum computation. This type of the control looks as more realistic because we can realize the tunneling by means of laser impulses. 

\subsection{Computations controlled by tunneling}

To prove the universality of the proposed simplified scheme of control on fermionic computations we must make one technical preparation, namely, to establish some different correspondence between Hilbert space of qubits and Fock space of occupation numbers. This correspondence will be different from the natural correspondence we spoke earlier. 

We fix some division of the set of energy levels to two parts and choose some one-to-one correspondence between these parts. For the determinacy we can take the $k$-th level down from Fermi level $\epsilon_F$ and agree that it corresponds to the $k$-th level up from $\epsilon_F$. We denote $j$-th level don from Fermi border by the standard letter, and the $j$-th level up from this level by this letter with the stroke $j'$. We call the first level the $j$-th the lower level and the second level the $j$-th upper level. Fock space $\cal F$ can be then represented as 

$$
{\cal F}={\cal F}_1 \bigotimes {\cal F}_2 
\bigotimes\ldots\bigotimes {\cal F}_k
$$

 where each ${\cal F}_j$ corresponds to $j$-th pair of corresponding energy levels. We consider the subspace $F_j$ in ${\cal F}_ j$, which is generated by two following vectors. The first will be: "$j'$-th level is occupied, $j$-th free", the second will be: "$j$-th level is occupied, $j'$-th is free". 
We denote them by $|1\rangle_j$ and $|0\rangle_j$ correspondingly. 

We will work with the subspace 
$$
F=F_1 \bigotimes F_2 \bigotimes\ldots\bigotimes F_k
$$

 in Fock space ${\cal F}$. We define the function $\theta$, which maps our Hilbert space $\cal H$ to $F$ by the following definition on basic states: 
$$
\theta (|\xi_1 ,\xi_2 \ldots\xi_n \rangle )=|\xi_1 \rangle_1 \bigotimes |\xi_2 \rangle_2 \bigotimes\ldots\bigotimes |\xi_n \rangle_n
$$
 where all 
$\xi_j$ are zeroes and units. Then the function $\theta$ establishes the non-standard correspondence between Hilbert and Fock spaces (see the picture A2).

One qubit state in Hilbert space corresponds to two-qubit state in the usual identification with qubits (one level - one qubit).
We will see that this identification better fits to our aims than the natural. Now all is ready for the representation of unitary operators in Hilbert space in terms of operators acting in the space of occupation numbers. We consider an arbitrary Hermitian operator $H$ in two-dimensional Hilbert space of one qubit states $\cal H$. It has the form $H_0
+H_1$, where
$$
H_0 =\left(
\begin{array}{ccc}
&d_1 &0\\
&0 &d_2
\end{array}
\right) ,
H_1=\left(
\begin{array}{ccc}
&0 &d\\
&\bar d &0
\end{array}
\right) .
$$

\setlength{\unitlength}{0.6pt}
\begin{picture}(720,420)(0,90)
\put(100,500){$\xi\in{\cal H}$}
\put(247,500){$|0\rangle_2$}
\put(332,500){$|1\rangle_2$}
\put(520,500){$|010\rangle$}
\multiput(200,230)(0,36){7}{\multiput(0,0)(17,0){10}{\line(1,0){7}}\multiput(260,0)(15,0){14}{\line(1,0){5}}}
\put(255,410){\circle{10}}
\put(340,266){\circle{10}}
\put(500,374){\circle{10}}
\put(536,266){\circle{10}}
\put(572,446){\circle{10}}
\put(340,410){\circle*{10}}
\put(500,302){\circle*{10}}
\put(536,410){\circle*{10}}
\put(255,266){\circle*{10}}
\put(572,230){\circle*{10}}
\put(100,338){$\theta (\xi )$}
\put(200,338){\line(1,0){170}}
\put(460,338){\line(1,0){215}}
\put(700,338){$\epsilon_F$}
\put(150,230){$3$}
\put(150,266){$2$}
\put(150,302){$1$}
\put(150,338){$0$}
\put(150,374){$1'$}
\put(150,410){$2'$}
\put(150,446){$3'$}
\put(0,150){Picture A2. \ \ Correspondence between Fock and Hilbert spaces}
\end{picture}

It can be straightforwardly verified that for operators $\tilde H_0 = d_1 a^+_k 
a_k +d_2 a^+_{k'} a_{k'}$ and $\tilde H_1 = d a^+_k a_{k'} +\bar d a^+_{k'} a_k$ 
(external field and tunneling) the following equalities take place: $ \tilde H_i \theta 
=\theta H_i$ for $i=0,1$. Using the linearity $\theta$, we find $(\tilde H_0 
+\tilde H_1 )\theta =\theta H$.
Now we consider one qubit unitary operator $U$ in Hilbert space. It has the form $e^{-iH}$ for Hamiltonian $H$ (we have chosen the appropriate unit system to get rid of Plank constant and the time). Due to the linearity of $\theta$ and the equation $\theta^{-1} H^s \theta =(\theta ^{-1}H\theta )^s$ for integer $s$ we find that for any one qubit unitary operator $U$ we can effectively find the corresponding Hamiltonian in Fock space containing only the external field and the tunneling, which makes the diagram A from the picture A3 closed.

We take up two qubit transformations in Hilbert space. Since all diagonal matrices commute, for all diagonal transformations in the spaces ${\cal F}_k 
\bigotimes {\cal F}_j$ м can effectively find the corresponding diagonal operator in Hilbert space, which makes the diagram B from the picture A3 closed.

Now all is ready for the transfer of the trick from the work \cite{OF} with one qubit control to Fock space. The combination of diagrams from the picture A3 gives the diagram from the picture A4.

\begin{picture}(700,380)(0,150)
\put(64,330){\vector(1,0){256}}
\put(424,330){\vector(1,0){266}}
\put(64,220){\vector(1,0){256}}
\put(424,220){\vector(1,0){266}}
\put(50,234){\vector(0,1){86}}
\put(330,234){\vector(0,1){86}}
\put(410,234){\vector(0,1){86}}
\put(700,234){\vector(0,1){86}}
\put(37,275){$\theta$}
\put(317,275){$\theta$}
\put(397,275){$\theta$}
\put(687,275){$\theta$}
\put(43,213){$\cal H$}
\put(323,213){$\cal H$}
\put(403,213){$\cal H$}
\put(693,213){$\cal H$}
\put(43,323){$\tilde F$}
\put(323,323){$\tilde F$}
\put(403,323){$\tilde F$}
\put(693,323){$\tilde F$}
\put(80,340){ext. field + tunneling}
\put(480,340){$\tilde F
$ diagonal}
\put(80,200){one-qubit on $\cal H$}
\put(480,200){$\cal H$ diagonal}
\put(150,150){A}
\put(550,150){B}
\put(120,90){Picture A3.\ \ Correspondence of operators in }
\put(180, 65){Fock and Hilbert subspaces. $\tilde F=F_j \bigotimes F_k .$}
\end{picture}

\begin{picture}(700,490)
\put(90,320){$\cal F$ diag}
\put(250,320){f+t}
\put(580,320){$\cal F$ diag}
\put(90,130){diag}
\put(250,130){one qubit}
\put(580,130){diag}
\put(40,300){\vector(1,0){125}}
\put(220,300){\vector(1,0){125}}
\put(550,300){\vector(1,0){125}}
\put(40,150){\vector(1,0){125}}
\put(220,150){\vector(1,0){125}}
\put(550,150){\vector(1,0){125}}
\put(30,165){\vector(0,1){115}}
\put(197,165){\vector(0,1){115}}
\put(700,165){\vector(0,1){115}}
\put(390,300){.\ .\ .}
\put(390,150){.\ .\ .}
\put(20,297){F}
\put(190,297){F}
\put(693,297){F}
\put(20,142){$\cal H$}
\put(187,142){$\cal H$}
\put(685,142){$\cal H$}
\put(100,80){Picture A4.\ \ Correspondence of computations in}
\put(180,55){Fock and Hilbert spaces}
\end{picture}

Let the diagonal part of Hamiltonian of interaction in Fock space be fixed and act permanently in the non-controlled mode. We then can find the corresponding diagonal interaction in Hilbert space, making closed the "diagonal" parts of diagrams from the picture A4. By virtue of the result of theh work \cite{OF} we can find one-qubit transformations realizing the control on the arbitrary quantum algorithm in Hilbert space, in the form of the lower sequence of transformations in the diagram. Al last, we can find the control of the form "field + tunneling" on the state in Fock space making closed the whole diagram. We note that all operators of creation and annihilation considered in the whole Fock space are non-local due to the multiplier $(-1)^{\s_j}$, which depends on a given state.   

For the diagonal operator $a_j^+ a_j a_k^+ a_k$ and the external field these multipliers compensate each other. The tunneling operator $a_j^+ a_{j'}$ in the space $F$ brings the multiplier $(-1)^{\s'}$ where $\s'=\sum\limits_{s+j}^{j'-1}n_s =j'-j$, which does not depend on the given state $|\bar n\rangle\in F$, because for such state exactly the half of levels between $j$ and $j'$ are occupied by fermions. The sign we can factorize from all states, and ignore.

We thus obtain the universal quantum computer on states in the space of occupation numbers, controlled by the external field and the tunneling only.

\section{Lecture 8. Implementation of quantum computing on optical cavities}

The physical implementation of quantum computing is a separate big topic that requires a physical foundation for building gates. Here we give an example of such an implementation on atomic excitations in optical cavities. The physical part is described in the framework of finite-dimensional models of quantum electrodynamics in optical cavities - the Jaynes-Cummings-Hubbard model.

\subsection{The Jaynes-Cummings Model}{\label{JC}}

The difficulty of experimental accounting for the electromagnetic field is that its excitation quanta travel at the speed of light, so that as soon as a photon appears, it will cover most of the distance from the Earth to the Moon in a second. The method of photon retention is ideologically simple: it is necessary to place mirrors reflecting the photon opposite each other so that it runs between them and does not fly far away for a sufficiently long time. 

\begin{figure}[h]
\centering
\includegraphics[scale=0.7]{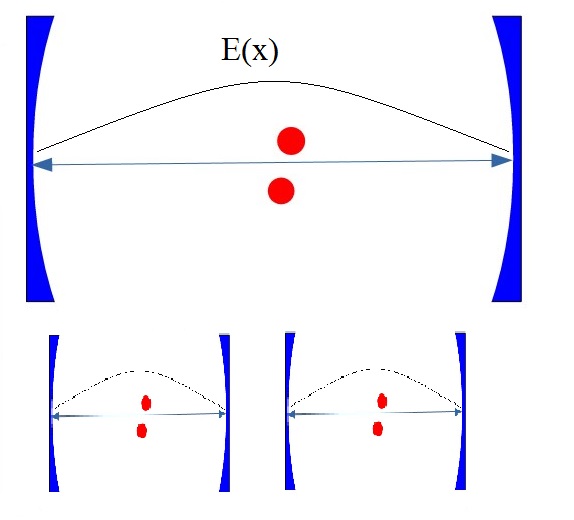}
 \caption{Optical cavity with atoms}
\label{fig:cavity}
\end{figure}

This device is called a Fabry-Perot interferometer (see Figure \ref{fig:cavity}). Two mirrors form a kind of cavity, or resonator, into which a photon can be launched with a laser, and extracted with a mirror with variable reflectivity; such a mirror is called a Pockels cell. A voltage can be applied to the cell, then it will begin to reflect the photon that has fallen on it; when the voltage is turned off, it becomes transparent and the photon passes through it freely. The side walls of the cavity are also made of a light-reflecting material. 

\begin{figure}[h]
\centering
\includegraphics[scale=0.95]{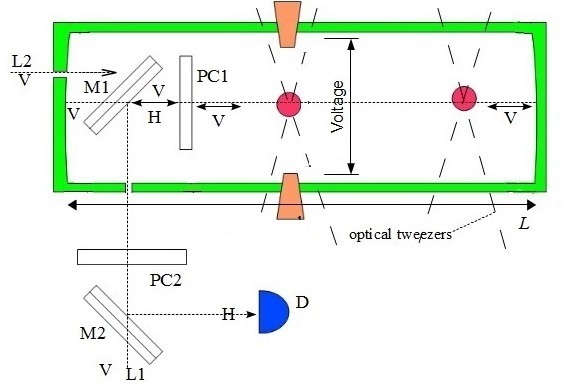}
 \caption{Manipulation with atoms in optical cavity}
\label{fig:resonator1}
\end{figure}

Possible operation with atoms in the cavity is shown at the Figure \ref{fig:resonator1}.

If an atom is placed in such an optical cavity, it will be able to interact with the field inside the cavity, if for some energy levels of its electrons, which we denote by $ |0 \rangle$ - conditionally basic, and $|1\rangle$ - conditionally excited levels, the transition energy between these levels $ \Delta E=\hbar\w$ is such that $\w$ very accurately approximates the frequency of a photon in the cavity. At the same time, when a photon is absorbed, the atom passes from the ground state $|0\rangle$ to the excited $|1\rangle$, and vice versa, during the reverse transition of the atom, a photon is emitted. 

The complete cycle of interaction of an atom with a field, including the absorption of a photon by an atom and its subsequent emission, is called a Rabi oscillation. In order for one such oscillation to occur, for example, in the rubidium atom $Rb^{85}$, where the transition frequency between the levels is approximately $\w_{Rb}\approx 10^{10}\ sec^{-1}$, the photon must be held in the cavity for a long enough time, so that during this time it manages to be reflected from the mirrors several tens of thousands of times. Therefore, the mirrors must be of very high quality; they are made of a superconducting material, for example, niobium, and they function at a very low temperature of liquid helium.

But the quality of mirrors is not enough, since a photon can leak out of the cavity due to its interference nature. This nature follows from the principle of interference, which we briefly described in the first chapter. I recommend the reader to refer to the detailed explanation of the operation of this principle in relation to photons, given in R. Feynman's book \cite{Fe1}.

In order for the photon to remain in the cavity for a long time, it is necessary that the electric field created by it constructively interferes with itself inside the cavity, and destructively-outside it. This is provided by the length of the cavity - the distance between the mirrors $L$. The length must be a multiple of the half-wave of the photon length, that is, $L=k\la/2$, where $\la=2\pi c/\w$, $k=1,2,...$.

An atom placed in the cavity interacts with a field with an interaction energy $g$, which is calculated by the formula
\begin{equation}
\label{fieldatom}
g=\sqrt{\frac{\hbar\w}{V}}dE(x),
\end{equation}
where $V$ is the effective volume of the cavity (the volume where the photo is present), $d$ is the absolute value of the $\bar d $ dipole transition between $|0 \rangle$ and $|1\rangle$, $E (x)$ is the factor of the location of the atom inside the cavity. The goal is to make $g$ as large as possible, for the fastest possible manifestation of the properties of the interaction of light and matter. Therefore, the length of the cavity should be chosen so that $k=1$, and $L=\la/2$. In this case, the constructive interference of the photon's electric field inside the cavity is maximal, and its intensity is distributed along the sinusoid, so that $E (x)=sin (\pi x)/L$.

The dipole transition is calculated by the formula $\bar d = e\int\limits_{R^3}\psi_0^*\bar r\psi_1d\bar r$, where $\psi_0,\ \psi_1$ are the wave functions of the electron states in the ground and excited states inside the atom, depending on the three-dimensional vector $\bar r$, $e$ - electron charge (we suppose that the transition between atomic states is connected with one electron). 

The actual output of the formula \eqref{fieldatom} can be found in the book \cite{FeH}. The constant $g$, generally speaking, is complex, but it is possible by multiplying the basis vector $|1 \rangle$ by a suitable complex number $e^{i\phi}$ to achieve that $g$ is a real non-negative number, which we will assume in the future.

Thus, the interaction of the field with the atom inside the cavity is the interaction of the atom with the quantum harmonic oscillator described in the Appendix of \cite{Oz0}. The conditional "coordinate" of the $ x $ field with an accuracy of constants is expressed in terms of the operators of the birth of $a^+$ and the destruction of $a$ of a photon in the field as $x=a^++a$.

We introduce, similarly to field operators, the operators $\s^+,\s$ - atomic excitation and relaxation operators; we get a set of field operators and atoms of the form
\begin{equation}
\label{operators}
\begin{array}{lll}
&a:\ |n\rangle_{ph}\rightarrow \sqrt{n}|n-1\rangle_{ph}, &a^+: |n\rangle_{ph}\rightarrow \sqrt{n+1}|n+1\rangle_{ph},\\
&\s:\ |0\rangle_{at}\rightarrow 0,\ &|1\rangle_{at}\rightarrow |0\rangle,\\
&\s^+:\ |0\rangle_{at}\rightarrow |1\rangle_{at},\ &|1\rangle_{at}\rightarrow 0,
\end{array}
\end{equation}
so the number of photons $n=0,1,2,...$, and the number characterizing the atomic excitation takes only two values $0$ or $1$, and the relaxation $\s$ an atom already in the ground state $|0\rangle_{at}$ leads to the destruction of the state as such (zero in the state space), and the excitation of the already excited state of the atom gives the same result; in other cases, the operators act naturally.

We introduce a conditional "coordinate" of the atomic excitation $X$ by analogy with the field "coordinate": $X=\s^++\s$. Let the interaction of the field with the atom be denoted by $G (x,X)$, where $x,\ X$ are the conditional "coordinates" of the field and the atomic excitation, respectively. Decomposing this function into a Taylor series, we see that the lowest interaction term containing both coordinates has the form $gxX=g(a^++a)(\s^++\s)$. This is called the dipole approximation of the interaction of an atom and a field. It is valid if the size of the atom is significantly smaller than the wavelength of the photon; this assumption is fulfilled in most practically important cases, for example, in chemistry.

If we take into account the following terms in the Taylor expansion of the function \newline $G (x,X)$, we will get higher terms in the interaction approximation; we will not deal with them.

The eigenvalues of the atom and the field are given by the formulas $E_{at}=\hbar\w\s^+\s$, $E_{ph}=\hbar\w a^+a$. We omit the energy of the vacuum state $\hbar\omega/2$, since in this case it does not play a role. 

The listener is provided to check that adding a constant to the Hamiltonian, that is, the transition from $H$ to $H+cI$, leads only to the appearance of an additional phase multiplier of the form $e^{-ic t/\hbar}$ in solving the Schrodinger equation, which has no physical meaning and disappears when passing to the Schrodinger equation for the density matrix. Summing them with the interaction energy, we obtain the Jaynes-Cummings Hamiltonian for a two-level atom in an optical cavity:

\begin{equation}
\label{JC}
H_{JC}=\hbar\w a^+a+\hbar\w \s^+\s+g(a^++a)(\s^++\s).
\end{equation}

It is quite difficult to solve the Cauchy problem for the Schrodinger equation with such a Hamiltonian. The fact is that the interaction contains the terms $a\s$ and $a^+\s^+$. which individually do not save energy. This means that for a potentially infinite matrix of the operator $H_{JC}$ there are no finite-dimensional invariant subspaces, and one has to deal with infinities, which is difficult and essentially incorrect.

Fortunately, this difficulty is avoided for most applications where the interaction force $g$ is small compared to the excitation energy $\hbar\w$ of the atom. If $g/\bar\w\ll 1$, the non-energy-conserving terms can be discarded, and the Hamiltonian will take a much more convenient form

\begin{equation}
\label{TC}
H_{JC}^{RWA}=\hbar\w a^+a+\hbar\w \s^+\s+g(a^+\s+a\s^+).
\end{equation}
This is the so-called rotating wave approximation RWA, its output can be found in the suplementing materials.

Our physical system is composite. It consists of two parts: a field and an atom. Let's agree to denote the basic states by writing out first the number of photons in the field, and then the atomic excitation: $|n,m \rangle$, so that $n=0,1,2,...$, $m=0,1$, and omit the subscripts $ph$ and $at$. When writing operators, we usually accept the following agreement: if an operator acting on another element of the composite system is not specified, it is assumed to be an identical operator: $I_{at}$ or $I_{ph}$. Thus, for example, the entry $a^+a$ should be interpreted as $a^+a\otimes I_{at}$, and the entry $a\s^+$ - either as $a\otimes\s^+$, or as a matrix product of $a\otimes I_{at}\cdot I_{ph}\otimes\s^+$. Check that both paths give the same result.

For the Hamiltonian $H_{JC}^{RWA}$, the space of quantum states decays into a direct sum of invariant two-dimensional subspaces ${\cal H}_n$, each corresponding to the energy $E_n=\hbar\w n$, and is generated by the vectors $|n, 0 \rangle,\ |n-1,1\rangle$. The Hamiltonian bounded by ${\cal H}_n$ has the form
\begin{equation}
\label{2_}
H_n=\left(\begin{array}{lll}
&\hbar\w n&g\sqrt{n}\\
&g\sqrt{n}&\hbar\w n
\end{array}
\right).
\end{equation}

\begin{figure}
\centering
\includegraphics[scale=1.0]{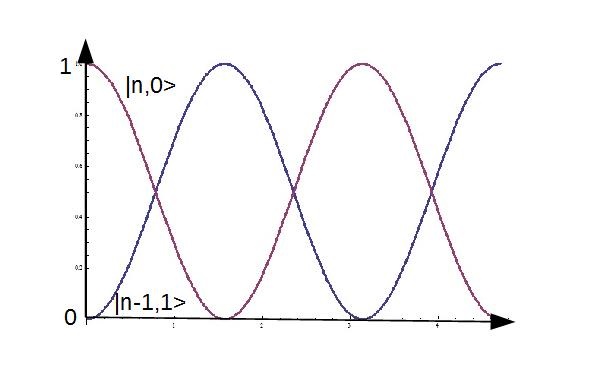}
 \caption{Rabi oscillations between populations of states $|n,0\rangle$ и $|n-1,1\rangle$.}
\label{fig:rabi}
\end{figure}

The expression \eqref{2_} says that the states $|n, 0 \rangle$ and $|n-1,1\rangle$ pass into one another during evolution, and their population changes according to a sinusoidal law (see Figure \ref{fig:rabi}).

From this it can be seen that the populations of the states $|n, 0 \rangle$ and $|n-1,1\rangle$ alternate, oscillating in the opposite phase. If,at one of the vertices of the population graph of the state $|n, 0\rangle$, photons are somehow extracted from the cavity (this is done using a special optical mirror-a Pokkels cell), then the atom will remain in the cavity in the ground state. This important remark will be useful to us in the future, when designing the coCSign quantum gate.

\subsection{The Tavis-Cummings-Hubbard model}

If we connect two cavities with a waveguide consisting of an optical fiber, along which photons from one cavity can move to another cavity, we get a more complex model, which is described by the Tavis-Cummings-Hubbard Hamiltonian of the form
\begin{equation}
\label{TCH}
H_{TCH}=\sum\limits_{i=1}^mH^i_{TC}+\sum\limits_{1\leq i<j\leq m}\mu_{ij}(a_i^+a_j+a_ia_j^+).
\end{equation}

\begin{figure}
\centering
\includegraphics[scale=0.7]{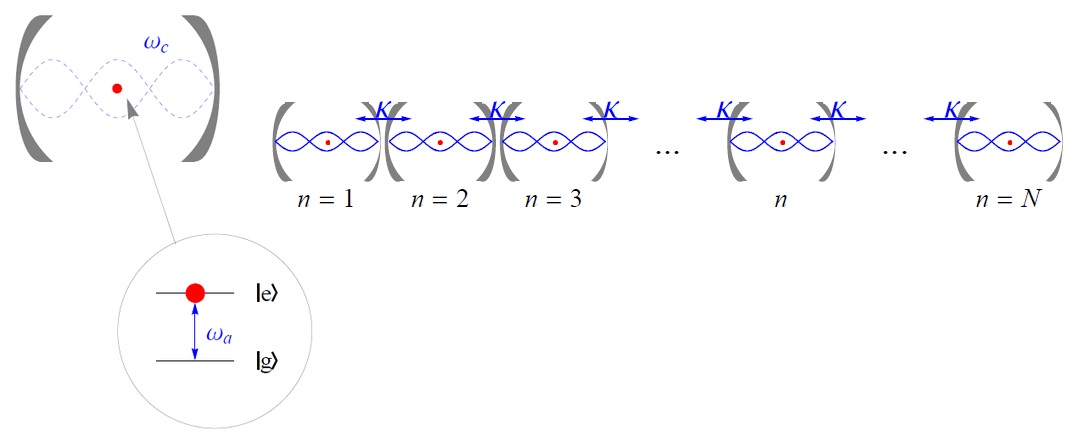}
 \caption{The Jaynes-Cummings-Hubbard model. The TCH model differs only in the presence of several atoms in the cavities}
\label{fig:JCH}
\end{figure}

This model is closer to reality than the single-band model, since it already allows the possibility for photons to become distinguishable by being in different cavities (see Figure \ref{fig:JCH}).

\subsection{Entangling gate in JCH model}

Quantum computing is an intrusion of quantum theory into the field of complex processes, where the operation of its basic laws has not yet been studied. Therefore, the construction of the simplest schemes of computations, in which the quantum laws would appear as clearly as possible, is an urgent task. The dark place here is decoherence, which occurs due to the interaction of charges and the field, the quanta of which closely connect the quantum computer with the environment. This makes it necessary to account for and control, or even explicitly use photons in quantum protocols.

Photons as information carriers make it possible to use linear optical devices to implement single-qubit gates, but the design of entangling operations is difficult, since photons do not directly interact with each other. There is a popular KLM scheme (see \cite{KLM}), where measurements are used as an interaction ersatz, and its improvement \cite{DC} with teleportation (see \cite{tele}), which significantly increases its efficiency, as well as a number of variants of this scheme for atoms (see, for example, \cite{Po}). However, the use of classical probabilistic schemes in practical implementation places increased requirements on the efficiency, at least theoretically, of quantum gates on single particles. The use of classical probability obscures the main question of a quantum computer: how does coherence work for complex systems of different particles?

The most basic methods are more suitable here, the main of which are optical cavities with several atoms, the interaction of which with a single-mode field is clearly described from the first principles (for the capabilities of this type of devices, see, for example, \cite{Re}). Thus, the CNOT gate was implemented using the external - vibrational-degrees of freedom of the atom (see \cite{Wi}).

 However, the essence of quantum computing is not the coherent behavior of a single qubit, but the scaling of a Feynman quantum processor that implements the theoretical possibilities of unitary dynamics in the entire Hilbert state space, giving, for example, the Grover \cite{Gr} algorithm on the same equipment as the Shor \cite{Sh} algorithm. The use of external factors to demonstrate the dynamics of individual atoms and the field is useful for individual atoms, but the inevitable interference introduced will certainly affect the scaling.

Therefore, the value is represented by gate implementation schemes that use minimal means, which are well described from the first principles. One of these schemes is proposed in the article X.Azums \cite{A}, where dual-rail states of single photons are used as qubits. In this scheme, the interaction of photons with atoms is used only to perform an entangling transformation $CSign$, which requires two optical cavities; two beam splitters and phase shifters are also required.

We will describe a simplification of the Azuma scheme, where only one cavity is used, and the beam splitters are replaced by a time shift for photons entering it. We will have asynchronous states of atoms in Rabi oscillations as logical cubes. This scheme can be modified for purely photonic carriers, with a time shift that determines the value of the qubit. However, atoms as information carriers have the advantage that they are much easier to control, as well as the photons they emit. 

The advantage of the proposed scheme is its simplicity. The disadvantage is the same as in the \cite{A} scheme-the dependence on the time accuracy of the operation of the Pokkels cell or its analog, the operating time of which must be made significantly less than the time of the Rabi oscillation of the atom in the cavity.

For technical reasons, we will implement the $coCSign gate:\ |x, y\rangle\rightarrow (-1)^{x(y\oplus 1)} |x, y\rangle$, changing sign at a single state $|10\rangle$, related to the $CSign $ gate, which is implemented in \cite{A}; there is no difference, since $coCSign=CSign\sigma_x (y)$, and single-qubit gates are implemented by linear optical devices.

\subsection{Calculation of phase shifts \label{pha}}

The core of this scheme is an optical cavity with one two-level atom with an energy gap $ \hbar\omega$ between the main $ |0\rangle$ and the excited $|1\rangle$ levels, where $\omega$ coincides with the frequency of a photon of a certain mode held in the cavity. The interaction constant $g$ between an atom and a field is assumed to be small: $g/\hbar\omega\ll 1$ (in practice, this ratio should be no more than $10^{-3}$) for the possibility of applying the RWA approximation, in which the Jaynes-Cummings Hamiltonian of the "atom+field" system (\cite{JC}) has the form
\begin{equation}
H=H_{JC}=H_0+H_{int};\ H_0=\hbar\omega a^+a+\hbar\omega\sigma^+\sigma,\ H_{int}=g(\a^+\sigma+a\sigma^+),
\label{HamJC}
\end{equation}
where $a, a^+$ are the operators of photon annihilation and creation, $\sigma,\sigma^+$ are the relaxation and excitation of the atom. We will write the basic states of the atom and the field in the form $|n\rangle_{ph}|m\rangle_{at}$, where $n=0,1,2,...$ is the number of photons in the cavity, $m=0,1$ is the number of atomic excitations. We will have $n=0,1,2$. We will consider several cavities, and supply the cavity operators $i$ with the subscript $i$, so that the total Hamiltonian will be equal to the sum of $ \sum\limits_iH_i$; the interaction of atoms with the field $H_{int}$ in all regions will then be equal, respectively, to the sum of $\sum\limits_iH_{int\ i}$.

The Hamiltonian will change during the execution of the $coCSign$ gate: a term of the form $H_{jump}=\nu (a_ia^+_j+a_ja^+_i)$ will be added to its term $H_{int}$, meaning the transition of a photon from the cavity $i$ to the cavity $j$ and vice versa, but the energy of independent atoms and the field $H_0$ will not change (the Jaynes-Cummings-Hubbard model JCH). Therefore, the phase raid associated with $H_0$ will be common to all states, and it can be ignored. 

Next, we will consider the phase raid relative to either the identical operator $I$, or relative to $\sigma_x$, since all the operations discussed below are reduced to either the first or the second with a phase change, so that the phase raid when applying, for example, $ - i\sigma_x$ will be $ - \frac{\pi}{2}$.

Leta $\tau_1=\pi\hbar/g,\ \tau_2=\pi\hbar/g\sqrt{2}$ -are periods of Rabi oscollations for the total energy $\hbar\omega$ and $2\hbar\omega$ correspondingly. Operators $U_t=e^{-\frac{i}{\hbar}Ht}$, induced by the evolution in the important time instants will depend on the total energy of the cavity. If it equals $\hbar\omega$, in the basis $|\phi_0\rangle=|1\rangle_{ph}|0\rangle_{at}, \ |\phi_1\rangle=|0\rangle_{ph}|1\rangle_{at}$, we have:

\begin{equation}
U_{\tau_1/2}=-i\sigma_x,\ U_{\tau_1}=-I,\ U_{2\tau_1}=I,
\label{phase_add}
\end{equation}
where $\sigma_x$ is the Pauli matrix, and similar relations with the replacement of $\tau_1$ by $\tau_2$ for the total energy of the cavity $2\hbar\omega$.

When a photon is moved from the cavity $j$ to the cavity $i$ and vice versa, which is realized by the simultaneous inclusion of Pokkels cells or similar devices in these cavities, the addition of $H_{jump}$ to the interaction of $H_{int}$ is realized, which, in the absence of atoms in the cavities, implements exactly the same dynamics as the Rabie oscillations,but with the period $\tau_{jump}=\pi\hbar/\nu_{i, j}$. We will assume that $ \nu\gg g$, so that it is possible to move a photon from a cavity to a cavity so that the atom does not affect this process at all, so that the phase incursion can be calculated using formulas similar to \eqref{phase_add}. 

As noted in \cite{A}, this is difficult to implement in an experiment, but there are reasons to consider it a technical difficulty. If this condition is met, the phase gain for the $\sigma_x$ operator applied to the photons of the two cavities will be $-\pi/2$, as well as for half of the Rabie oscillation.

Due to the incommensurability of the periods of the Rabi oscillations $\tau_1$ and $\tau_2$ we can choose such natural numbers $n_1$ and $n_2$ that the approximate equality will be fulfilled with high accuracy

\begin{equation}
2n_2\tau_2\approx 2n_1\tau_1+\frac{\tau_1}{2},
\label{noncommon}
\end{equation}
which will be the basis for the nonlinear phase shift required for the implementation of $coCSing$.

\subsection{Implementation of coCSign}

The state of the qubit $|0\rangle$ is realized in our model as the state of the optical cavity of the form $|0\rangle_{ph}|1\rangle_{at}$, and the state of the qubit $|1\rangle$ is realized as $|1\rangle_{ph}|0\rangle_{at}$. Thus, for the state $|10 \ rangle$, which needs to invert the phase, has the form $|10\rangle_{ph}|01\rangle_{at}$, where the first photon qubit belongs to the cavity $x$, and the second to the cavity $y$. Note that after a time $\tau_1/2$, zero and one change places with the phase raid, which is included in $H_0$, and therefore is ignored.

\begin{figure}
\centering
\includegraphics[scale=0.60]{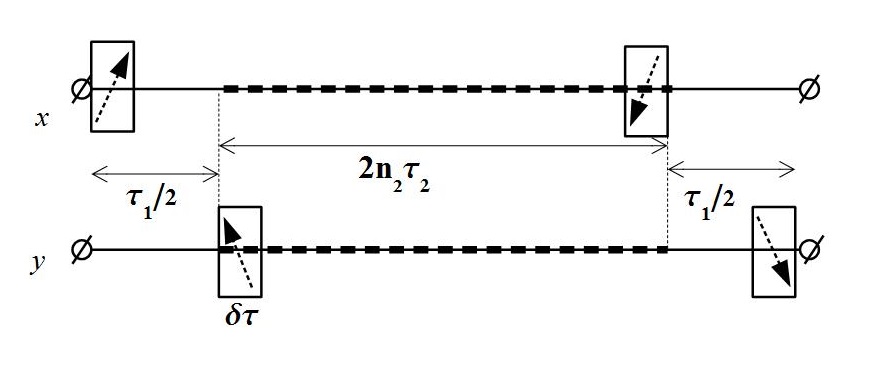}
\caption{The sequence of operations when implementing the com Sign gate: $|x, y\rangle\rightarrow (-1)^{x(y\oplus 1)}|x, y\rangle$ on asynchronous atomic excitations in optical cavities, $\delta\tau=\tau_{jump}/2$}
\label{fig:1}
\end{figure}

\begin{figure}
\centering
\includegraphics[scale=0.60]{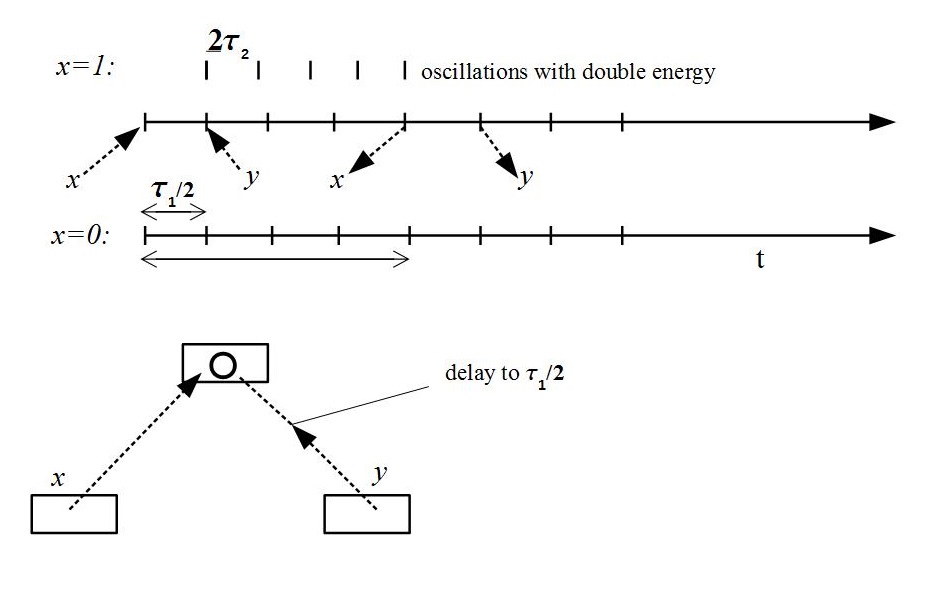}
\caption{Implementation of the gate coCSign}
\label{fig:coCSign}
\end{figure}

The sequence of operations implementing $coCSign$ is shown in Figure \ref{fig:1}, and the participating cavities are shown in Figure \ref{fig:coCSign}. First, we launch a photon from the cavity $x$ into an auxiliary cavity with an atom in the ground state and a vacuum state of the field, then, with a delay of $ \tau_1/2$, a photon from the cavity $y$, then, after a time of $2n_2\tau_2$, we move a photon from the auxiliary cavity to the cavity $x$, then, after a time of $\tau_1/2$, we move a photon from the auxiliary cavity to the cavity $y$. It follows from our choice of photon travel times that at these moments there will be either one photon or none in the participating cavities, so the inclusion of Pokkels cells on a small time interval $ \delta \tau=\pi\hbar/2\nu\ll \tau_1$ will give exactly the movement of photons.

It follows from the previous calculations that at the energy of the central cavity $ \hbar\omega$ (initial states $|00\rangle,\ |11\rangle$), the phase incursion during the photon transfer back and forth will be $-\pi$, and in interaction with the atom $ - \pi$, so the total phase incursion will be zero, as in the case of zero energy of the central cavity (initial state $|01\rangle$). For the energy $2\hbar\omega$ - in the case of $x=1, y=0$, the transfer of two photons will give zero, and the interaction will give $-\frac{\pi}{2}-\frac{\pi}{2}=-\pi$, which was required.

In the article \cite{SOL}, a calculation is given, from which it follows that to achieve satisfactory accuracy of such entangling gates on nonlinearity in cavities, it is enough to take the numbers of incommensurable periods $n_1, n_2$, not exceeding several tens, which corresponds to the number of observed Rabi oscillations in optical cavities.

\subsection{Implementation of single-qubit gates}

For quantum computing, in addition to the coCSign entangling gate, one-qubit gates are also needed. We will show how two gates can be implemented: the phase rotator $|x\rangle\rightarrow e^{i\phi x}|x\rangle$ and the Hadamard operator $H: |x\rangle\rightarrow \frac{1}{\sqrt 2}(|0\rangle+(-1)^x|1\rangle)$.

First, we note that the logical somersaults defined by us differ only in the time of the appearance of an explicit photon in the cavity. Let two waveguides, 1 and 2, be connected to the cavity. Using the fast switching on and off of the Pokkels cell, as above, we can direct the photon along the waveguide 1, if the logical qubit is zero, and along the waveguide 2, if it is equal to one.

The phase shifter changes the phase of the logical somersault, increasing it by the angle $\phi$, if and only if it is equal to one. For such a phase change, it is enough to extend the waveguide 2, into which the photon will fall, if the logical qubit is equal to one. The excess length is wound on the coil, so that at the output we will again have the same photons, but the phase shift along the waveguide 2 will be equal to $\phi$. Since the period of the Rabi oscillations $ \tau$ significantly exceeds the photon wavelength, such an elongation of the photon path in the second waveguide will not affect the determination of logical qubits in any way.

\begin{figure}
\centering
\includegraphics[scale=0.60]{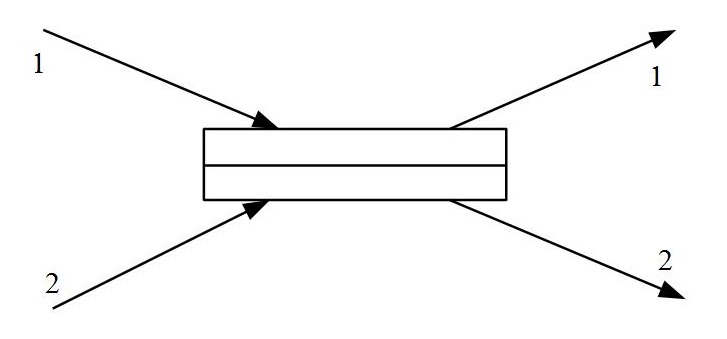}
\caption{The beam splitter.}
\label{fig:BC}
\end{figure}

Now let's move on to the Hadamard gate $H$. To implement it, we use a linear beam splitter, shown in the figure \ref{fig:BC}. This device implements the conversion of photons in waveguides 1 and 2 of the form:

\begin{equation}
\begin{array}{ll}
|n_1m_2\rangle=\frac{1}{\sqrt{n!m!}}(a_1^+)^n(a_2^+)^m|0_10_2\rangle&\rightarrow\\
&\frac{1}{\sqrt{n!m!}}[\frac{1}{\sqrt 2}(a_1^++a_2^+)]^n[\frac{1}{\sqrt 2}(a_1^+-a_2^+)]^m|0_10_2\rangle,
\end{array}
\end{equation}
where the subscript denotes the waveguide number. For $n=1, m=0$ or $n=0, m=1$, that is, for one logical qubit, this transformation will exactly give the Hadamard operator.

Thus, one-qubit gates necessary for the implementation, for example, the Grover algorithm, can be made on optical cavities, within the framework of the Jaynes-Cummings-Hubbard model.
The main difficulty is in the speed of operation of the Pokkels cell, which seems to be technically surmountable.

The advantage of the proposed gate implementation scheme is its simplicity and the possibility of accurately following the theoretical JCH model, which, despite the mentioned technical difficulty, inspires optimism in terms of scalability and comparison of the theory of a quantum computer with experiments on a large number of qubits.

\subsection{Density matrix}

So far, we have considered either unitary evolution, or measurement is an idealized scheme. In reality, quantum memory is closely surrounded by various particles and the electromagnetic field of different modes. For example, an atom in a cavity interacts with a photon for a limited time, since the lifetime of a photon in a cavity is limited: it flies out of it outside. Upon contact with the environment, the state of $|\Psi\rangle$, which we will call pure in the future, deteriorates in a specific way: it turns into a mixed state, which can no longer be described as a state vector of $|\Psi\rangle$, but can only be described as a density matrix. We will study it now.

Consider the matrix $|\Psi\rangle\langle\Psi|=\rho_\Psi$ associated with the state $|\Psi\rangle$ of the form \eqref{state}. This is the Landau density matrix of a given state. It is Hermitian and its rank and trace are equal to one. These three conditions on the matrix, in turn, mean that it has the form $\rho_\Psi$ for some vector $|\Psi\rangle$ of the form \eqref{state}. {\it A listener is invited to prove these statements independently.} On the diagonal of the density matrix there are probabilities $p_j$, and the non-diagonal terms, called coherence, symbolize the quantum properties of a given state $|\Psi\rangle$: if it is basic, it has no quantum properties, it is classical.

When the basis is changed, the density matrix is transformed according to the matrix law: $\rho_\Psi\rightarrow T\rho_\Phi T^*$, where $T$ is a unitary transition operator to a new basis.

In the continuous case, the density matrix is a function of the form \newline $\bar\Psi(r_1, t)\Psi(r_2,t)$, where $r_1,r_2$ is a pair of possible positions of the particle.

The Born rule is the only link between quantum formalism and experiments. We cannot extract any information about the state in which a given particle is located other than by making a measurement over it in some pre-selected basis.

To determine the amplitudes of the$ \la_j $ state of $|\Psi\rangle$, it is necessary to make many measurements on many particles equally prepared in this state. Moreover, after the measurement, the state of the particle changes irreversibly, so that, generally speaking, we cannot use the same particle.

It may seem that if we measure a particle in any basis, then after repeated measurement we will get the same result, since the result of the projection of the state on the pre-selected basis vector $|j\rangle$ has the form $|j\rangle\langle j|$ (prove it!) and its square is equal to the same projection. But this is a gross mistake characteristic of Copenhagen physics, in which formalism exists separately from reality. In fact, the measurement cannot exist separately from the so-called unitary dynamics, (see below, the solution is the Schrodinger equation) as well as unitary dynamics - without measurement. By measuring, for example, a coordinate, we will inevitably give the particle such an impulse that it will fly away from our laboratory, so we will have to re-measure the coordinate at another particle.

The Schrodinger equation for the density matrix has the form
\begin{equation}
\label{Sh-den}
i\hbar\dot{\rho}=[H,\rho]=H\rho-\rho H;
\end{equation}
it is equivalent to the usual Schrodinger equation and is easily derived from it. Thus, the solution of the equation \eqref{Sh-den} is a unitary dynamics in the Hilbert space of states, that is, the dynamics in the absence of decoherence, when we assume the absence of the influence of the environment measuring the system.

\subsection{An open quantum system. The quantum master equation \label{Liuvil}}

What happens when the system in question comes into contact with an environment that does not have long-term memory, but is able to cause measurements of some part of the system? This question is of great practical importance. For example, if an atom emits a photon, and this photon flies away, being in an entangled state with an atom, the measurement of this photon will automatically lead to the appearance of a mixed state of the atom, that is, it will cause decoherence, in which the atom must be described by a density matrix.

We may not even know what happens to the emitted photon; maybe no one is watching it, but it will be reflected from a distant mirror and fly back to us again - all the same, if it is not nearby, we must consider the state of the atom we have as mixed. A photon, once emitted and not measured by anyone, will arrive again-well, the atom-photon system''it will be in a pure state again, and if another photon arrives instead of ours, which got into someone's detector, then we will have a density matrix of the mixed state of the composite system. 

That is, we can determine whether a photon has been measured only when it arrives to us again. If we arrange an experiment so that a photon arrives at us at each repetition, we can, by changing the basis, determine by tomography whether there was a measurement, that is, whether it was the same photon that once flew out of our atom, or another: when detecting, the photon disappears.

However, this method is statistical. With its help, we can only check whether there is a systematic measurement of outgoing photons in the same type of experiments, or there are no measurements, and all the photons are reflected from the mirror, flying back to us. For a specific case, it is impossible to draw such a conclusion: the conclusions of the quantum theory are always only statistical.

The time change of the density matrix of a system interacting with a stationary environment that does not have long-term memory is described by a generalization of the Schrodinger equation for the density matrix, which is called the Kossakovsky - Lindblad - Glauber - Sudarshan quantum master equation:
\begin{equation}
\label{L}
i\hbar\dot{\rho}=[H,\rho]+i{\cal L}(\rho),\ \ {\cal L}(\rho)=\sum\limits_{j=1}^{N^2-1}\g_j(A_j\rho A_j^+-\frac{1}{2}\{ A_j^+A_j,\rho\})
\end{equation}
where the operators $A_j$ are called decoherence factors, and must, together with an identical operator, form an orthonormal basis in the $N^2$- dimensional Liouville space of operators of size $N\times N$, in which the scalar product is defined by the formula $\langle A|B\rangle = tr(A^+B)$. Here, following the tradition, we denote the conjugate operator by a cross, and the non-negative numbers $\g_j$ are the intensities of the decoherence factor $A_j$.

This equation is a generalization to the quantum case of the main Markov equation $ \dot{P}=AP$ for the probability distribution $P$; if a given dynamics of probability distributions is considered in random processes, that is, the dynamics of the main diagonal of the density matrix, then in quantum physics the entire density matrix is considered, and the physical causes of such dynamics are investigated.

The numerical solution of the equation \eqref{L} can be carried out by the Euler method. The fact is that the main term in the right part of $[H,\rho]$ corresponds to the unitary dynamics; this dynamics does not increase the magnitude of the error, so there are no pathological cases of its rapid growth and, as a rule, there is no need to use more accurate methods of the Runge-Kutta type. The solution can be represented as a sequence of steps, each of which corresponds to the time $t_j$, begins with the density matrix $ \rho (t_j)$ and consists of two actions:

1. The unitary dynamics of the density matrix is calculated

$$\tilde\rho(t_{j+1})=\rho(t_j)+\frac{1}{i\hbar}[H,\rho(t_j)]dt.
$$

2. The action of the Lindblad superoperator ${\cal L}$ on the $\tilde\rho(t_{j+1})$: 

$$
\rho(t_{j+1})=\tilde\rho(t_{j+1})+\frac{1}{\hbar}{\cal L}(\tilde\rho(t_{j+1}))dt.
$$

The density matrix $ \rho (t)$ at any given time must be positive definite, Hermitian, and have a unit trace. The last two conditions, in the presence of random errors, can be easily provided by switching from a slightly corrupted matrix $\rho (t)$ to a corrected matrix $(\rho (t)+\rho^+(t))/tr (\rho (t))$. To ensure positive certainty in case of random errors, it is possible to calculate the eigenvalues once, for example, in 20 steps, and then, when a small negative value appears, correct these values by redistributing the error to all other eigenvectors.

\section{Lecture 9. The complexity of quantum system ad the accuracy of its description}

\subsection{Introduction}

Our understanding of quantum theory has evolved greatly since its inception. If until about 80-90 years of the 20th century, as a rule, simple, from the classical point of view, systems were studied: individual atoms, molecules or ensembles consisting of identical particles that could be reduced to separate independent simple objects, then in recent decades the focus of research has shifted towards more complex systems. In particular, the relevance of microbiology and virology has also aroused physicists ' interest in studying objects related to living things, for example, the DNA molecule, which can no longer be attributed to simple systems.

Meanwhile, quantum theory, which is the basis of our understanding of the microcosm, and, therefore, an accurate understanding of complex systems, has a very rigid and well-defined mathematical apparatus based on the matrix technique. The predictions of quantum mechanics have always proved to coincide very precisely with experiments on simple systems that are traditional for physics, but for complex systems this theory meets a fundamental obstacle. The very procedure of obtaining theoretical predictions requires such unimaginable computational resources that we will never have them at our disposal.

If for simple systems the procedure of computation the quantum state had no relation to the physics of its evolution and was only a technical technique, then for complex systems the situation is different. Here, the computation process is the main part of the definition of the quantum state itself, and therefore should be considered as a physical process, and the device that implements this computation is an integral part of any experiment with complex systems at the quantum level.

This computing device is an abstract computer that simulates the evolution of the complex system under consideration. Thus, all restrictions on this computer, following from the theory of algorithms, have the status of physical laws; and these laws have absolute priority over physical laws in the usual sense in the case of complex systems and processes.

This is a new situation that did not exist in classical physics, where the procedure for obtaining theoretical predictions was not very complicated. In any case, the complexity there has almost always been within the reach of classical supercomputers, which are created mainly to cover processes from the point of view of classical physics - by the number of particles in the system under consideration. In quantum mechanics, the complexity increases exponentially with the number of particles, and the classical way of computing becomes unacceptable. This was strictly proved by the discovery of theoretically possible (from the point of view of the standard - Copenhagen quantum theory) processes that cannot be modeled on any classical supercomputers - the so-called fast quantum algorithms (\cite{Sh}, \cite{Gr}). 

The attempt to circumvent the complexity barrier with the help of a quantum computer proposed by R. Feynman (\cite{Fey}) has given us a lot to understand the microcosm and some interesting applications, for example, in cryptography and metrology. However, this attempt did not solve the main problem: the scaling of a fully functional quantum computer is very questionable due to decoherence. Decoherence occurs as a result of spontaneous measurements of the states of the simulated system from the environment, which is traditionally considered in the framework of the concept of an open quantum system in contact with the environment (see \cite{BP}), so that the influence of the environment is reduced to uncontrolled measurements of the state of the original system.

Thus, decoherence is a fundamental factor that cannot be eliminated with the help of mathematical techniques, such as error correction codes (they begin to really work only for a quantum computer with more than a hundred qubits). If we set the task of modeling complex systems at the quantum level, decoherence should be embedded in the quantum formalism itself, and not introduced into it as an extraneous influence. The deviation from the linear unitary law of evolution resulting from decoherence must be naturally justified mathematically.

There is a complexity barrier to matrix formalism. 

The more precisely we know the amplitudes of the quantum state, the simpler it should be. If the state is complex, we will not be able to determine its amplitudes exactly. Given a system of $n$ particles - qubits, to apply quantum mechanics we try to learn their state as precise as possible. 

A quantum state $|\Psi\rangle$ cannot be the state of a single system of qubits; the wave vector is a characteristic of a huge number of equally prepared such systems. Thus, the state of $| \Psi\rangle $ characterizes a certain imaginary apparatus that produces exactly the same ensembles consisting of $n$ qubits.

The accuracy of the quantum state is the accuracy of determining its amplitudes of the basic states using measurements: the more copies of the state we have, the more accurately we can determine these amplitudes by simultaneously measuring all these copies.

Let there be a system of $n$ qubits about which we think that it is in the quantum state $|\Psi\rangle$.  We will call the accuracy of this state the maximum possible number of such equally prepared ensembles. The accuracy of a state is thus the maximum possible number $A$ of copies of this state that can be available to us at the same time, when we can measure them. 

We arrange such $n$ qubit ensembles $S_j$ having the same states in the form of a sequence of $nA$ qubits of the form
$$
S_1,S_2,...,S_A,\ \ S_j=(s_1^j,s_2^j,...,s_n^j),
$$
thus obtaining the memory of some anstract Main Computer (MC), from which we can learn what this state is. The memory of MC cannot be unlimited, hence there is the constant  $Q$, such that  
\begin{equation}
\label{1}
An\leq Q.
\end{equation}

But the number $n$ of qubits cannot be the exact measure of the complexity of the state $|\Psi\rangle$, even if these qubits are all entangled. To define the real complexity we have to use the so called canonical transformation, which can radically reduce the number $n$ of qubits without changing the state $|\Psi\rangle$ substantionally. 

For the right definition of complexity of a state $|\Psi\rangle$ the following inequality takes place:

\begin{equation}
\label{und}
AC\leq Q.
\end{equation} 

We will argue in favor of such a ratio of complexity and accuracy in the general case, for complex particles; in particular, we will show that the quantization of the amplitude makes it possible to introduce a certain determinism into the quantum formalism, the nature of which is not reduced to the classical one.

\section{The Main Computer}

A complex system is a system whose behavior cannot be reduced to independent particles. Predictions of the behavior of such a system can be obtained only by relying on quantum mechanics and computer ideology, since analytical techniques do not work here. Therefore, we must introduce the concept of a Main Computer (MC), which adequately represents real processes for us - an abstract computing device, the laws of which will be for complex systems (and only for them!) have priority over physical laws, the scope of which is unconditionally limited to simple systems and processes.

A physical prototypes of a Main Computer can have only limited power; such devices will be able to adequately represent the processes traditionally related to chemistry, as well as to those areas of physics in which quantum methods are successfully applied, for example, to electrodynamics. Nuclear physics does not yet belong to such processes, and its complexity radically exceeds the electrodynamics  complexity (see \cite{Oz}).

Thus, we have, within the framework of this limitation of the computational capabilities of the MC, a balance between the complexity and accuracy of the representation of the state vector, which must be found specifically in each specific case.

\begin{figure}
\centering
\includegraphics[scale=0.57]{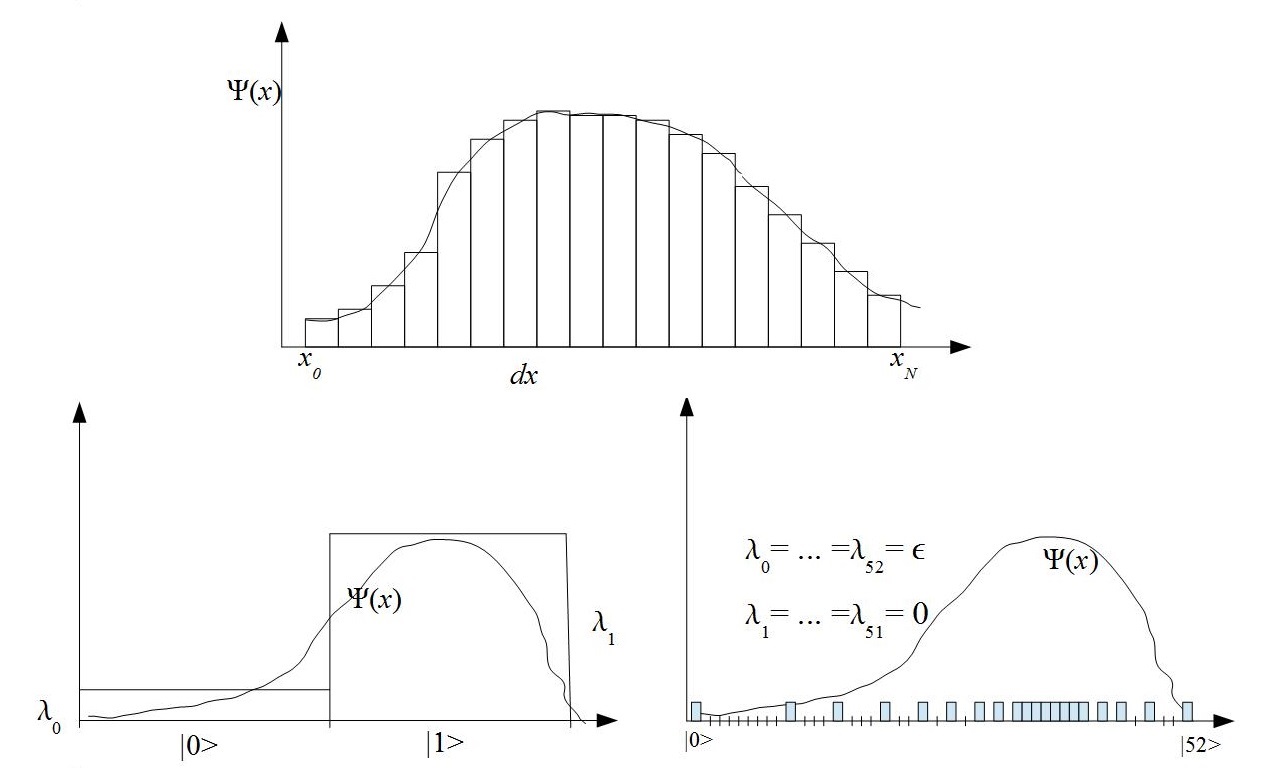}
\caption{Representation of the state vector. Curves represent the hypothetic wave function $|\Psi\rangle$, predicted by Copenhagen quantum theory. Rectangles denote the information about it, which we can obtain via MC. Bottom left - the main computing resource captures accuracy: $|\Psi\rangle=\lambda_0|0\rangle+\lambda_1|1\rangle $; idealized case where we reduce the number of basic states to 2, as for a particle in two hole potential - this gives the satisfactory agreement with experiment.  The top - the resource is divided equally between accuracy and complexity. The best agreement with experiments; there is the typical area of application of quantum mechanics. Bottom right - the main resource is captured by complexity: $|\Psi\rangle=\sum\limits_{j=0}^{N-1}\varepsilon |j\rangle,\ \varepsilon \in\{ 0,\epsilon\}$. Our knowledge is limited by only one basic state, which we extract from the single measurement. Here we have to follow the trajectory of one quantum of amplitude (see below). }
\label{fig:discr1}
\end{figure}

For one qubit we can find its amlitudes with the most possible precision. For the simple systems with low complexity, which was in the focus of attention of the physics of 20th century, the possible precision was equal to the accuracy of experimental results. For such systems with  of intermediate complexity we were able to determine the amplitudes more accurately. For the more complex systems, as the prototypes of quantum computer, we already meet difficulties that is called the decoherence.
For an extremally complex system $A=1$ and we have only one sample of it; we can make only one measurement, which means that we can receive only one basic state.  The figure \ref{fig:discr1} represents these cases.

\subsection{Complexity of the quantum state}

In order to give a correct definition of the complexity of the quantum state of a system of $ n $ particles, we must first consider the canonical transformation - the main method of reducing such complexity.

Any coordinate on a unit segment is a real number represented as its binary expansion $2^{-l}\sum\limits_{j=0}^{l-1}a_j2^j$ with an accuracy of $2^{-l}$, where $a_j=0.1$ are the values of $l$ qubits representing this coordinate\footnote{To represent a coordinate on any other segment, we need to apply a suitable linear transformation, for example, on the segment $[-2^{l/2}, 2^{l/2}]$ the approximate qubit representation looks like this: $2^{-l/2} \sum\limits_{j=0}^{l-1}a_j2^j-2^{l/2-1}$.} By ordering the qubit values lexicographically, we get a standard ordering of basis vectors, in which any operator will have a certain matrix.

Let the classical state of the particle $i$ be the real vector $x_i$. Then the classical state of the system of $n$ particles will be a vector of the form $ \bar x=(x_1, x_2,..., x_n)$. Let $\bar x'$ and $\bar x"$ be two vectors with non-empty sets of coordinates such that their Cartesian product coincides with $\bar x$. This means that we have split the set of particles into two non-empty subsets $X'$ and $X"$, so that the given vectors are the sets of coordinates of these subsets. Let $H (\bar x)$ be the Hamiltonian of this system, having the form

\begin{equation}
\label{ham}
H(\bar x)=H_1(\bar x')+H_2(\bar x'');
\end{equation}
here, we assume by default that $H(\bar x')$ is $H(\bar x')\otimes I (\bar x")$, that is, on particles not included in the first subset, this term of the Hamiltonian acts as an identical operator, and similarly with the second term. Then we call the Hamiltonian $H$ reducible. Let $X'$ be the maximum subset of the components of the vector $\bar x$ in terms of the number of elements, such that the equality \eqref{ham} holds, and the Hamiltonian $H_1(\bar x')$ is not reducible. Then $X'$ is called the kernel of the given Hamiltonian.

We will assume by default that any quantum evolution begins with the basic state of the system under consideration. Since \eqref{ham} implies the equality $exp(-\frac{i}{\hbar}H)=exp(-\frac{i}{\hbar})H_1\otimes exp (- \frac{i}{\hbar}) H_2$, we see that the core of the Hamiltonian is the maximum set of particles whose states in quantum evolution with the Hamiltonian $H$ can be entangled; we denote the number of particles in this set by $ \nu(H)$, and call it the {\it naive} complexity of this Hamiltonian.

Consider the transformation of the coordinates of the particles of the form

\begin{equation}
\label{trans}
q_i=q_i(x_1,x_2,...,x_n),\ i=1,2,...,n;
\end{equation}
denote $\bar q =(q_1, q_2,..., q_n)$ and let $H_q=H(x_1(\bar q),x_2(\bar q),..., x_n(\bar q))$ is the original Hamiltonian written in terms of the new variables $q_i$, $\bar x=\bar x(\bar q)$ is the inverse of \eqref{trans}. We will introduce virtual particles with coordinates $q_1, q_2,..., q_n$, which we will call quasiparticles.

The classical state of the source system is a set of specific values $x_1,x_2,..., x_n$. Then each classical state will correspond to the classical state of the same system, obtained using the formulas \eqref{trans}. The basis vector of a Hilbert space passes into another basis vector of the same space. The standard ordering of the basis vectors, according to the qubit representation of coordinates, will pass to another ordering, that is, the coordinate transformation \eqref{trans} is a permutation of the basis vectors of the Hilbert space of quantum states.

In this case, the qubits representing the coordinate values will already have a different, new meaning. In the new coordinates, for quasiparticles, the Hamiltonian will have a different form $H_q$. We will call the coordinate transformation \eqref{trans} canonical if $ \nu (H_q)$ is minimal. In this case, the transition to quasiparticles will mean a reduction in the complexity of the original Hamiltonian. So, the canonical transformation is a permutation of the bisis vectors that minimizes the complexity of the Hamiltonian.

\subsection{Example: the system of interacting harmonic oscillators}

We will consider the concept of a quasiparticle, within the framework of the so-far ordinary quantum theory, on the model problem of a system of interacting harmonic oscillators. This task was chosen because of its special importance for everything further. For one oscillator, the Hamiltonian has the form $H=\frac{p^2}{2m}+m\w^2 q^2/2$; its eigenfunctions have the form

\begin{equation}
\Psi_n(x)=\frac{1}{\sqrt{2^nn!}}\left(\frac{m\w}{\pi h}\right)^{1/4}exp(-m\w x^2/2h)H_n(x\sqrt{m\w/h})
\label{osci}
\end{equation}
where Hermite polynomials $H_n=(-1)^ne^{x^2}\frac{d^n}{dx^n}e^{-x^2}$, and the corresponding eigenvalues of the energy $E_n=h\w (n+1/2)$.

Consider a system of $ N $ harmonic oscillators interacting with each other according to Hooke's law. Such a system can be, for example, a chain of positive metal ions in a Paul trap. The Coulomb interaction between them, if we consider small fluctuations near the equilibrium position, gives quadratic potentials, that is, we can approximately assume that the force between the ions obeys Hooke's law.

Let $u_n$ denote the deviation of the oscillator $n$ from its equilibrium position. The Hamiltonian of such a system has the form
$$
H=\sum\limits_{n=1}N(\frac{p_n^2}{2m}+\kappa u_n^2)-\kappa\sum\limits_{n=1}^Nu_nu_{n+1}
$$
which is obtained from subtracting the Hooke forces $ - \kappa (u_n-u_{n+1})$ over all $n$, with the reduction of similar terms (so it turns out $u_nu_{n+1}$) and neglecting boundary effects.

The struggle against entanglement is the struggle against interaction. The essence of the canonical transformation is to get rid of the interaction member $u_nu_{n+1}$. This term should not be for quasiparticles - these new "particles" should be independent of each other. It turns out that the Fourier transform is suitable for this purpose, but not over the wave function, as it was during the transition from the coordinate to the pulse basis in the Hilbert state space, but over the amplitudes $u_n$ of oscillator oscillations themselves.

The role of the $r$ coordinate here will be played by $n$ - the number of the oscillator, and the role of the psi function will be played by $u_n$. The values of this psi function are also called amplitudes, so there is complete agreement with linguistics here. However, the fact that the "coordinate" of $n$ and the "amplitude" of $u_n$ are $r$ and $\Psi (r)$ is not just a funny analogy. He talks about the nature of the quantum amplitude and its deep connection with the role representation of real particles. Mechanical oscillatory amplitudes $u_n$ cannot be complex, like quantum amplitudes. And the classical oscillation equation describing the dynamics of oscillators also cannot be a prototype of the Schrodinger equation.

Therefore, oscillators are not suitable as a basis for a complete model of quantum dynamics in the general case. However, they are suitable as a model of the electromagnetic field, the interaction with which gives a complete picture of the dynamics of particles, which turns out to be the dynamics of "particles + fields". The field turns out to be a refuge of determinism. Quantum chaos is connected precisely with the fact that the amplitudes there will be complex; this is due to the chaotic nature of the particles themselves, and this cannot be overcome in the same way as in the case of the deterministic world of oscillators. This means that we will have to quantize the field, that is, the system of oscillators, which will, in fact, be the propagation to the field of the internal stochastic behavior that was originally characteristic of matter particles.

So, the canonical transformation in our case looks like this:
\begin{equation}
u_n=\frac{1}{\sqrt N}\sum\limits_qU_qe^{-iqnd},
\label{can}
\end{equation}

and the reverse to it:

\begin{equation}
U_q=\frac{1}{\sqrt N}\sum\limits_nu_ne^{iqnd},
\label{can-1}
\end{equation}

where $d=2\pi/N$. Using the definition of impulse $p_n=\frac{h}{i}\frac {\partial}{\partial u_n}$ and the differentiation rules, we can derive formulas for impulse transformations of the form

\begin{equation}
p_n=\frac{1}{\sqrt N}\sum\limits_qP_qe^{iqnd},
\label{canp}
\end{equation}

and the reverse to it:

\begin{equation}
P_q=\frac{1}{\sqrt N}\sum\limits_np_ne^{-iqnd},
\label{canp-1}
\end{equation}

where $P_n=\frac{h}{i}\frac{\partial}{\partial U_n}$. The fact that we have allowed complex amplitudes here is not significant, because we will now return to the real numbers, which could not be done if we had a psi function instead of $u_n$.

The canonical transformation is linear, and translates any small cube of division of the configuration space into an equally small parallelepiped, so that instead of one division, another will arise. Moreover, our transformation of the form (\ref{can}) will even be orthogonal, that is, as we will see below, the cubes will turn into cubes. Therefore, we can always represent it as a permutation of the basis vectors of the Hilbert space.

We also shift the origin of coordinates for $q$ so that this parameter, which replaces $n$, takes values from a symmetric interval. Then instead of $q+q'=N$ , we will write $q+q'=0$. The inequality relation is induced from the old set $1,2,\ldots, N$ so that the pairs $q> - q$ are almost half (we neglect the edge effect everywhere)

By rewriting the Hamiltonian in the new coordinates, we have:
$$
\begin{array}{ll}
H=&\sum\limits_{n=1}^N\frac{1}{2mN}(\sum\limits_{q,q'}P_qP_{q'}e^{1(q+q')nd})+\frac{K}{N}\sum\limits_qU_qU_{q'}\\
\ &-\frac{K}{N}\sum\limits_{q,q'}(U_qU_{q'}e^{-iqnd}e^{-iq'(n+1)d})=\\
&=\frac{1}{2mN}\sum\limits_qP_qP_{-q}-\frac{K}{N}\sum\limits_{q,q'}U_qU_{q'}(\sum
\limits_{n=1}^Ne^{-ind(q+q')})e^{-iq'd}+\frac{K}{N}\sum\limits_qU_qU_{q'}=\\
&=\frac{1}{2mN}\sum\limits_qP_qP_{-q}-\frac{K}{N}\sum\limits_{q}U_qU_{-q}e^{+iqd}+\frac{K}{N}\sum\limits_{q}U_qU_{-q}=\\
&=\frac{1}{2mN}\sum\limits_qP_qP_{-q}+\frac{2K}{N}\sum\limits_{q>-q}U_qU_{-q}(1-cos(qd)).
\end{array}
$$
Here $K=m\w^2/2$, the standard formula for summing the exponential geometric progression was used, which gives 0 in the case of $q\neq q'$, and the pairs $q,-q$ were ordered, so that we explicitly wrote out only half in which $q>q'$ - hence the coefficient 2 in the last term.

Now let's move on once again to the new variables, this time - real:
$$
\begin{array}{ll}
&U_q=X_q+iY_q,\ X_q=\frac{U_q+U_{-q}}{2},\ Y_q=\frac{U_q-U_{-q}}{2i};\\
&X_q=\frac{1}{\sqrt N}\sum\limits_nu_ncos(qnd),\ Y_q=\frac{1}{\sqrt N}\sum\limits_nu_nsin(qnd),\\
&\frac{\partial}{\partial U_q}=\frac{\partial}{\partial X_q}\frac{1}{2}+\frac{\partial}{\partial Y_q}\frac{1}{2i},\\
&\frac{\partial}{\partial U_{-q}}=\frac{\partial}{\partial X_q}\frac{1}{2}-\frac{\partial}{\partial Y_q}\frac{1}{2i},\\
&\frac{\partial^2}{\partial U_q\partial U_{-q}}=\frac{1}{4}\left(\frac{\partial^2}{\partial X_q^2}+\frac{\partial^2}{\partial Y_q^2}\right).
\end{array}
$$
Finally we get
$$
H=-\frac{1}{4mN}\sum\limits_{q>q'}\left(\frac{\partial^2}{\partial X_q^2}+\frac{\partial^2}{\partial Y_q^2}\right)+\frac{2K}{N}\sum\limits_{q>-q}(X_q^2+Y_q^2)(1-cos(qd)).
$$

We see that in the new coordinates, our system is a set of independent harmonic mass oscillators $ \tilde m=2m$, with a new coefficient $\tilde K=2K(1-cos (qd))$ and frequencies
\begin{equation}
\tilde\w_q=\sqrt{\frac{2K}{m}(1-cos(qd)}
\label{frictions}
\end{equation}

\subsection{Quasi particles}

\begin{figure}
\centering
\includegraphics[height=0.5\textwidth]{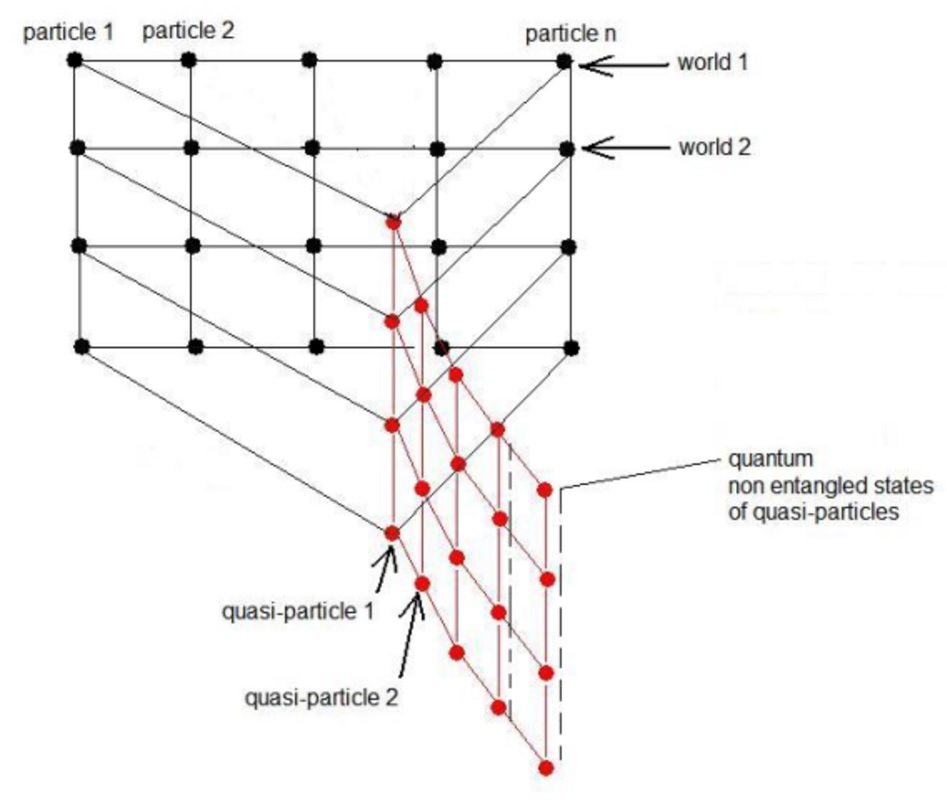}
\caption{Quantum representation of quasiparticles}
\label{fig:quasiparticles}
\end{figure}

The transition to the description of evolution in the form of quasiparticles has the form $H=\tau^{-1} H_q \tau$ where $\tau$ is the transformation of the transition to quasiparticles under the canonical transformation. Then the representation of the evolution operator is $exp (- \frac{i}{\hbar}Ht)=\tau^{-1}exp(-\frac{i}{\hbar}H_qt)\tau$ requires less computational resources than the direct calculation of $exp (- \frac{i}{\hbar}Ht)$, since the main resource is spent on the kernel, which for a quasi-partial representation will have a minimum size. Quantum representation of quasiparticles is shown at the figure \ref{fig:quasiparticles}

An example is a chain of interacting harmonic oscillators, for which the Fourier transform of their coordinates $x_i$ is canonical, and the kernel of the quasiparticle representation-in terms of phonons - will generally consist of a single particle, that is, the phonons are all independent. Here we have the maximum reduction of the kernel. 

A simpler example would be the Hamiltonian $H$ of a closed chain of 4 interacting qubits, which is reduced by the canonical transformation $CNOT$ to a completely reduced Hamiltonian of the form
 $H_q=\s_x^{(1)}\otimes I_2+I_1\s_x^{(2)}$: 
$$
H=
\begin{pmatrix}
&0&1&0&1\\
&1&0&1&0\\
&0&1&0&1\\
&1&0&1&0
\end{pmatrix}
=CNOT
\begin{pmatrix}
&0&1&1&0\\
&1&0&0&1\\
&1&0&0&1\\
&0&1&1&0
\end{pmatrix}
CNOT.
$$
Note that the permutation of the basis vectors, which is a canonical transformation, must be an entangling and simultaneously disentangling operator in the state space, since it reduces the kernel of the Hamiltonian. So, for example, the $CNOT $ operator applied repeatedly to the state $|00...0\rangle+|11...1\rangle$ completely untangles this state.

The opposite example is given by the Tavis-Cummings Hamiltonian for $n$ two-level atoms interacting with the resonant mode field in the optical cavity. Here there is a basic state of the field and atoms of the form $ |n \rangle_{ph}|00...0 \rangle_{at}$, such that the column of the Hamiltonian matrix corresponding to this state consists of the numbers $g\sqrt{n}$ and one number $n\hbar\w$, and such a column is the only one. No Hamiltonian of the form \eqref{ham} can have this property even at $n=2$, so there is no non-identical canonical transformation for the Tavis-Cummings Hamiltonian.

The quantum complexity of the Hamiltonian $H$, denoted by $c(H)$, is the minimal naive complexity of the operators $ \tau^{-1} H \tau$ over all possible permutations $\tau $ of basis vectors. For the above examples, the complexity of the Hamiltonian is 1, that is, it can be completely reduced by the canonical transformation.

The complexity of the quantum state $|\Psi\rangle$is determined in a similar way. Its {\it naive} complexity is $\nu(|\Psi\rangle)$ is defined as the number of particles in the maximum tensor divisor $|\Psi_1\rangle$ of the state $|\Psi\rangle=|\Psi_1\rangle\otimes |\Psi_2\rangle$.

The quantum complexity $C(|\Psi\rangle)$ of the state $| \Psi\rangle$ is the minimal naive complexity of the state $\tau |\Psi\rangle$ over all permutations $\tau$ of particles. In other words, the quantum complexity of $C(|\Psi\rangle)$ of a state is the logarithm of the kernel length of that state in the canonical representation.

For example, the quantum complexity of the generalized GHZ state $ |GHZ\rangle=\frac{1}{\sqrt 2}(|00...0\rangle+|11...1\rangle)$ is equal to 1, since it can be untangled by successive CNOT operators. If we start from the basic state $|\Psi(0)\rangle$ in the canonical representation of the Hamiltonian$ H $, then an evolution with $H$ will only contain states with a complexity of no more than $C(H)$. 

The quantum complexity of the Hamiltonian has a clear algorithmic meaning. Suppose we model quantum evolution on a classical supercomputer with an unlimited parallelization resource. Then, for the reduced Hamiltonian, you can parallelize such a simulation by assigning separate nodes of the supercomputer to simulate separate groups of variables. Thus, the node that will be assigned the most difficult task will be the one that will be assigned the kernel of the Hamiltonian. We can say that the classical complexity of the Hamiltonian is the memory size of a single node of a classical supercomputer that can simulate quantum evolution with a given Hamiltonian. The quantum complexity of the Hamiltonian is the logarithm of this value.

\subsection{Amplitude granularity}

Here we show the possible way how to prolong quantum formalism over  the border $Q$. This is the kind of quantum determinism that is not reducible to the known quasiclassical effects.

The qubit representation of the classical coordinates and impulses determines the grain size of the amplitudes. For any expansion of the quantum state $|\Psi\rangle$ over an arbitrary orthonormal set of basis states $|\psi_j\rangle$ of the form 
\begin{equation}
\label{state}
|\Psi\rangle = \sum\limits_{j\in J}\la_j|\psi_j\rangle,\ \la_j\neq 0,
\end{equation}
the amplitudes of $\la_j$ must be bounded from below modulo some nonzero constant.

To preserve the principle of linearity in the region where the nonzero value of this constant does not play a role, we must assume that any amplitude has the form 
\begin{equation}
\label{quanta}
\la_j = \e n_j+i\e m_j
\end{equation}
where $n_j,m_j\in Z$ are integers, $\e>0$ is a constant that is the quantum of the amplitude. This restriction of the matrix formalism entails the rejection of the absolute equivalence of any bases in the space of quantum states, which also manifests itself only at sufficiently small amplitudes $ \la_j$, and, accordingly, large sets of coherent states $J$.

However, modeling does not provide for such equivalence of bases; it is only an algebraic technique that cannot be experimentally tested for complex systems.
The smallest modulo possible nonzero amplitude is thus $\e$. This restriction is very well consistent with the probabilistic nature of the state vector, since to determine the value of $|\la_j|^2$ with an accuracy $ \delta$, it is necessary to conduct about $1/\delta$ measurements of equally prepared samples of the original system; for complex systems with small amplitudes of $\la_j$, this will be possible only if the minimum probability $\e^2$ is separated from zero to obtain the most unlikely outcome.

You can determine the value of $ \e$ by "smearing" the amplitude over as many basic classical states of the system as possible, for which the presence in them can simply be detected by measurement. If all the amplitudes are equal to: $\la_j=\e$, we get for the total number of coherent basis states the estimate $|J|=[1/\e^2]$ - as the integer part of the inverse square of the amplitude quantum. In the \eqref{state} expansion, the amplitudes of $\la_j$ have no physical dimension, the dimension has the basis states of $|\psi_j\rangle$. The number of qubits whose possible quantum states are physically realizable is equal to the constant $q$, for which we get the expression $Q=log_2([1/\e^2])$.

We can now formulate a hypothesis about the relation of the form "accuracy - complexity" in the final form:
\begin{equation}
C(|\Psi\rangle) A(|\Psi\rangle) \leq Q,
\label{main}
\end{equation}
 and $q$ is the maximum number of completely entangled qubits that cannot be disentangled by any permutation of the basis states.

Consider, as an example, the state of a set of $n$ qubits of the form 
\begin{equation}
\label{GSAstate}
|\Psi_{GSA}(t)\rangle= \a\sum\limits_{j\neq j_0,0\leq j<N}|j\rangle+\b|j\rangle,
\end{equation}
where $\a=cos(t)/\sqrt{N-1},\ \b=sin(t)$ for some $t$, and $N=2^n$. The quantum complexity of this state is $ n$ if $t\neq k\pi/2$ for no integer $k$. Indeed, this superposition has the property that all its basic components, except for exactly one, have the same non-zero amplitude, and one has a different amplitude from them.

This property is preserved under any permutation of the basis states, that is, under any quasiparticle representation. But if the state is reducible, then it should have the form $\la_1|i_1\rangle+\la_2|j_2\rangle+...)\otimes(\la_3|j_3\rangle+\la_4|j_4\rangle+...)$ for some basis $|j_i\rangle$, and such a superposition cannot contain exactly 2 amplitude values for any basis states, since there must either be at least 3 different non-zero amplitude values of the components, or it must have only two different non-zero amplitude values that correspond to two groups of basis states containing an equal number of terms. Both of these possibilities are excluded for states of the form \eqref{GSAstate}.

Note that the discrete representation of amplitudes in the form of \eqref{quanta} allows us to naturally include state measurements in its unitary evolution. The hard contact of the system in the state $|\Psi\rangle=\sum\limits_{j\in J}\la_j|j\rangle$ with the measuring device means the inclusion of the states of this device in the system, that is, the transition to the state of the extended system of the form $|\Psi_{ex}\rangle=\sum\limits_{j\in J,\nu_j\in{\cal N}_j}\mu_{j,\nu_j}|j,\nu_j\rangle$, where, under the condition of the applicability of quantum mechanics, all $\mu_{j,\nu_j}$ must be minimal, which means that the quantum of the amplitude of $\epsilon$is equal. Since contact with the environment is a unitary evolution, we have $|\la_j|^2=\sum\limits_ {\nu_j\in{\cal N}_j}|\mu_{j,\nu_j}|^2$ that is, the measurement will be an arbitrary choice of one of the states of the $\nu_j$ meter and we get the standard urn scheme from probability theory.

\subsection{Experimental finding of a constant $Q$}

You can find the approximate value of $Q$ by the building the state of the form \eqref{GSAstate}. These states have the memory of a quantum computer when implementing the Grover GSA algorithm with a single target state $|j_0 \rangle$ (see \cite{Gr}). Let $N=2^n$. We put $t_0=arcsin(1/\sqrt{N})$. We start with  
$$
|\Psi(0)\rangle=\frac{1}{\sqrt N}\sum\limits_{j=0}^{N-1}|j\rangle=(\frac{1}{\sqrt 2}(|0\rangle+|1\rangle)^{\otimes n}, 
$$
and the complexity of this state is 1; this is the initial state for the GSA algorithm. As soon as the first step of the algorithm is performed, $t$ becomes equal to $t_0$ and we get a state of complexity $N$, of the form \eqref{GSAstate}. Already at the first step of the algorithm, the complexity jumps from one to the maximum total number $n$ of qubits. If $n=Q$, at the first step of the GSA algorithm, we will go beyond the limits of acceptable states with ampitudes of the form \eqref{quanta}, which have the property \eqref{main}.

Thus, we can estimate the constant $Q$ from above, increasing the value of $n$ to the limit when the GSA stops working correctly. Here, by correct operation, we mean the possibility to raise the amplitude of the target state by an amount of the order of $1/ \sqrt{2}$ compared to the others, which can be fixed by quantum tomography, since the amplitudes of other states will have the order of $1/\sqrt N$. For a more rough estimate, fixing the jump in the amplitude by $1/\sqrt N$ compared to the main mass is also suitable, but this is only possible for small $n$, not exceeding 20.

\begin{figure}
\centering
\includegraphics[height=0.7\textwidth]{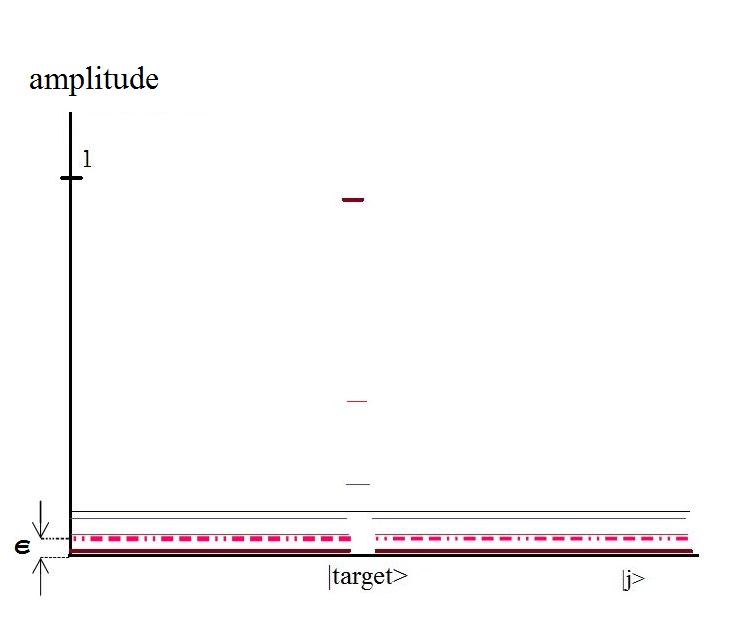}
\caption{Limitations on the work of GSA in terms of amplitude quanta}
\label{fig:GSA2}
\end{figure}

The Figure \ref{fig:GSA2} shows the border of the grover algorithm work in terms of amplitude quanta.

The question of what will happen to the real state if the amplitudes calculated according to standard quantum theory become less than $\e$ is formally open. However, it is natural to assume that small amplitudes should simply disappear, with a corresponding renormalization of the remaining state. This means that implementing GSA near the boundary of $n\approx Q$ we will get the target state very quickly, much faster than when implementing GSA in a normal model. However, this will only happen with an ideal implementation of GSA; in practice, the amplitudes of the main mass of states in $|\Psi_{GSA}(t)\rangle$ in \eqref{GSAstate} cannot be exactly the same, so that zeroing will not occur simultaneously for all states, which can greatly distort the picture.

For the experimental detection of such an effect, it is necessary to fine-tune the gates so precisely as to achieve the maximum possible dimension  $Q$ of the quantum kernel. Below we will discuss the estimates for this constant.

Let's consider two processes: the transition of the states of an electron in an atom $Rb^{85}$ and the decay of an unstable nucleus $He^6$. The first process is described by quantum electrodynamics quite accurately, a complete quantum description of the second is not yet available. 

We will proceed from the criterion of accurate drawing of the wave function, when each step of its computer description requires one new basic state. This follows from the speed of the quantum walk, in which the wave front propagates at a linear speed (in contrast to the classical walk, in which the speed is proportional to the square root of time). Let $t$ be the total time of the process, $dt$ be the step of the computer description of this process in time, then the number of basic states necessary for "accurate drawing" of the process will be $N=t/dt$. The values of $t$ are determined experimentally, and $dt$ is found from the energy-time uncertainty ratio.

For the Rabi oscillation of the rubidium atom, which occurs with the emission of a photon with a wavelength of approximately $ 1.4 \ cm$, we have:
$$
\omega \approx 10^{10}\ sec^{-1}, \ E_{QED}=\hbar\omega\approx 10^{-17}, \ dt\approx\hbar/E_{QED}=10^{-10}. 
$$
Given the time of the Rabi oscillation $t\approx 10^{-6}\ sec$, we get $N=t/dt\approx 10^{4}$. Then $Q\geq 10^{4}< 2^{14}$ and for a good representation of this process on the basis of quantum theory, it is enough to work GSA on 14 qubits, which seems real.

Now let's consider the decay of the nucleus of the helium isotope:
$$
He^6\rightarrow He^5 + n \rightarrow He^4 + 2n
$$
(in this rough approximation, we take into account only the nucleons). The characteristic energy value will be about $10\ Mev\approx 10^{-5}\ erg$, and the energy-time uncertainty ratio will give $dt\approx 10^{-22}\ sec$. The whole process takes about $ 1.6 \ sec$, from where $N=t/dt\approx 10^{22}\approx 2^{73}$, and if quantum mechanics can be continued to nuclear processes such as the decay of the helium - 6 isotope to a stable isotope 4, GSA should work well already at $73$ qubits.

The decay of the helium 6 isotope is a very complex process from the point of view of quantum mechanics. We can only consider its last stage, when one neutron is split off from the stable helium 4 nucleus (see \cite{Masl}). It takes about $10^{-11}\ sec$. For it, estimates similar to the above will give about $36 $ qubits of a reliable implementation of GSA, which is less realistic, but the corresponding value of $Q\approx 36$ can already be verified by experiments on GSA.

Thus, the acceptance of the amplitude grain hypothesis directly connects the question of the applicability of quantum theory to real microprocessors and the implementation of GSA. The implementation of GSA thus becomes a central issue of quantum theory and the theory of complex systems as such.

\subsection{The grain of the amplitude as the cause for the measurements}

The measurement of the quantum state $| \ Psi\rangle=\sum\limits_{j\in J}\la_j|j\rangle$ is a random variable that takes the values $|j\rangle,\ j\in J$ with probabilities $|\la_j|^2$. Physically, it begins with the contact of the original system with the meter, that is, a unitary transformation of the form
\begin{equation}
|\Psi\rangle|\bar 0\rangle_{meas}\rightarrow \sum\limits_{j\in J}\la_j|j\rangle\sum\limits_{i_j\in I_j}\mu_{i_j}|i_j\rangle_{meas}.
\label{meas}
\end{equation}

If the number of elements in each set of states of the meter $I_j$ is very large, so that the amplitudes of all specific states of the meter become approximately equal to the grain $\epsilon$: $\sum\limits_{i_j\in I_j}\mu_{i_j}\approx\epsilon$, then due to the unitarity of the transition \eqref{meas}, the numbers of elements of the sets $I_j$ become proportional to the probabilities $|\la_j|^2$ getting the result of $|j\rangle$ in the dimension.
Therefore, measurement means choosing at random from the urn scheme, giving the probabilities of the outcome in accordance with the Born rule.

So, the presence of a grain removes a strange barrier between unitary dynamics and the collapse of the wave vector, which is a hindrance for modeling dynamics, since the use of a quantum basic equation requires a quadratic increase in memory resources in computer modeling compared to unitary dynamics, since it is necessary to store a density matrix in memory instead of a state vector.

In the limiting case, with a complete collapse at each step of evolution, we will get a completely deterministic description of the dynamics if there is some superiority of the amplitude of a single state $|j_{tar} \rangle$ over all others, as in Grover's algorithm.
The basic states of $|j \rangle$, in this case, generally speaking, must be nonlocal due to the fundamental quantum nonlocality; we have already discussed it above. This form of description of dynamics is quite unusual, but it is most likely that it will be adequate for real complex systems that are the subject of a quantum computer project.

\subsection{Equilibrium states}

We will begin by describing the classes of states for which the amplitude quantization is introduced in the most visual way.

For a complex number $z=a+ib,\ a,b\in R$, we introduce the notation $\{ z\}=|a|+|b|$.
For the quantum state $|\psi\rangle$, we define $\{ \psi\}=\sum\limits_{i=0}^{N-1}\{\langle i|\psi\rangle\}$.

Let $A$ be a linear operator, and $|j \rangle$ be some basis vector, $j\in\{ 0,1,..., N-1\}$. Define $|a_j\rangle=A|j\rangle$. We call the state vector $| \Psi\rangle$ equilibrium with respect to the operator $A$, if all the numbers $\{a_j\}$ are the same for all the base components $|j\ rangle$ that are included in it with nonzero amplitudes.

As an example, consider the Hamiltonian of a one-dimensional particle moving in the potential $V$: $H=\frac{p^2}{2m}+V$. We will reduce the matrix of this Hamiltonian, assuming that there are no too long transitions of a given particle in space. Then the equilibrium states in the coordinate basis for this Hamiltonian will be exactly the states $| \Psi\rangle$, all the basis components of which have the same potential.

An important class of multiparticle equilibrium states are connected states. Here is an example of such a state. 
Consider $k$ of two-level atoms in an optical cavity holding photons with the transition energy between the ground and excited levels of the atoms. We choose a basis consisting of vectors of the form $|n\rangle_{ph}|m_1,m_2,..., m_k\rangle_{at}$, where $n$ is the number of photons in the cavity, $m_j\in\{ 0,1\rangle$ is the state of the atom $j$, ground and excited. Let $g_j,\ j=1,2,..., k$ be the forces of the interaction of atoms with the field. Then the dynamics of the system of atoms and the field under the condition $g_{j}/\hbar\omega\ll 1$, where $ \ omega$ is the frequency of the cavity, will obey the Schredinger equation with the Tavis-Cummings Hamiltonian in the RWA approximation:
\begin{equation}
H_{TC}^{RWA}=\hbar\omega (a^+a+\sum\limits_{j=1}^k\sigma^+_j\sigma_j)+a^+\bar\sigma+a\bar\sigma^+,\ \ \bar\sigma=\sum\limits_{j=1}^kg_j\sigma_j,
\label{exampleHam}
\end{equation}
where $a, a^+$ are the standard field operators of photon annihilation and creation, and $\sigma_{j},\sigma_j^+$ are the atomic relaxation and excitation operators of the atom $j$. The connected states in such a system will be for $k=2$ only for $g_1=g_2$, and this will be either one of the basic states, or the states $|n\rangle_{ph} (\a|10\rangle_{at}+ \b|01\rangle_{at})$, of which in all but the singlet state $\b=-\a$ the atoms will interact with the field. All such states will be in equilibrium.

The general definition of connectivity looks like this.

Let $H$ be a Hamiltonian in the state space of $n$ qubits. If a qubit is associated with a real or virtual two-level particle, $H$ can be, for example, the Tavis-Cummings Hamiltonian or some modification of it. Let $S_n$ be a group of permutations of qubits that are naturally extended to operators on the entire space of quantum states ${\cal H}$, namely: on the basis states, the permutation $\eta\in S_n$ acts directly, and $\eta\sum\limits_j|j\rangle=\sum\limits_j\eta|j\rangle$.

Denote by $G_H$ the subgroup $S_n$ consisting of all permutations of qubits $ \tau$ such that $[H,\tau]=0$. Let $A\subseteq\{ 0,1,..., 2^n-1\}$ be a subset of the basis states of the $n$ - qubit system. Its linear shell $L (A)$ is called a connected subspace with respect to $H$ if for any two states $|i \rangle,\  |j\rangle\in A$ there exists a permutation of qubits $\tau\in G_H$ such that $\tau (i)=j$. The state of a $|\Psi\rangle$ $n$ - qubit system is called connected with respect to $H$ if it belongs to a connected subspace with respect to $H$, and $H|\Psi\rangle\neq is 0$.

The connectedness of a state means that all its nonzero components are obtained from one another by permutations of those particles that behave in the same way with respect to a given Hamiltonian. 
The above example of the state $ |n\rangle_{ph}(\a|10\rangle_{at}+ \b|01\rangle_{at})$ will obviously be connected, since the permutation of atoms interacting equally with the field does not change the Hamiltonian. The states of the form $|n\rangle_{ph}(\a|10\rangle_{at}+ \b|01\rangle_{at}+c|00\rangle_{at}+d|11\rangle_{at})$, for non-zero values of the amplitudes $a,b,c, d$ will not be connected. 
\bigskip

{\it Lemma.

If $|\Psi\rangle=\sum\limits_j\la_j|j\rangle$ is connected with respect to $H$, then any two columns of the matrix $H$ with numbers $j_1,\ j_2$, such that $\la_{j_1}$ and $\la_{j_2}$ are nonzero, differ from each other only by the permutation of elements. The same is true for the unitary evolution matrix $U_t=exp(-\frac{i}{\hbar}Ht)$. }
\bigskip

Indeed, for such basis states $j_1$ and $j_2$, according to the definition of $H$ - connectivity, there exists $\tau\in G_H$ such that $j_2=\tau(j_1)$. The columns numbered $j_1,\ j_2$ consist of the amplitudes of the states $H|j_1\rangle$ and $H|j_2\rangle$, respectively. From the commutation condition, we have $\tau H|j_1\rangle=H\tau |j_1\rangle=H|j_2\rangle$, and this just means that the column $j_2$ is obtained from the column $j_1$ by the permutation of elements induced by $\tau$. Moving on to the evolution matrix $U_t$, we see that the switching relation $\tau U_t|j_1\rangle=U_t\tau |j_1\rangle=U_t|j_2\rangle$ will hold for it as well, which is what is required. Lemma is proven.

It follows from Lemma that the states connected with respect to the Hamiltonian $H$ are equilibrium with respect to $H$ and with respect to the evolution operator $U_t=e^{-\frac{i}{\hbar}Ht}$ corresponding to this Hamiltonian.

\subsection{Amplitude quanta and determinism\label{quantampl}}

The exact description of the dynamics, even in the classical framework, has some degree of nondeterminism or stochasticity (see \cite{det1},\cite{det2}). The advantage of the quantum language lies in the precise limitation of this stochasticity to the state vector, which, according to the main thesis of Copenhagen mechanics, provides an exhaustive description of the dynamics of microparticles.

The possibility of introducing determinism into quantum theory, which interested researchers at the beginning of the history of quantum physics, has not lost the interest of researchers (see \cite{det3}) and is again becoming relevant for complex systems; for example, for systems of extreme complexity, the DNA molecule determines the trajectory of its owner with an accuracy unattainable in physical experiments. This higher type of determinism must have an analog for the simple systems described by standard quantum theory.

We describe a possible specific form of determinism at the level of amplitude quanta.

Our goal is to show that if the state $|\Psi\rangle$ is equilibrium with respect to the evolution operator $U_t$, then the amplitudes of all the basic states in $|\Psi\rangle$ can be divided into small portions - quanta of amplitude, so that for each quantum its trajectory will be uniquely determined when $U_t$ acts on a previously fixed time interval $t$, in particular, it will be uniquely determined, and with which other quantum of amplitude it will contract when summing the amplitudes to obtain the subsequent states.

This fact is also true for the arbitrary operator $A$, for which we will formulate the quantization of the amplitude. 

Let $| \Psi\rangle$ be an arbitrary equilibrium state with respect to $A$, the decomposition of which in terms of the basis has the form

\begin{equation}
\label{connected_state}
|\Psi\rangle=\sum\limits_j\la_j|j\rangle.
\end{equation}

We introduce the important concept of the amplitude quantum as a simple formalization of the transformation of a small portion of the amplitude between different base states when multiplying the state vector by the matrix $A$. Let $T=\{ +1,-1,+i,-i\}$ be a set of 4 elements, which are called amplitude types: real positive, real negative, and similar imaginary ones. The product of types is defined in a natural way: as a product of numbers. The quantum of the amplitude of the size $\varepsilon>0$ is called a list of the form
\begin{equation}
\label{quanta}
\kappa=(\varepsilon,id, |b_{in}\rangle, |b_{fin}\rangle, t_{in},t_{fin}),
\end{equation}
where $|b_{in}\rangle,\ |b_{fin}\rangle$ are two different basic states of the system of atoms and photons, $id$ is a unique identification number that distinguishes this quantum from all others, $t_{in}, t_{fin}\in T$. Transition of the form $|b_{in}\rangle\rightarrow\ |b_{fin}\rangle$ is called a state transition, $t_{in}\rightarrow t_{fin}$ is called a type transition. Let's choose the identification numbers so that if they match, all the other attributes of the quantum also match, that is, the identification number uniquely determines the quantum of the amplitude. In this case, there must be an infinite number of quanta with any set of attributes, except for the identification number. Thus, we will identify the quantum of the amplitude with its identification number, without specifying this in the future. Let's introduce the notation:
$$
t_{in}(\kappa)=t_{in},\ t_{fin}(\kappa)=t_{fin},\ s_{in}(\kappa)=b_{in},\  s_{fin}(\kappa)=b_{fin}.
$$

The transitions of states and types of amplitude quanta actually indicate how a given state should change over time, and their choice depends on the choice of $A$; the size of the amplitude quantum indicates the accuracy of a discrete approximation of the action of this operator using amplitude quanta.


The set of $ \theta $ quanta of the amplitude of the size $\varepsilon$ is called the quantization of the amplitude of this size, if the following condition is met:

{\bf Q}. In the set $\theta$, there are no such quanta of the amplitude $\kappa_1$ and $\kappa_2$ that their state transitions are the same, $t_{in}(\kappa_1)=t_{in}(\kappa_2)$ and $t_{fin}(\kappa_1)=-t_{fin}(\kappa_2)$, and there are no such quanta of amplitude $\kappa_1$ and $\kappa_2$ that $s_{in}(\kappa_1)=s_{in}(\kappa_2)$ and $t_{in}(\kappa_1)=-t_{in}(\kappa_2)$.

\bigskip

The condition {\bf Q} means that during the transition described by the symbol " $ \rightarrow$", the total value of the amplitude quantum cannot contract with the total value of a similar amplitude quantum, and also that the amplitude quanta do not contract with each other directly in the initial state record.

Quantization of the $\theta $ amplitude sets a pair of quantum states
\begin{equation}
\label{theta}
|\theta_{in}\rangle=\sum\limits_j\la_j|j\rangle, \ |\theta_{fin}\rangle=\sum\limits_i\mu_i|i\rangle, 
\end{equation}
according to the natural rule: for any basis states $|j\rangle, \ |i\rangle$, the equalities must be satisfied
 \begin{equation}
\la_j=\langle j|\theta_{in}\rangle=\varepsilon\sum\limits_{\kappa\in\theta:\ s_{in}(\kappa)=j}t_{in}(\kappa),\ \ \ \ 
\mu_i=\langle i|\theta_{fin}\rangle=\tilde\varepsilon\sum\limits_{\kappa\in\theta:\ s_{fin}(\kappa)=i}t_{fin}(\kappa),
\label{shift_}
\end{equation}
where $ \tilde\varepsilon$ is some normalization coefficient, so that the state $|\theta_{fin}\rangle$ has a unit norm, and $|\theta_{in}\rangle$ has an arbitrary nonzero one. The coefficient $\tilde\varepsilon$ does not have to coincide with $\varepsilon$, because when quantizing the amplitude, the usual norm of the state vector, in general, is not preserved; if we took $\tilde\varepsilon=\varepsilon$, then the value we entered $\{|\Psi\rangle\}$ at the transition $|\theta_{in}\rangle\rightarrow |\theta_{fin}\rangle$ could this is exactly due to the fact that some quanta of the amplitude "contract" with each other in the second sum from the formula \eqref{shift_}.

Let's fix the dimension $dim({\cal H})$ of the state space, and we will make estimates (from above) of the considered positive quantities: the time and the size of the amplitude quantum up to an order of magnitude, considering all constants to depend only on the independent constants: $dim({\cal H})$ and from the minimum and maximum absolute values of the elements of the matrix $A$. At the same time, the term "strict order''will mean the evaluation of both the top and bottom positive numbers that depend only on independent constants.

To quantize the amplitude $\theta$ and the numbers $i, j$ of the basis states via $n_{i,j}(\theta)$ denote the number of elements of the set ${\cal N}_{i,j}(\theta)=\{ \kappa\in\theta:\ s_{in}(\kappa)=j,\ s_{fin}(\kappa)=i\}$. 

Let $\theta (\varepsilon)$ be some function that maps some sequence of positive numbers - values of $ \varepsilon $ converging to zero, into quantizations of the amplitude of the size $\varepsilon$. This function will be called parametric quantization of the amplitude.

The parametric quantization of the amplitude $\theta (\varepsilon)$ is called consistent with the operator $A$ if there exists a function $c (\varepsilon)$ such that
for any basis states $i,j$
\begin{equation}
с(\varepsilon) n_{i,j}(\theta(\varepsilon) )\rightarrow\{ \la_j\} \{\langle i|A|j\rangle\}\ (\varepsilon\rightarrow 0)
\label{agreement}
\end{equation}
for $\la_j$, defined in \eqref{shift_}.

If $A$ is the evolution operator $U_t$, then having a parametric quantization of the amplitudes $\theta(\varepsilon)$ consistent with $A$ is a completely non-trivial property of the quantum states $|\theta_{in}(\varepsilon)\rangle$, which says that it is possible to introduce a hidden parameter corresponding to the dynamics set by the evolution matrix $U_t$, and making the quantum evolution $U_t$ deterministic. Such a parameter is the quantum of the amplitude $\kappa\in\theta (\varepsilon)$, where the accuracy of the deterministic description is determined by the value $\varepsilon$.

\bigskip

{\it The amplitude quantization theorem. 

Let $A$ be an arbitrary matrix. For every equilibrium state $|\Psi\rangle$ with respect to $A$, there exists a parametric quantization of the amplitudes $\theta (\varepsilon)$, consistent with the operator $A$, such that

\begin{equation}
\label{in_fin}
|\theta_{in}(\varepsilon)\rangle\rightarrow |\Psi\rangle,\ |\theta_{fin}(\varepsilon)\rangle\rightarrow A|\Psi\rangle,\ \ (\varepsilon\rightarrow 0). 
\end{equation}
}

{\it Proof. } 

Let we be given the equilibrium state with respect to $A$ $ | \Psi\rangle=\sum\limits_j\la_j|j\rangle$ and the number $\varepsilon>0$. For $|j\rangle$ with nonzero $\la_j\neq 0$ let
\begin{equation}
\la_j=\langle j|\Psi\rangle\approx sign_{re}( \underbrace{\varepsilon+\varepsilon+\ldots +\varepsilon}_{M_j})+sign_{im}i( \underbrace{\varepsilon+\varepsilon+\ldots +\varepsilon}_{N_j} ),
\label{quanta_expansion}
\end{equation}
where $sign_{re}\varepsilon M_j+sign_{im}i\varepsilon N_j\approx \la_j$ is the best approximation of the amplitude of $\la_j$ with the accuracy of $\varepsilon$; $M_j,\ N_j$ are natural numbers, $sign_{re\ (im)}=\pm 1$. Thus, the first tendency relation from \eqref{in_fin} will be satisfied, and the second relation must be satisfied if the parametric quantization is consistent with the Hamiltonian.

Let us approximate each element of the evolution matrix in the same way as we approximate the amplitudes of the initial state:
\begin{equation}
\label{hamapp}
\langle i|A|j\rangle\approx \pm(\underbrace{\varepsilon+\varepsilon+...+\varepsilon}_{R_{i,j}})\pm i (\underbrace{\varepsilon+\varepsilon+...+\varepsilon}_{I_{i,j}}),
\end{equation}
where $R_{i,j},\ I_{i, j}$ are natural numbers; the real and imaginary parts are exactly $\varepsilon$ each, and the signs before the real and imaginary parts are chosen based on the fact that this approximation should be as accurate as possible for the selected $\varepsilon$.

The amplitudes of the resulting state $A|\Psi\rangle$ are obtained by multiplying all possible expressions \eqref{quanta_expansion} by all possible expressions \eqref{hamapp}:

\begin{equation}
\label{mult}
\la_j\langle i|A|j\rangle\approx (sign_{re}M_{j}\varepsilon+i\ sign_{im}N_{j}\varepsilon )(\pm R_{i,j}\varepsilon\pm i\ I_{i,j}\varepsilon).
\end{equation}

We will expand the brackets in the right part of the expression \eqref{mult}, but we will not make abbreviations. Each occurrence of the expression $\varepsilon^2$ in the amplitudes of the resulting state after opening the brackets in the right part of \eqref{mult} will be obtained by multiplying a certain occurrence of $\varepsilon$ in the right part of \eqref{quanta_expansion} by a certain occurrence of $\varepsilon$ in the right part of \eqref{hamapp}. The problem is that the same occurrence of $\varepsilon$ in \eqref{quanta_expansion} corresponds to not one, but several occurrences of $\varepsilon^2$ in the result, and therefore we can not match the amplitude quanta directly to the occurrences of $\varepsilon$ in \eqref{quanta_expansion}.

How many occurrences of $\varepsilon^2$ in the amplitudes of the state $A|\Psi\rangle$ from the result of opening the brackets in \eqref{mult} correspond to one occurrence of $\varepsilon$ in the approximation of the amplitude $\la_j=\langle j|\Psi\rangle$ of the state $|\Psi\rangle$? This number - the multiplicity of the given occurrence of $\varepsilon$ - is equal to $ \sum\limits_i(R_{i,j}+I_{i,j}).$ These numbers can be different for an arbitrary operator $A$ and the state $|\Psi\rangle$. However, since $| \Psi\rangle$ is equilibrium with respect to $A$, $ \sum\limits_i(R_{i,j}+I_{i, j})$ will be the same for different $j$.

We introduce the notation $ \nu=\sum\limits_i(R_{i,j}+I_{i, j})$ - this is the number of occurrences of $\varepsilon$ in any column from the expansion of the matrix \eqref{hamapp}; this number $\nu$ has the order $1/\varepsilon$ at $\varepsilon\rightarrow 0$.

\bigskip

Denote by $Z_{i,j}$ the set of occurrences of the letter $\varepsilon$ in the right part of the expression \eqref{hamapp}, and let $Z_j=\bigcup_iZ_{i, j}$. Then the number of elements in the set $Z_j$ will be equal to $ \nu$.

Consider the smaller value of the amplitude quantum: $\epsilon=\varepsilon/\nu$. We substitute in the expression \eqref{quanta_expansion} instead of each occurrence of $\varepsilon$ its formal decomposition of the form $\varepsilon=\overbrace{\epsilon+\epsilon+\ldots +\epsilon}^\nu$, obtaining the decomposition of the amplitudes of the initial state into smaller numbers:

\begin{equation}
\label{refined_expansion}
\begin{array}{ll}
\la_j=&\langle j|\Psi\rangle\approx sign_{re}( \underbrace{\overbrace{\epsilon+\epsilon+\ldots +\epsilon}^\nu+\overbrace{\epsilon+\epsilon+\ldots +\epsilon}^\nu+\ldots +\overbrace{\epsilon+\epsilon+\ldots +\epsilon}^\nu}_{M_j})+\\
&sign_{im}i( \underbrace{\overbrace{\epsilon+\epsilon+\ldots +\epsilon}^\nu+\overbrace{\epsilon+\epsilon+\ldots +\epsilon}^\nu+\ldots +\overbrace{\epsilon+\epsilon+\ldots +\epsilon}^\nu}_{N_j}).
\end{array}
\end{equation}

Let $W^j_1,W^j_2,..., W^j_{M_j+N_j}$ be the sets of occurrences of the letter $\epsilon$ in the right part of the expression \eqref{refined_expansion}, marked with upper curly brackets. In each of these sets of$ \ nu $elements, as in the previously defined sets of$Z_j$. Therefore, we can construct for each such set $W^j_s$ a one-to-one mapping of the form $\xi:\ W_s^j\rightarrow Z_j$. For each occurrence of $\varepsilon$ in \eqref{quanta_expansion}, its descendants are naturally defined - the occurrences of $\epsilon$ in \eqref{refined_expansion}; the descendants for each occurrence will be $\nu$.

We define the quantization of the amplitudes $\theta=\theta (\epsilon)$ so that the $ id $ of the  amplitude quanta $\kappa\in\theta$ will simply be the occurrences of $ \epsilon$ in the expansions of \eqref{refined_expansion} for all $j$. Define, as required in \eqref{quanta}, the initial state and initial type of this quantum as the state and type of this occurrence. It remains to determine the transitions of states and types. This definition is given in the following natural way.

\begin{figure}
\centering
\includegraphics[height=0.5\textwidth]{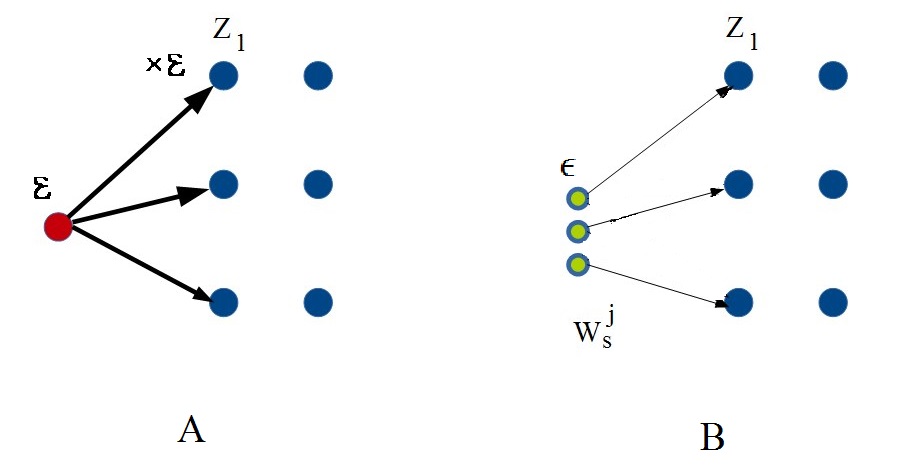}
\caption{A. Multiplying the state vector by the matrix. The contribution of each occurrence of $\varepsilon$ is multiplied by $\varepsilon$.
B. $\theta$- initial state shift. The size of the $\epsilon $ amplitude quantum is of the order of
 $\varepsilon^2$. }
\label{fig:nps}
\end{figure}

Each pair of the form $(w_s^j,\xi(w_s^j))$, where $w_s^j\in W_s^j$, will correspond to the transition of states and the transition of types in a natural way. Namely, the state transition will have the form $j\rightarrow i$ for such $i$ that $\xi(w_s^j)\in Z_{i,j}$; the transition of the types $t_{in}\rightarrow t_{fin}$ is defined so that $t_{in}$ is the type of occurrence\footnote{The type of occurrence is determined naturally after opening the parentheses, for example, for the occurrence of $...-i\epsilon ...$ the type is $-i$.} $w_s^j$, and the type $t_{fin}$ is the product of the occurrence type $t_{in}$ by the occurrence type $\xi(w_s^j)$. The sets $W^j_s$ do not intersect at different pairs $j,s$, so we consider all occurrences of the letter $\epsilon$ in the right part of\eqref{refined_expansion} to be the domain of the function definition $\xi$ (see Fig. \ref{fig:nps}).

Now let the transition of states and types for a given quantum $\kappa\in\theta$ correspond to the mapping $\xi$ in the sense defined above. The condition {\bf Q} will be met, since there are no reducing terms in the expression for the matrix element \eqref{hamapp}. So we defined the quantization of the amplitude.

Due to our definition of the function $ \xi$, the distribution of amplitudes in the state $|\theta\Psi\rangle$ will be approximately proportional to the distribution of amplitudes in the state $A|\Psi\rangle$, and the accuracy will increase indefinitely with the decrease of $\varepsilon$ to zero. In order to determine the value of the function $c (\epsilon)$ necessary for consistency of $\theta$ with the operator $A$, we calculate the contribution of each occurrence of $\varepsilon^2$ to the right side of the equality \eqref{mult} and compare it with the contribution of the corresponding letter $\epsilon$ to $ |\theta\Psi\rangle$.

\begin{figure}
\centering
\includegraphics[height=0.5\textwidth]{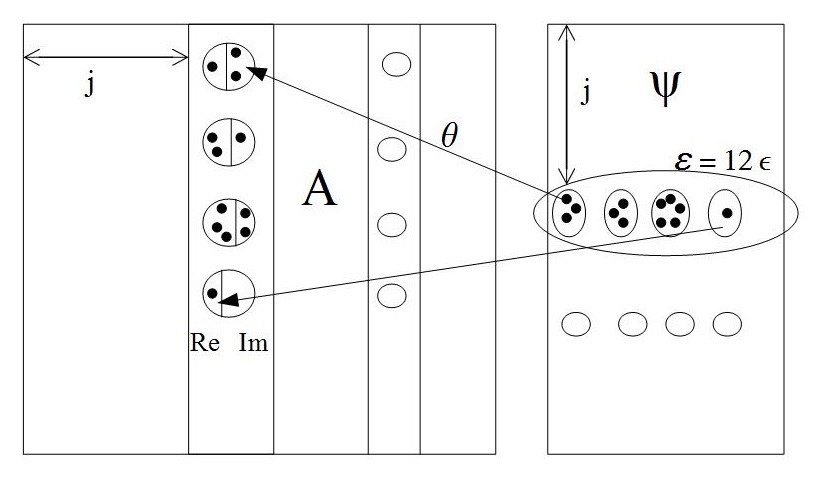}
\caption{Determinism of trajectories when multiplying the state vector $|\Psi\rangle$ by the matrix $A$. Each quantum of the ampitude passes into a certain quantum of the amplitude of the resulting state, there are no branches. }
\label{fig:matrix}
\end{figure}

Let's fix any transition of the types $t_{in}\rightarrow t_{fin}$ and the transition of the states $s_{in}\rightarrow s_{fin}$. We will call the occurrence of $\varepsilon^2$ in the result of opening brackets in \eqref{mult} corresponding to these transitions if $j=s_{in},\ i=s_{fin}$, and this occurrence is obtained by multiplying the occurrence of $\varepsilon$ of type $t_{in}$ in the first factor of the right side of \eqref{mult} by the occurrence of $\varepsilon$ in the second factor of type $t'$, so that $t_{in}t'=t_{fin}$. For each such occurrence, $\varepsilon^2 $ corresponds to exactly one quantum of the amplitude of the size $ \epsilon$ from the quantization of the amplitude defined above via the function $\xi$, which has the same transitions of states and types: this quantum corresponds to the occurrence of $\epsilon$, which is translated by a one-to-one mapping $\xi$ to this occurrence $\varepsilon^2$ (see Fig. \ref{fig:matrix}).

So, the occurrences of $\varepsilon^2$ in \eqref{mult} are in one-to-one correspondence with the occurrences of $\epsilon$ in \eqref{refined_expansion}. How many quanta of amplitude will the transition of states of the form $ |j \rangle\rightarrow |i\rangle$have? In each bracket of \eqref{refined_expansion} such quanta will be approximately $\{\langle i|A|j\rangle\}/\varepsilon$ pieces. And the total number of brackets is about $\{\la_j\}/\varepsilon$ pieces. Therefore, the number $n_{i, j}$ will be approximately equal to $\{\la_j\}\{\langle i|A|j\rangle\}/\varepsilon^2$, and the accuracy will grow indefinitely at $\varepsilon\rightarrow 0$. So, $c(\epsilon)=\varepsilon^2$. 

Note that if $A=0$, you can take $c (\varepsilon)=0$ and any quantization of the amplitude will be suitable. 

The theorem is proved.

Note that if we abandon the condition {\bf Q} and the condition of the equilibrium state $|\Psi\rangle$ with respect to $A$, we can also define matrix determinism, only we need to introduce the reducing terms $\varepsilon - \varepsilon$ into the matrix elements; then the formal entries of the amplitudes for all columns of $A$ will contain the same number of terms, and the reasoning will be valid, but interference will now occur not only between the descendants of different base states, as for equilibrium states. $|\Psi\rangle$, and also between descendants of the same state.

Note also that a quantum computation in which only gates of the form $CNOT$ and Hadamard are used has the property that for any gate the number of elements in all $Z_j$ will be the same, so that for such calculations determinism is provided with interference only between images of different basis states. This is, in particular, the Grover GSA algorithm.

Moreover, for promising implementations of quantum gates on photons (see, for example, \cite{A}), the states resulting from the implementation of gates are connected, so that interference in the course of such quantum calculations also has the property noted above.

\subsection{Conclusion}

We have argued for limiting the standard matrix formalism of quantum theory in the field of complex systems, as an upper bound on the possible number of qubits that can have an irreducible entangled state. This approach does not contradict any experiments with multi-bit systems, but allows you to build models of such systems on existing supercomputers. This restriction also applies to the equivalence of bases in the state space, and implies the existence of a minimum nonzero amplitude in the superposition. The size of this quantum of amplitude can be approximately found in experiments on the implementation of Grover's algorithm. The quantization of the amplitude makes it possible to introduce some type of determinism into quantum theory, which is not reduced to a quasi-classical approximation.

\section{Lecture 10. Quantum nonlocality, Bell inequality and distributed quantum computing}


The scheme of the experiment proving the presence of quantum instantaneous action at a distance was proposed by J. Bell in the early 60s (\cite{Bell},\cite{Bell2}); the experiments themselves were conducted for the first time in the 1980s by A. Aspec and A. Zeilinger (\cite{Aspe}, \cite{Zei}, as well as references in \cite{Gen} and \cite{Gen2}). In these experiments, the entanglement is manifested not at extreme distances, as in a hydrogen molecule, but at distances of several hundred kilometers.

In the experiment, states of the form $| \Psi\rangle =\frac{1}{\sqrt 2}|00\rangle +\frac{1}{\sqrt 2}|11\rangle$ are obtained for photons that are detected at distances of several hundred kilometers between them. Let's imagine that the first photon is detected by the observer Alice, and the second one is detected by Bob. The experimental conditions are such that Alice has two possibilities to choose a detector, that is, a measurement basis in the state space of her qubit, and Bob also has two possibilities.

Namely, Alice can choose the eigenvectors of the Hermitian operator $\sigma_x$ or $\sigma_z$, and Bob: $(\sigma_x +\sigma_z)/\sqrt 2$ or $(\sigma_x-\sigma_z)/\sqrt 2$, respectively. Since all the listed operators have only $1$ or $-1$ eigenvalues, we will assume that Alice got the value $X$ or $Y$, and Bob got $a$ or $b$, respectively, in the above order. For example, we can assume that $1$ means that a photon with horizontal polarization is detected, and $-1$ - with vertical (relative to the corresponding position of the detector). This is equivalent to each participant of the experiment choosing {\it observable} from two possibilities, for each-with a probability of 1/2.

We define the random variable $ \xi$ as the product of the measurement results of Alice and Bob, taken with a minus sign, in the case when the detector selections were $Y$ and $b$, respectively, and the product of the results with a plus sign in all other cases. The value of such a value is obtained by simply multiplying and changing the sign accordingly, after Alice and Bob have found out which orientation of the detectors each of them has chosen; during the measurement itself, they do not agree on their choice.

Alice and Bob get one by one pairs of biphotons in the state $\frac{1}{\sqrt 2}(|00\rangle+|11\rangle)$ and perform their tests with randomly selected observables, drawing up a protocol of experiments. Then they come together and calculate the value of the random variable $ \ xi$, which is equal to the product of the values of the observed Alice and Bob, if their choice of detectors was: $Y,a$ or $X,b$ or $Y,a$, and the product of these values with the opposite sign, if the choice was $Y,b$.

At a superficial glance, it may seem that $X,Y,a, b$ are random variables that can be operated on as with ordinary numbers. Let's temporarily accept this point of view, and conduct some simple calculation of the mathematical expectation $E$ of the value $\xi$. We will take $X$ and $Y$ out of brackets in the expression
\begin{equation}
E=Xa+Xb+Ya-Yb,
\label{bell_}
\end{equation}
which will be obtained if we add up all the possible results of calculations $\xi$. Then it turns out that in one bracket there is 0, and in the other there is a number modulo 2. Then we can estimate $E$ as $|E|= 2/4=1/2$, since all four choices of detector orientations are equally probable. Naturally, for random variables $X,Y,a, b$, we will have exactly the same inequality, and it does not matter whether they are dependent or not. So, for the mathematical expectation $M (\xi)$ values $\xi$ we get the inequality
\begin{equation}
M(\xi)\leq 1/2,
\label{Bell_}
\end{equation}
which is called Bell inequality.

And now let's calculate $M(\xi)$ for the observables we have written out using the quantum mechanical rule $\langle A\rangle_{\Psi} = tr(\rho_{\Psi}A)$ definitions of the average value (eigenvalues) of the Hermitian operator $A$ in the state $\Psi$. It follows from the definition of the mathematical expectation for the eigenvalues of the Hermitian operator $A$: $\langle A\rangle = \langle\Psi |A|\Psi\rangle$. The listener is asked to a) prove that the mathematical expectation of the observed $A$ in the state $| \ Psi\rangle$ is calculated using this formula, and b) the formula given in the text, based on this.

Simple calculation (It is necessary to consider all the cases of the orientation of the detectors, and for each to make a reduced density matrix of our state, and then apply the full probability rule) will show that $M (\xi)=\frac{1}{4}2\sqrt 2$ (the $1/4$ multiplier occurs everywhere due to the equal probability of choosing all 4 combinations of detectors). This is exactly what is detected in the experiment, which we will return to later. What is the matter? Where did we make a mistake in reasoning? Obviously, there is only one possibility to make it: the assumption that the results of Alice and Bob's measurements are expressed as random variables $X,Y,a,b$ was not strictly used by us, due to the fact that we did not apply the definition of a random variable. Now we will fill this gap, and see how this will lead us to a new understanding of the meaning of the experiment with two entangled photons.

Let's consider the experiment more strictly. To do this, we recall the basic concepts of the Kolmogorov probability theory. It includes 3 objects: probability space, random variables, and their numerical characteristics. First, we define the central concept: the set of elementary outcomes. This is (we always have a finite) set
$$
\Omega =\{\w_1, \w_2,\ldots, \w_k\},
$$
each element of which reflects the entire essence of the world, which plays a role for the experiment under consideration. This means that by selecting any element $\w_j\in\Omega$, we automatically select the outcome of any experiment from the set under consideration, including the position of the detector, the state of all elementary particles in it, as well as all parameters that we don't even know, but which determine what the outcome of the experiment will be. The fact that we do not know the structure of $\Omega$ does not play any role. We should still consider this object explicitly if we are talking about probabilities.

Due to the inflexible form of the "ban on hidden parameters" in Copenhagen quantum mechanics, we cannot even consider the approximation of the $\Omega$ set within its framework. Thus, an accurate consideration of quantum mechanical problems for many particles must go beyond the limits of the Copenhagen quantum theory. Going beyond the limits does not mean breaking laws, but considering entities that are not available in the Copenhagen theory.

You can put it differently. Standard problems of quantum theory involving the use of the apparatus of wave functions and projections should not use hidden parameters, that is, they should not concern the probabilistic structure of the wave function. The standard ones include problems about the behavior of a single quantum particle or reduced to them.

However, our problem about entangled photons is not standard, since it already concerns two particles. Although the entangled state of two particles can in some sense be reduced to a single-particle one, but the observation operators used by Alice and Bob are significantly different, and therefore we are dealing with a substantially non-single-particle quantum state. Sometimes the arsenal of Copenhagen theory is enough to solve such problems, but our case clearly does not belong to this category, and therefore we must use probability theory to find a solution, considering the set of elementary outcomes $\Omega$. The limitation of the standard formalism here is that we must consider this set to be finite; although in this case this will not affect the conclusions in any way.

On the set $S$ of all subsets of $\Omega$ is necessary to define the so-called likelihood function of the form $P:\ S\ar [0,1]$ that satisfies the axioms of probability:
$P(A\cup B)=P(A)+P(B)$ for disjoint $A,B\in S$, $P(\emptyset )=0$, $P(\Omega )=1$.
Define $P$ is very simple: $P(A)$ is the quotient of the number of all elements of $A$ on $l$. This is sometimes called the frequency determination of probability.

A random variable is any function of the form
$$
\xi :\ \Omega \ar R.
$$

We can calculate the mathematical expectation of a random variable $ \xi$ using the standard formula $M(\xi )=\sum\limits_{x\in R} xP(\{\w\in\Omega\ |\ \xi (\w )=x\})$. 

It should be said at once that it is pointless to look for "explanations" of the experiment with biophotons without resorting to the strict definition of probability given above. The only mathematically accurate formulation of the concept of "probability" follows from the above definition.
Now let's look at how this arsenal is applied to the situation under consideration.

From the point of view of quantum mechanics, the state $|\Psi\rangle$ of the two photons under consideration represents a single vector in the Hilbert space of states. This means that there is such a space of elementary outcomes $ \Omega$ that all the quantities $X,Y,a, b$ are random variables over this space, that is, functions of elementary outcomes: $X(\w ), Y(\w ), a(\w ), b(\w )$. 

Now we must reformulate the experimental conditions in the language of probability theory. We have the following situation. Alice and Bob, independently of each other and completely randomly, each choose any one state of the detector from the two possibilities available to it, immediately after which each photodetector detects a photon that has fallen into it (a specific measurement result is 1 if the state $|\e_1\rangle$ is obtained from the basis of the eigenvectors of the operator of the observed, and $-1$ - if $|\e_2\rangle$, the choice of these alternatives is based on the characteristics of the detector).

This means that Alice's choice of detector orientation is included in some object $\w_1$, and Bob's choice of detector orientation is included in the object $\w_2$, so that the elementary random outcome of one experiment $\w\in\Omega$ has the form $(\w_1,\w_2)$. If the photons have any hidden parameters, then we consider the parameters of the photon that arrived to Alice to be included in $\w_1$, and for the photon that arrived to Bob - in $\w_2$. Thus, we must assume that $\Omega=\Omega_1\times \Omega_2$, where the sets $\Omega_1$ and $\Omega_2$ correspond to the choices of Alice and Bob, respectively.

This assumption expresses the so-called freedom of will in both participants of the experiment. The absence of free will would simply mean that the choice of, say, Alice, would automatically determine Bob's choice. In real experiments, the question of orientation is solved not by people, but by electronics, based on such events that, from the point of view of common sense, must be independent (for example, streams of extraneous photons from different regions of outer space). The free will of the participants in the experiment is a necessary assumption if we are engaged in science.

Now consider that there are random variables $X, Y,a, b$. Since $\w_1$ automatically determines the orientation of Alice's detector, we denote by $\Omega_1^X$ such a subset of $\Omega_1$ that corresponds to the orientation of the detector $X$, and similarly we denote the subsets corresponding to $Y, a, b$. In this case, $\Omega_1$ will be the sum of disjoint subsets of $\Omega_1^X$ and $\Omega_1^Y$, and $\Omega_2$ will be the sum of also disjoint $\Omega_2^a$ and $\Omega_2^b$.

We must assume that the result of Alice's detection is a random variable $\xi_1(\w_1,\w_2)$, and the result of Bob's detection is a random variable $\xi_2(\w_1,\w_2)$, so that the overall result is the Cartesian product $\xi=(\xi_1,\xi_2)$, and $X$ is the restriction of the function $\xi(\w_1,\w_2)$ on the domain $\w_1\in\Omega_1^X$, $Y$ there is a restriction of the function $\xi(\w_1,\w_2)$ on the domain $\w_1\in\Omega_1^Y$, $a$ there is a restriction of the function $\xi$ on the domain $\w_2\in\Omega_2^a$, and $b$ - on the domain $\w_2\in\Omega_2^b$. In order to make the values $X,Y, a, b$ defined on the entire set of elementary outcomes $ \ Omega$, we will add them to zero in those areas where they are not explicitly defined by us.

Let's define the random variable $\xi$ as follows:
$$
\xi(\w_1,\w_2)=\left\{
\begin{array}{lll}
& \xi_1(\w_1,\w_2)\xi_2(\w_1,\w_2),\ \ &\text{если}\ \ \w_1\notin\Omega_1^Y\ \text{или}\ \ \w_2\notin\Omega_2^b,\\
&- \xi_1(\w_1,\w_2)\xi_2(\w_1,\w_2),\ \ &\text{если}\ \ \w_1\in\Omega_1^Y\ \ \text{и}\ \ \w_2\in\Omega_2^b.
\end{array}\right.
$$

We then have: $\xi = Xa+Xb+Ya-Yb$. 

Let's calculate its expectation by the given definition, choosing the frequency definition of probability. We will obtain

$$
\begin{array}{ll}
M(\xi)=&\frac{1}{k}\sum\limits_{\w_1,\w_2} X(\w_1,\w_2)a(\w_1,\w_2)+X(\w_1,\w_2)b(\w_1,\w_2)\\
&+Y(\w_1,\w_2)a(\w_1,\w_2)-Y(\w_1,\w_2)b(\w_1,\w_2).
\end{array}
$$

Note that we are using what many people call realism. This means that we have the right to repeatedly use a limited number of letters $\w_j$ so that any combinations of them will correspond to real experiments on detecting photons. In another way, this can be formulated as freedom of will when choosing from a finite set of reality options. 

It is not difficult to make sure that it is impossible to do with this expression for the expectation as it was done above with numbers when proving the Bell inequality \eqref{Bell_}, due to the presence of arguments for random variables. Indeed, since the result of measuring one of the participants depends on the elementary outcomes for both of them, we would have to write another expression instead of the expression \eqref{bell_}: $E=Xa+X'b+Ya'-Y'b'$, and we would not be able to put the common factors out of brackets, that is, our naive reasoning would be incorrect.

However, let's assume that, in addition to the obvious realism for us, we also have a so-called locality. In short, locality means that the result of Alice's measurement does not depend in any way on the orientation of Bob's detector and vice versa. We will discuss the physical meaning of locality below. Formally, locality means that $X$ and $Y$ depend only on $\w_1$, and $a$ and $b$ depend only on $\w_2$. Then we can do the same trick with the expression for the mathematical expectation as when proving Bell's inequality. Namely, we will group all the terms of a large sum into groups of 4 types
$$
X(\w_1)a(\w_2')+X(\w_1)b(\w_2)+Y(\w_1')a(\w_2')-Y(\w_1')b(\w_2),
$$
consisting of non-zero terms, so that within each group it will be possible to take $X$ and $Y$ out of brackets, and just as above to prove that this group does not exceed 2. Since 4 different $\w$ are involved in the group, we get that the mathematical expectation of $\xi$ does not exceed $1/2$. That is, locality leads to the fulfillment of Bell's inequality. Thus, we came to the conclusion that the nonlocality of quantum mechanics follows from the experiment on detecting biphotons.

Now let's look at the nonlocality in more detail. It means that the random variables related to Bob depend not only on its component of the elementary outcome, but also on the component belonging to Alice, and vice versa, that is, all the outcomes of $X,Y,a, b$ depend on both $\w_1$ and $\w_2$.

How can this be implemented? Just like this: there is some object $ \tilde\w$ that travels from Alice to Bob and back, carrying information about the other half of the elementary outcome of the corresponding experiment. If this object $\tilde\w $ obeys the restriction of relativism, and cannot move faster than light, then we can deduce restrictions on the times of emission of a biphoton by the source and the times of detection of the arrival of each of the photons by Alice and Bob.

Let $ \Delta t$ be the natural uncertainty of the moment of emission of a biphoton by the source, about which we assume that all biphotons whose emission time lies outside this range do not play any role for obtaining statistics in this experiment. The presence of such an interval is a direct consequence of the energy-time uncertainty ratio. Now we assume that the clocks of Alice, Bob, and the source of the biphoton are exactly synchronized, and we enter the value $ \delta t$ equal to the difference between the moment of triggering the detector and the moment of choosing its position (that is, the choice between $X$ and $Y$ and between $a$ and $b$). Then, if the material object $ \tilde\w$, which carries information about the other half of the elementary outcome, obeys relativism, then the following inequality must be satisfied
\begin{equation}
\Delta t+\delta t \geq d/c,
\label{rel}
\end{equation} 
where $d$ is the distance between Alice or Bob, and the source of biphotons, $c$ is the speed of light.

Experiments show that this inequality is violated for biphotons detected at distances of several hundred kilometers, which has absolutely fundamental consequences for quantum theory. Indeed, the violation (\ref{rel}) says that $\tilde\w$ it cannot be a hidden parameter of any of the photons. In real experiments, as a rule, they do not check the violation (\ref{rel}), but directly prevent the reverse transfer of information by the photons themselves, exposing the stubs after their passage.

That is, $ \tilde\w$ directly transfers information about the orientation of the detectors from Alice to Baba or vice versa. This effect is commonly called "quantum nonlocality"; it directly follows from the standard quantum formalism, but in fact, it makes it necessary to move from the narrow Copenhagen framework to a post-quantum theory, in which the random outcomes of $ \omega$ should have a real meaning, and not serve only a formal goal - mathematical consistency. You can get acquainted with various points of view on quantum nonlocality, for example, by reading articles from \cite{Kh1}.

Quantum theory is fully consistent with the principle of relativism, according to which no information can move at a speed exceeding the speed of light. Formally, this is expressed in the fact that the statistics of Alice's measurements does not depend in any way on whether Bob changes his qubit or not. That is, with the help of an entangled quantum state, it is impossible to transfer the information generated by the participants of the experiment to each other. But we have just found out that this restriction does not apply to information about elementary outcomes in specific experiments about measuring quantum states when they are put together!

There is only one conclusion to be drawn from this. There is a kind of administrative system, the interaction with which determines the reality. This interaction exactly corresponds to the user's interaction with the computer. The user, that is, the experimenter, determines the conditions (the position of the detectors), after which the administrative system working with elementary outcomes outputs the result of the experiment. At the same time, the time spent by the administrative system to coordinate the conditions set by various users is not real physical time.

We are using programming terminology here, in which the administrative system means a very specific thing that should be included in the post-quantum formalism, and therefore should not cause any other associations. The non-locality of the elementary outcomes of $ \ omega$ suggests that these outcomes can find real meaning precisely for complex systems and processes affecting large spatial regions. For simple systems, for example, for a single atom or even a molecule, non-locality itself does not play a big role: it manifests itself in a rather subtle experiment described by us, and its effect for simple systems is even less than relativistic corrections.

However, quantum long-range operation allows you to create amazing information exchange protocols, one of which we will consider below.

\subsection{An example of quantum superiority in distributed computing with one-way control}

The construction of a quantum computer is a complex and multifaceted process, and an important role in it is played by limited models of quantum computing, for example, quantum branching programs (\cite{Ab}) or biochemistry modeling programs (\cite {Do}.) The advantage of quantum methods may not be in speeding up calculations in the usual sense, but in using individual elements of quantum nature to obtain the final gain as the result obtained.

Here we demonstrate how breaking the Bell inequality can help improve the efficiency of some distributed computing. The example that we will give is artificially constructed and is intended only to illustrate the possibility of practical use of the amazing property of quantum nonlocality; moreover, the effect of this use is not too great. However, this example has similarities with the biological process of growth of complex molecules with a linear organization of the primary structure, and therefore it suggests that the search for further applications of quantum nonlocality can be fruitful.

\begin{figure}
\centering
\includegraphics[height=0.5\textwidth]{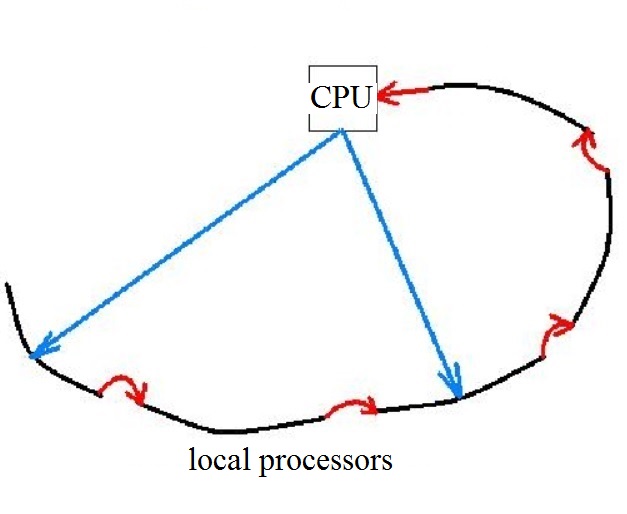}
\caption{One-way control scheme}
\label{fig:odn}
\end{figure}

\subsection{One-way control }

\begin{figure}
\centering
\includegraphics[height=0.5\textwidth]{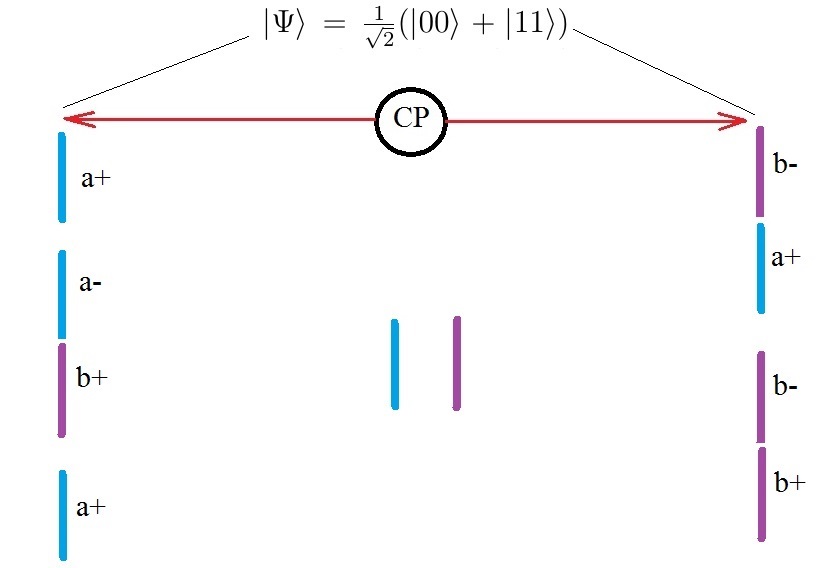}
\caption{Distributed synthesis of chains}
\label{fig:fig4_old}
\end{figure}

We show how this goal can be achieved using a violation of Bell's inequality. Let's consider a model of distributed computing with one-way control, where all computing devices are divided into a central processor (CPU) and remote peripheral devices that can directly receive commands from the CPU. The reverse transfer of information from peripheral devices to the central processor does not occur directly, but only in the form of a sequential transfer through a chain of peripheral devices that locally interact with each other, as shown in the figure \ref{fig:odn}.

In the example that we will analyze, the use of entangled photon states in the control gives an increase in the quality of the calculation result, exceeding the result of classical control by about $1.138$ times. This is the task of synthesizing two remote chains consisting of separate links, carried out on two peripheral devices.

The central processor sends a signal to two peripheral processors, each of which is responsible for the corresponding subsystem of the entire system. For example, the CPU solves the problem of synthesis on one subsystem of a certain polymer $ A $, which has a special activity, and, at the same time, the problem of synthesis of another polymer $ B$, which suppresses (or, conversely, intensifies) this activity, already on another subsystem. The CPU sends the corresponding signal to both subsystems, and switches to other tasks, for example, to synthesize another pair of polymers $ A'$ and $ B'$.

What would happen if the peripheral processors began to send signals to each other directly? Let's say we have $ m $ subsystems, each of which is controlled by its own processor. For the correct addressing of signals between all possible pairs (of the order of $ m^2 $), we would have to load the CPU with this work. The CPU would have to wait for the time $ cD $ for each pair, where $ D $ is the distance between the peripheral processors, $ c $ is the speed of light, before switching to the next task.

Еif $ m $ is large enough (in real bio-systems, this number is very large), such a calculation scheme based on addressing signals through the CPU would lead to a fatal control delay, which would make the entire scheme unusable.

Thus, we come to the need for one-way control, when the CPU sends signals to peripheral processors immediately, without waiting for a response from them. The feedback information is sent to the CPU not directly, but through a chain of intermediaries, as in a cellular automaton. This form of information processing organization can be effective in living organisms, since in them the central nervous system, which plays the role of the CPU, should be free from routine work on managing metabolism.

\subsection{Quantum bi-photonic signals}

\begin{figure}
\centering
\includegraphics[scale=0.8]{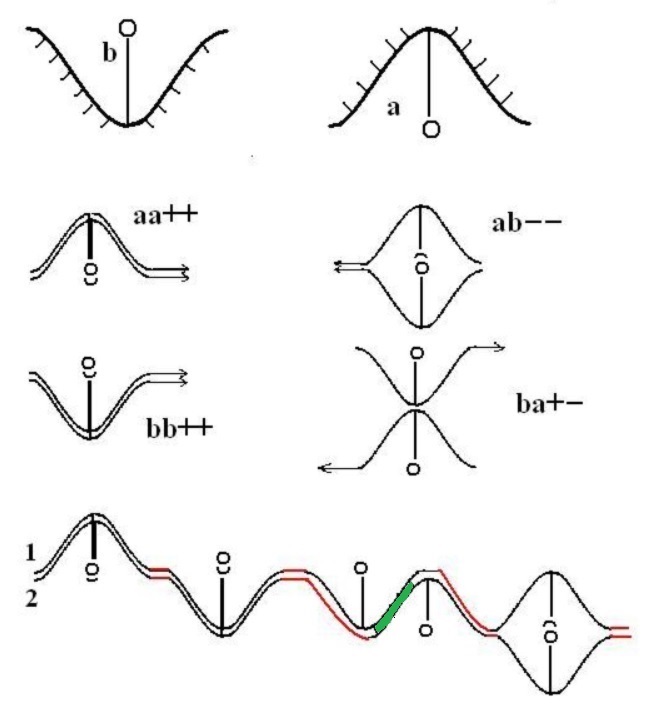}
 \caption{The superposition of two polymers. The arrows indicate the direction of bond stretching (red) between adjacent monoblocks during polymer synthesis. Overlays of the form $aa++(--), ab++ (--), bb++ (--), ba+-(-+)$ give gluing, other gluing does not give. At the bottom, all pairs give glues.}
\label{fig:polimers}
\end{figure}

In this situation, the use of CPU biphotons (entangled states of photons) gives an advantage compared to a purely classical CPU. To demonstrate this, we will consider the following abstract problem. Suppose you want to synthesize two polymer molecules, the chemical structure of which has the form $C_1=(c_1^1,c_2^1,...,c_M^1),\ C_2=(c_1^2,c_2^2,...,c_M^2)$,
so, what they consist of monoblocks two types $ a $ and $ b $: $c_i^j\in\{ a,b\}$ (see Figure \ref{fig:polimers}).

The quality of such a mutual assembly of two polymers is checked by superimposing finished chains on each other: the first $C_1$ on the second $C_2$, and the quality criterion is the degree of gluing of these chains.
Each monoblock has an external (convex) and internal (concave) surface, where the latter is equipped with a special ball located in its center. In a fixed position, two monoblocks can stick together in one of the following cases: 1) their surfaces or half of the surfaces are completely aligned by vertical displacement, or 2) their central balls are at the same point with such a shift, as shown in the figure \ref{fig:polimers}.

The physical structure of the polymer, on which the gluing depends, is determined not only by the sequence of monoblocks in the chain; the gluing also depends on an additional option: their exact location relative to each other in the chain. Neighboring monoblocks in the polymer are connected by a flexible bond, which can either give up by $dx$, which is a quarter of the length of the monoblock, or stretch to the same length. In these cases, we will say that the monoblock is shifted backward or forward, respectively, relative to the equilibrium position of the connection. 

During the synthesis, monoblocks are installed with these restrictions and their positions are fixed. Then the two chains are superimposed on each other and for each pair of overlapping monoblocks, the presence of gluing is established. It follows from the accepted restriction that if in such a pair of overlapping monoblocks they were shifted to one side, they are glued together in the same way as if there were no shifts; and if in different ones, the resulting shift is half the length of the monoblock.
After that, the number of glued pairs of superimposed monoblocks is calculated and this number is considered a numerical characteristic of the assembly quality of a pair of chains.

The synthesis of chains occurs as a sequential attachment of a new monoblock to each of the existing chains - the one that first appeared at the assembly point of one and the other chain. Monoblocks are taken from the environment surrounding the growth points, where they are in chaotic motion and both types are distributed equally. In this case, you can move the newly attached monoblock either backwards or forwards by a distance of $ dx $. We will denote the forward shift by $ + $, the backward shift by $ - $. Each $j$ - th pair of monoblocks in both chains, superimposed on each other after synthesis, thus correspond to the four $ c_j^1c_j^2s_j^1s_j^2$, where the last two terms are shifts $ s_j^{1,2} \in \{ +, - \} $.

It follows from our rules (see Figure \ref{fig:polimers}) that the gluing corresponds to pairs of superimposed monoblocks of the form: $aa++ (--),\ ab++(--), bb++(--), ba+ - ( -+)$, whereas pairs of a different type:
$aa+ - (-+), ab+ - (-+), bb+ - (-+), ab++ (--)$ do not give gluing. Note the asymmetric behavior of monoblocks of the type $a$ and $b$: pairs of $a b$ and $ba$ are glued together in different ways with the same engines. This asymmetry looks like an asymmetry in the Bell inequality, which will give us an increase in the quality of the resulting gluing with biphoton control compared to classical control.

We assume that the growth of the polymer $ C_1 $ occurs at one point, and the growth of $ C_2 $ occurs at another, and these points are separated by a large distance (for example, they occur in different countries). The task is to organize this synthesis so that the number of non-glued pairs of superimposed monoblocks is minimal, or, in other words, so that the number of glues is maximum. 

A similar problem may arise when modeling the synthesis of a gene and an antigen in different living cells. We can create an information channel for controlling them from one center; however, with a large distance between the assembly points, such control can slow down the assembly process itself, which is a separate problem for real polymers that goes beyond our model.

We will show how to use biophoton control of the simultaneous synthesis process to obtain quantum superiority.

So, to minimize critical (non-glued) pairs of monoblocks, we use CPU signals in the form of EPR states $ | \Psi \rangle = \frac{1}{\sqrt 2} (| 00 \rangle + | 11 \rangle)$, and calculate the number of critical pairs that occur during such control. If the control were classical, in the notation from the previous paragraph we would have Bell's inequality
\begin{equation}
E(a_1b_2+b_1b_2+a_1a_2-b_1a_2)\leq 2.
\label{Bell}
\end{equation}
  
Let's accept the following agreement. The subscript indicates the assembly point (polymer number) 1 or 2. The letter $ a $ or $ b $ indicates the type of monoblock attached to the chain, the sign corresponds to the direction of the shift of this monoblock, as we agreed. The result of joining monoblocks at both assembly points is determined if, for the lower index 1 and 2, we have, first, the letter $ a $ or $ b $, and secondly, the shift sign $ + $ or $ - $. The letter $ a $ or $ b $ always determines the type of monoblock closest to the assembly point at the moment.

The CPU operating on the principles of classical physics can thus control the assembly only by choosing the shift sign $ + $ or $ - $ at both assembly points. The CPU selects these signs simultaneously, so that no waiting for the signal to pass between the assembly points can slow down the process: information about the sign appears at both points simultaneously, and just at the moment when it is needed. If we allowed a time delay, it would be possible to make the assembly generally perfect, avoiding critical pairs altogether.

For the classical type of correlation between the choice of signs, we have Bell's inequality. For each step of the process, we keep the criticality index $ Cr = + 1 $, if the overlap of the corresponding monoblocks is not critical (there is a gluing), and $ Cr = -1 $ otherwise. We are interested in the resulting number of necrotic overlays along the entire length of the chains of synthesized polymers: $ NonCr $; our goal is to make this number the maximum.

For one pair of monoblocks, we have $No nC r=\frac{1}{2}(1+Cr)$. Since all combinations of $aa, ab, ba, bb$ for both synthesis points have the same probabilities of $1/4$, for the average value of $E (Cr)$ of the criticality index we have
\begin{equation}
E(Cr)=\frac{1}{4}(a_1b_2+b_1b_2+a_1a_2-b_1a_2),
\label{cr}
\end{equation}
where the letter $a$ or $b$ with an index denotes a random variable corresponding to the choice of the type of monoblock with the sign $\pm1$, depending on the shift sign chosen for it.

For classical control due to Bell's inequality for $E(Cr)$ of the form \eqref{cr} the average number of critical overlays satisfies the inequality
$$
E(NonCr)\leq \frac{1}{2}(1+\frac{2}{4})=\frac{3}{4}=0.75. 
$$

In the case of quantum biphoton control, the situation will be different. Here we cannot consider $a_{1}$ and $b_{1}$ as random variables defined on separate sets of elementary outcomes for $a_2,\ b_2$, that is, the evaluation of \eqref{Bell} will not follow from the obvious expression $a (X+Y)+b(X-Y)\leq 2$ for the numbers $a,b, X, Y=\pm 1$; here we should write $a_1b_2+b_1b'_2+a'_1a_2-b'_1a'_2$ instead of the left side of the inequality \eqref{Bell}, which will make this inequality incorrect.

For biphoton control, our random variables are defined on the same set of elementary outcomes, we do not have Bell's inequality and must count the probabilities directly using the Born rule.

Let us assume that for each of the assembly points we have a photodetector that can be instantly oriented in accordance with the observables that we associate with $ a $ and $ b $. For the first and second assembly points, let these observables have the form:
\begin{equation}
\label{ob}
\begin{array}{lll}
&a_1=\sigma_x,\ &b_1=\sigma_z,\\
&a_2=\frac{1}{\sqrt 2}(\sigma_x-\sigma_z),\ &b_2=\frac{1}{\sqrt 2}(\sigma_x+\sigma_z)
\end{array}
\end{equation}
accordingly. Here we do not consider the interesting question of the practical implementation of such observables.

Let's assume that the type of the current monoblock determines the position of the detector for both points and the shift sign of the monoblock is the value of the corresponding observable. Since all combinations of monoblock types $ aa, ab, ba, bb $ are equally probable, we can use the formula \eqref{cr} for the average value of the criticality index.

Now we have: $E=E(a_1b_2+b_1b_2+a_1a_2-b_1a_2)=E(a_1b_2)+E (b_1b_2)+E(a_1a_2)-E (b_1a_2)$. Using the definition of observables \eqref{ob} and applying the rule for calculating the averages $ \ langle A\rangle_{\psi}=tr(A\rho_{\psi})$ for all observables $A$ taken from \eqref{ob}, we will find $E= 2 \sqrt{2} $ and for the average value of the number of non-critical overlays (glues) we will get the value of $ E (NonCr) =
\frac{1}{2} (1+ \frac{2 \sqrt{2}}{4}) \approx 0.85 $. So, the use of EPR pairs of photons in the assembly control gives a significant gain in quality - a little more than $ 1.138$ for such a formulation of the problem.
 

\end{document}